\documentclass[openright,a4paper,12pt]{memoir}

\chapterstyle{veelo}
\setsecnumdepth{subsection}
\usepackage{layout}
\usepackage{blindtext}
\usepackage{graphicx}
\usepackage[english]{babel}
\usepackage[a4paper]{geometry}
\usepackage{amssymb}
\usepackage{amsmath}
\usepackage{tabularx}
\usepackage{multirow}
\usepackage{rotating}
\usepackage{mathabx}
\usepackage{wasysym}
\usepackage{textcomp}
\usepackage{xifthen}
\newboolean{chapterbib}
\setboolean{chapterbib}{false}
\ifthenelse{\boolean{chapterbib}}{\usepackage[sectionbib]{chapterbib}    \usepackage[sectionbib]{natbib}}{\usepackage{natbib}}
\bibpunct{(}{)}{;}{a}{}{,} 
\usepackage{color}
\usepackage{xspace}
\usepackage{bbm}
\usepackage{cancel}
\usepackage{float}

\usepackage[%
pdfauthor={Tilman Birnstiel},
pdftitle={The Evolution of Gas and Dust in Protoplanetary Accretion Disks},
pdfstartview=FitH,
linkcolor=black,
anchorcolor=black,
citecolor=black,
filecolor=black,
menucolor=black,
urlcolor=black,
colorlinks=true]{hyperref}
\newcommand{\e}[1]{\ensuremath{\times 10^{#1}}}
\newcommand{\Del}{\mathrm{d}}
\newcommand{\del}{\partial}
\newcommand{\ddel}[2]{\frac{\partial #1}{\partial #2}}
\newcommand{\half}{\frac{1}{2}}
\newcommand{\St}{\text{St}\xspace}
\newcommand{\rhos}{\ensuremath{\rho_\text{s}}\xspace}
\newcommand{\amax}{\ensuremath{a_\text{max}}\xspace}
\newcommand{\Sc}{\text{Sc}}
\newcommand{\uf}{\ensuremath{u_\text{f}}\xspace}
\newcommand{\csound}{\ensuremath{c_\mathrm{s}}\xspace}
\newcommand{\Hp}{\ensuremath{H_\mathrm{p}}\xspace}
\newcommand{\Hd}{\ensuremath{H_\mathrm{d}}\xspace}
\newcommand{\mpr}{\ensuremath{m_\mathrm{p}}\xspace}
\newcommand{\kb}{\ensuremath{k_\mathrm{b}}\xspace}
\newcommand{\Ok}{\ensuremath{\Omega_\mathrm{k}}\xspace}
\newcommand{\sigb}{\ensuremath{\sigma_\mathrm{B}}\xspace}
\newcommand{\tauros}{\ensuremath{\tau_\mathrm{R}}\xspace}
\newcommand{\taupla}{\ensuremath{\tau_\mathrm{P}}\xspace}
\newcommand{\kapros}{\ensuremath{\kappa_\mathrm{R}}\xspace}
\newcommand{\kappla}{\ensuremath{\kappa_\mathrm{P}}\xspace}
\newcommand{\rhodust}{\ensuremath{\rho_\mathrm{d}}\xspace}
\newcommand{\rhogas}{\ensuremath{\rho_\mathrm{g}}\xspace}
\newcommand{\Tm}{\ensuremath{T_\mathrm{mid}}\xspace}
\newcommand{\tvisc}{\ensuremath{t_\mathrm{visc}}\xspace}
\newcommand{\Siggas}{\ensuremath{\Sigma_\mathrm{g}}\xspace}
\newcommand{\Sigdust}{\ensuremath{\Sigma_\mathrm{d}}\xspace}
\newcommand{\nug}{\ensuremath{\nu_\mathrm{g}}\xspace}
\newcommand{\dx}{\ensuremath{\mathrm{d}}}
\newcommand{\N}{\ensuremath{\mathcal{N}}}
\newcommand{\Fmm}{\ensuremath{{F_\text{1mm}}}\xspace}
\newcommand{\alphamm}{\ensuremath{\alpha_\text{1-3mm}}\xspace}
\newcommand{\betamm}{\ensuremath{\beta_\text{1-3mm}}\xspace}
\newcommand{\alphat}{\ensuremath{\alpha_\text{t}}\xspace}
\newcommand{\aBT}{\ensuremath{a_\mathrm{BT}}\xspace}
\newcommand{\aonetwo}{\ensuremath{a_{12}}\xspace}
\newcommand{\ugas}{\ensuremath{u_\text{gas}}\xspace}
\newcommand{\asett}{\ensuremath{a_\mathrm{sett}}\xspace}
\newcommand{\Rey}{\ensuremath{\mathrm{Re}}\xspace}
\newcommand{\mf}{\ensuremath{m_\text{f}}\xspace}
\newcommand{\sighyd}{\ensuremath{\sigma_{\mathrm{H}_2}}\xspace}
\definecolor{mygray}{gray}{.5}

\defcitealias{Ricci:2010p9423}{R10}
\defcitealias{Brauer:2008p215}{B08a}
\begin{document}
    \pagestyle{empty}
    \vspace*{0.63in}
    \parbox[t][2in][t]{\textwidth}{
    \large
    \centering
    \textsc{{\Large Dissertation}\\
    submitted to the\\
    Combined Faculties for the Natural Sciences and for Mathematics\\
    of the Ruperto-Carola University of Heidelberg, Germany\\
    for the degree of\\
    Doctor of Natural Sciences}
    }
    \vspace*{\fill}
    \begin{center}
        \textsc{Put forward by}\\[0.5cm]
        \begin{tabular}{ll}
            \textsc{M.Sc.}                    &   \textsc{Tilman Birnstiel}\\
            \textsc{Born in}                  &   \textsc{Miltenberg am Main}\\
            \textsc{Date of oral examination} &   \textsc{18.10.2010}
        \end{tabular}
    \end{center}
    \movetooddpage
    \vspace*{0.63in}
    \begin{center}
        \textsc{\large The Evolution of Gas and Dust in Protoplanetary Accretion Disks}
        \vfill
        \begin{tabular}{ll}
            Referees:&  Prof. Dr. Cornelis P. Dullemond\\
            &           Prof. Dr. Ralf S. Klessen
        \end{tabular}
    \end{center}
    \movetooddpage
    \vspace*{\fill}
    \begin{center}
        \textsc{\large Zusammenfassung}
    \end{center}
    Obwohl Staub nur etwa ein Prozent der Masse einer protoplanetaren Scheibe ausmacht, ist er doch ein wichtiger Bestandteil von chemischen Modellen, Modellen zur Planetenentstehung oder in der Modellierung von Strahlungstransport und Beobachtungen. Der anf\"angliche Wachstumsprozess von sub-$\mu$m gro\ss{}en Staubpartikeln bis hin zu Planetesimalen, sowie der radiale Transport von Staub in den Gasscheiben um junge Sterne ist das Thema dieser Arbeit. Radiale Drift, Sedimentation, turbulenter Transport sowie Koagulation, Fragmentation und Erosion bestimmen die zeitliche Entwicklung von zirkumstellarem Staub.
    
    Wir gehen dieses Problem von drei verschiedenen Richtungen an: analytische Berechnungen, numerische Simulationen und Vergleich zu Beobachtungen. Wir beschreiben die physikalischen und numerischen Konzepte, mit denen unser Modell die Entwicklung des Staubes \"uber Millionen von Jahren in einer zeitabh\"angigen, viskosen Gasscheibe simuliert. Wir vergleichen die simulierten Staubgr\"o\ss{}enverteilungen mit unseren analytischen Vorhersagen und leiten daraus ein einfaches Rezept zur Bestimmung von station\"aren Gr\"o\ss{}enverteilungen ab. Mit dem vorgestellten Modell ist es uns m\"oglich zu zeigen, dass Fragmentation erkl\"aren kann, warum zirkumstellare Scheiben f\"ur mehrere Millionen Jahre reich an Staub sein k\"onnen. Des Weiteren befassen wir uns mit dem Problem, vor das uns Beobachtungen stellen, n\"amlich dass Staubpartikel von der Gr\"o\ss{}e einiger Millimeter in gro\ss{}en Abst\"anden von ihrem Zentralstern nachgewiesen wurden. Unter der Annahme, dass radiale Drift ineffektiv ist, k\"onnen wir einige beobachtete spektrale Indices und Fl\"usse im mm-Wellenl\"angenbereich reproduzieren. Lichtschw\"achere Quellen deuten darauf hin, dass das Staub-zu-Gas Verh\"altnis oder die Opazit\"aten geringer sind als weithin angenommen.  
    \vspace*{\fill}
    \movetooddpage
    \vspace*{\fill}
    \begin{center}
        \textsc{\large Abstract}
    \end{center}
Dust constitutes only about one percent of the mass of circumstellar disks, yet it is of crucial importance for the modeling of planet formation, disk chemistry, radiative transfer and observations. The initial growth of dust from sub-$\mu$m sized grains to planetesimals and also the radial transport of dust in disks around young stars is the topic of this thesis. Circumstellar dust is subject to radial drift, vertical settling, turbulent mixing, collisional growth, fragmentation and erosion.

We approach this subject from three directions: analytical calculations, numerical simulations, and comparison to observations. We describe the physical and numerical concepts that go into a model which is able to simulate the radial and size evolution of dust in a gas disk which is viscously evolving over several million years. The resulting dust size distributions are compared to our analytical predictions and a simple recipe for obtaining steady-state dust size distributions is derived. With the numerical model at hand, we show that grain fragmentation can explain the fact that circumstellar disks are observed to be dust-rich for several million years. Finally, we investigate the challenges that observations present to the theory of grain evolution, namely that grains of millimeter sizes are observed at large distances from the star. We have found that under the assumption that radial drift is ineffective, we can reproduce some of the observed spectral indices and fluxes. Fainter objects point towards a reduced dust-to-gas ratio or lower dust opacities. 
    \vspace*{\fill}
    \movetooddpage
\movetooddpage
\setlength{\epigraphwidth}{8.2cm}
\epigraphtextposition{flushleftright}
\vspace*{\fill}
\epigraph{The light dove, cleaving in free flight the thin air, whose resistance it feels, might imagine, that her movements would be far more free and rapid in airless space.}
{\textit{Critique of pure reason (1787), Introduction, III.}\\ \textsc{-- Immanuel Kant}}
\vspace*{\fill}
\cleardoublepage
\pagestyle{Ruled}
\renewcommand{\sectionmark}[1]{\markright{\thesection\ #1}}
\tableofcontents
\chapter{Introduction}\label{chapter:intro}
The average density in the universe today is about one atom per cubic meter \citep{Schneider:2008p10345}. Nevertheless, we are living on an over-density of 29 orders of magnitude, known to us as the Earth. Even though the question how our world was created is almost as old as mankind itself, we are still far away from being able to answer it. Today, astronomers are able to draw a coarse picture of how stars and their planetary systems are ``produced'' but at the same time,   many more questions have been discovered which have not been posed before.

This chapter tries to give a very brief overview over the historical aspects and the nowadays widely believed scenario of star and planet formation, which sets the scene for the following chapters of this thesis. 

\section{Historical perspective}
The question about the origin of the Earth and our solar system has always been particularly attractive as it is one of the keys to constraining the uniqueness of our own existence. The earth was believed to be the center of the universe, until 400 years ago in January 1610, Galileo Galilei pointed his telescope to Jupiter \citep{Drake:2003p10346}. By the discoveries of the moons of Jupiter and the phases of Venus, he did not only become the father of modern astronomy but also did he degrade the Earth to be just one planet amongst others, orbiting the Sun. This hypothesis was soon accepted by other thinkers of his time as evidence was accumulating.

%
\begin{figure}[t!hb]
  \centering
    \includegraphics[height=0.6\hsize]{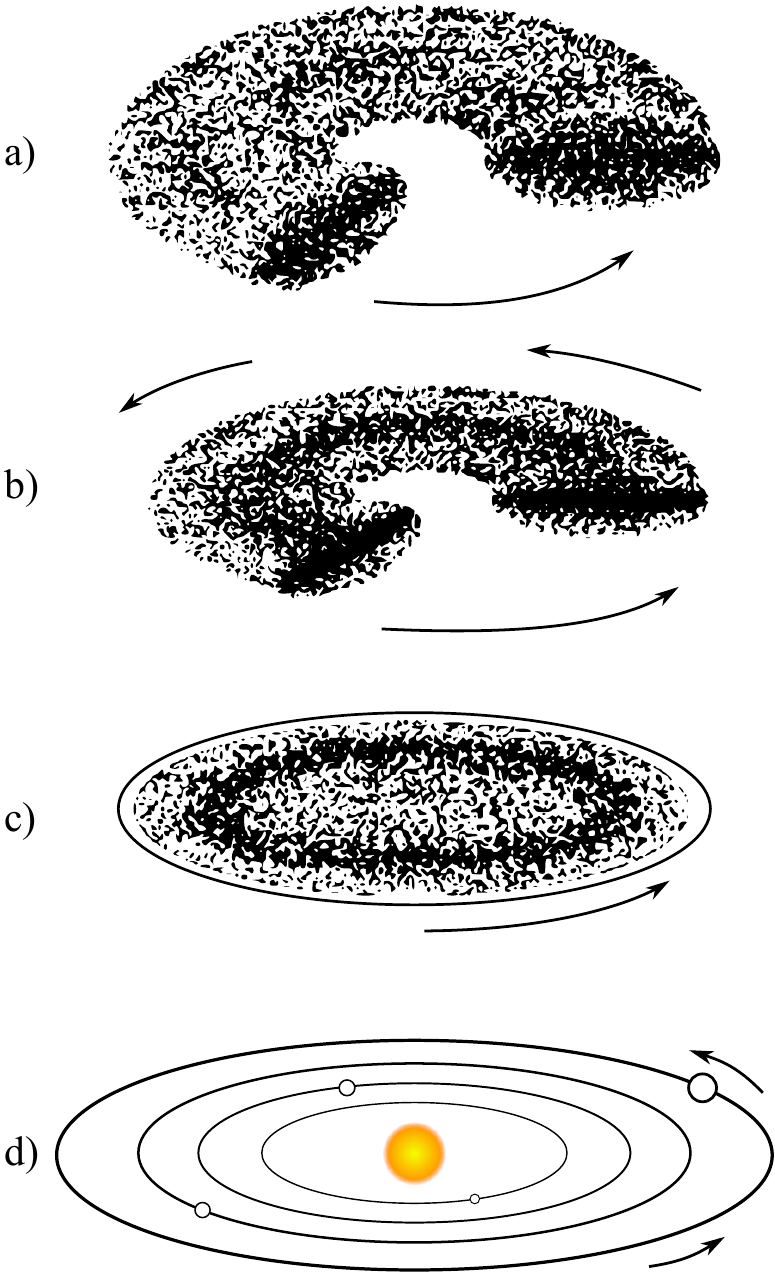}
  \caption{The nebula hypothesis, first suggested by Laplace. In this picture of solar system formation, the initial cloud contracts (a) the contraction causes the forming disk to spin up due to angular momentum conservation (b). Once the centrifugal force equals the gravitational force of the star, the disk fragments locally into rings (c) and finally, planets condense out of the rings (d).}
  \label{fig:intro:nebula}
\end{figure}

While previous theories about the world were mostly mystic or philosophical, the Age of Enlightenment and the observations of Galilei and Kepler led to more mechanical and more realistic philosophies. One of the first of such philosophies about the solar system was put forward by Descartes in 1644. He imagined the solar system and the universe in total as a system of vortices, thus implying that each star itself could host a planetary system like our own. While the idea that the solar system was born out of a large vortex was not too far from what we know today, his theory still lacked a mathematical basis. He writes \citep[see,][]{Gaukroger:2006p10348}:
\begin{quote}
``Finally, we see that, although these whirlpools always attempt a circular motion, they practically never describe perfect circles, but sometimes become too great in width or in length. Thus we can easily imagine that all the same things happen to the planets; and this is all we need to explain all their remaining phenomena.''
\end{quote}
Nevertheless, things were a bit more complicated than Descartes imagined and his ideas were quickly swept away by the Newtonian/Keplerian system which seemed to explain perfectly the motions of the planets by a mathematical formalism. Yet it did not explain the origin of the solar system.

During the 18th century, two scenarios were developed, the first of which was later known as the \emph{tidal theory}, first put forward by the french naturalist Georges Buffon, the other one was called the \emph{nebula hypothesis} which was proposed by Pierre Simon de Laplace in 1796 \citep{Murdin:2001p10349}.
In Buffons scenario, a close encounter between the sun and a comet should have stripped off material which later condenses to form the earth and the other planets. This theory implied that the formation of the solar system was a particularly rare event which was consistent with the fact that no evidence for other planetary systems was known at that time.

For Laplace, on the other hand, the very uniform motion of the planets and their satellites proved that the system was created by a more general law of nature. Thus he proposed that the solar atmosphere once extended much further out through the whole solar system, initially being hot and luminous, and then cooled and contracted (see Fig.~\ref{fig:intro:nebula}). In his view, matter would spin up due to angular momentum conservation until it condenses locally, forming rings around the sun out of which in turn the planets condense. Other defenders of this scenario include Immanuel Kant and Friedrich Wilhelm Herschel who already observed nebulae which seemed to resemble the nebulae predicted by this theory. In this scenario, planet formation is a natural, universal by-product of star formation which would mean that each star could in principle have a planetary system.

This theory became more and more accepted and few people at that time still considered theories of interaction. In 1905, T. C. Chamberlin and F. R. Moulton successfully revived Buffons one-century-old dualistic theory \citep[see][]{Brush:1996p10354}, arguing that the Earth could not have formed out of hot gas because its atmosphere would have evaporated. They suggested that the solar system formed through a close encounter of the sun with another star which induces tidal bulges which are then thrown off into the surrounding space. However, unlike in Buffons scenario, the planets did not ``condense'' out of this matter, in their view, but were assembled out of small cold particles which they called \emph{planetesimals} (from 'infinitesimal planets').
But also the tidal theory had its flaws: Henry N. Russel showed that the expelled gas would be far too hot to condense into planets and that it is impossible to expel enough material to large enough radii to explain the existence of the gas giants \citep{Murdin:2001p10349}.

During the first and the beginning of the second half of the twentieth century, many different scenarios and theories were discussed, amongst them the externally-triggered cloud collapse model of A. G. W. Cameron, the theory of magnetic breaking of Hannes Alfv\'{e}n and Fred Hoyle or the Bondi-Hoyle-Lyttleton cloud capture scenario. One of the most influencing ones was by Carl Friedrich von Weizs\"acker who revisited the idea of Descartes' vortices on a more solid basis. He was also the first to realize that a turbulent disk would transport mass inside and angular momentum outside \citep{Bodenheimer:2006p11151}.

The subject of star and planet formation experienced an ever increasing impetus since the second half of the last century. \citet{Larson:1969p2574}, \citet{Penston:1969p2601}, and \citet{Shu:1977p843} derived first solutions of the collapse of self gravitating spheres, while the development of computers enabled astronomers to carry out simulations of the proposed scenarios which could not be computed analytically. Today, it is generally accepted that stars form out of molecular clouds with turbulence being both triggering as well as regulating the formation of stars \citep[see, e.g.][]{McKee:2007p2813}.
 
If a cloud core exceeds a certain critical mass, called the \emph{Jeans mass}\footnote{The Jeans mass can very simplistically be derived by equating the thermal and the gravitational energy of a sphere of constant density. More sophisticated calculations like Jeans' original analysis as well as more recent work derive a critical mass with the same dependencies and only slightly different coefficients \citep[e.g.,][]{Larson:1985p10355}.}, it becomes gravitationally unstable and will collapse, unless other stabilizing effects, such as turbulence, rotational energy or magnetic fields are strong enough to stabilize the cloud. These effects also gain importance as the cloud collapses: rotational energy can cause binary formation and/or the the formation of a circumstellar disk \citep{Larson:2003p3025}. Additionally, magnetic fields play a role by driving outflows and transporting angular momentum.

These facts show how diverse and complicated the formation of stars is, already on larger scales. Things will not become simpler if we focus on smaller scales now. Large parts of our current basic understanding of disk evolution and planet formation has to be contributed to a few influential papers: \citet{LyndenBell:1974p1945} and \citet{Shakura:1973p4854} pioneered the theory of accretion disks. These disks are an important aspect of star formation and at the same time the birth places of planets. \citet{Safronov:1969p11177} and independently \citet{Goldreich:1973p11184} envisioned a scenario of planetesimal formation which is (with modifications) still discussed today.

\section{Theoretical concepts}
Circumstellar disks are the birth places of planets, thus understanding the mechanism which determine their shape and structure is the key to finding out how planets can form. The evolution of the gas and the dust in circumstellar disks, the growth and fragmentation of the precursors of planet formation and some observational implications thereof are the topic of this thesis. In the following, we will discuss the most important theoretical concepts of circumstellar disks and planet formation and also some observational constrains which together provide the framework of this thesis.

\subsection{Structure and evolution of circumstellar disks}
Already the most basic observations of our solar system contain valuable hints about the formation of stars and planets: most of the mass (about 99.87\%) of the solar system is contained in the sun, which consists almost entirely of hydrogen and helium. In contrast to this, the angular momentum of Jupiter already exceeds the angular momentum of the sun by about two orders of magnitude \citep[e.g.,][]{Armitage:2009p10482}. The typical angular momentum of a one solar mass molecular cloud core \citep[see][]{Goodman:1993p8362} is another 3-4 orders of magnitude larger than the total angular momentum of the solar system (assuming a sphere of constant density in solid-body rotation). This tells us that which ever mechanism produced our solar system out of such a cloud core needs to transport angular momentum outside while most mass in transfered inwards. This is known as the \emph{angular momentum problem} of star formation.

A collapse of a cloud with non-vanishing angular momentum will form a disk. This can be understood if one imagines each parcel of gas to orbit around the core center maintaining its angular momentum. If the total angular momentum of the previously mentioned sphere of constant density is about $J \sim 10^{54}$ g cm$^2$ s$^{-1}$ (assuming a density of $n = 10^{5}$ cm$^{-3}$ and the ratio of rotational over gravitational energy $\beta = 0.02$), then the parcel of gas with an average specific angular momentum $l = J/M_\odot$ will hit the equator at a radius of about $r = \frac{l^2}{G\,M_\odot} \sim 200$~AU \citep{Ulrich:1976p856}. At the mid-plane, it will collide with a symmetrically moving parcel of gas. This will produce a so called accretion shock where the kinetic energy is dissipated and thus a disk is formed. In reality, the physics of a cloud core collapse is  much more complicated because the rotational support can produce multiple stars/disks and turbulence and magnetic fields also play an important role during the collapse. The formation of a disk is however a well established outcome, also of more realistic simulations of cloud collapse \citep[see,][for state-of-the-art simulations]{Inutsuka:2010p11359}. The disks around new-born stars are usually called circumstellar disks, accretion disks (implying an inward flux of mass) or protoplanetary disks (implying that planets will form inside of them). Due to the geometry of a disk, the vertical sound-crossing time scale is much shorter than the radial one, thus it is usually a good approximation to treat the vertical structure to be in an equilibrium, while the radial evolution is happening on larger time scales.

A first order vertical structure of the gas disk can be derived by assuming the disk to be isothermal (vertical temperature profile $T(z) = $ const.) and thin (i.e., the height above the mid-plane $z$ is small compared to the distance to the star $r$). Then, the vertical component of the stellar gravity needs to be balanced by a vertical pressure gradient,
\begin{equation}
\csound^2 \frac{\mathrm{d}\,\rhogas}{\mathrm{d}z} = -\rhogas\,z \, \frac{G\,M_\star}{r^3},
\end{equation}
where $\csound$ is the isothermal sound speed, $G$ the constant of gravity, \rhogas the gas density, and $M_\star$ the mass of the central star. From this we derive by integration a Gaussian density profile
\begin{equation}
\rhogas = \rho_0 \, \exp\left(-\frac{z^2}{2\, \Hp^2}\right),
\end{equation}
where we denoted the mid-plane density as $\rho_0$ and defined the pressure scale height \Hp to be the ratio of the sound speed over the Kepler frequency $\Omega$.

Having a first expression for the vertical structure, we can now focus on the radial structure and its evolution. Depending on the desired resolution/complexity, this can be done in 1, 2 or 3 dimensions. However, in this thesis, we are interested in (parameter studies of) the long-time evolution of these disks and multidimensional calculations are computationally too expensive to simulate several million years of disk evolution (especially if grain growth is considered as we will discuss later on). Therefore -- and also to get an analytical feeling of the disk evolution -- we will now consider the 1+1D approach, where an axisymmetric disk is assumed to be in a vertical equilibrium as discussed above. We are thus only interested in the time evolution of the surface density, which is the vertically integrated column density, defined as
\begin{equation}
\Siggas(r) = \int_{-\infty}^{\infty} \rhogas(r,z) \, \dx{z}.
\end{equation}

If there is viscosity acting within the disk (we will cover this question further below), then two neighboring sheets of gas will interact through friction due to the differential rotation. The gas is approximately orbiting in Keplerian rotation. Therefore, an inner sheet of gas is rotating faster than an outer one and viscous friction thus accelerates the outer sheet while the inner sheet is decelerated. This mechanism obviously transports angular momentum outside and at the same time allows matter to flow inwards.

The mathematical description of this process is called the \emph{theory of accretion disks} and was first derived by \citet{LyndenBell:1974p1945}. They derived the radial velocity of the gas due to mass and angular momentum conservation in the presence of a viscosity $\nu$ to be
\begin{equation}
v_\mathrm{gas}(r) = - \frac{3}{\Siggas \sqrt{r}} \, \frac{\partial}{\partial r} \left(\Siggas \, \nu \, \sqrt{r}\right).
\label{eq:intro:vgas}
\end{equation}
The evolution of the surface density is then given by the vertically integrated mass conservation equation, and can be written as
\begin{equation}
\frac{\del \Siggas}{\del t} = \frac{3}{r} \, \frac{\del}{\del r} \left[r^{1/2} \, \frac{\del}{\del r}\left(\nu \Siggas r^{1/2}\right) \right]
\label{eq:intro:ssd}
\end{equation}
\citep[for a derivation, see also][]{Pringle:1981p805}.
This equation can now be solved numerically for a given viscosity and initial condition (see Chapter~\ref{chapter:model} and \ref{chapter:algorithm}), but it is instructive to discuss some general properties of this equation analytically, which we will do in the following.

Substituting $x = 2\,\sqrt{r}$ and $u = \Siggas\, \sqrt{r}$ in Eq.~\ref{eq:intro:ssd}, we can derive a diffusion equation
\begin{equation}
\frac{\del u}{\del t} = D \frac{\del^2 u}{\del x^2}
\end{equation}
with diffusion constant $D=12\nu/x^2$ (assuming a constant $\nu$). The diffusion constant defines the time on which diffusion acts to smooth out gradients in the concentration (in this case: \Siggas). A viscous disk with size $r$ will therefore evolve on the \emph{viscous timescale} given by
\begin{equation}
\tvisc = \frac{x^2}{D} \approx \frac{r^2}{\nu}.
\label{eq:intro:tvisc}
\end{equation}
From the observed live time of circumstellar disks, which is about 3~million years (see Section~\ref{sec:intro:observations}), we can now get constraints on the viscosity of these disks. It turns out that the viscous time scale of a 100~AU disk (with typical values for the temperature and surface density) is of the order of $10^{13}$ years if we consider only molecular viscosity. This is much longer than the age of the universe. To explain the observed disk lifetimes, another mechanism of angular momentum transport is needed and the \emph{anomalous viscosity} provided by turbulence is typically believed to be solution (although gravitational instability and disk winds may also play a role).

The source of the turbulence is still unknown. Keplerian disks are linearly stable to axisymmetric perturbations \citep[e.g.,][]{Safronov:1969p11177}, but a number of other possible sources have been discussed, such as the convective instability \citep[today discarded due to the fact that it would transport angular momentum inside instead of outside, see][]{Stone:1996p10488}, the gravitational instability \citep{Toomre:1964p1002,Gammie:2001p6435}, or the baroclinic instability \citep{Klahr:2003p10848,Lesur:2010p10824} to name a few of them. However none of these has been as successful as the magnetorotational instability (MRI), discovered by \citet{Balbus:1991p4932}, because it draws its energy directly from the Keplerian shear flow \citep{Balbus:1998p7450}. Therefore it has been shown to be a very reliable and effective mechanism of angular momentum transport which is effective enough to explain the observed accretion rates. Still, it requires two conditions in order to operate in circumstellar disks: outwardly decreasing differential rotation and an ionization fraction of at least $n_\mathrm{ion}/n_{neutral} \gtrsim 10^{-13}$ \citep[see][]{Balbus:1998p7450}.

While the former condition is always the case in (nearly) Keplerian disks, the latter condition may be questionable. A certain level of ionization is provided by cosmic rays, high energy irradiation of the central star, and by decay of radioactive nuclei within the disk. However dense enough regions and/or the large scale magnetic field line structure may shield some parts of the disk \citep[e.g.,][]{Gammie:1996p1515,Sano:2000p9889,Turner:2009p11341}. In this case, there will be a zone where MRI shuts off (called \emph{dead zone}) surrounded by an MRI-active layer. Since the vertically averaged  angular momentum transport in layered (i.e., partly-dead) regions is not as effective as in active regions, material will accumulate in the dead parts of the disk \citep{Gammie:1996p1515}. These \emph{layered} disks can become unstable and produce outbursts \citep[thought to explain the FU Orionis objects, e.g.][]{Armitage:2001p993} but stationary solutions are also possible \citep{Terquem:2008p7731}.

Coming back to the evolution of the disks surface density evolution, we note that there are possible mechanisms providing the necessary anomalous viscosity, but it is still unknown which mechanism operates and how effective it really transports angular momentum. It has therefore been proven very helpful to parametrize our ignorance in the following way, proposed by \citet{Shakura:1973p4854}: turbulent viscosity is provided by eddies with a certain size $L$ and velocity $V$. The largest reasonable values in a circumstellar disk are the scale height \Hp and the sound speed \csound, respectively. The viscosity can therefore be written as a fraction of a product of these quantities,
\begin{equation}
\nu = \alphat \, \csound \, \Hp,
\label{eq:intro:viscosity}
\end{equation}
where the dimensionless parameter $\alphat$ specifies the amount of turbulence in the disk and is therefore usually bound between $0<\alphat<1$. Typical values of $\alphat$ from simulations of MRI-active disks are in the range of $10^{-3}$ to a few times $10^{-2}$ \citep[e.g.,][]{Johansen:2005p8425,Dzyurkevich:2010p11360}. For a given value of $\alphat$, we can now calculate the viscosity if the temperature profile is known. Realistic models need to take into account the effects of external irradiation, viscous heating, optical depth, and shadowing. Thus multidimensional radiative transfer is in principle necessary to derive a realistic temperature structure, however we are only interested in some scaling relations of the surface density in this introductory chapter and will therefore assume a power-law profile for the temperature
\begin{equation}
T(r) \propto r^{-q},
\end{equation}
and also for the surface density
\begin{equation}
\Siggas(r) \propto r^{-p}.
\end{equation}
Postulating a steady-state of the surface density ($\frac{\del \Siggas}{\del t}=0$) and using the power-laws for temperature and surface density in Eq.~\ref{eq:intro:ssd}, we derive
\begin{equation}
p+q = -\frac{3}{2}.
\end{equation}
This means that for a given temperature profile, a disk will evolve towards an equilibrium power-law with index $p$. The time scale of this evolution is given by Eq.~\ref{eq:intro:tvisc}. From this result, we can also derive the accretion rate of the disk: the product rule applied to Eq.~\ref{eq:intro:vgas} gives
\begin{equation}
v_\mathrm{gas}(r) = - \frac{3\nu}{2\,r} - \frac{3}{\Siggas }\frac{\partial \left(\Siggas\nu\right)}{\partial r}.
\end{equation}
Inserting this equation into the definition of the mass accretion rate,
\begin{equation}
\dot M(r) = -2\pi \, r \, \Siggas \, v_\mathrm{gas},
\end{equation}
results in two contributions to the mass flux,
\begin{equation}
\dot M(r) = 3\pi \,\Siggas\,\nu + 6\pi \,r\frac{\del\left(\Siggas \, \nu\right)}{\del r}.
\end{equation}
We can see now that this mass accretion rate is constant in a steady state disk since the product of $\Siggas\cdot\nu$ is constant with radius and that the mass in a non-stationary disk will be rearranged to approach a steady state.
Eq.~\ref{eq:intro:ssd} can also be solved analytically in a less hand-waving way, which can be found in \citet{LyndenBell:1974p1945}. Their self-similar solution of the disk evolution shows similar features. In general, a viscous disk (with power-laws for temperature and viscosity) evolves in the following way: the disk profile in the inner parts of the disk \citep[defined by a characteristic radius, see][for example]{Hartmann:1998p664} follows the same power-law derived above, while in the outer parts, the surface density decreases exponentially with radius. The mass accretion rate causes the mass of the disk to gradually decrease. In order to achieve this, a small fraction of the mass needs to ``absorb'' the angular momentum. The outer parts of the disk therefore spreads while the inner parts accrete onto the star. The final outcome of this evolution is the state of lowest energy: almost all the mass is in the central object while almost all angular momentum is in a small fraction of the mass at a large radius. This arrangement is similar to what is observed in our solar system. 

\subsection{The dynamics of dust}
So far we have only talked about the gas disk evolution. The canonical value of the dust-to-gas ratio in the interstellar medium used in the literature is only 1\%, thus one might be tempted to neglect it. However, not only is this small fraction of the total mass the material out of which terrestrial planets, asteroids and the cores of giant planets are made, it is also an important ingredient to many aspects of circumstellar disks. Almost all the opacity of the disk is provided by grains. They do therefore determine most observable properties of disk. Additionally, they are important for chemical surface reaction (e.g., formation of water) or for photoelectric heating which strongly influences the temperature in the upper layers of the disk \citep[and also the FUV photoevaporation, see][for example]{Gorti:2009p8414} and also for the ionization fraction, which detemins the coupling of the disk to the magnetic field and thus its MRI activity \cite[e.g.,][]{Sano:2000p9889,Turner:2009p11341}. All these facts tell us how important the radial and also the size distribution of dust in circumstellar disks is. In this subsection we will introduce the most important aspects wich determine the evolution of circumstellar dust.

As noted above, the dust mass in circumstellar disk is believed to be (initially) only about 1\% of the gas mass since the disk material stems from the interstellar medium. Therefore, the gas is the dynamically dominating phase. This can only change if some transport mechanism effectively accumulates the dust relative to the gas until the mass fraction approaches unity. We will therefore begin by discussing the transport of dust in both the vertical and the radial direction. Analyzing the interplay between gas and dust motion is beyond the scope of this chapter \citep[for details, see][and references therein]{Chiang:2010p11570}. We will rather discuss the steady-state solutions of the dust velocity without considering back reactions to the gas. 

While the gas reacts to pressure forces, the dust only feels the gravitational forces acting on it and the drag force by which it is coupled to the gas motion. 
The drag force is given by
\begin{equation}
\boldsymbol{F}_\mathrm{drag} = - \frac{m}{\tau_\mathrm{s}}\boldsymbol{\Delta u},
\label{eq:intro:dragforce}
\end{equation}
where $m$ is the mass of the particle, $\boldsymbol{\Delta u}$ is the particle velocity relative to the gas and $\tau_\mathrm{s}$ is the stopping time, which will be discussed in more detail in Chapter~\ref{chapter:model}. This coupling to the gas causes dust particles to move towards the region of highest pressure. If we consider a particle at a height $z$ above the mid-plane, then its vertical motion would be an oscillation around the mid-plane if it were not coupled to the gas. The drag force damps this motion and thus the particle will settle towards the mid-plane. Equating the vertical component of the gravitational force $F_\mathrm{g} = - m \,\Ok^2\,z$ and the drag force (Eq.~\ref{eq:intro:dragforce}) yields the terminal settling velocity
\begin{equation}
u_\mathrm{sett} = - \tau_\mathrm{s} \, \Ok^2 \,z.
\label{eq:intro:v_sett}
\end{equation}
It turns out to be very practical to define a dimensionless stopping-time of the particle called the \emph{Stokes number} \St. It describes the coupling of the particle to the gas motion and is in our context defined as the ratio of the stopping time and the orbital time,
\begin{equation}
\St = \tau_\mathrm{s} \,\Ok.
\end{equation}
This way, particles with different structure or composition but the same \St will behave dynamically entirely the same.

As noted before, the gas disk is most probably turbulent. The distribution of dust is thus mixed along with the turbulent gas motion, again depending on the coupling to the gas. The vertical distribution of dust is now determined by the two counteracting mechanisms of turbulent mixing and vertical settling. We can derive a steady-state distribution from the diffusion-advection equation of the dust concentration \citep[e.g.,][]{Dubrulle:1995p300}
\begin{equation}
\frac{\del \rhodust}{\del t} = -\nabla \left[u_\mathrm{sett}\,\rhodust - D_\mathrm{d} \, \rhogas \nabla\left(\frac{\rhodust}{\rhogas}\right) \right].
\end{equation}
Postulating a steady-state ($\del \rhodust/\del t = 0$), using Eq.~\ref{eq:intro:v_sett} and integration results in a Gaussian distribution of the dust concentration with a scale height
\begin{equation}
h_\mathrm{d} = \Hp\,\sqrt{\frac{D_\mathrm{d}}{\St}}.
\end{equation}
This leads to the question what the diffusion constant $D_\mathrm{d}$ is. \citet{Youdin:2007p2021} showed that the dust diffusion coefficient is in good approximation given by
\begin{equation}
D_\mathrm{d} = \frac{D_\mathrm{g}}{1+\St^2}.
\end{equation}
It is then typically assumed that the gas diffusivity $D_\mathrm{g}$ is equal to the viscosity $\nu$. \citet{Johansen:2005p8425} found that in MRI turbulence, radial diffusion is as strong as turbulent viscosity, while the vertical diffusivity is slightly weaker.

The dust-gas coupling also influences the radial distribution of the dust, again, by turbulent mixing and ``radial settling'' towards the star but also by co-moving it along with the radial gas flux. These topics will extensively be discussed in Chapter~\ref{chapter:model}. Especially the radial drift has important implications for planet formation, as we will see in the following: larger bodies (i.e., $\St \gg 1$) are not significantly affected by the gas drag (because of their small surface to mass ratio) and thus move on Keplerian orbits. In contrast to this, very small particles ($\St \ll 1$) are so tightly coupled to the gas that they are basically frozen-in to the gas. In between these two limits, there is a critical regime around a Stokes number of unity: these particles are marginally coupled, thus they are not frozen-in to the gas but are orbiting at Keplerian velocity. However unlike the much larger bodies, they are still significantly influenced by the gas drag. The fact that the gas is orbiting at sub-Keplerian velocity due to its pressure support induces a relative velocity between the dust particles and the gas. The dust particles loose angular momentum because of this head wind and consequently spiral inwards. The resulting radial velocity can be written as
\begin{equation}
u_\mathrm{drift} \simeq \frac{\St}{1+\St^2}\,\frac{\Hp}{r} \, \left(\frac{\del \ln P}{\del \ln r}\right)\,\csound,
\end{equation}
as derived by \citet{Weidenschilling:1977p865} and \citet{Nakagawa:1986p2048} (here, $P = \rhogas\,\csound$ is the gas pressure). For typical disk models, the time scale of this drift can be very short, in the order of only a few local orbits \citep[e.g.,][]{Brauer:2007p232}. This drift poses a fundamental problem for all theories which assume hierarchical growth from small dust particles to planetesimals and will be discussed in further detail in Chapter~\ref{chapter:model}. The basic ideas of grain growth and planet(-esimal) formation will be introduced in the next section.
    
\subsection{From dust to planets}
    The growth from sub-$\mu$m sized monomers to larger aggregates is inevitably the first stage of the formation of planets. Small dust aggregates are sticky because of their large surface to mass ratio and inter-molecular binding forces (e.g. van der Waals forces or hydrogen bonds). This fact is proven by both experimental \citep[e.g.,][]{Heim:1999p11580,Poppe:2000p9447} and theoretical work \citep[e.g.,][]{Dominik:1997p9440}.    
    In order to assemble larger aggregates, these sub-$\mu$m sized dust particles need to collide. There are several mechanisms which could provide relative motion of the dust particles:
    \emph{Brownian motion}, the random motion caused thy the thermal motion of the gas, leads to random velocities between small particles. The smaller the particle, the more it is influenced by Brownian motion.
    \emph{Radial, azimuthal, and vertical motion} of the dust (as discussed in the previous section) is also size dependent, thus, particles with different sizes will have different velocities which leads to relative motion between grains of different sizes.
    Different sized dust grains couple to different sized eddies of the \emph{turbulent motion} of the gas. The relative particle velocities induced by turbulence have been derived by \citet{Ormel:2007p801}.
    
    The relative velocities together with the efficient sticking of small dust aggregates leads to a fast growth of dust; sizes of around centimeters can be reached within less than 1000~years at a few AU, as was shown by many authors \citep[][to name a few]{Weidenschilling:1980p4572,Weidenschilling:1995p11583,Nakagawa:1981p4533,Tanaka:2005p6703,Dullemond:2005p378,Brauer:2008p215}. However, two problems arise: particles of decimeter sizes at a few AU (or millimeter sizes at around 100~AU) have a Stokes number of unity and, once they are formed, are quickly lost towards the star. Additionally, the relative velocities due to drift and turbulence also reach their maximum at these sizes and are large enough to fragment or erode the particles \cite[e.g.,][]{Brauer:2008p215,Blum:2008p1920}. Additionally, it is questionable if the efficient sticking assumption which is often used is still valid in this size regime \citep{Youdin:2010p10939}. Aggregates of mm size do not seem to stick at any velocity, but rather bounce or fragment \citep{Blum:2008p1920}. This obstacle might be overcome by the collisions of aggregates of very different sizes \citep[e.g.,][]{Wurm:2005p1855,Teiser:2009p7785}.
    
    The \emph{gravitational instability} of the dust is an attractive scenario which basically ``jumps'' over this problematic size regime. In the original idea, developed by \citet{Safronov:1969p11177} and independently by \citet{Goldreich:1973p11184}, growth and vertical settling of dust particles forms a dust layer at the mid-plane of the disk which becomes unstable to its own gravity and thus collapses to form planetesimals. \citet{Weidenschilling:1980p4572} showed that this scenario does not hold since the thin dust layer is unstable to the Kelvin-Helmholtz instability because of the shear at the boundary between the dust and the gas. It turns out that mid-plane turbulence is both a curse and a cure \citep{Youdin:2010p10939}: it may on average stir up concentrations of solids, but it also does create random over-densities. The streaming instability \citep{Youdin:2005p11585} is particularly effective in creating dust over-densities, it can even create gravitationally bound clumps, as shown by \citet{Johansen:2007p4788}. However, this scenario still requires particles of around decimeter sizes, underlining the importance of the initial ``sticking'' growth of dust.  

Leaving these problems aside, a new mode of growth is reached, once particles have grown to sizes where gravity dominates the accretion of solid bodies \citep[however, gas drag can still influence the accretion process, see][]{Ormel:2010p11591}. The gravitational attraction effectively increases the cross section \citep{Lissauer:1993p11600}
\begin{equation}
\sigma_\mathrm{grav} = \pi\,a^2\,\left(1+\left(\frac{u_\mathrm{esc}}{\Delta u}\right)^2 \right),
\end{equation}
where $\Delta u$ is the relative velocity of the planetesimals and $u_\mathrm{esc}$ is the escape velocity of the two colliding bodies. The accretion rate in a system where gravitational focusing is important ($u_\mathrm{esc}\gg\Delta u$) scales superlinearly with mass \citep{Ormel:2010p11595}. Under these conditions, it can be shown \citep[e.g.][]{Armitage:2009p10482} that the ratio of the masses of two bodies growth with time. This is called \emph{runaway growth} and will continue until the runaway bodies have reached sizes of a few 100~km \citep{Wetherill:1989p11603,Thommes:2003p11622,Ormel:2010p11601}. At this point, the largest body gravitationally stirs up the distribution of smaller bodies and the accretion rate then becomes again sub-linearly with mass \citep[e.g.][]{Ida:1993p11613,Kokubo:2000p11616,Thommes:2003p11622}. Ultimately, the growth of the largest body (called the \emph{oligarch}) is stalled once it has accreted all material within its gravitational reach, the so-called \emph{feeding zone}. The resulting protoplanets typically have about the mass of Mars in the terrestrial zone \citep{Thommes:2006p11628}.

Unless the disk mass is scaled up by a factor of 10, this scenario fails to explain the formation of Earth and Venus and also the formation of the cores of gas giant planets in situ: according to \citet{Pollack:1996p11635}, a core of about 10~$M_\Earth$ is needed to start runaway gas accretion and thus produce gas giants. Possible solutions to this include the migration of small bodies by aerodynamic drag as well as the migration of the planets and cores. Scattering between the cores is also an attractive scenario as it could possibly explain the formation of Uranus and Neptune which are too massive to be explained by core accretion at their current location in the disk \citep{Thommes:2003p11622}.

An entirely different approach to planet formation is the \emph{disk instability} scenario: while the disk is typically gravitationally stable in the inner parts, this might be debatable in the outer parts where the disk is colder and the orbital frequency is lower. This can be seen from Toomre's stability criterion \citep{Toomre:1964p1002} which states that disk becomes gravitationally unstable if the stability parameter
\begin{equation}
Q = \frac{\csound\,\Ok}{\pi \, G \, \Sigma},
\label{eq:intro:toomreQ}
\end{equation}
becomes smaller than unity. Whether the disk really fragments can however not be seen just from this parameter. Before collapse sets in (i.e., $1<Q<2$), the disk develops spiral arms which tend to stabilize the disk by rearranging the surface density \citep{Gammie:2001p6435}. Additionally, the disk must be able to cool quickly enough so that a contracting clump is not sheared apart by the Keplerian rotation. Several authors \citep[e.g.,][]{Boss:1997p11650,Boss:2000p11660,Boss:2007p11741,Rice:2005p11667,Meru:2010p11987} have produced very different results which depend highly on the efficiency of cooling in the disk (and thus also on the numerical scheme). Today it seems unlikely that disk instability alone can explain giant planet formation since the simulations typically produce planets at the high end of the mass scale \citep{Rafikov:2005p11787}, however some of the planets detected by direct imaging (see Sect.~\ref{sec:intro:exoplanets}) are candidates for this formation scenario.
    
\section{Observational constraints}\label{sec:intro:observations}
After having introduced the basic concepts theoretically, it is time to provide some proof from the observational side which led to our current understanding (and also the problems) of disk evolution and planet formation. In this section, we will discuss the most important constraints, derived from observations, ordered by size: from observations of gas and tiny dust grains, to the meteorites in our own system, and finally to the observed population of planets in and beyond our solar system.

\begin{figure}[tbh]
  \centering
  \resizebox{0.65\hsize}{!}{\includegraphics{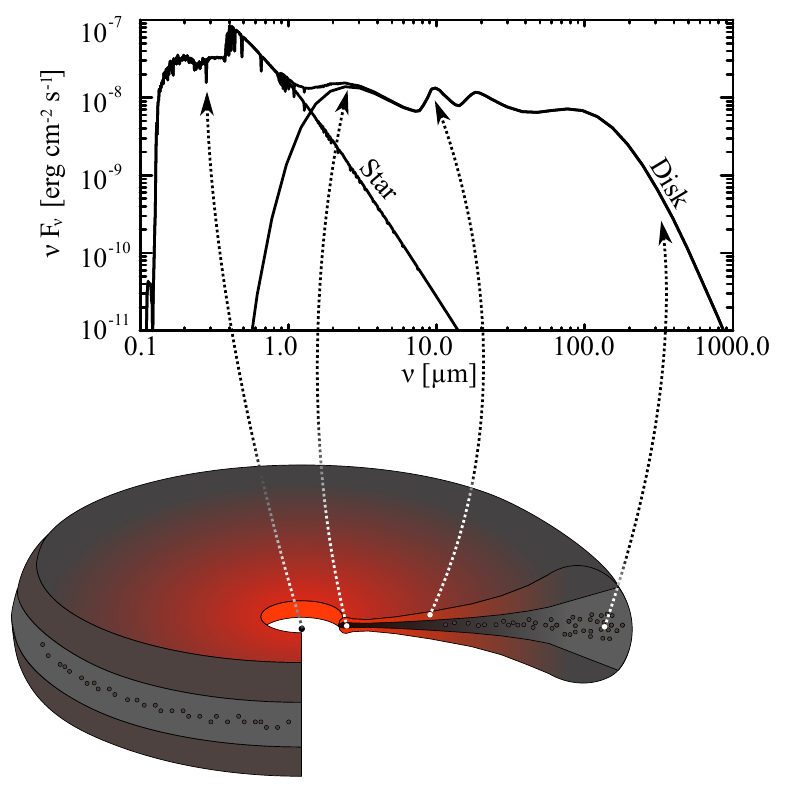}}
  \caption{Origin of the spectral energy distribution: at larger wavelength, the disk emission outweighs the stellar spectrum. Generally, the shorter wavelength emission comes from hotter, inner regions of the disk. Large wavelength (sub-millimeter) probe the dust mass in the outer, optically thin regions. Figure adapted from \citet{Dullemond:2007p336}.}
  \label{fig:intro:diskSED}
\end{figure}

\subsection{Disk observations}
Circumstellar disks can be detected by several techniques, each tracing specific features, temperatures and/or regions of the disk. Combining this knowledge enables us not only to proof the existence of disks but also to probe the physics which is happening inside of them.

Most of the disk mass is expected to reside in gas (because disks are made out of interstellar material), however this component is the hardest to detect as it is often frozen out, especially in the outer, cold parts of the disk and in the mid-plane.

Another gas feature is a strong H$\alpha$ excess, which is is believed to be due to an accretion flow onto the stellar surface \citep[see][]{Bouvier:2007p10356}. Traditionally, T Tauri stars are divided into classical T Tauri stars (CTTS) and weak-lined T Tauri stars (WTTS), depending on the strength of the emission lines. The accreting material has to come from circumstellar material and thus CTTS are thought to be younger objects which still have an accretion disk around them. Indeed, the accretion rate (typically of the order of $10^{-8}$ solar masses per year) and the fraction of young stars which have detectable disks (see below) seem to correlate (in the same way) with the age of the system \citep[e.g.,][]{Calvet:2000p1083,Haisch:2001p10426,SiciliaAguilar:2006p10359,SiciliaAguilar:2006p10361}, indicating a typical lifetime of a gaseous circumstellar disk to be around 3~million years. Only few gaseous disks have been detected around stars older than 6~million years but dust can continue to exist in debris disks, which are probably the remnants of planet(-esimal) formation, similar to the asteroid belts in our own solar system.

The accretion signature and the observed lifetimes of disks show that some mechanism must efficiently transport angular momentum outwards and mass inwards. In general, the transition from accreting to non-accreting seems to be very rapid, in the order of $10^5$ years \citep{Simon:1995p11800,Wolk:1996p11850,Andrews:2005p3779}. This is in contradiction with simple viscous evolution which predicts a gradually fading disk \citep[e.g.,][]{Hueso:2005p685}. Photoevaporative dissipation of the disk \citep{Hollenbach:1994p1186,Clarke:2001p969,Alexander:2006p136} or planets blocking the accretion of the disk \citep[e.g.,][]{Calvet:2002p10424} are some of the frequently suggested scenarios. However at every age some sources do not show any signs of accretion. The reason for this is yet unknown.

Much easier to detect than the gas is the dust which re-radiates the stellar irradiation in the infrared (IR). This emission occurs as a long-wavelength excess above the stellar photospheric flux, where different wavelength correspond to different temperatures and thus different regions in the disk. The disk geometry and the according spectral energy distribution (SED) is schematically shown in Fig.~\ref{fig:intro:diskSED}. In fact this IR excess was the first observational evidence for disks around young stars \citep[e.g.,][]{Strom:1989p9475,Beckwith:1990p3768}.

The emission of warm dust in the inner disk can be used as a tracer for grain growth at smaller sizes: the silicate 10~$\mu$m emission feature is prominent for small dust sizes (around sub-$\mu$m) but disappears as the grain size is increased beyond a few $\mu$m. \citet{vanBoekel:2003p8117} found evidence for grain growth at these aggregate sizes as one would expect from both theory and laboratory experiments \citep{Blum:2000p8099,Blum:2000p8110}.

While the near- and mid-IR traces mostly the warm/hot inner regions and the surface layers of the optically thick dust disk, observations at millimeter wavelength probe the cold outer regions of the disk. Radiation at these wavelength comes from the far end of the disk SED, which is close to the Rayleigh-Jeans limit. The disk SED can be approximated by \citep[e.g.,][]{Beckwith:1990p3768,Andrews:2007p3380}
\begin{equation}
\nu \, F_\nu \approx \nu \, \kappa_\nu\, B_\nu(T) \, M_\mathrm{d}/d^2,
\label{eq:intro:SED}
\end{equation}
if the disk is optically thin and the emission is from an isothermal region. Here $\nu$ denotes the frequency, $\kappa_\nu$ the opacity (per gram of dust), $B_\nu(T)$ the Planck function at a characteristic temperature $T$, $M_\mathrm{d}$ the dust mass, and $d$ the distance to the source. Eq.~\ref{eq:intro:SED} shows that under these conditions, the SED is proportional to the product of dust mass and opacity. For an assumed opacity law, the dust mass can be determined. The total mass of the disk can then roughly be estimated of one assumes the dust-to-gas ratio to be the same as in the interstellar case. Disk masses were found to be ranging from 0.005 to 0.14 solar masses spread out over sizes of between 20 to several hundred AU \citep[e.g.][]{Andrews:2009p7729}. The resolution of observations at these wavelength has reached a state where the emission can be radially resolved \citep[e.g.][]{Isella:2010p9438}. The Atacama Large Millimeter Array (ALMA) will be able to resolve the emission down to a few AU for the most nearby disks.

In the framework of viscous disks and $\alpha$-turbulence (see Eqs.~\ref{eq:intro:ssd} and \ref{eq:intro:viscosity}), spatially resolved observations can be used to constrain important parameters of the theoretical disk models. Results indicate for example that the turbulence parameter should be in the range of $\alphat = 5\e{-4}$--Ð$8\e{-2}$ \citep{Andrews:2009p7729}, which compares well to values predicted by MRI simulations.

\begin{figure}[tbh]
  \centering
  \resizebox{0.75\hsize}{!}{\includegraphics{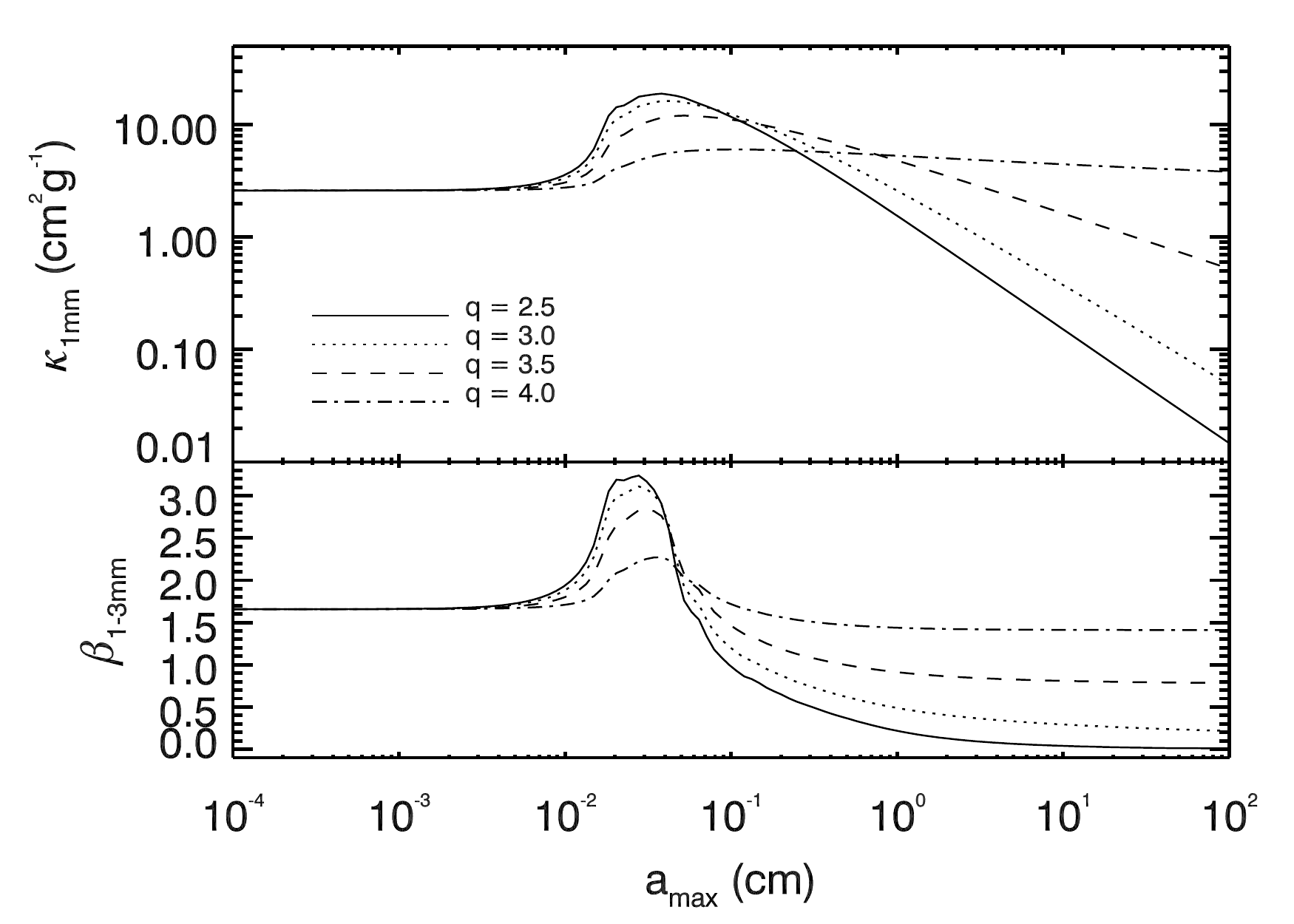}}
  \caption{Dust opacity (upper panel) and the resulting opacity index for different dust size distributions. The parameters of the dust size distribution are the maximum grain sizes \amax and the (negative) power-law index $q$ (see Eq.~\ref{eq:intro:dust_distri}).}
  \label{fig:intro:opacity}
\end{figure}

Additionally, the mm-wavelength regime is also an important probe of grain growth: Eq.~\ref{eq:intro:SED} shows that the spectral dependency of the SED is given by the dust opacity and the Planck function. If we know the mid-plane temperature (or are sufficiently deep in the Rayleigh-Jeans regime), measuring the slope of the SED (by observations at two or more different wavelengths) provides a direct measure of the spectral index of the dust opacity law $\kappa_\nu \propto \lambda^{-\beta}$. $\beta$ depends on the dust size distribution (and the opacity model). For a power-law dust size distribution,
\begin{equation}
n(a) \propto
\left\{\begin{array}{ll}
a^{-q}  &   \text{for } a<\amax \\
0       &   \text{else }
\end{array}\right.,
\label{eq:intro:dust_distri}
\end{equation}
we can now compare which parameters $q$ and \amax are possible for a certain measured value of $\beta$ (see Fig.~\ref{fig:intro:opacity}). For example, a $\beta$ value of unity can only be reached by a distribution which has a maximum grain size \amax of around a few millimeters.

Several authors \citep[e.g.][]{Testi:2003p3390,Wilner:2005p3383,Rodmann:2006p8905,Ricci:2010p9423} detected emission which indicates mm- or cm-sized dust particles at several tens of AU. On the other hand, \citet{Brauer:2007p232} has shown that (for reasonable disk parameters) this distribution of grains should drift towards the star on a short time scale. A solution to this problem is yet unknown, but particle clumping in streaming instabilities or spiral density waves could significantly decrease the drift time scale.    

\subsection{Meteorites}
Additional constrains are coming from a quite different direction, namely meteorites:
they provide ample information with a precision unmatched by any astronomical observation about the formation time scales of solar system bodies. The largest class of meteorites, the \emph{chondrites} are extraordinarily diverse, but they are \emph{undifferentiated} which means that their parent body has not been molten which would have differentiated it into different phases. While the oldest known rocks on earth are ``only'' about $4280$ million years old \citep{Oneil:2008p10905} \citep[some zircon minerals are as old as $4404$ million years, see][]{Wilde:2001p10906}, meteorites can be significantly older: the age of \emph{calcium-aluminium rich inclusions} (CAIs), which can be found in chondrites, was determined to be 4568.3 million years, with an uncertainty of only 700~000~years \citep{Burkhardt:2008p10908}. This way, meteorites are the ``fossils of the planet formation process'', conserving hints about the structure and the chemical composition of the early solar system.

\begin{figure}[tbh]
  \centering
  \resizebox{0.65\hsize}{!}{\includegraphics{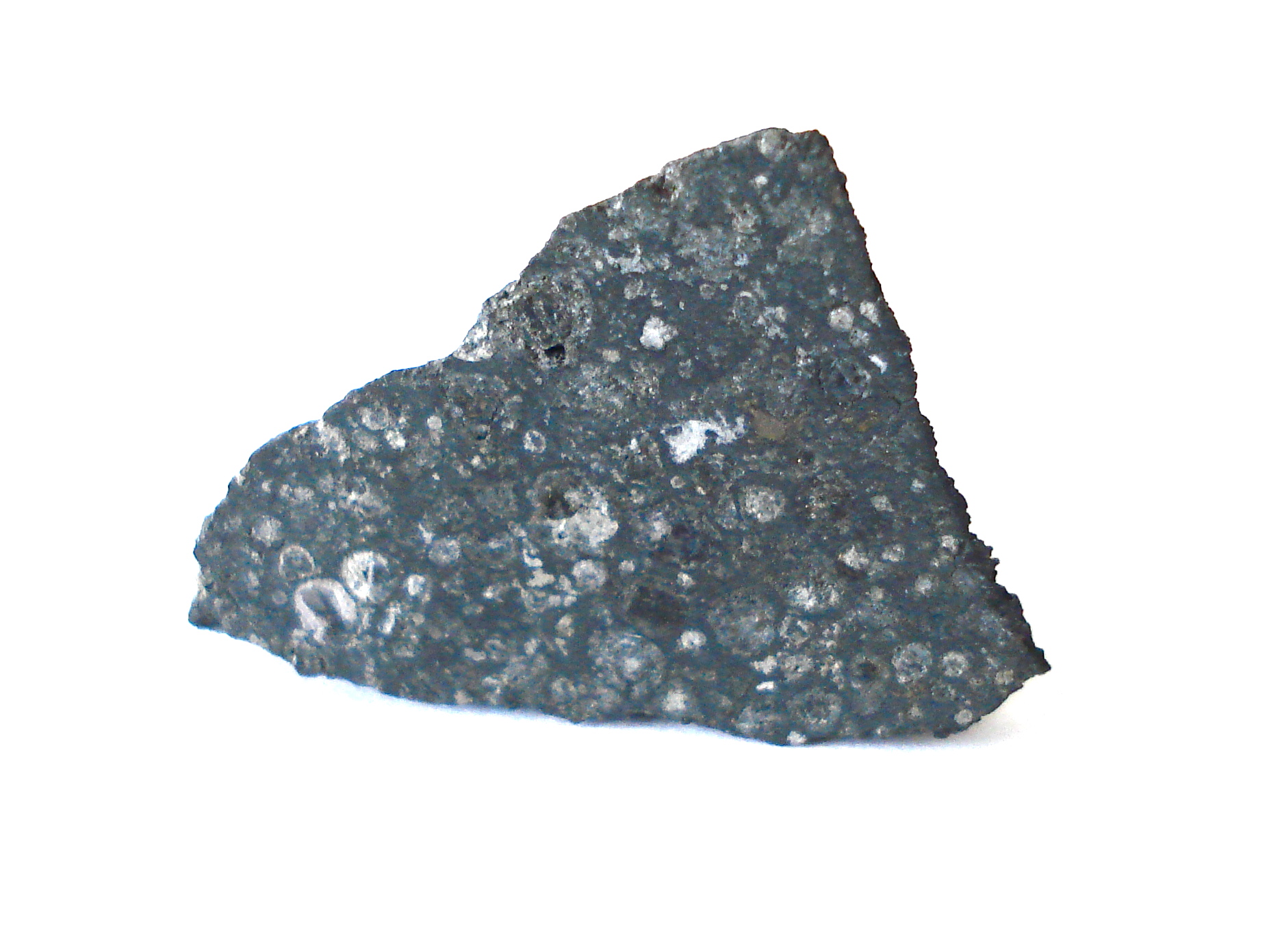}} 
  \caption{A fragment of the Allende meteorite.}
  \label{fig:intro:allende}
\end{figure}

Most chondrites consist of three main components: Firstly, the CAIs, up to cm-sized particles, which are considered to be the first condensates in the solar system. Secondly, \emph{chondrules}, small, mostly spherical silicate crystals with diameters in the range of 50~$\mu$m to several mm \citep{Scott:2007p10936} comprise more than 70\% by volume of the ordinary chondrites \citep{Youdin:2010p10939}. Thirdly, the space between CAIs and chondrules is filled with the \emph{matrix}, which consists of sub-$\mu$m sized grains of similar minerals. A cut through a fragment of the famous Allende meteorite is shown in Fig.~\ref{fig:intro:allende}.

The decay of radioactive nuclei can be used to infer both the ages and the formation time-scales of these components \citep[for a review, see][]{Russell:2006p11856}. Results indicate that CAIs are the oldest components and were formed during a time interval of about 0.25~million years. Chondrules were formed during an extended period of around 2~million years, starting 1--2 million years after the formation of CAIs \citep{Kleine:2009p10853}. Today, it is believed that chondrules also formed earlier than this, but these early chondrules have been incorporated into parent bodies which differentiated due to the heat from $^{26}$Al-decay. Dating of iron meteorites indicate that they have formed within $<1$ million years after CAI formation, they are therefore often thought to be core material of the early parent bodies which were differentiated by the $^{26}$Al-heating.

Despite the ample laboratory data (or perhaps due to it) there is no satisfying theory which explains the formation of these bodies. Still two major points that have been learned from meteoritic data should be mentioned: firstly, apart from some very volatile materials, the composition of chondritic meteorites is strikingly similar to the composition of the sun. This tells us that meteorites (and thus the planets) formed from the same material as the sun itself. Secondly, the constraints on the lifetime of the solar nebula are in very good agreement with the lifetimes of circumstellar disks. 

\subsection{Exoplanets}\label{sec:intro:exoplanets}
The most important observational constraint is that planets do exist, not only in our own solar system.
The basic concepts behind the most successful planet-searching techniques are rather simple: the \emph{radial velocity method} measures the ``wobble'' of the star induced by the common motion around the center of mass, while the \emph{transit method} uses the dip in the light curve to proof the presence of an extrasolar planet. Together, these two methods are responsible for more than 93\% of all exoplanet detections. Still, it is the high precision of the measurements that is needed, which makes detections of exoplanets such a challenging task. Therefore it has taken 40 years after the first proposal of the radial velocity method by \citet{Struve:1952p10743} that an extrasolar planetary companion (around a neutron star) could be detected by \citet{Wolszczan:1992p10543}. Detections of planets around solar-type stars followed a few years later \citep{Mayor:1995p10615}.

Today, the number of extrasolar planets is ever increasing (unlike the number of planets in the solar system, which has decreased), proving that planet formation is indeed a common by-product of star formation. Extrapolation of recent discoveries suggest that about 20\% of all stars have planets orbiting within 20~AU \citep{Cumming:2008p11954}. Population synthesis models predict an even higher rate: about 30-40\% of all FGK stars should host planets \citep{Mordasini:2009p11966}.

\begin{figure}[tbh]
  \centering
  \resizebox{0.75\hsize}{!}{\includegraphics{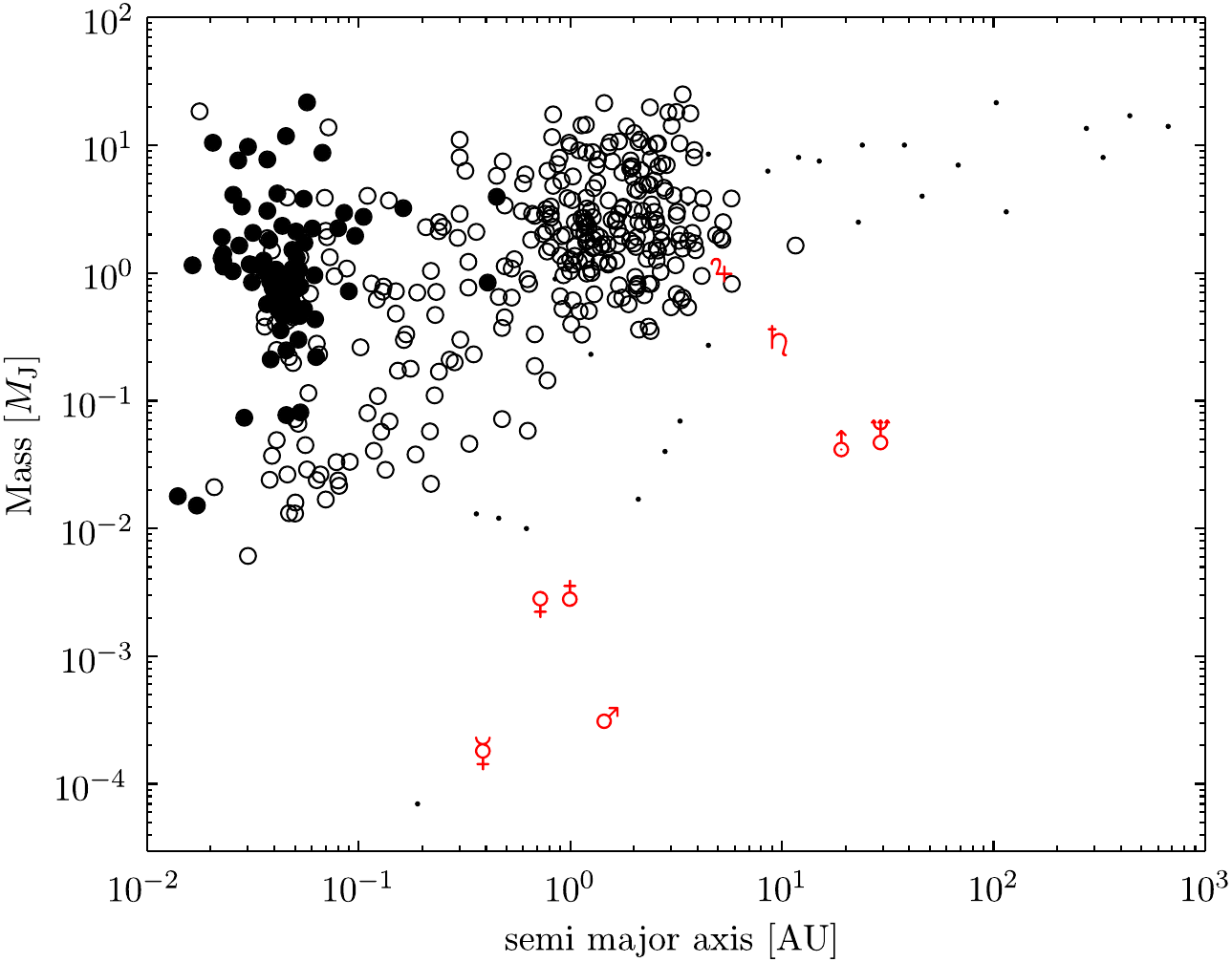}}
  \caption{The mass of the currently known planets and exoplanets as function of their semimajor axis. The solid black circles (\textbullet) denote transiting planets, the empty circles (\textopenbullet) denote radial velocity detections, and all other detections (pulsar timing, microlensing, and direct imaging) are displayed as black dots ($\cdot$). The red symbols represent the solar system planets.}
  \label{fig:intro:planets}
\end{figure}

Fig.~\ref{fig:intro:planets} shows the mass of the currently \footnote{Data of July 22, 2010 from \url{exoplanet.eu}} known planets as function of their semimajor axis. The known population of exoplanets suggests that stars with higher metallicity are more likely to host planets \citep{Fischer:2005p11968}, which is in general in agreement with the core accretion scenario.

Recently, the first \emph{direct imaging} detections of extrasolar planets have been reported \citep[e.g.,][]{Kalas:2008p11975,Marois:2008p11978}, which have typically several Jupiter masses and are located at large distances from the star (hundreds of AU). Some of these are thought to have formed by disk instability \citep{Meru:2010p11987}.   

\section{Summary}
Summarizing all of this, problems with planet formation seem to come in all sizes: smaller particles, as discussed above, face the problem that growth can be halted by charging of the grains \citep[see][]{Okuzumi:2009p7473}, by fragmentation and erosion (see Chapter~\ref{chapter:dustretention}), or by bouncing \citep[see][]{Zsom:2010p9746}. If larger grains are formed, they seem to be lost quickly to the star due to the radial drift. And even if one can explain the formation of planetary cores, these bodies still face the problem of migration due to interaction with the disk. Additionally, planet formation needs to happen relatively quickly (i.e., a few million years), otherwise the gas disk, which is needed for the formation of gas giant planets, will have disappeared. Thus, we can see that we are able to understand the formation of the solar system and the planets far better than Descartes, however we are still far away from a final, consistent answer.    

\section{About this thesis}
This thesis focuses on the very first stages of planet formation, namely the growth of sub-$\mu$m particles to planetesimals along with the evolution of the radial distribution of these particles. Dust is an important ingredient in circumstellar disks, not only because terrestrial planets and the cores of giant planets are made out of it, but also it is the main source of opacity in disks (thus, important for interpreting observations), it influences the coupling of the disk to magnetic fields by influencing the ionization fraction (important for the viscous evolution of the whole disk) and it is an important catalyst in disk chemistry, especially for the formation of water. Therefore, we build a numerical code which is able to trace the radial and size evolution of dust suspended in a gaseous disk which itself is evolving due to viscous transport of angular momentum.

\paragraph{Chapter~\ref{chapter:model}} describes the physics of growth and radial transport of dust as well as the physics of the gas disk which dominates the dynamics of the dust. This is the first work which takes into account the disk build-up phase and subsequent gas disk evolution in calculations of the dust evolution. We investigate how the dust-to-gas ratio (which is an important parameter for both observations of disks and the theory of planetesimal formation) evolves over time if coagulation and fragmentation are taken into account. We also revisit the barriers to planet formation by sticking collisions, namely the radial drift and the fragmentation barrier. We show that radial variations in the tensile strength of particles (caused by compositional changes due to the presence of ices) can cause a strong enhancement of dust inside the snow line. The fragmentation velocity turns out to be one of the most important parameters determining the particle shape as well as the dust disk shape and lifetime.

The outward spreading of the gas disk is able to carry dust along. Whether and to which extent this effect works strongly depends upon the maximum particle size and thus also on the critical fragmentation velocity of the gains. There is always a radius where even the smallest grains decouple from the spreading gas disk. Future observations (e.g., with ALMA) might be able to detect the different outer radii of the gas and the dust disks. 

\paragraph{Chapter~\ref{chapter:algorithm}} discusses the numerical implementation of the model. The radial transport of the gas and the evolution (in size and radial position) of dust are solved using implicit integration schemes. We also show test cases which validate the results of our numerical simulations by comparison to analytic estimates.

\paragraph{Chapter~\ref{chapter:distri}} focuses on the grain distribution in the scenario of an equilibrium between coagulation and fragmentation. We generalize previous analytical work on the slope of steady-state grain size distribution by taking into account both fragmentation and coagulation at the same time. These results not only validate the distributions we derived in the previous chapters but they also give us the opportunity to understand the influence of parameters like the temperature or the gas surface density on the size distribution of grains in circumstellar disks. We also derive a fitting recipe which can readily be used for further modeling such as grain surface chemistry in disks, dust-dependent MRI simulations or radiative transfer modeling.

\paragraph{Chapter~\ref{chapter:dustretention}} investigates the the fact that disks are generally observed to be dusty for several million years and how this can be understood if grain fragmentation is considered. We show that grain fragmentation at impact velocities similar to what is found in experimental studies causes grains to be continuously fragmented to small enough sizes to avoid the radial drift regime and to keep the optical depth as high as it is observed.

\paragraph{Chapter~\ref{chapter:mmobs}} is an application of our previously derived steady state dust size distributions to the field of millimeter observations. A grid of different input parameters is used to calculate the dust distribution in steady-state. From this, we can predict the spectral indices and flux levels at millimeter wavelength and compare this to an existing sample of observations. We show that typical values can explain the low spectral indices and also the flux levels of a subset of the observations. We propose several possibilities which further reduce the fluxes while keeping the spectral indices low enough.

\paragraph{Chapter~\ref{chapter:outlook}} summarizes our findings and discusses future prospects of this thesis which are planned or already in progress.
    
\ifthenelse{\boolean{chapterbib}}
{
    \clearpage
    \bibliographystyle{aa}
    \bibliography{/Users/til/Documents/Papers/bibliography}
}
{}
\chapter{Evolution of gas and dust in protoplanetary disks}\label{chapter:model}
\chapterprecishere{\centering From \citealp*[][ A\&A, 513, 79]{Birnstiel:2010p9709}}
\begin{abstract}
{Current models of the size- and radial evolution of dust in protoplanetary
disks generally oversimplify either the radial evolution of the disk (by focussing at one single radius or by using steady state disk models) or they assume particle growth to proceed monodispersely or without fragmentation. Further studies of protoplanetary disks -- such as observations, disk chemistry and structure calculations or planet population synthesis models -- depend on the distribution of dust as a function of grain size and radial position in the disk.}
{We attempt to improve upon current models to be able to investigate how the initial conditions, the build-up phase,  and the evolution of the protoplanetary disk influence growth and transport of dust.}
{We introduce a new model similar to \citet{Brauer:2008p215} in which we now include the time-dependent viscous evolution of the gas disk, and in which more advanced input physics and numerical integration methods are implemented.}
{We show that grain properties, the gas pressure gradient, and the amount of turbulence are much more influencing the evolution of dust than the initial conditions or the build-up phase of the protoplanetary disk. We quantify which conditions or environments are favorable for growth beyond the meter size barrier. High gas surface densities or zonal flows may help to overcome the problem of radial drift, however already a small amount of turbulence poses a much stronger obstacle for grain growth.}
\end{abstract}
%
%

\section{Introduction}\label{sec:model:introduction}
The question of how planets form is one of the key questions in modern astronomy today. While it has been studied for centuries, the problem is still far from being solved. The agglomeration of small dust particles to larger ones is believed to be at least the first stage of planet formation. Both laboratory experiments \citep{Blum:2000p8110} and observations of the 10~$\mu$m spectral region \citep{Bouwman:2001p8118,vanBoekel:2003p8117} conclude that grain growth must take place in circumstellar disks. The growth from sub-micron size particles to many thousand kilometer sized planets covers 13 orders of magnitude in spatial scale and over 40 orders of magnitude in mass. To assemble a single 1 km diameter planetesimal requires the agglomeration of about $10^{27}$ dust particles. These dynamic ranges are so phenomenal that one has to resort to special methods to study the growth of particles though aggregation in the context of planet(esimal) formation.

A commonly used method for this purpose makes use of particle size distribution functions. The time dependent evolution of these particle size distribution functions has been studied by \citet{Weidenschilling:1980p4572}, \citet{Nakagawa:1981p4533}, \citet{Dullemond:2005p378}, \citet{Brauer:2008p215} (hereafter \citetalias{Brauer:2008p215}) and others. It was concluded that dust growth by coagulation can be very quick initially (in the order of thousand years to grow to centimeter sized aggregates at 1 AU in the solar nebula), but it stalls around decimeter to meter size due to what is known as the ``meter size barrier'': a size range within which particles achieve large enough velocities to undergo destructive collisions and fast radial inward drift toward the central star \citep{Weidenschilling:1977p865,Nakagawa:1986p2048}.

While the existence of this meter size barrier (ranging in fact from a couple of centimeters to tens of meters at 1 AU) has been known for over 30~years, a thorough study of this barrier, including all known mechanisms that induce motions of dust grains in protoplanetary disks, and at all regions in the disk, for various conditions in the disk and for different properties of the dust (such as sticking efficiency and critical fragmentation velocity), has been only undertaken recently \citepalias[see][]{Brauer:2008p215}. It was concluded that the barrier is indeed a very strong limiting factor in the formation of planetesimals from dust, and that special particle trapping mechanisms are likely necessary to overcome the barrier. 

However, this work was based on a static, non-evolving gas disk model. It is known that over the duration of the planet formation process the disk itself also evolves dramatically \citep{LyndenBell:1974p1945,Hartmann:1998p664,Hueso:2005p685}, which may influence the processes of dust coagulation and fragmentation. It is the goal of this work to introduce a combined disk-evolution and dust-evolution model which also includes additional physics: we include relative azimuthal velocities, radial dependence of fragmentation critical velocities and the Stokes-drag regime for small Reynolds numbers.

The aim is to find out what the effect of disk formation and evolution is on the process of dust growth, how the initial conditions affect the final outcome, and whether certain observable signatures of the disk (for instance, its degree of dustiness at a given time) can tell us something about the physics of dust growth.

Moreover, this model will serve as a supporting model for complementary modeling efforts such as the modeling of radiative transfer in protoplanetary disks (which needs information about the dust properties for the opacities) and modeling of the chemistry in disks (which needs information about the total amount of dust surface area available for surface chemistry). In this chapter we describe our model in quite some detail, and thus provide a basis for future work that will be based on this model.

Furthermore, additional physics, such photoevaporation or layered accretion can be easily included, which we aim to do in the near future.

As outlined above, this model includes already many processes which influence the evolution of the dust and the gas disk. However, there are several aspects we do not include such as back-reactions by the dust through opacity or collective effects \citet{Weidenschilling:1997p4593}, porosity effects \citep{Ormel:2007p7127}, grain charging \citep{Okuzumi:2009p7473} or the ``bouncing barrier'' \citet{Zsom:2010p9746}.

This chapter is organized as follows: Section~\ref{sec:model:model} will describe all components of the model including the radial evolution of gas and dust, as well as the temperature and vertical structure of the disk and the physics of grain growth and fragmentation. In Section~\ref{sec:model:results} we will compare the results of our simulations with previous steady-state disk simulations and review the aforementioned growth barriers. As an application, we show how different material properties inside and outside the snow line can cause a strong enhancement of dust within the snow line. Section~\ref{sec:model:discussion} summarizes our findings.
A detailed description of the numerical method along with results for selected test cases can be found in Chapter~\ref{chapter:algorithm}.
\section{Model}\label{sec:model:model}
The model presented in this work combines a 1D viscous gas disk evolution code and a dust evolution code, taking effects of radial drift, turbulent mixing, coagulation and fragmentation of the dust into account. 
We model the evolution of gas and dust in a vertically integrated way. The gas disk is viscously evolving after being built up by in-falling material from a collapsing molecular cloud core.

The radial distribution of grains is subject to gas drag, radial drift, and turbulent mixing. To which extend each effect contributes, depends on the grain/gas coupling of each grain size. By simultaneously modeling about 100--200 different grain sizes, we are able to follow the detailed evolution of the dust sub-disk being the superposition of all sizes of grain distributions.

So far, the evolution of the dust distribution depends on the evolving gas disk but not vice versa. A completely self consistently coupled code is a conceptually and numerically challenging task which will be the subject of future work.

\subsection{Evolution of gas surface density}\label{sec:model:gas}
Our description of the viscous evolution of the gas disk follows closely the models described by \citet{Nakamoto:1994p798} and \citet{Hueso:2005p685}. In this thesis we shall therefore be relatively brief and put emphasis on differences between those models and ours.

The viscous evolution of the gas disk can be described by the continuity equation,
\begin{equation}
\frac{\del \Siggas }{\del t} - \frac{1}{r}\frac{\del}{\del r}\left( \Siggas \, r \, u_\mathrm{g}\right) = S_\mathrm{g},
\label{eq:model:ssd}
\end{equation}
where the gas radial velocity $u_\mathrm{g}$ is given by \citep[see][]{LyndenBell:1974p1945}
\begin{equation}
u_\mathrm{g} = - \frac{3}{\Siggas\sqrt{r}} \frac{\del}{\del r} \left( \Siggas \nu_\mathrm{g} \sqrt{r} \right).
\label{eq:model:u_gas}
\end{equation}
$\Siggas = \int_{-\infty}^{\infty} \rhogas(z) \dx z$ is the gas surface density, $r$ the radius along the disk mid-plane and \nug the gas viscosity. The source term on the right hand side of Eq.~\ref{eq:model:ssd}, denoted by $S_\mathrm{g}$ can be either infall of material onto the disk or outflow.

The molecular viscosity of the gas is too small to account for relevant accretion onto the star, the time scale of viscous evolution would be in the order of several billion years. 
Observed accretion rates and disk lifetimes can only be explained if turbulent viscosity drives the evolution of circumstellar disks. Therefore \citet{Shakura:1973p4854} parameterized the unknown viscosity as the product of a velocity scale and a length scale. The largest reasonable values for these scales in the disk are the pressure scale height $\Hp$
\begin{equation}
\Hp = \frac{\csound}{\Ok}
\end{equation}
and the sound speed $\csound$. Therefore the viscosity is written as
\begin{equation}
\nug = \alpha \: \csound \: \Hp,
\end{equation}
where $\alpha$ is the turbulence parameter and $\alpha \leq 1$.

Today it is generally believed that disks transport angular momentum by turbulence, however the source of this turbulence is still debated. The magneto-rotational instability is the most commonly accepted candidate for source of turbulence \citep{Balbus:1991p4932}. Values of $\alpha$ are expected to be in the range of $10^{-3}$ to some $10^{-2}$ \citep[see][]{Johansen:2005p8425,Dzyurkevich:2010p11360}. Observations confirm this range with higher probability for larger values \citep[see][]{Andrews:2009p7729}.

If the disk becomes gravitationally unstable, large scale mechanisms of angular momentum transport such as through the formation of spiral arms come into play. The stability of the disk can be described in terms of the Toomre parameter \citep{Toomre:1964p1002}
\begin{equation}
Q = \frac{\csound \Ok}{\pi \,G \, \Siggas}.
\end{equation}
Values below a critical value of $Q_\text{cr} = 2$ describe a weakly unstable disk, which forms non-axisymmetric instabilities like spiral arms. $Q$ values below unity lead to fragmentation of the disk.

The effect of these non-axisymmetric structures is to transport angular momentum outward and rearranging the surface density in the disk so as to counteract the unstable configuration. This mechanism is therefore to a certain extent self-limiting. Values above $Q_\text{cr}$ are not influenced by instabilities, values below $Q_\text{cr}$ form instabilities which increase $Q$ until the disk is marginally stable again. Our model approximates this mechanism by increasing the turbulence parameter $\alpha$  according to the recipe of \cite{Armitage:2001p993},
\begin{equation}
\alpha(r) = \alpha_0 + 0.01  \left(  \left(\frac{Q_\text{cr}}{\min(Q(r),Q_\text{cr})}\right)^2 - 1  \right),
\label{eq:model:alpha_of_Q}
\end{equation}
where $\alpha_0$ is a free parameter of the model which is taken to be $10^{-3}$, unless otherwise noted.

During the time of infall onto the disk, we use a constant, high value of $\alpha = 0.1$ to mimic the effective redistribution of surface density during the infall phase which also increases the overall stability of the disk. Once the infall stops, we gradually decrease the turbulence parameter on a timescale of 10~000 years until it reaches its input value.

Eq. \ref{eq:model:ssd} is a diffusion equation, which is stiff. This means, one faces the problem that the numerical step of an explicit integration scheme goes $\propto \Delta r^2$ (where $\Delta r$ is the radial grid step size) which would make the simulation very slow. One possible solution to this problem is using the method of implicit integration. This scheme keeps the small time scales of diffusion i.e. the fast modes in check. We are not interested in these high frequency modes, but they would become unstable if we used a large time step. With an implicit integration scheme (see Section~\ref{sec:algorithm:alg_advdif}) the time step can be chosen larger without causing numerical instabilities, thus increasing the speed of the computation.

\subsection{Radial evolution of dust}\label{sec:model:dust}
If the average dust-to-gas ratio in protoplanetary disks is in the order of $10^{-2}$ (as found in the ISM), then the dust does not dynamically influence the gas, while the gas strongly affects the dynamics of the dust.

Thermally, however, the dust has potentially a massive influence on the gas disk evolution through its opacity, but we will not include this in this work. Therefore the evolution of the gas disk can, in our approximation, be done without knowledge of the dust evolution, while the dust evolution itself {\em does} need knowledge of the gas evolution.

There might be regions, where dust accumulates (such as the mid-plane of the disk or dead-zones and snow-lines) and its influence becomes significant or even dominant but we will not include feedback of such dust enhancements onto the disk evolution in this work.

For now, we want to focus on the equations of motion of dust particles under the assumption that gas is the dominant material by mass. The interplay between dust and gas can then be described in terms of a dimensionless coupling constant, the \textit{Stokes number} which is defined as
\begin{equation}
\St = \frac{\tau_\text{s}}{\tau_\text{ed}},
\label{eq:model:ST_general}
\end{equation}
where $\tau_\text{ed}$ is the eddy turn-over time and $\tau_\text{s}$ is the stopping time.

The stopping time of a particle is defined as the ratio of the momentum of a particle divided by the drag force acting on it. There are four different regimes for the drag force which determine the dust-to-gas coupling \citep[see][]{Whipple:1972p4621,Weidenschilling:1977p865}. Which regime applies to a certain particle, depends on the ratio between mean free path $\lambda_\text{mfp}$ of the gas molecules and the dust particle size $a$ (i.e. the Knudsen number) and also on the particle Reynolds-number $Re = {2 a u/\nu_\mathrm{mol}}$ with
$\nu_\mathrm{mol}$ being the gas molecular viscosity
\begin{equation}
\nu_\mathrm{mol} = \half \, \bar u \, \lambda_\text{mfp},
\end{equation}
$\bar u$ the mean thermal velocity. The mean free path is taken to be
\begin{equation}
\lambda_\text{mfp} = \frac{1}{n\:\sigma_{\mathrm{H}_2}}
\end{equation}
where $n$ denotes the mid-plane number density and $\sigma_{\mathrm{H}_2} =2 \e{-15}$~cm$^2$.

The different regimes\footnote{It should be noted that ``Stokes regime'' refers to the regime where the drag force on a particle is described by the Stokes law -- this is not directly related to the Stokes number.} are
\begin{equation}
\tau_\text{s} = 
\left\{
\begin{array}{lll}
\frac{\rho_\text{s}\: a}{\rho_\text{g}\:\bar u}&        \text{for }  \lambda_\text{mfp}/a\gtrsim\frac{4}{9} &       \text{Epstein regime}\\
\\
\frac{2 \rho_\text{s}\: a^2}{9 \nu_\mathrm{mol} \:\rho_\text{g}}&        \text{for } Re<1& \text{Stokes regime 1}\\
\\
\frac{2^{0.6}\:\rho_\text{s}\:a^{1.6}}{9 \nu_\mathrm{mol}^{0.6} \: \rho_\text{g}^{1.4}\:u^{0.4}} & \text{for } 1<Re<800& \text{Stokes regime 2}\\
\\
\frac{6 \rho_\text{s}\: a}{\rho_\text{g}\:u}&       \text{for } Re>800& \text{Stokes regime 3}\\
\end{array}\right.
\label{eq:model:stopping_time}
\end{equation}
Here $u$ denotes the velocity of the dust particle with respect to the gas, $\bar u = c_\text{s} \sqrt{{\pi}/{8}}$ denotes the mean thermal velocity of the gas molecules, $\rho_\text{s}$ is the solid density of the particles and $\rho_\text{g}$ is the local gas density.

For now, we will focus on the Epstein regime. To calculate the Stokes number, we need to know the eddy turn-over time. As noted before, our description of viscosity comes from a dimensional analysis. We use a characteristic length scale $L_\text{c}$ and a characteristic velocity scale $V_\text{c}$ of the eddies. This prescription is ambiguous in a sense that it does not specify if angular momentum is transported by large, slow moving eddies or by small, fast moving eddies, that is
\begin{equation}
\nug = (\alpha^{q} V_\text{c}) \cdot (\alpha^{1-q} L_\text{c}).
\end{equation}

This is rather irrelevant for the viscous evolution of the gas disk, since all values of $q$ lead to the same viscosity, but if we calculate the turn-over-eddy time, we get
\begin{equation}
\tau_\text{ed} = \frac{2\pi L_\text{c}}{V_\text{c}} = \alpha^{1-2q}\: \frac{1}{\Ok}.
\end{equation}
The Stokes number and therefore the dust-to-gas coupling as well as the
relative particle velocities strongly depend on the eddy turnover time and
therefore on $q$ . In this work $q$ is taken to be 0.5
\citep[following][]{Cuzzi:2001p2167,Schrapler:2004p2394} which leads to a
turn-over-eddy time of
\begin{equation}
\tau_\text{ed} = \frac{1}{\Ok}.
\end{equation}
The Stokes number then becomes
\begin{equation}
\St = \frac{\rho_\text{s}\cdot a}{\Siggas} \cdot \frac{\pi}{2} \qquad \text{for } a<\frac{9}{4}\lambda_\text{mfp}.
\label{eq:model:ST_epstein}
\end{equation}
The overall radial movement of dust surface density $\Sigdust$ can now be described by an advection-diffusion equation,
\begin{equation}
\ddel{\Sigdust}{t} + \frac{1}{r} \ddel{}{r} \Bigl( r \, F_\text{tot} \Bigr) = 0,
\end{equation}
where the total Flux, $F_\text{tot}$ has contributions from a diffusive and an advective flux.
The diffusive part comes from the fact that the gas is turbulent and the dust couples to the gas. The dust is therefore turbulently mixed by the gas. Mixing counteracts gradients in concentration, in this case it is the dust-to-gas ratio of each size that is being smoothed out by the turbulence. The diffusive flux can therefore be written as
\begin{equation}
F_\text{diff} = - D_\text{d} \: \ddel{}{r} \left(\frac{\Sigdust}{\Siggas}\right) \cdot \Siggas.
\end{equation}
The ratio of gas diffusivity $D_\text{g}$ over dust diffusivity $D_\text{d}$ is called the Schmidt number. We follow \cite{Youdin:2007p2021}, who derived
\begin{equation}
\Sc \equiv \frac{D_\text{g}}{D_\text{d}} = {1+\St^2},
\end{equation}
and assume the gas diffusivity to be equal to the turbulent gas viscosity \nug.

The second contribution to the dust flux is the advective flux,
\begin{equation}
F_\text{adv} = \Sigdust \cdot u_\text{r},
\end{equation}
where $u_r$ is the radial velocity of the dust. There are two contributions to it,
\begin{equation}
u_\text{r} = \frac{u_\text{g}}{1 + \St^2} - \frac{2 u_\text{n}}{\St + \St^{-1}}.
\label{eq:model:u_r_dust}
\end{equation}
The first term is a drag term which comes from the radial movement of the gas which moves with a radial velocity of $u_\mathrm{g}$, given by Eq.~\ref{eq:model:u_gas}. Since the dust is coupled to the gas to a certain extend, the radially moving gas is able to partially drag the dust along.

The second term in Eq. \ref{eq:model:u_r_dust} is the radial drift velocity with respect to the gas. The gas in a Keplerian disk does in fact move sub-keplerian, since it feels the force of its own pressure gradient which is usually pointing inwards. Larger dust grains do not feel this pressure and move on a keplerian orbit. Therefore, from a point of view of a (larger) dust particle, there exists a constant head wind, which causes the particle to loose angular momentum and to drift inwards. This depends on the coupling between the gas and the particle and is described by the second term in Eq. \ref{eq:model:u_r_dust}.
$u_n$ denotes the maximum drift velocity of a particle,
\begin{equation}
u_\text{n} = - E_\mathrm{d} \cdot \frac{ \ddel{P_g}{r}}{2 \: \rho_\text{g} \: \Ok},
\label{eq:model:u_eta}
\end{equation}
which has been derived by \citet{Weidenschilling:1977p865}. Here, we introduced a radial drift efficiency parameter $E_\mathrm{d}$. This parameter describes how efficient the radial drift actually is, as several mechanisms such as zonal or meridional flows might slow down radial drift. This will be investigated in Section~\ref{sec:model:drift_barrier}.

Putting all together, the time dependent equation for the surface density of one dust species of mass $m_i$ is given by
\begin{equation}
\ddel{\Sigdust^i}{t} + \frac{1}{r}\ddel{}{r}
\left\lbrace r \cdot \left[ \Sigdust^i \cdot u_r^i -  D^i_\text{d} \cdot \ddel{}{r} \left( \frac{\Sigdust^i}{\Siggas}\right) \cdot \Siggas \right] \right\rbrace = S_\mathrm{d}^i,
\label{eq:model:dustequation}
\end{equation}
where we have included a source term $S^i_\mathrm{d}$ on the right hand side which can be positive in the case of infall or re-condensation of grains and negative in the case of dust evaporation or outflows.
This source term does not include the sources caused by coagulation and fragmentation processes (see Section~\ref{sec:model:smolu}). All sources will be combined into one equation later which is implicitly integrated in an un-split scheme (see Chapter~\ref{chapter:algorithm}).

Note that both, the diffusion coefficient and the radial velocity depend on the Stokes number and therefore on the size of the particle.

\begin{figure}[thb]
  \centering
  \resizebox{\hsize}{!}{\includegraphics{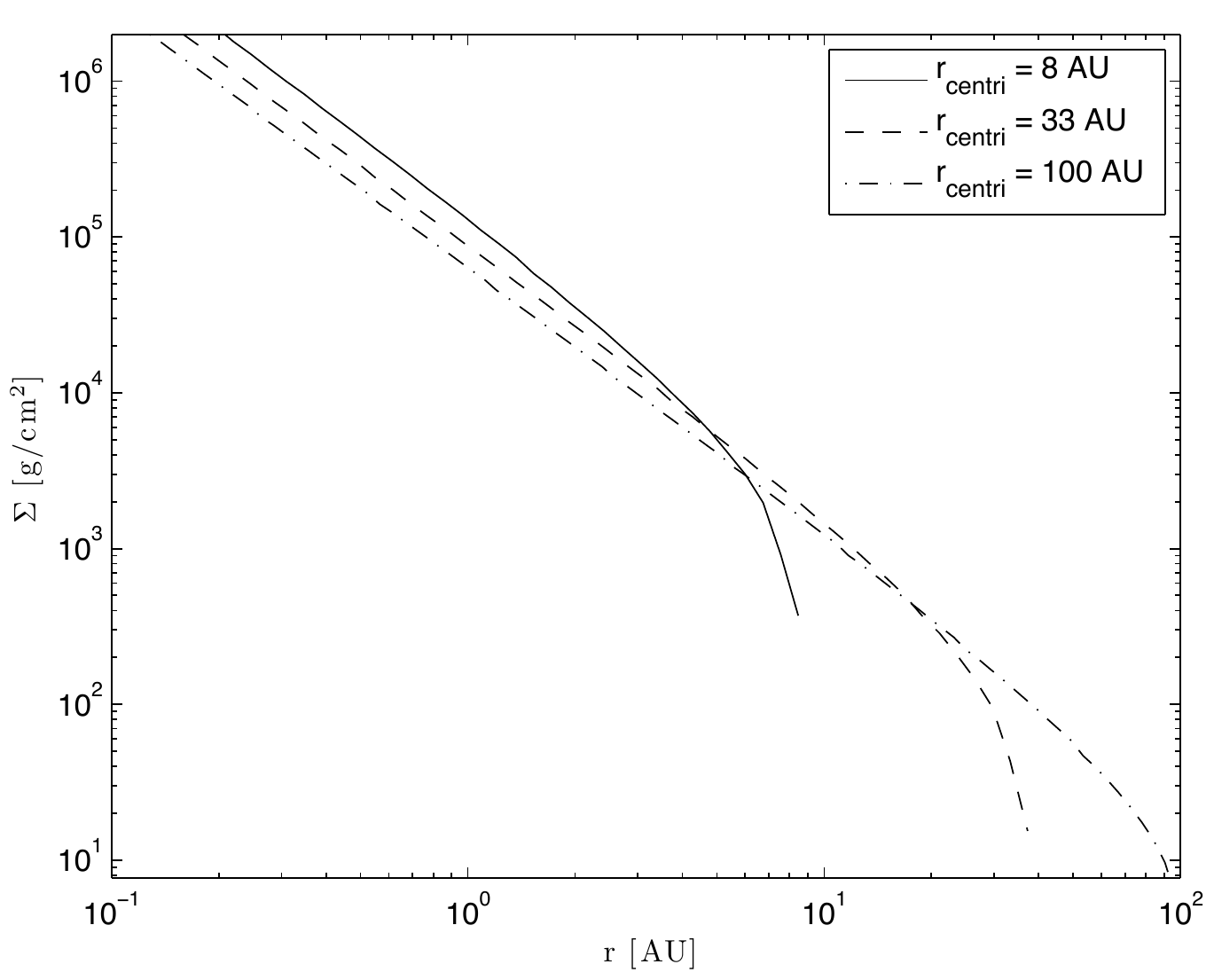}}
  \caption{Total amount of in-fallen surface density as function of radius according to the Shu-Ulrich infall model (see Section~\ref{sec:model:collapse}) assuming a centrifugal radius of 8~AU (solid line), 33~AU (dashed line), and 100~AU (dash-dotted line).}
  \label{fig:model:mass_loading}
\end{figure}

\subsection{Temperature and vertical structure}\label{sec:model:temperature}
The vertical structure can be assumed as being in hydrostatic equilibrium at all times if the disk is geometrically thin ($\Hp(r)/r\ll r$) and the vertical sound crossing time is much shorter than the radial drift time scale of the gas.
The isothermal vertical density structure is then given by
\begin{equation}
\rhogas(z) = \rho_0 \: \exp\left( - \half \: \frac{z^2}{\Hp^2} \right),
\end{equation}
where
\begin{equation}
\rho_0 = \frac{\Siggas}{\sqrt{2\:\pi} \: \Hp}.
\label{eq:model:rho_midplane}
\end{equation}
The viscous heating is given by \cite{Nakamoto:1994p798}
\begin{equation}
Q_+ = \Siggas \: \nug \left( r\: \ddel{\Ok}{r} \right)^2,
\end{equation}
were $\nug$ denotes the turbulent gas viscosity and $\Ok$ the Kepler frequency.

\citet{Nakamoto:1994p798} give a solution for the mid-plane temperature with an optically thick and an optical thin contribution,
\begin{equation}
{\Tm^4} = \frac{9}{8 \sigb} \left(   \frac{3}{8}{\tauros} + \frac{1}{2\: {\taupla}} \right)    \Siggas \: \alpha \: \csound^2\: \Ok + T_\text{irr}^4
\label{eq:model:t_mid}
\end{equation}
where we used $\nug = \alpha\, \csound\, \Hp$ and the approximation $\tau_\text{R/P} = \kappa_\text{R/P} \,\Siggas/2$. \kapros and \kappla are Rosseland and Planck mean opacities which will be discussed in the next section.

$T_\text{irr}$ contains contributions due to stellar or external irradiation.
Here, we use a formula derived by Ruden \& Pollack \citep[see][App. B]{Ruden:1991p1806},
\begin{equation}
T_\text{irr} =  T_\star \cdot \left[ \frac{2}{3 \pi} \left(\frac{R_\star}{r}\right)^3 + \half\: \left(\frac{R_\star}{r}\right)^2 \: \left(\frac{\Hp}{r}\right) \: \left(\frac{d\ln \Hp}{d\ln r} - 1 \right) \right]^{\frac{1}{4}},
\end{equation}
with a fixed $\mathrm{d ln}\Hp/\mathrm{d ln} r = 9/7$, following \citet{Hueso:2005p685}.

The main source of opacity is the dust. Due to viscous heating, the temperature will increase with surface density. If the temperature rises above $\sim 1500$ K, the dust (i.e. the source of opacity) will evaporate, which stops the disk from further heating until all dust is vaporized. To simulate this behavior in our model, we calculate a gas temperature (assuming a small, constant value for gas opacity) in the case where the dust temperature rises above the evaporation temperature. Then $T_\text{mid}$ is approximated by
\begin{equation}
T_\text{mid} = \max(T_\text{gas},T_\text{evap.}),
\end{equation}
only if $T_\text{mid}$ from Eq.~\ref{eq:model:t_mid} would be larger than $T_\text{evap}$.

This is a thermostat effect: if $T$ rises above 1500 K, dust will evaporate, the opacity will drop and the temperature stabilizes at $T=1500$ K. Once even the very small gas opacity is enough to get temperatures $>1500$ K, all the dust is evaporated and the temperature rises further.

\subsection{Opacity}\label{sec:model:opacities}
In the calculation of the mid-plane temperature we use Rosseland and Planck mean opacities which are being calculated from a given frequency dependent opacity table. The results are stored in a look up table and interpolated during the calculations. The opacity table is for a mixture of 50\% silicates and 50\% carbonaceous grains.

Since these are dust opacities, we convert them from \emph{cm$^2$/g dust} to \emph{cm$^2$/g gas} by multiplying the values with the dust-to-gas ratio $\epsilon_0$, which is a fixed parameter in our model, taken to be the canonical value of 0.01. This assumes that the mean opacity of the gas is very small and that the dust-to-gas ratio does not change with time. To calculate the opacity self-consistently, the total mass of dust and the distribution of grain sizes has to be taken into account, meaning that the dust evolution has a back reaction on the gas by determining the opacity. For now, our model does not take back-reactions from dust to gas evolution into account. Only in the case where the temperature rises above 1500 K, the drop of opacity due to dust evaporation is considered, as described above.

\subsection{Initial infall phase: cloud collapse}\label{sec:model:collapse}
The evolution of the protoplanetary disk also depends on the initial infall phase which builds up the disk from the collapse of a cold molecular cloud core. This process is still not well understood. First similarity solutions for a collapsing sphere were calculated by \cite{Larson:1969p2574} and \cite{Penston:1969p2601}. \cite{Shu:1977p843} found a self similar solution for a singular isothermal sphere. The Larson \& Penston solution predicts much larger infall rates compared to the inside-out collapse of Shu ($\dot m_\text{in} \approx 47\: c_\text{s}^3/G$ and $0.975\: c_\text{s}^3/G$ respectively).

More recent work has shown that the infall rates are not constant over time, but develop a peak of high accretion rates and drop to smaller accretion rates at later times. The maximum accretion rate is about $13\: c_\text{s}^3/G$ if opacity effects are included \citep[see][]{Larson:2003p3025}. Analytical, pressure-free collapse calculations of \cite{Myers:2005p4950} show similar behavior but with a smaller peak accretion rate of $\dot m_\text{in} = 7.07 \csound^3/G$. They also argue that the effects of pressure and magnetic fields will further increase the time scales of cloud collapse.

This initial infall phase is important since it provides the initial condition of the disk and also influences the whole simulation by providing a source of small grains and gas at larger distances to the star during later times of evolution.

It should be noted that several groups perform 3D hydrodynamic simulations of
collapsing cloud cores which show more complicated evolution
\citep[e.g.,][]{Banerjee:2006p8491,Whitehouse:2006p8532}. However, to be able to study general trends of the infall phase, we use the Shu-model since it is adjustable by a few parameters whose influences are easy to understand. In this model the collapse proceeds with an infall rate of $\dot m_\text{in}=0.975 \: c_\text{s}^3/G$ which stays constant throughout the collapse.

We assume the singular isothermal sphere of mass $M_\text{cloud}$ to be in solid body rotation $\Omega_\text{s}$. If in-falling material is conserving its angular momentum, all matter from a shell of radius $r_\text{s}$ will fall onto the star and disk system (with mass $m_\mathrm{cent}(t)$) within the so called centrifugal radius,
\begin{equation}
r_\text{centr}(t) = \frac{\Omega_\text{s}^2\: r_\text{s}^4}{G \: m_\text{cent}(t)},
\label{eq:model:r_centri}
\end{equation}
where $G$ is the gravitational constant and $r_\text{s} = 0.975\cdot c_\text{s}\:t /2$. The path of every parcel of gas can then be described by a ballistic orbit until it reaches the equatorial plane. The resulting flow onto the disk is
\begin{equation}
\dot\Sigma_\text{d}(r,t) = 2\;\rho_1(r,t) \cdot u_\text{z}(r,t),
\end{equation}
where
\begin{equation}
u_\text{z}(r,t) = \sqrt{\frac{G \: m_\text{cent}(t)}{r}} \cdot \mu,
\end{equation}
and
\begin{equation}
\rho_1(r,t) = \frac{\dot m_\mathrm{in}}{8 \pi \sqrt{G \: m_\text{cent}(t) \: r^3}} \cdot  \frac{r}{r_\text{centr}(r,t)} \cdot \frac{1}{\mu^2}
\end{equation}
as described in \cite{Ulrich:1976p856}.
Here, $\mu$ is given by
\begin{equation}
\mu = \sqrt{1-r/r_\text{centr}(r,t)}.
\end{equation}
The centrifugal radius can therefore be approximated by (cf. \citet{Hueso:2005p685})
\begin{equation}
  r_\mathrm{centri}(t) \simeq
  1.4 \left( \frac{\Omega_\mathrm{s}}{10^{-14} \mathrm{~s}^{-1}} \right)^2
  \left(\frac{m_\mathrm{cent}(t)}{M_\odot}\right)^3
  \left(\frac{\csound}{3\times 10^4 \mathrm{~cm~s}^{-1}}\right)^{-8}\mathrm{ AU}.
\end{equation}

We admit that this recipe for the formation of a protoplanetary disk is perhaps oversimplified. Firstly, as shown by \citet{Visser:2009p9087}, the geometrical thickness of the disk changes the radial distribution of in-falling matter onto the disk surface, because the finite thickness may ``capture'' an in-falling gas parcel before it could reach the midplane. Secondly, star formation is likely to be messier than our simple single-star axisymmetric infall model. And even in such a simplified scenario, the Shu inside-out collapse model is often criticized as being unrealistic. However, it would be far beyond the scope of this chapter to include a better infall model. Here we just want to get a feeling for the effect of initial conditions on the dust growth, and we leave more detailed modeling to future work.

\subsection{Grain growth and fragmentation}\label{sec:model:growth_and_frag}
\begin{figure*}[t]
\centering
\resizebox{\hsize}{!}{\includegraphics{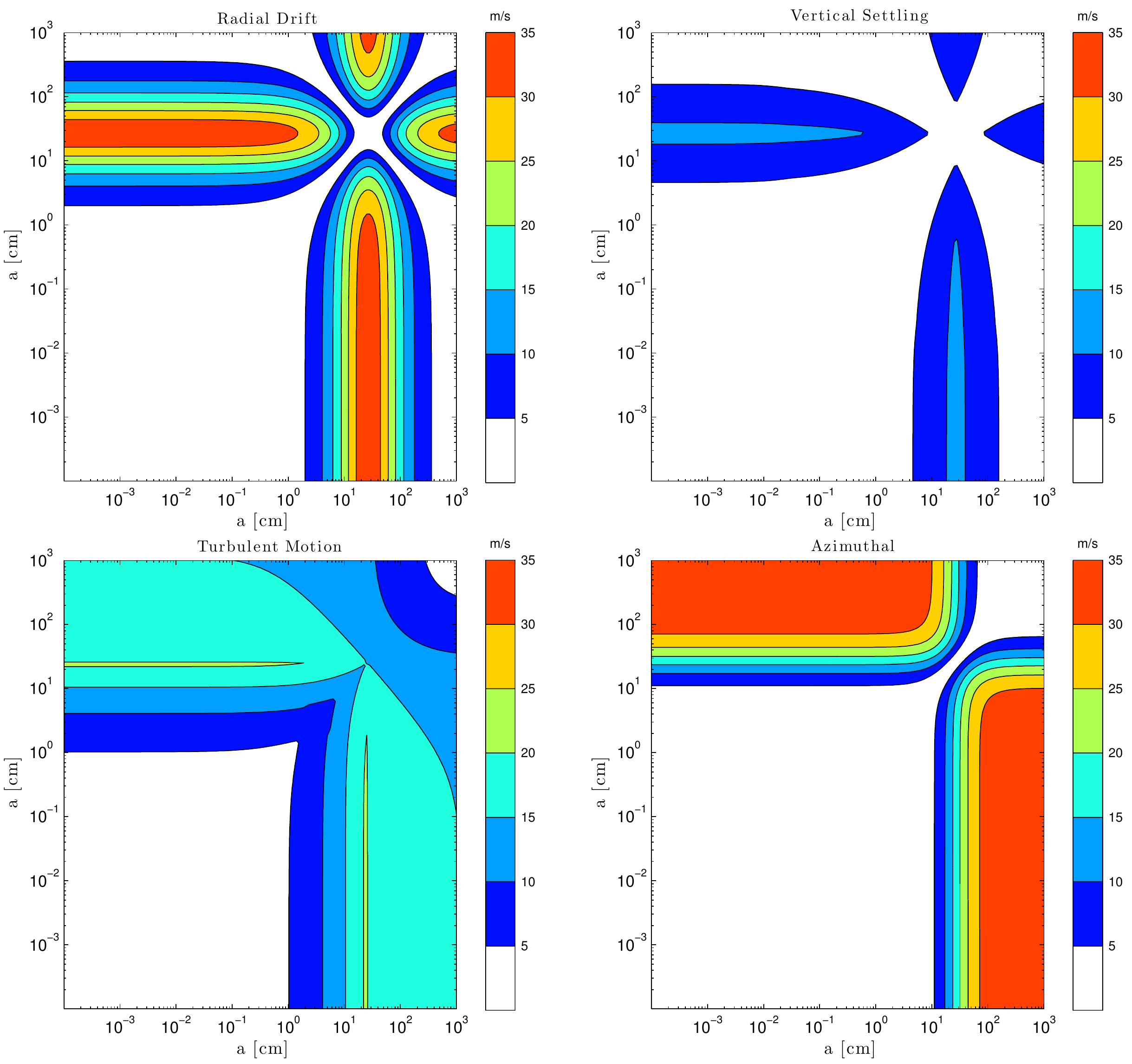}}
\caption{Sources of relative particle velocities considered in this model (Brownian motion velocities are not plotted) at a distance of 10~AU from the star. The turbulence parameter $\alpha$ in this simulation was $10^{-3}$. It should be noted that relative azimuthal velocities do not vanish for very large and very small particles.}
\label{fig:model:rel_vel}
\end{figure*}
The goal of the model described in this thesis is to trace the evolution of gas and dust during the whole lifetime of a protoplanetary disk, including the grain growth, radial drift and turbulent mixing.

Here, the problem arises that radial drift and coagulation ``counteract'' each
other in the regime of $\St=1$ particles: coagulation of smaller sizes restores the population around $\St=1$, whereas radial drift preferentially removes these particles. To be able to properly model this behavior, the time step has to be chosen small enough if the method of operator splitting is used.

The upper limit for this time step can be very small. If larger steps are used the solution will ``flip-flop'' back and forth between the two splitted sub-steps of motion and coagulation, and the results become unreliable. A method to allow the choice of large time steps while preserving the accuracy is to use a non-splitted scheme in which the integration is done ``implicitly''. \citetalias{Brauer:2008p215} already use this technique to avoid flip-flopping between coagulation and fragmentation of grains, and we refer to that paper for a description of the general method. What is new in the current work is that this implicit integration scheme is extended to also include the radial motion of the particles. So now radial motion, coagulation and fragmentation are done all within a single implicit integration scheme. See Chapter~\ref{chapter:algorithm} for details.

\subsubsection{Smoluchowski equation}\label{sec:model:smolu}
The dust grain distribution $n(m,r,z)$, which is a function of mass $m$, distance to the star $r$ and height above the mid-plane $z$, describes the number of particles per cm$^3$ per gram interval in particle mass. This means that the total dust density in g cm$^{-3}$ is given by
\begin{equation}
\rho(r,z)=\int_0^\infty n(m,r,z) \cdot m \; \dx{m}.
\end{equation}

With this definition of $n(m,r,z)$, the coagulation/fragmentation at one point in the disk can be described by a general two-body process,
\begin{equation}
\begin{split}
\frac{\del}{\del t} n(m,r,z) =& \iint_0^\infty M(m,m',m'') \times\\
&\phantom{ \iint_0^\infty}\times n(m',r,z) \cdot n(m'',r,z) \,\dx m' \,\dx m'',
\end{split}
\label{eq:model:smolu}
\end{equation}
where $M(m,m',m'')$ is called the kernel. In the case of coagulation and fragmentation, this is given by
\begin{equation}
\begin{split}
M(m,m',m'')     =   &\phantom{+}\half K(m',m'') \cdot \delta(m'+m''-m)\\
                &- K(m',m'') \cdot \delta(m''-m)\\
                &+\half L(m',m'') \cdot S(m,m',m'')\\
                &- L(m',m'') \cdot \delta(m-m'').
\end{split}
\label{eq:model:combined_kernel}
\end{equation}
For better readability, the dependency of $M$ on radius and height above the mid-plane was omitted here. The combined coagulation/fragmentation kernel consists of four terms containing the coagulation kernel $K$, the fragmentation kernel $L$ and the distribution of fragments after a collision $S$.

The first two terms in Eq. \ref{eq:model:combined_kernel} correspond to gain (masses $m'$ and $m-m'$ coagulate) and loss ($m$ coagulates with $m'$) due to grain growth.

The third term describes the fragmentation of masses $m$ and $m'$, governed by the fragmentation kernel $L$ and the fourth term describes the fact that when masses $m'$ and $m''$ fragment, they distribute some of their mass via fragments to smaller sizes.

The coagulation and fragmentation kernels will be described in section
\ref{sec:model:coag_kernel}, the distribution of fragments, $S$, will be described in
the next section.

To be able to trace the size and radial evolution of dust in a combined way, we need to express all contributing processes in terms of the same quantity. Hence, we will formulate the coagulation/fragmentation equation in a vertically integrated way. The vertical integration can be done numerically \citepalias[as in][]{Brauer:2008p215}, however coagulation processes are most important near the mid-plane, which allows to approximate the kernels as being vertically constant (using the values at the mid-plane). If the vertical dust density distribution of each grain size is taken to be a Gaussian (see Section~\ref{sec:model:rel_velocities}, Eq.~\ref{eq:model:h_dust}), then the vertical integration can be done analytically, as discussed in Section~\ref{sec:algorithm:alg_coag}. Unlike the steady-state disk models of \citetalias{Brauer:2008p215} which have fixed surface density and temperature profiles, we need to recompute the coagulation and fragmentation kernels (which are functions of surface density and temperature) at every time step. Therefore this analytical integration also saves significant amounts of computational time.

We therefore define the vertically integrated dust surface density distribution
per logarithmic bin of grain radius, $\sigma(r,a)$ as
\begin{equation}
\sigma(r,a) = \int_{-\infty}^\infty n(r,z,a) \cdot m\cdot a\; \dx{z},
\label{eq:model:def_sigma}
\end{equation}
where $n(r,z,a)$ and $n(r,z,m)$ are related through $m=4\pi/3 \rhos a^3$.
The total dust surface density at any radius is then given by
\begin{equation}
\Sigma_\mathrm{d}(r) = \int_0^\infty \sigma(r,a)\; \mathrm{dln}a.
\end{equation}

Defining $\sigma(r,a)$ as in Eq.~\ref{eq:model:def_sigma} makes it a grid-independent density unlike the mass integrated over each numerical bin. This way, all plots of $\sigma(r,a)$ are meaningful without knowledge of the size grid that was used. Numerically, however we use the discretized values as defined in Chapter~\ref{chapter:algorithm}.

In our description of growth and fragmentation of grains, we always assume the
dust particles to have a constant volume density meaning that we do not trace the evolution of porosity of the particles as this is currently computationally too expensive with a statistical code as used in this work. This can be achieved with Monte-Carlo methods as in \citet{Ormel:2007p7127} or \citet{Zsom:2008p7126}, however these models have do not yet include the radial motion of dust and therefore cannot trace the global evolution of the dust disk.

\subsubsection{Distribution of fragments}\label{sec:model:distr_of_fragments}
The distribution of fragments after a collision, $S(m,m',m'')$, is commonly described by a power law,
\begin{equation}
n(m) \text{d}m \propto m^{-\xi} \text{d}m.
\label{eq:model:frag_powerlaw}
\end{equation}

The value $\xi$ has been investigated both experimentally and theoretically.
Typical values have been found in the range between 1 and 2, by both
experimental \citep[e.g.,][]{Blum:1993p4324,Davis:1990p7995} and theoretical
studies \citep{Ormel:2009p8002}. Unless otherwise noted, we will follow \citetalias{Brauer:2008p215} by using the value of $\xi=1.83$.

In the case where masses of the colliding particles differ by orders of magnitude, a complete fragmentation of both particles is an unrealistic scenario. More likely, cratering will occur \citep{Sirono:2004p8225}, meaning that the smaller body will excavate a certain amount of mass from the larger one. The amount of mass removed from the larger one is parameterized in units of the smaller body $m_\text{s}$,
\begin{equation}
m_\text{out} = \chi \: m_\text{s}.
\end{equation}
The mass of the smaller particle plus the mass excavated from the larger one is then distributed to masses smaller than $m_\text{s}$ according to the distribution described by Eq. \ref{eq:model:frag_powerlaw}. In this work, we follow \citetalias{Brauer:2008p215} by assuming $\chi$ to be unity, unless otherwise noted.

\subsubsection{Coagulation and fragmentation kernels}\label{sec:model:coag_kernel}
The coagulation kernel $K(m_1,m_2)$ can be factorized into three parts,
\begin{equation}
K(m_1,m_2) = \Delta u(m_1,m_2) \: \sigma_\mathrm{geo}(m_1,m_2) \: p_\text{c},
\end{equation}
and, similarly, the fragmentation kernel can be written as
\begin{equation}
L(m_1,m_2) = \Delta u(m_1,m_2) \: \sigma_\mathrm{geo}(m_1,m_2) \: p_\text{f}.
\end{equation}
Here, $\Delta u(m_1,m_2)$ denotes the relative velocity of the two particles, $\sigma_\mathrm{geo}(m_1,m_2)$ is the geometrical cross section of the collision and $p_\text{c}$ and $p_\text{f}$ are the coagulation and fragmentation probabilities which sum up to unity. In general, all these factors can also depend on other material properties such as porosity, however we always assume the dust grains to have a volume density of $\rhos=1.6$~g~cm$^{-3}$.

The fragmentation probability is still not well known and subject of both theoretical \citep{Paszun:2009p8871,Wada:2008p4903} and experimental research \citep[see][]{Blum:2008p1920,Guttler:2010p9745}. In this work, we adopt the simple recipe
\begin{equation}
p_\text{f} = \left\{
\begin{array}{ll}
0&                              \text{if } \Delta u < \uf - \delta u\\
\\
1&                              \text{if } \Delta u > \uf\\
\\
1-\frac{\uf-\Delta u}{\delta u}&    \text{else}
\end{array}
\right.
\end{equation}
with a transition width $\delta u$ and the fragmentation speed \uf as free parameter which is assumed to be 1 m~s$^{-1}$, following experimental work of \citet{Blum:1993p4324} and theoretical studies of \citet{Leinhardt:2009p5282}.

\subsubsection{Relative particle velocities}\label{sec:model:rel_velocities}
The different sources of relative velocities considered here are Brownian motion, relative radial and azimuthal velocities, turbulent relative velocities and differential settling. These contributions will be described in the following, an example of the most important contributions is shown in Figure~\ref{fig:model:rel_vel}.

\textit{Brownian motion}, the thermal movement of particles, dependents on the mass of the particle. Hence, particles of different mass have an average velocity relative to each other which is given by
\begin{equation}
\Delta u_\text{BM} = \sqrt{\frac{8 k_\text{B} \: T (m_1 + m_2)}{\pi \: m_1 \: m_2 }}.
\end{equation}
Particle growth due to Brownian relative motion is most effective for small particles.

\textit{Radial drift}, as described in section \ref{sec:model:dust} also induces relative velocities since particles of different size are differently coupled to the gas. The relative velocity is then
\begin{equation}
\Delta u_\text{RD} = \left| u_\text{r}(m_1) - u_\text{r}(m_2) \right|,
\end{equation}
where the radial velocity of the dust, $u_\text{r}$ is given by Eq. \ref{eq:model:u_r_dust}.

\textit{Azimuthal relative velocities} are induced by gas drag in a similar way as radial drift. However while only particles (plus/minus 2 orders of magnitude) around \St=1 are significantly drifting, relative azimuthal velocities do not vanish for encounters between very large and vary small particles (see Figure~\ref{fig:model:rel_vel}). Consequently, large particles are constantly suffering high velocity impacts of smaller ones. According to \citet{Weidenschilling:1977p865} and \citet{Nakagawa:1986p2048}, the relative azimuthal velocities for gas-dominated drag are
\begin{equation}
 u_\varphi = \left| u_\mathrm{n} \cdot \left( \frac{1}{1+\St_1^2} - \frac{1}{1+\St_2^2} \right) \right|,
 \label{eq:model:dv_az}
\end{equation}
where $u_\mathrm{n}$ is defined by Eq.~\ref{eq:model:u_eta}.

\textit{Turbulent motion} as source of relative velocities is discussed in detail in \cite{Ormel:2007p801}. They also derive closed form expressions for the turbulent relative velocities which we use in this work.

\textit{Differential settling} is the fifth process we consider contributing to relative particle velocities. \cite{Dullemond:2004p390} constructed detailed models of vertical disk structure describing the depletion of grains in the upper layers of the disk. They show that the equilibrium settling speed for particles in the Epstein regime is given by
$u_\text{sett} = - z \: \Ok \: \St$ which can be derived by equating the frictional force $F_\text{fric} = - m \: u / t_\text{fric}$ and the vertical component of the gravity force, $F_\text{grav} = - m \: z \: \Ok^2$.
To limit the settling speed to velocities smaller than half the vertically projected Kepler velocity, we use
\begin{equation}
u_\text{sett} = - z \: \Ok\:\min\left(\St,\half\right)
\label{eq:model:u_settling}
\end{equation}
for calculating the relative velocities.

Since we do not resolve the detailed vertical distribution of particles, we take the scale height of each dust size as average height above the mid-plane, which gives
\begin{equation}
\Delta u_\text{DS} = \left| h_i \cdot \min(\St_i,1/2)  -  h_j  \cdot \min(\St_j,1/2)  \right| \: \cdot \Ok.
\end{equation}

\begin{figure}[thb]
\centering
\resizebox{0.6\hsize}{!}{\includegraphics{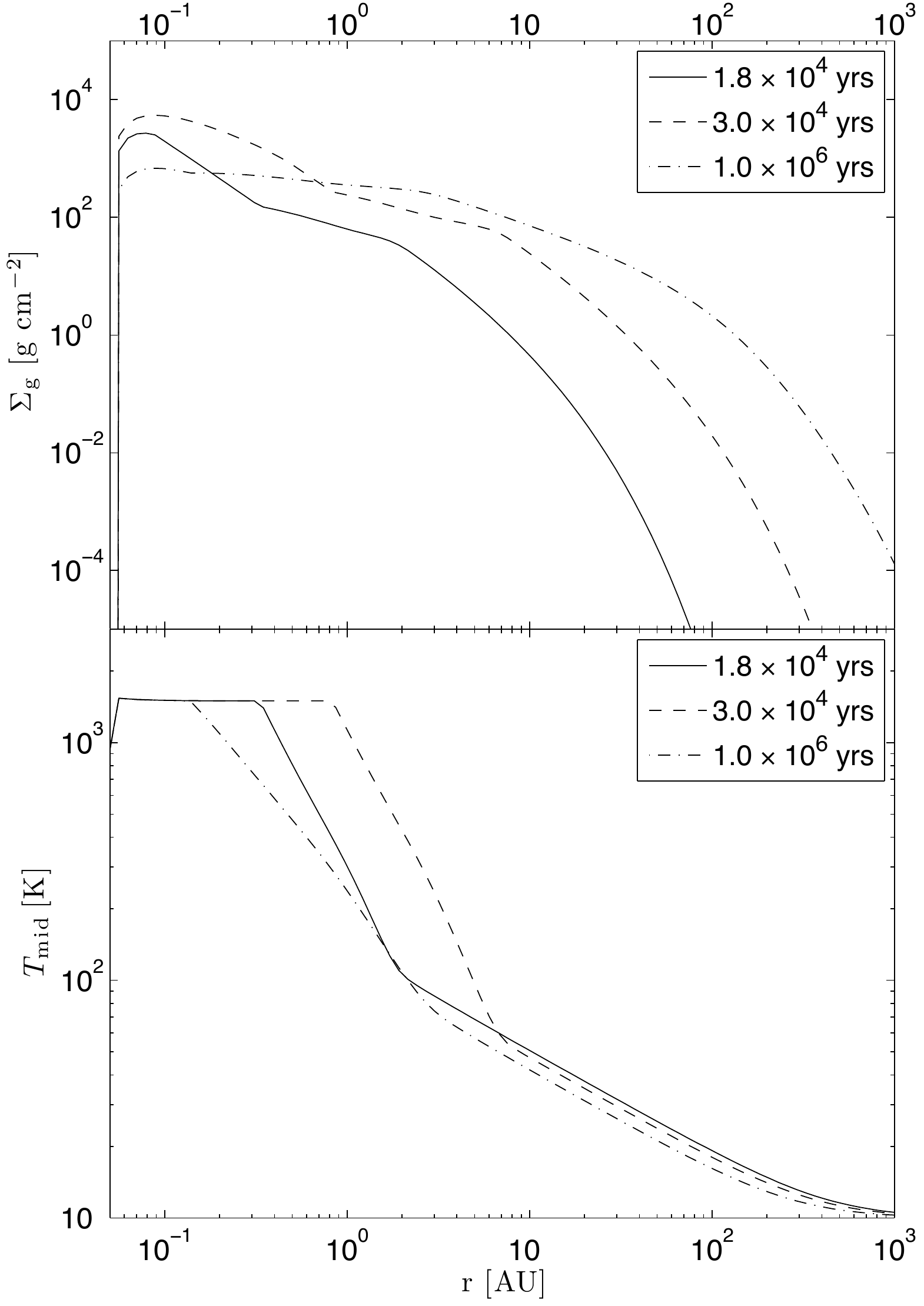}}
\caption{Evolution of disk surface density distribution (top) and midplane temperature (bottom) of the fiducial model described in
\ref{sec:model:visc_evol_of_gas_disk}.}
\label{fig:model:gas_T_snapshots}
\end{figure}

The dust scale height is calculated by equating the time scale for settling,
\begin{equation}
t_\text{sett} = \frac{z}{u_\text{sett}}
\end{equation}
and the time scale for stirring \citep{Dubrulle:1995p300,Schrapler:2004p2394,Dullemond:2004p390},
\begin{equation}
t_\text{stir} = \frac{z^2}{D_\text{d}}.
\end{equation}

By limiting the vertical settling velocity as in Eq. \ref{eq:model:u_settling} and by constraining the dust scale height to be at most equal to the gas scale height, one can derive the dust scale height to be
\begin{equation}
h_\text{d} = \Hp \cdot \min\left(1, \sqrt{\frac{\alpha}{\min(\St,1/2)(1+\St^2)}} \right).
\label{eq:model:h_dust}
\end{equation}

This prescription is only accurate for the dust close to the mid-plane, however most of the dust (and hence most of the coagulation/fragmentation processes) take place near the mid-plane, therefore this approximation is accurate enough for our purposes.


\section{Results}\label{sec:model:results}

\subsection{Viscous evolution of the gas disk}\label{sec:model:visc_evol_of_gas_disk}
We will now focus on the evolution of a disk around a T Tauri like star.
We assume the rotation rate of the parent cloud core to be $7\e{-14}$~s$^{-1}$,
which corresponds to 0.06 times the break up rotation rate of the core.
The disk is being built-up from inside out due to the Shu-Ulrich
infall model, with the centrifugal radius being 8~AU.
The parameters of our fiducial model are summarized in
Table~\ref{tab:model:model_parameters}.

\begin{figure}[thb]
  \centering
\resizebox{\hsize}{!}{\includegraphics{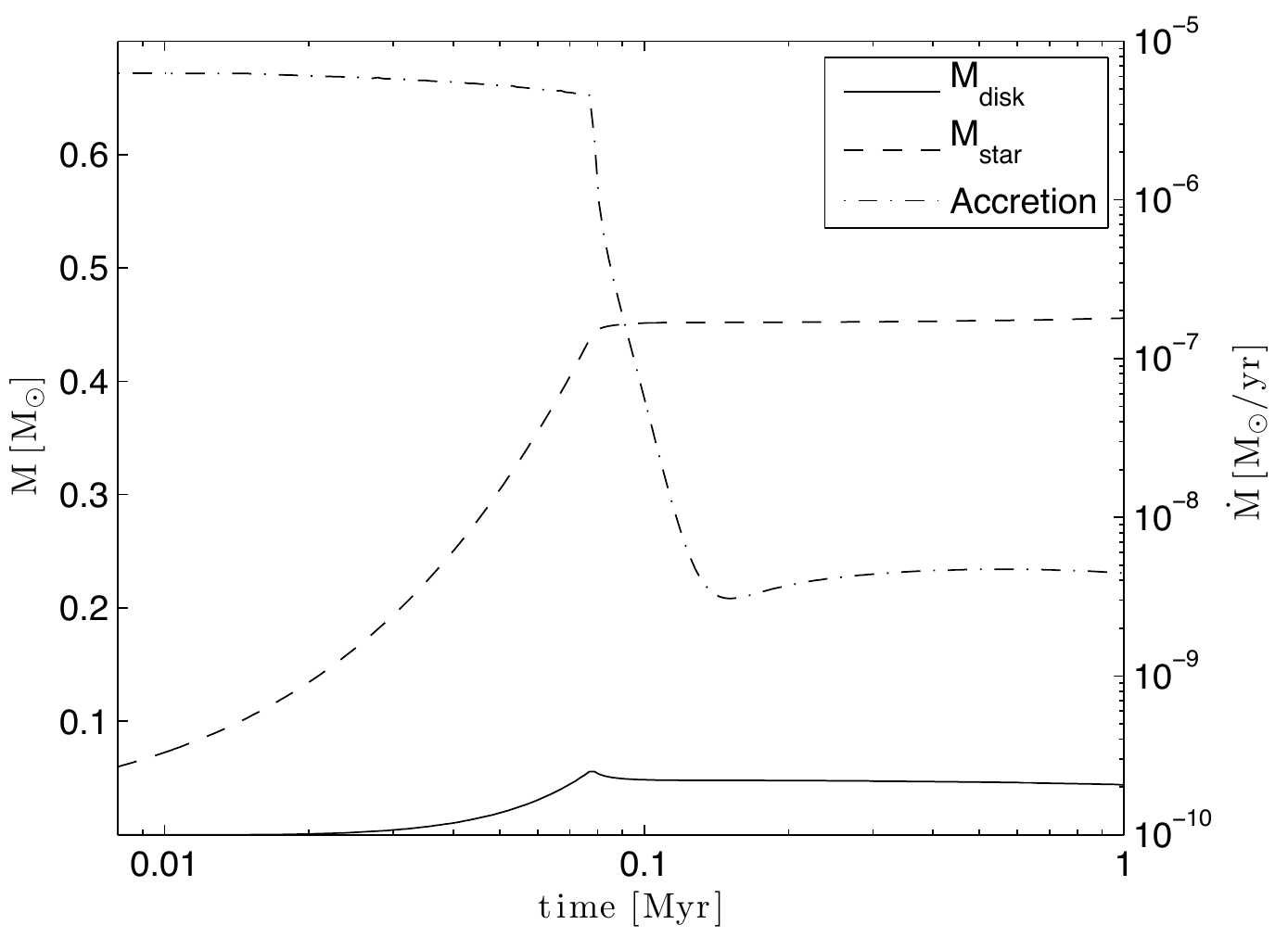}}
\caption{Evolution of disk mass and stellar mass (solid and dashed line on left scale respectively) and accretion rate onto the star (dash-dotted line on right scale). Adapted from Figure~5 in \citet{Hueso:2005p685}.}
\label{fig:model:accretion}
\end{figure}

Figure~\ref{fig:model:gas_T_snapshots} shows how the gas surface
density and the mid-plane temperature of this model evolve as the disk gets
built up, viscously spreads and accretes onto the star. It can be seen that
viscous heating leads to a strong increase of temperature at small radii. This
effect becomes stronger as the disk surface density increases during the infall
phase. After the infall has ceased, the surface density and therefore also the
amount of viscous heating falls off.

This effect also influences the position of the dust
evaporation radius, which is assumed to be at the radius where the dust
temperature exceeds 1500~K. This position moves outwards during the infall
(because of the stronger viscous heating described above). Once the infall
stops, the evaporation radius moves back to smaller radii as the large surface
densities are being accreted onto the star.

Figure~\ref{fig:model:accretion} shows the evolution of accretion rate onto the star, stellar mass
and disk mass. The infall phase lasts until
about 150\,000~years. At this point, the disk looses its source of gas and
small-grained dust and the disk mass drops off quickly until the disk has 
adjusted itself to the new condition. This also explains the sharp drop of the
accretion rate. The slight increase in the accretion rate afterwards comes from
the change in $\alpha$ after the infall stops (see Section~\ref{sec:model:gas}).
\citet{Hueso:2005p685} find a steeper, power-law decline of the accretion rate
after the end of the infall phase because their model does not take the effects
of gravitational instabilities into account.

\subsection{Fiducial model without fragmentation}\label{sec:model:fiducial_model_nofrag}

\begin{figure*}[htp]
\centering
\includegraphics[height=0.69\textheight]{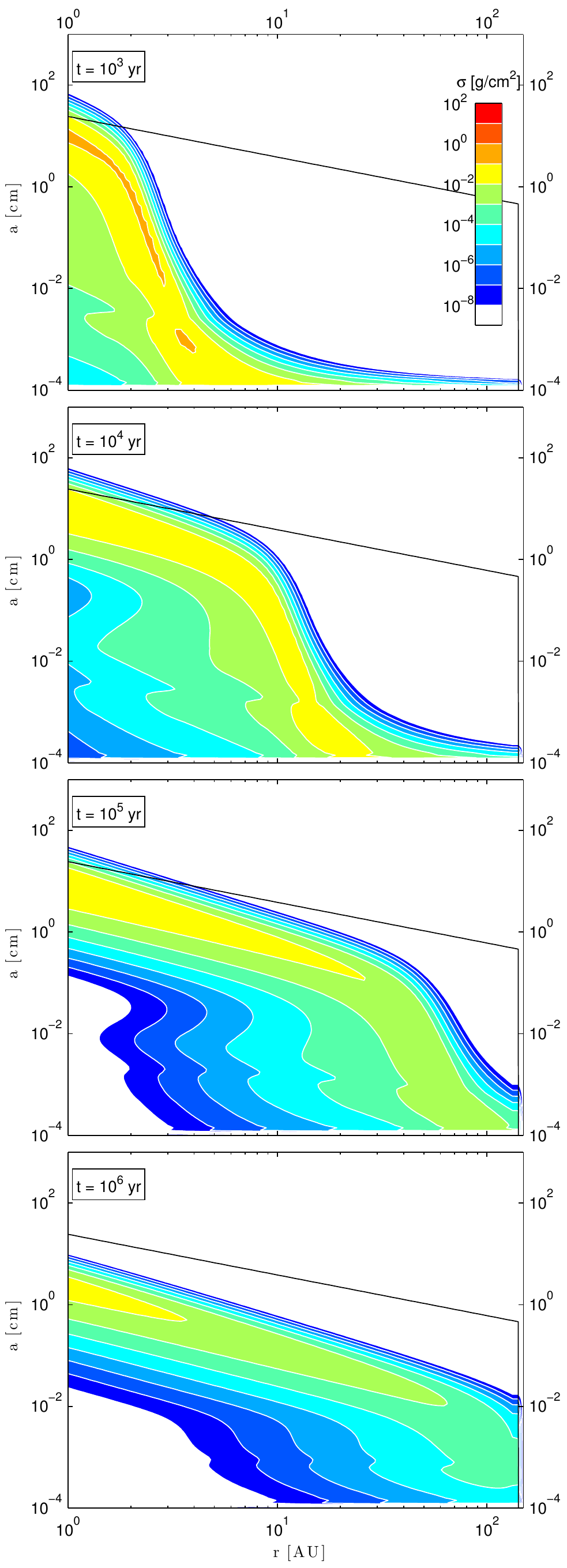}
\includegraphics[height=0.69\textheight]{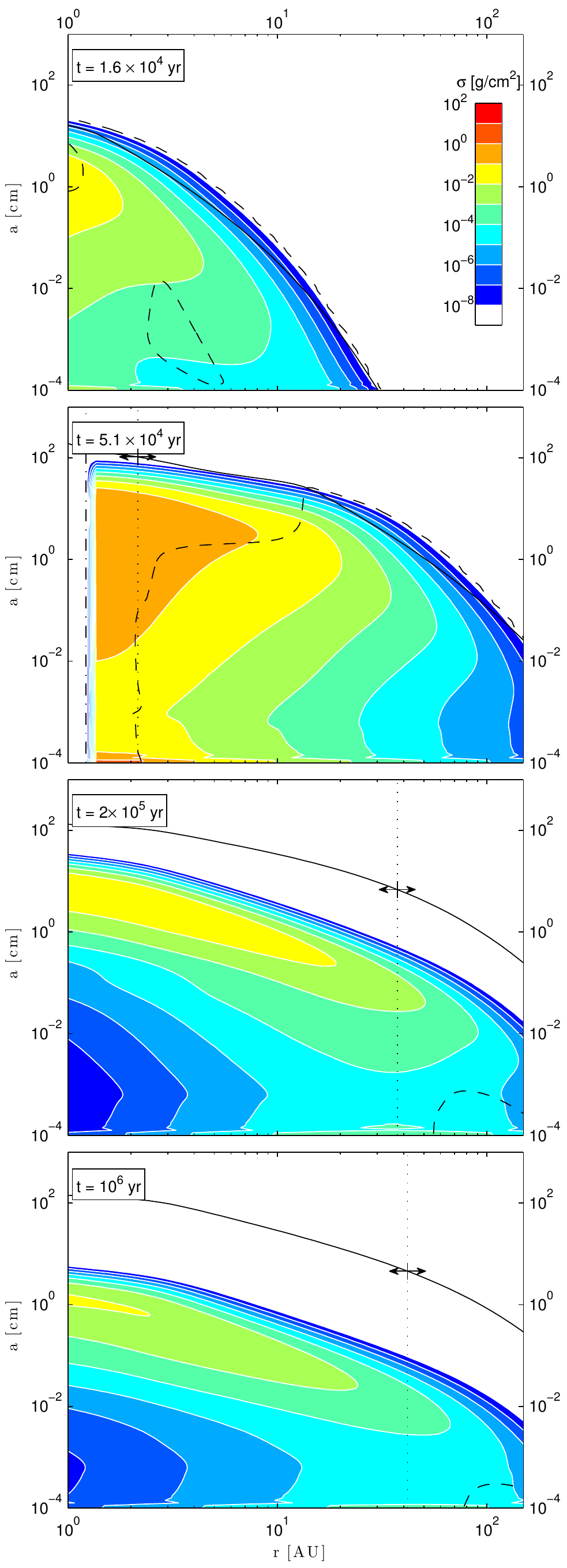}
\caption{Snapshots of the vertically integrated dust density distributions (defined in Eq.~\ref{eq:model:def_sigma}) of a steady state disk as in \citetalias{Brauer:2008p215} (left column) and of an evolving disk (fiducial model, right column). No coagulation is calculated within the evaporation radius (denoted by the dash-dotted line), fragmentation is not taken into account in both simulations. The solid line shows the particle size corresponding to a Stokes number of unity.
Since $a_\mathrm{St=1} \propto \Siggas$ (see Eq.~\ref{eq:model:ST_epstein}) this curve
in fact has the same shape as $\Siggas(r)$, so it reflects, as a ``bonus'', what the gas disk looks like. The radius dividing the evolving disk into accreting and expanding regions is marked by the dotted line and the arrows. Particles which are located below the dashed line have a positive flux in the radial direction due to coupling to the expanding gas disk and turbulent mixing (particles within the closed contour in the upper right plot have an inward pointing flux).}
\label{fig:model:snapshots_fid}
\end{figure*}

\begin{table}
    \centering
    \caption[]{Parameters of the fiducial model.}
    \label{tab:model:model_parameters}
    \begin{tabular}{llll}
    \hline
    \noalign{\smallskip}
    parameter                       & symbol            & value     & unit\\
    \noalign{\smallskip}
    \hline
    \noalign{\smallskip}
    turbulence parameter            & $\alpha$            & $10^{-3}$   & -\\
    irradiation angle               & $\varphi$           & $0.05$      & -\\
    cloud mass                      & $M_\text{cloud}$    & $0.5$       & $M_\odot$\\
    effect. speed of sound in core  & $\csound$           & $3\e{4}$    & cm~s$^{-1}$\\
    rotation rate of cloud core     & $\Omega_\text{s}$   & $7\e{-14}$  & s$^{-1}$\\
    solid density of dust grains    & $\rhos$             & $1.6$       & g/cm$^3$\\
    stellar radius                  & $R_\star$           & $2.5$       & $R_\odot$\\
    stellar temperature             & $T_\star$           & $4000$      & K\\
    \noalign{\smallskip}
    \hline
    \end{tabular}
\end{table}
We will now focus on the dust evolution of the disk. This fiducial simulation includes only grain growth without fragmentation or
erosion. All other parameters as well as the evolution of the gas surface
density and mid-plane temperature are the same as discussed in the previous section. The evolution of this model is visualized in Figure~\ref{fig:model:snapshots_fid}.

The contour levels in Figure~\ref{fig:model:snapshots_fid} show the vertically
integrated dust surface density distribution per logarithmic bin of grain
radius, $\sigma(r,a)$, as defined in Eq.~\ref{eq:model:def_sigma}.
The left column of Figure~\ref{fig:model:snapshots_fid} shows the results of dust
evolution for a steady state (i.e. not viscously evolving) gas disk as described
in \citetalias{Brauer:2008p215}.

The right column shows the evolution of the dust density distribution of the fiducial model, in which the gas disk is gradually built up through infall from the parent molecular cloud core, and the gas disk viscously spreads and accretes matter onto the star. The solid line marks the grain size corresponding to \St=1 at the given radius. This grain size will be called $a_{\St=1}$ hereafter.  In the Epstein regime, $a_{\St=1}$ is proportional to the gas surface density of the disk, which can be seen from Eq.~\ref{eq:model:ST_epstein}.

There are several differences to the simulations of grain growth in steady-state disks, presented in \citetalias{Brauer:2008p215}: viscously evolving disks accrete onto the star by transporting mass inwards and angular momentum outwards. This leads to the fact that some gas has to be moving to larger radii because it is 'absorbing' the angular momentum of the accreting gas. This can be seen in numerical simulations, but also in the self similar solutions of \citet{LyndenBell:1974p1945}. They show that there is a radius $R_\pm$ which divides the disk between inward and outward moving gas. This radius itself is constantly moving outwards, depending on the radial profile of the viscosity.

The radius $R_\pm$ in the fiducial model was found to move from around 20~AU at the end of the infall to 42~AU at 1~Myr and is denoted by the dotted line in Figure~\ref{fig:model:snapshots_fid}.
Important here is that small particles are well enough coupled to the gas to be transported along with the outward moving gas while larger particles decouple from the gas and drift inwards. Those parts of the dust distribution which lie below the dotted line in Figure~\ref{fig:model:snapshots_fid} have positive flux in the radial direction due to the gas-coupling.

One might expect that the dotted and dashed lines always coincide for small grains as they are well coupled to the gas, however, it can be seen that this is not the case. The reason for this is that turbulent mixing also contributes to the total flux of dust of each grain size. The smallest particles in between the dotted line and the dashed line in the lower two panels of Figure~\ref{fig:model:snapshots_fid} do have positive velocities, but due to a gradient in concentration of these dust particles, the flux is still negative.

During the early phases of its evolution, a disk which is built up from inside out quickly grows large particles at small radii which are lost to the star by radial drift. During further evolution, growth timescales become larger and larger (since the dust-to-gas ratio is constantly decreasing) while only the small particles are mixed out to large radii.

\begin{figure*}[htb]
\centering
  \resizebox{0.95\hsize}{!}{\includegraphics{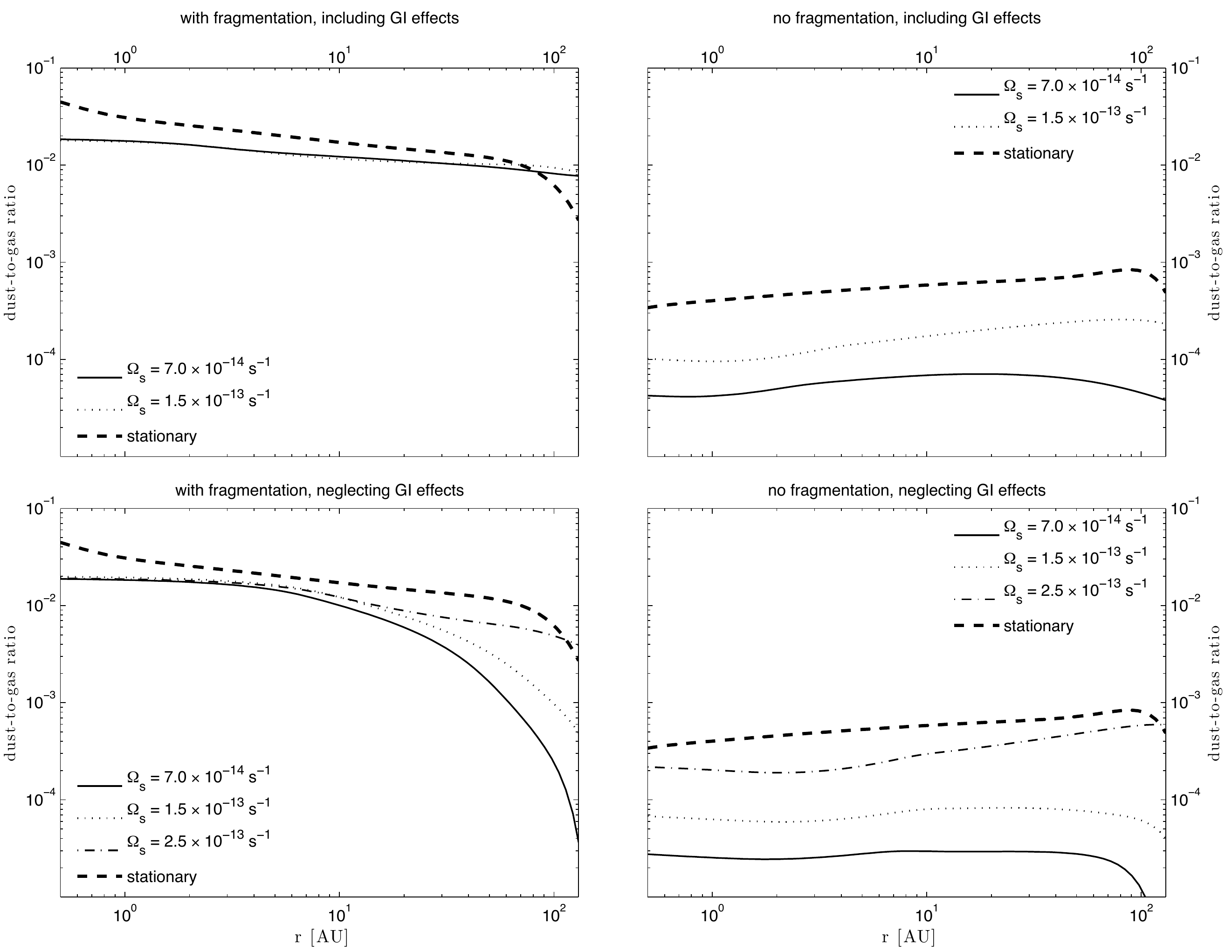}}
  \caption{Comparison of the radial dependence of the dust-to-gas ratio for the stationary simulations (thick lines) and the evolving disk simulations (thin lines). The four panels compare the results after 1~Myr of evolution with/without fragmentation (left/right column) and with/without effects of non-axisymmetric instabilities (top/bottom row). It can be seen that the dust-to-gas ratio of the evolving disk simulations is almost everywhere lower than the one of the stationary simulations. See Section~\ref{sec:model:influences_of_infall_model} for more details.}
  \label{fig:model:FGI}
\end{figure*}

The radial dependence of the dust-to-gas ratio after 1~Myr is shown in Figure~\ref{fig:model:FGI}.
These simulations show that the initial conditions of the stationary disk models (such as shown in \citetalias{Brauer:2008p215} and in the left column of Figure~\ref{fig:model:snapshots_fid}) are too optimistic since they assume a constant dust-to-gas ratio at the start of their simulation throughout the disk which is not possible if the disk is being built-up from inside out unless the centrifugal radius is very large (in which case the disk would probably fragment gravitationally) and grain fragmentation is included to prevent the grains from becoming large and start drifting strongly.
Removal of larger grains and outward transport of small grains lead to the fact that the dust-to-gas ratio is reduced by 0.5--1.5 orders of magnitude compared to a stationary model. This effect is also discussed in Section~\ref{sec:model:influences_of_infall_model}.

\subsection{Fiducial model with fragmentation}\label{sec:model:fiducial_model_frag}
The situation changes significantly, if we take grain fragmentation into account. As discussed in Section~\ref{sec:model:distr_of_fragments}, we consider two different kinds of outcomes for grain-grain collisions with relative velocities larger than the fragmentation velocity \uf: cratering (if the masses differ by less than one order of magnitude) and complete fragmentation (otherwise).


\begin{figure*}[h!tbp]
\centering
  \resizebox{0.4\hsize}{!}{\includegraphics{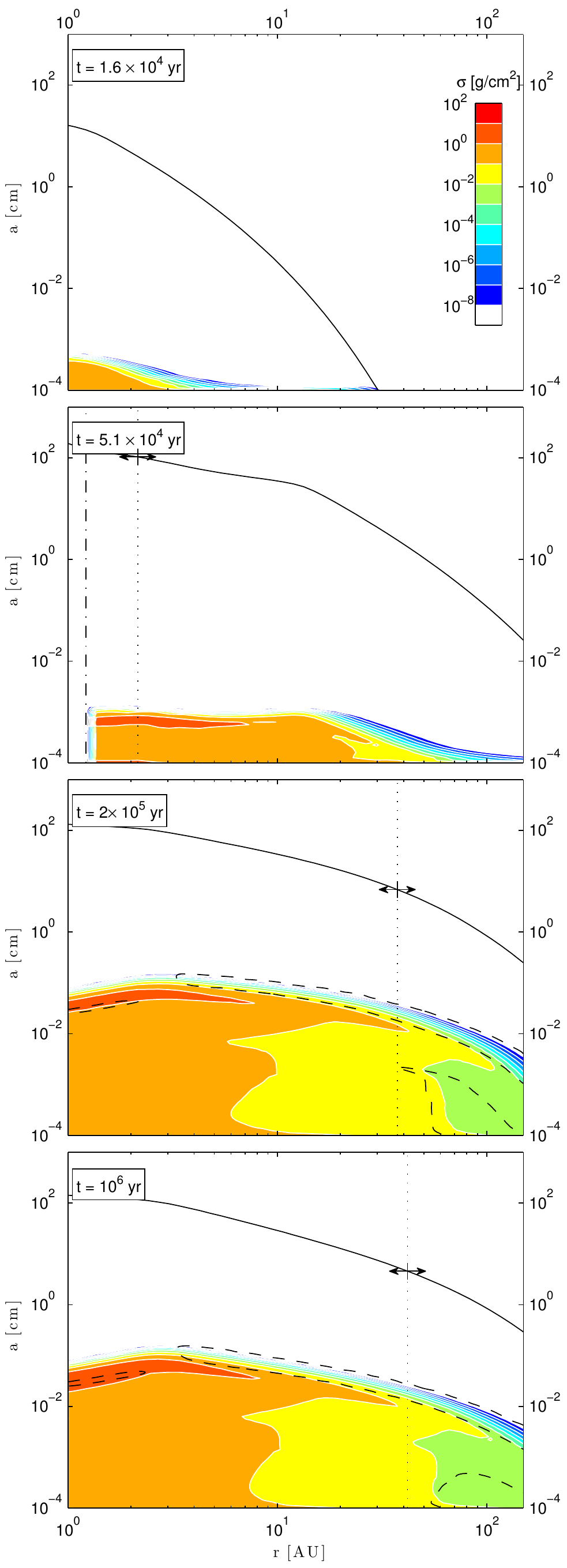}}
  \caption{Evolution of the dust density distribution of the fiducial model as in Figure~\ref{fig:model:snapshots_fid} but with fragmentation included. The dashed contour line (in the lower two panels) around the upper end of the size distribution and around small particles at $>60$~AU marks those parts of the distribution which have a positive (outward pointing) fluxes.}
  \label{fig:model:snapshots_frag}
\end{figure*}

For particles larger than about $\St\approx 10^{-3}$, relative velocities are dominated by turbulent motion (and to a lesser extend by vertical settling). Since the relative velocities increase with Stokes number (and therefore with grain size), we can calculate the approximate position of the fragmentation barrier by equating the assumed fragmentation velocity \uf with the approximate relative velocities of the particles,
\begin{equation}
\St_\text{max} \simeq \frac{\uf^2}{2\alpha\,\csound^2}
\label{eq:model:st_max}
\end{equation}
Particles larger than this size are subject to high-velocity collisions which will erode or completely fragment those particles. This is only a rough estimate as the relative velocities also depend on the size of the smaller particle and radial drift also influence the grain distribution. However Eq.~\ref{eq:model:st_max} reproduces well the upper end of the size distribution in most of our simulations and therefore helps understanding the influence of various parameters on the outcome of these simulations.

The evolution of the grain size distribution including fragmentation is depicted in Figure~\ref{fig:model:snapshots_frag}. The initial condition is quickly forgotten since particles grow on very short timescales to sizes at which they start to fragment. The resulting fragments contribute again to the growth process until a semi-equilibrium of growth and fragmentation is reached.

\begin{figure}[tbh]
  \resizebox{0.95\hsize}{!}{\includegraphics{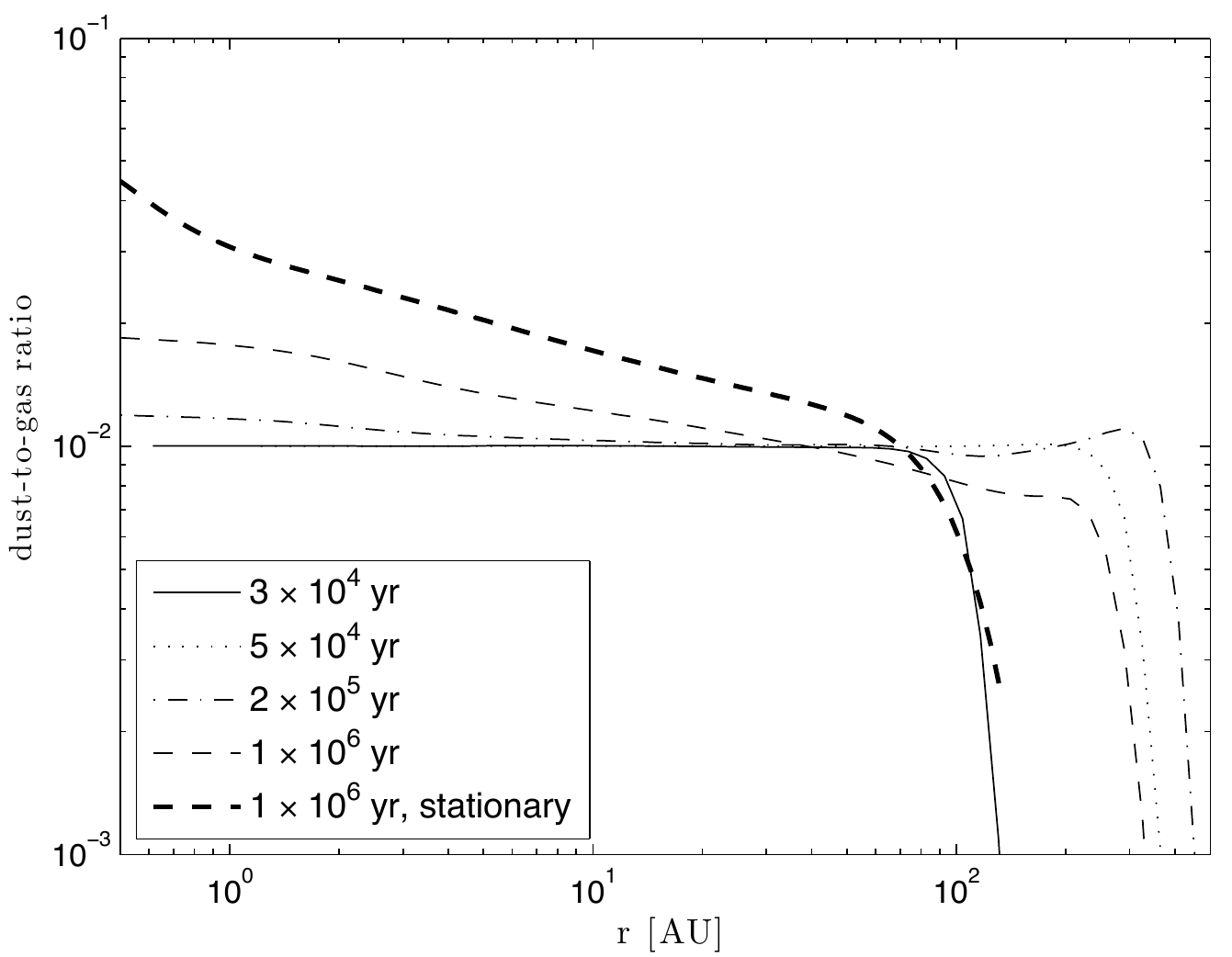}}
  \caption{Evolution of the radial dependence of the dust-to-gas ratio in the fiducial model including fragmentation with the times corresponding to the snapshots shown in Figure~\ref{fig:model:snapshots_frag}. The initial dust-to-gas ratio is taken to be~0.01. The thick dashed curve represents the result at 1~Myr of the static disk model for comparison.}
  \label{fig:model:fid_d2g_frag}
\end{figure}

It can be seen that particles stay much smaller than in the model without fragmentation. This means that 
they are less affected by radial drift on the one hand and better transported along with the expanding gas disk on the other hand. Consequently, considerable amounts of dust can reach radii of the order of 100~AU, as seen in Figure~\ref{fig:model:fid_d2g_frag}.

The approximate maximum Stokes number, defined in Eq.~\ref{eq:model:st_max}, is inversely proportional to the temperature (since relative velocities are proportional to \csound), which means that in regions with lower temperature, particles can reach larger Stokes numbers. By equating drift and drag velocities of the particles (cf. Eq.~\ref{eq:model:u_r_dust}), it can be shown that the radial velocities of particles with Stokes numbers larger than about $\alpha/2$, are being dominated by radial drift.

Due to the high temperatures below $\sim$5~AU (caused by viscous heating), $\St_\text{max}$ stays below this value which prevents any significant radial drift within this radius, particles inside 5~AU are therefore only transported along with the accreting gas. Particles at larger radii and lower temperatures can drift (although only slightly), which means that there is a continuous transport of dust from the outer regions into the inner regions. Once these particles arrive in the hot region, they get ``stuck'' because their Stokes number drops below $\alpha/2$. The gas within about 5~AU is therefore enriched in dust, as seen in Figure~\ref{fig:model:fid_d2g_frag}. The dust-to-gas ratio at 1~AU after 1~Myr is increased by 25\% over the value of in-falling matter, which is taken to be 0.01.

Figure~\ref{fig:model:fid_d2g_frag} also shows a relatively sharp decrease in the dust to gas ratio at a few hundred AU. At these radii, the gas densities become so small that even the smallest dust particles decouple from the gas and start to drift inwards.

The thick line in Figure~\ref{fig:model:fid_d2g_frag} shows as comparison the dust-to-gas ratio of the stationary disk model (cf. left column of Figure~\ref{fig:model:snapshots_fid}) after 1~Myr, which starts with a radially constant initial dust-to-gas ratio of 0.01.

\begin{figure}[thb]
  \includegraphics[width=0.95\textwidth]{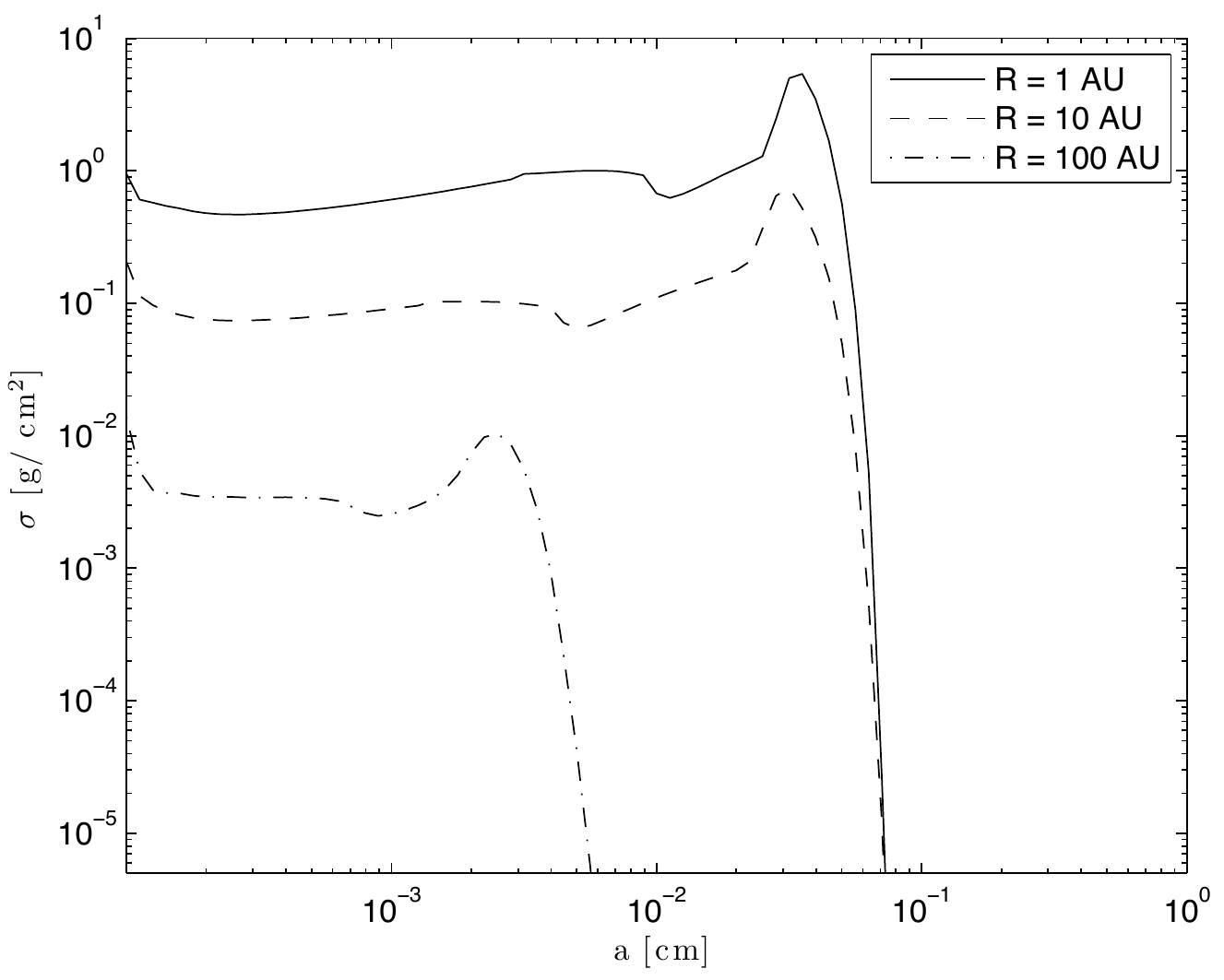}
  \caption{Vertically integrated (cf. Eq.~\ref{eq:model:def_sigma}) grain surface density distributions as function of grain radius at a distance of 1~AU (solid), 10~AU (dashed) and 100~AU (dot-dashed) from the star. These curves represent slices through the bottom panel of Figure~\ref{fig:model:snapshots_frag}.}
  \label{fig:model:distri_slices}
\end{figure}

Figure~\ref{fig:model:distri_slices} shows the semi-equilibrium grain surface density distribution at 1, 10 and 100~AU in the fiducial model with fragmentation at 1~Myr.
The exact shape of these distributions depends very much on the prescription of fragmentation and cratering. In general the overall shape of these semi-equilibrium distributions is always the same: a power-law or nearly constant distribution (in $\sigma$) for small grains and a peak at some grain size $a_\mathrm{max}$, beyond which the distribution drops dramatically. The peak near the upper end of the distribution is caused by cratering. This can be understood by looking at the collision velocities: the relative velocity of two particles increases with the grain size but it is lower for equal-sized collisions than for collisions with particles of very different sizes (see Figure~\ref{fig:model:rel_vel}). The largest particles in the distribution have relative velocities with similar sized particles which lie just below the fragmentation velocity (otherwise they would fragment). This means that the relative velocities with much smaller particles (which are too small to fragment the bigger particles but can still damage them via cratering) are above this critical velocity. This inhibits the further growth of the big particles beyond $a_\mathrm{max}$, causing a ``traffic jam'' close to the fragmentation barrier. The peak in the distribution represents this traffic jam.

\subsection{Influences of the infall model}\label{sec:model:influences_of_infall_model}
In the fiducial model without fragmentation, continuous resupply of material by infall is the cause why the disk has much more small grains than compared to the stationary disk model (cf. Figure~\ref{fig:model:snapshots_fid}), which relatively quickly consumes all available micrometer sized dust. The effect has already been found in \citet{Dominik:2008p4626}: if all grains start to grow at the same time, then the bulk of the mass grows in a relatively thin peak to larger sizes (see Figure~6 in \citetalias{Brauer:2008p215}). However if the bulk of the mass already resides in particles of larger size, then additional supply of small grains by infall is not swept up effectively because of the following: firstly, the number density of large particles is small (they may be dominating the mass, but not necessarily the number density distribution) and secondly, they only reside in a thin mid-plane layer while the scale height of small particles equals the gas scale height.

We studied how much the disk evolution depends on changes in the infall model.

For a given cloud mass, the so called centrifugal radius $r_\text{centr}$, which was defined in Eq.~\ref{eq:model:r_centri}, depends on the temperature and the angular velocity of the cloud. Both can be varied resulting in a large range of possible centrifugal radii reaching from a few to several hundred AU. Since the centrifugal radius is the relevant parameter, we varied only the rotation rate of the cloud core. We performed simulations with three different rotation rates which correspond to centrifugal radii of about 8 (fiducial model), 33~AU, and 100~AU. For each centrifugal radius, we performed two simulations: one which includes effects of gravitational instabilities (GI) -- i.e. increased $\alpha$ during infall and according to Eq.~\ref{eq:model:alpha_of_Q} -- and one which neglects them.

However for a centrifugal radius of 100~AU, too much matter is loaded onto the cold outer parts of the disk and consequently, the disk would fragment through gravitational instability. We cannot treat this in our simulations, hence, for the case of 100~AU, we show only results which neglect all GI effects.

The resulting dust-to-gas ratios are being shown in Figure~\ref{fig:model:FGI}.

Two general aspects change in the case of higher rotation rates: firstly, more of the initial cloud mass has to be accreted onto the star by going through the outer parts of the disk. Consequently, the disk is more extended and more massive than compared to the case of low rotation rate.

Secondly, as aforementioned the high surface densities in the colder regions at larger radii cause the disk to become less gravitationally stable.

If grain fragmentation is not taken into account in the simulations, both effects cause more dust mass to be transported to larger radii. Growth and drift timescales are increasing with radius and the dust disk with centrifugal radius of 33~AU (100~AU) can stay 5 (35)~times more massive than in the low angular momentum case after 1~Myr if GI effects are neglected.

If GI effects are included, matter is even more effectively transported outward, the dust-to-disk mass ratio for 8 and 33~AU is increased from 5 to 8.

However if fragmentation is included, it does not matter so significantly, where the dust mass is deposited onto the disk since grains stay so small during the build-up phase of the disk (due to grain fragmentation by turbulent velocities) that they are well coupled to the gas. Outwards of  $\sim 10$~AU (without GI effects) or of a few hundred AU (if GI effects are included), the gas densities become so small that even the smallest grains start do decouple from the gas. They are therefore not as effectively transported outwards. In these regions, the amount of dust depends on the final centrifugal radius while at smaller radii, turbulent mixing quickly evens out all differences in the dust-to-gas ratio (see left column of Figure~\ref{fig:model:FGI}).

It can be seen, that in all simulations, the dust-to-gas ratio is lower than in the stationary disk model. The trend in the upper right panel in Figure~\ref{fig:model:FGI} suggests that for a centrifugal radius of 100~AU and the enhanced radial transport by GI effects might have a higher dust-to-gas ratio than the stationary disk model. However in this case, the disk would become extremely unstable and would therefore fragment.

The reason for this is the following: to be able to compare both simulations, the total mass of the disk-star system is the same as in the stationary disk models. How the total mass is distributed onto disk and star depends on the prescription of infall. If a centrifugal radius of 100~AU is used, the disk becomes so massive that it significantly exceeds the stability criterion $M_\mathrm{disk}/M_\star \lesssim 0.1$.

\begin{figure}[h]
  \centering
  \resizebox{0.45\hsize}{!}{\includegraphics{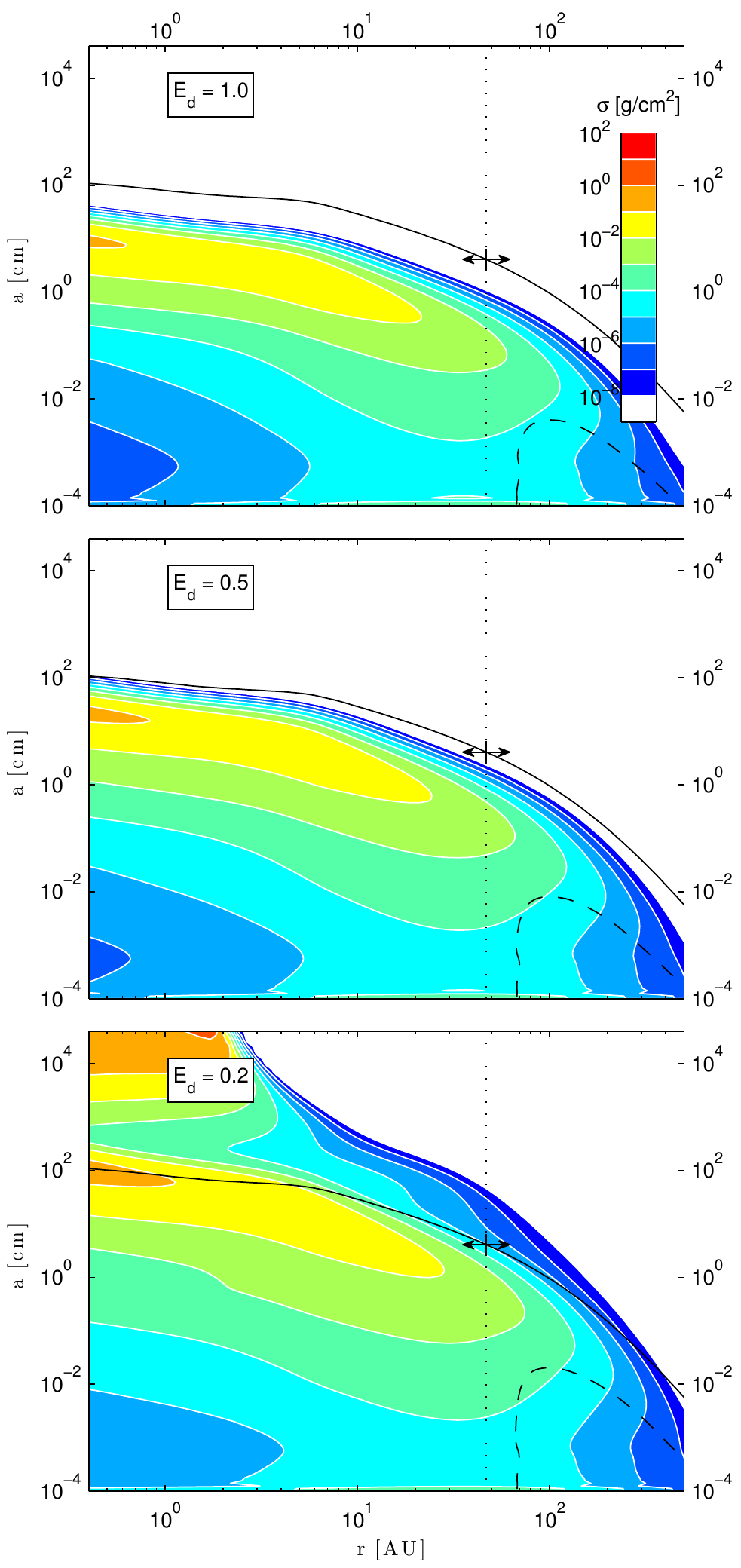}}
  \caption{Evolution of the dust surface density distribution of the fiducial model at 200\,000 years for different drift efficiencies $E_\text{d}$, without fragmentation. The solid line denotes the grain size $a_{\St=1}$ of particles with a Stokes number of unity. Gas outside of the radius denoted by the dotted line as well as particles below the dashed line have positive radial velocities. See section~\ref{sec:model:fiducial_model_nofrag} for more discussion.}
  \label{fig:model:snapshots_drift}
\end{figure}

\subsection{The radial drift barrier revisited}\label{sec:model:drift_barrier}
According to the current understanding of planet formation, several mechanisms seem to prevent the formation of large bodies via coagulation quite rigorously. The most famous ones -- radial drift and fragmentation -- have already been introduced above.
Radial drift has first been discussed by \citet{Weidenschilling:1977p865}, while the importance of the fragmentation barrier (which may prevent grain growth at even smaller sizes) was discussed in \citetalias{Brauer:2008p215}.
In the following, we want to test some ideas about how to weaken or to overcome these barriers apart from those already studied in \citetalias{Brauer:2008p215}.

\citetalias{Brauer:2008p215} has quantified the radial drift barrier by equating the timescales of growth and radial drift which leads to the condition
\begin{equation}
\frac{\tau_\text{g}}{\tau_\text{d}} = \frac{1}{\epsilon_0} \left( \frac{\Hp}{r} \right)^2 \leq \frac{1}{\gamma},
\label{eq:model:drift_barrier}
\end{equation}
where $\epsilon_0$ is the dust-to-gas ratio and $\tau_\text{d}$ and $\tau_\text{g}$ are the drift and growth timescales respectively. The  parameter $\gamma$ describes how much more efficient growth around \St=1 must be, so that the particles are not removed by radial drift. To overcome the drift barrier, obviously either particle growth must be accelerated, or the drift efficiency has to be decreased. \citetalias{Brauer:2008p215} have numerically measured the parameter $\gamma$ to be around 12. In other words, this means that the growth timescales have to be decreased (e.g. by an increased dust-to-gas ratio) until the condition in Eq.~\ref{eq:model:drift_barrier} is fulfilled.

However, there are other ways of breaking through the drift barrier. Firstly, the drift timescale for \St=1 particles also depends on the temperature (via the pressure gradient). A simple approximation from Eq.~\ref{eq:model:drift_barrier} with a 0.5~$M_\odot$ star and a dust-to-gas ratio of 0.01 gives
\begin{equation}
T < 103\text{ K } \left(\frac{r}{\text{AU}}\right)^{-1},
\end{equation}
which means that particles should be able to break through the drift barrier at 1~AU if the temperature is below $\sim$100~K (or 200~K for a solar mass star).
\citet{Dullemond:2002p399} have constructed vertical structure models of passively irradiated circumstellar disks using full frequency- and angle-dependent radiative transfer. They show that the mid-plane temperature of such a T Tauri like system at 1~AU can be as low as 60~K. Reducing the temperature by some factor reduces the drift time scale by the a factor of similar size which we will call the radial drift efficiency $E_\text{d}$ (cf. Eq.~\ref{eq:model:u_eta}).

Zonal flows as presented in \citet{Johansen:2006p7466} could be an alternative way of decreasing the efficiency of radial drift averaged over typical time scales of particle growth. \citet{Johansen:2006p7466} found a reduction of the radial drift velocity of up to 40\% for meter-sized particles.

Meridional flows \citep[e.g.,][]{Urpin:1984p1473,Kley:1992p7134} might also seem interesting in this context, however they do not directly influence the radial drift efficiency but rather reverse the gas-drag effect. This might be important for small particles (which, however are not strongly settling to the mid-plane) but for \St=1 particles, $\alpha$ would have to be extremely high to have significant influence: even $\alpha=0.1$ would result in a reduction of the particles radial velocity by approximately only a few percent.

Another possibility to weaken the drift barrier is changing its radial dependence. The reason for this is the following: particle radial drift is only a barrier if it prevents particles to cross the size $a_{\St=1}$. Since particles grow while drifting, the particle size corresponding to \St=1 needs to increase as well, to be a barrier. Otherwise drifting particles would grow (at least partly) through the barrier while they are drifting. If $a_{\St=1}$ is decreasing in the direction toward the star, then a particle that drifts inwards would have an increasing Stokes number even if the particle does not grow at all.

\begin{figure}[htb]
  \centering
  \resizebox{\hsize}{!}{\includegraphics{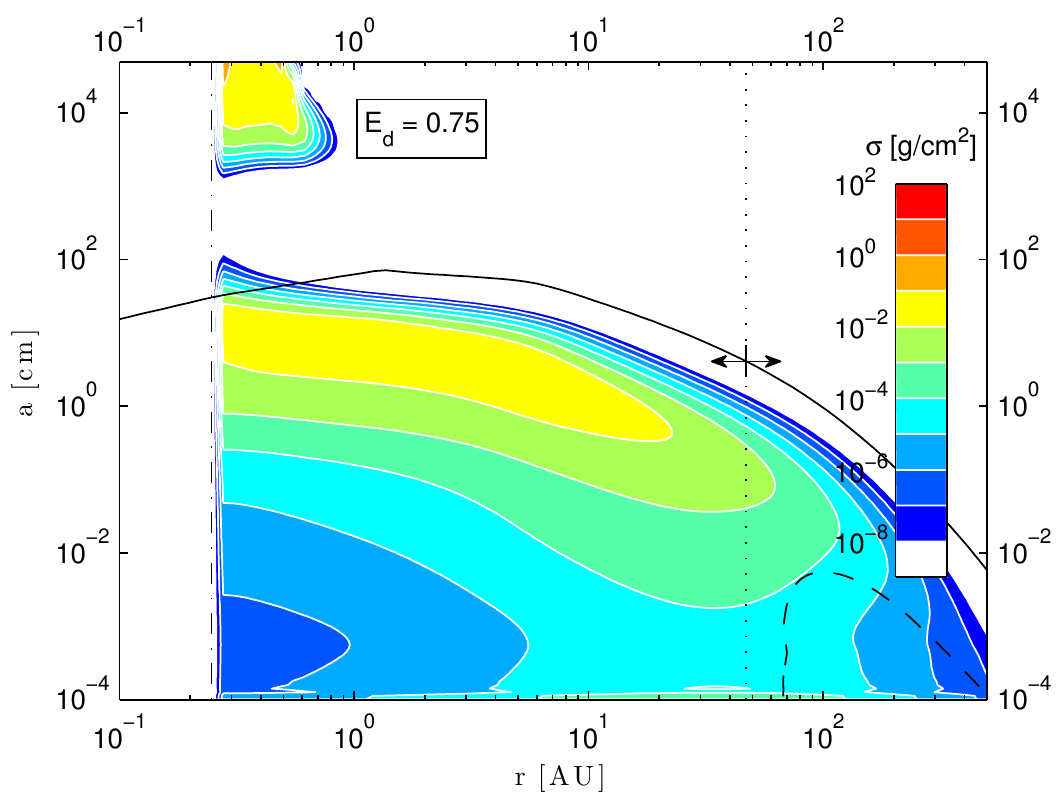}}
  \caption{Dust grain surface density distribution as in Figure~\ref{fig:model:snapshots_drift} at 200\,000 years but including the Stokes drag regime. The drift efficiency is set to $E_\text{d} = 0.75$ and fragmentation is not taken into account. It can be seen that $a_\mathrm{\St=1}$ (solid line) increases with radius until about 1~AU, which facilitates the break through the drift barrier.}
  \label{fig:model:snapshots_drift_ST}
\end{figure}

In the Epstein regime, the size corresponding to \St=1 is proportional to the gas surface density
\begin{equation}
a_\text{\St=1} = \frac{2 \Siggas}{\pi \rhos},
\end{equation}
meaning that a relatively flat profile of surface density (or even a profile with positive slope) is needed to allow particles to grow through the barrier. However, our simulations of the viscous gas disk evolution does not yield surface density profiles with positive slopes outside the dust evaporation radius.

To quantify the arguments above, we have performed simulations with varying drift efficiency $E_\text{d}$ to test how much the radial drift has to be reduced to allow break through. We have additionally included the first Stokes drag regime to see how the radial drift of particles is influenced by it.

Figure~\ref{fig:model:snapshots_drift} shows the grain surface density distribution after 200\,000 years of evolution for three different drag efficiencies. The most obvious changes can be seen in the region where the $a_{\St=1}$ line (solid line) is relatively flat: the grain distribution is shifted towards larger Stokes numbers. As explained above, particles can grow while drifting, which can be seen in the case of $E_\text{d}=0.5$. The Stokes number and size of the largest particles is significantly increasing towards smaller radii. However the radial drift efficiency has to be reduced by 80\% to produce particles which are large enough to escape the drift regime and are therefore not lost to the star.

\begin{figure*}[thb]
\centering
\resizebox{\hsize}{!}{\includegraphics{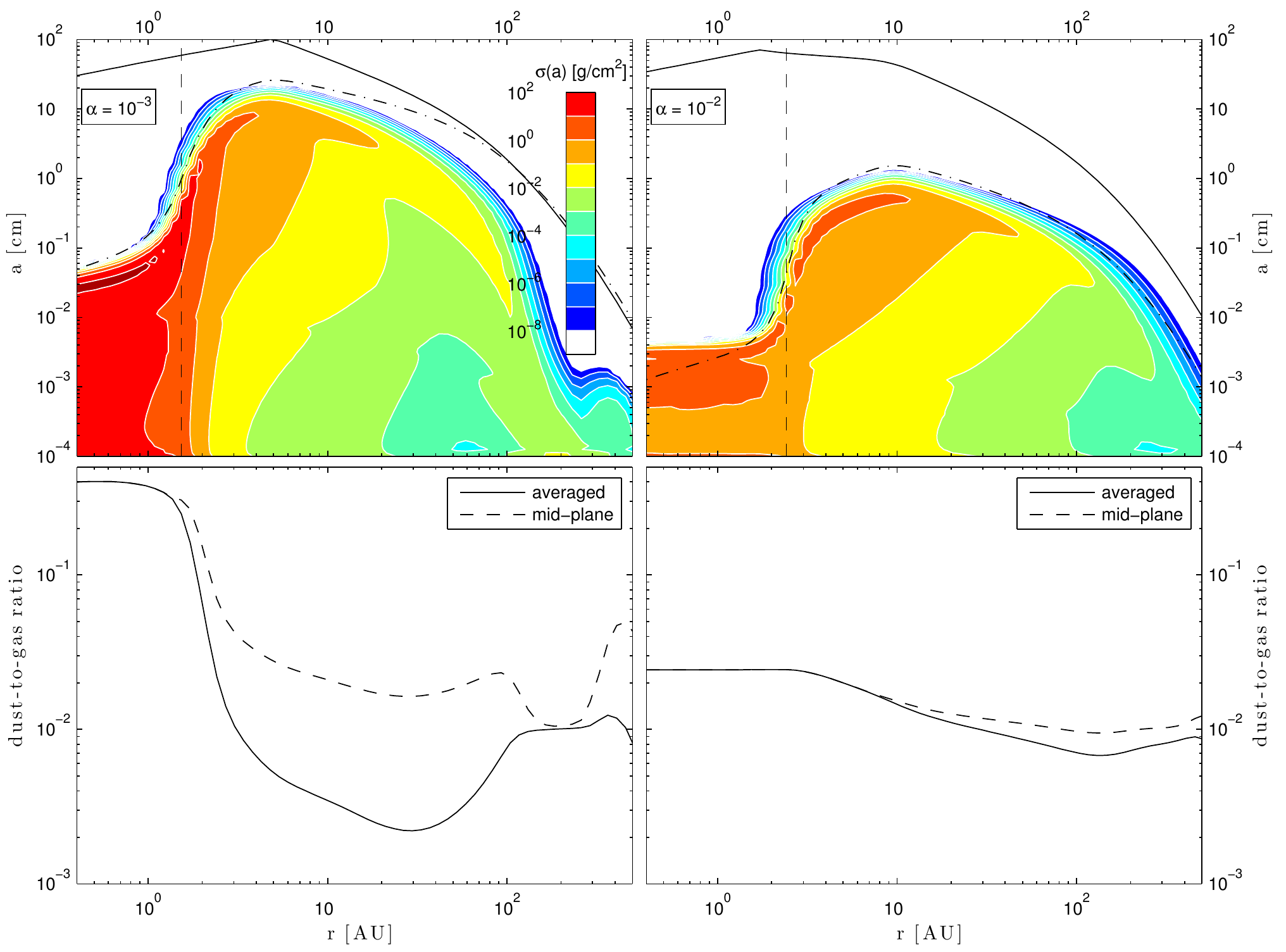}}
\caption{Dust surface density distributions (top row) and the according solid-to-gas ratio (bottom row) for the case of radius-dependent fragmentation velocity after 2\e{5} years of evolution. In the upper row, the vertical dashed line denotes the position of the snow line at the mid-plane, the solid line corresponds to $a_\mathrm{St=1}$ and the dot-dashed line shows the approximate location of the fragmentation barrier according to Eq.~\ref{eq:model:st_max}.
The snow line on the right plot lies further outside since viscous heating is stronger if $\alpha$ is larger.
In the bottom row, the solid line denotes the vertically integrated dust-to-gas ratio while the dashed line denotes the dust-to-gas ratio at the disk mid-plane.
The icy dust grains outside the snow line are assumed to fragment at a critical velocity of 10~m~s$^{-1}$ while particles inside the snow line fragment at 1~m~s$^{-1}$. The plots on the left and right hand side differ in the amount of turbulence in the disk ($\alpha = 10^{-3}$ and $10^{-2}$, respectively).}
\label{fig:model:snapshots_enhancement}
\end{figure*}

The situation changes, if the Stokes drag is taken into account: if gas surface densities are high enough for the dust particles to get into a different drag regime, a change in the radial dependency of $a_\text{\St=1}$ can be achieved. The Epstein drag regime for particles with Stokes number of unity is only valid if
\begin{equation}
\Siggas \lesssim 108\frac{\text{g}}{\text{cm}^2} \left(\frac{T}{\text{200 K}}\right)^{\frac{1}{4}} \left(\frac{R}{\text{AU}}\right)^{\frac{3}{4}}   \left(\frac{M_\star}{M_\odot}\right)^{-\frac{1}{4}}\left(\frac{\rhos}{1.6\frac{\text{g}}{\text{cm}^3}}\right)^{\frac{1}{2}},
\label{eq:model:sig_threshold}    
\end{equation}
otherwise, drag forces have to be calculated according to the Stokes drag law since the Knudsen number becomes smaller than 4/9 \citep[see][]{Weidenschilling:1977p865}. The Stokes number is then given by
\begin{equation}
\St = \frac{\sqrt{2\pi}}{9} \frac{\rhos\; \sigma_{\text{H}_2} \; a^2}{\mu\; \mpr} \: \Hp^{-1},
\end{equation}
with $\sigma_{\text{H}_2}$ being the cross section of molecular hydrogen. Interestingly, $a_{\St=1}$ is independent of $\Siggas$ and proportional to the square root of the pressure scale height which decreases towards smaller radii. This means that -- as long as the surface density is high enough -- it does not depend on the radial profile of the surface density. In this regime the radial drift itself could move particles over the drift barrier since drifting inwards increases the Stokes number of a particle without increasing its size.

Results of simulations which include the Stokes drag are shown in Figure~\ref{fig:model:snapshots_drift_ST}. In this case, particles can already break through the drift barrier if $E_\text{d} \lesssim 0.75$. This value and the position of the breakthrough depends on where the drag law changes from Epstein to Stokes regime and therefore on the disk surface density. As noted above, larger surface densities generally shift the position of regime change towards larger radii making it easier for particles to break through the drift barrier.

It should be noted that the physical way to avoid the Stokes drag regime is to decrease the surface densities, however we chose the same initial conditions for both cases -- with and without Stokes drag -- and just neglected the Stokes drag in the latter computations to be able to compare the efficiency factors independent of other parameters such as disk mass or temperature.

\subsection{The fragmentation barrier revisited}\label{sec:model:frag_barrier}

In the previous section, we have shown that several mechanisms could allow particles to break through the radial drift barrier, however fragmentation puts even stronger constraints on the formation of planetesimals.

As shown by \citet{Ormel:2007p801}, the largest relative velocities are of the order of
\begin{equation}
\Delta u_\text{max} \simeq \sqrt{2 \, \alpha} \, \csound.
\end{equation}
If particles should be able to break through the fragmentation barrier, then they need to survive these large relative velocities, meaning that $\Delta u_\text{max}$ has to be smaller than the fragmentation velocity of the particles, or
\begin{equation}
\frac{\uf}{\csound} \gtrsim \sqrt{2 \, \alpha}.
\end{equation}
This condition is hard to fulfill with reasonable fragmentation velocities, unless $\alpha$ is very small. E.g., for $\alpha = 10^{-5}$ and a temperature of 200 to 250~K, the fragmentation velocity needs to be higher than 4~m~s$^{-1}$, which could already seen in the simulations by \cite{Brauer:2008p212}, who have simulated particle growth near the snow-line.

Even in the case of very low turbulence, relative azimuthal velocities of large ($\St\gtrsim 1$) and small grains ($\St\ll 1$) are of the order of 30~m~s$^{-1}$, which means that large particles are constantly being 'sand-blasted' by small grains. The only way of reducing these velocities significantly is decreasing the pressure gradient (see Equations~\ref{eq:model:u_eta}~and~\ref{eq:model:dv_az}).

Another possibility to overcome this problem would be if high-velocity impacts of smaller particles would cause net growth, as has been found experimentally by \citet{Wurm:2005p1855} and \citet{Teiser:2009p7785}.

Taken together, these facts make environments as the inner edge of dead zones ideal places for grain growth \citep[see][]{Brauer:2008p212,Kretke:2007p697}: shutting of MRI leads to low values of $\alpha$, which are needed to reduce turbulent relative velocities and the low pressure gradients prevent radial drift and high azimuthal relative velocities.

\subsection{Dust enhancement inside the snow line}\label{sec:model:dustenh_in_snowline}

As we will discuss in Chapter~\ref{chapter:dustretention}, significant loss of dust by radial drift can be prevented by assuring that particles stay small enough and are therefore not influenced by radial drift. For typical values of $\alpha$ ($10^{-3}-10^{-2}$), this means that the fragmentation velocity must be smaller than about 0.5--5~m~s$^{-1}$. If particles have higher tensile strength, they can grow to larger sizes which are again affected by radial drift.

Typical fragmentation velocities for silicate grains determined both theoretically and experimentally are of the order of a few~m~s$^{-1}$ \citep[for a review, see][]{Blum:2008p1920}. The composition of particles outside the snow-line is expected to change due to the presence of ices. This can influence material properties and increase the fragmentation velocity \citep[see][]{Schafer:2007p7468,Wada:2009p8776}.

We have performed simulations with a radially varying fragmentation speed. We assume the fragmentation speed inside the snow-line to be 1~m~s$^{-1}$, outside the snow-line to be 10~m~s$^{-1}$. It should be noted that we do not follow the abundance of water in the disk or the composition of grains, we only assume particles outside the snow line to have larger tensile strength due to the presence of ice.
To be able to compare both simulations, we used the same 1~Myr old 0.09~$M_\odot$ gas disk around a solar mass star as initial condition.
The gas surface density profile of this disk was derived by a separate run of the disk evolution code. We used this gas surface density profile and a radially constant dust-to-gas ratio as initial condition for the simulations presented in this subsection. Apart from the level of turbulence, the input for both simulations is identical, the results are therefore completely independent of uncertainties caused by the choice of the infall model.

Results of the simulations are shown in Figure~\ref{fig:model:snapshots_enhancement}.
A one order of magnitude higher fragmentation velocity causes the maximum grain size to be about two orders of magnitude larger, which follows from Eq.~\ref{eq:model:st_max} since $\St_\text{max} \propto a_\text{max}$ in the Epstein regime (all particles in these simulations are small enough to be in the Epstein regime). This effect can be seen in Figure~\ref{fig:model:snapshots_enhancement}.
Particles outside the snow-line are therefore more strongly drifting inwards (because they reach larger Stokes numbers) where they are being pulverized as soon as they enter the region within the snow-line.

Strong drift outside the snow line and weaker radial drift inside the snow line cause the dust-to-gas ratio within the snow line to increase significantly (see bottom row of Figure~\ref{fig:model:snapshots_enhancement}): in the case of $\alpha = 10^{-3}$, the dust-to-gas ratio reaches values between 0.39 and 0.10 in the region from 0.2 to 1.9~AU.

Simulations for a less massive star (0.5~$M_\odot$) show the same behavior, however the increase in dust-to-gas ratio is not as high as for a solar mass star (dust-to-gas ratio of 0.27--0.20 from 0.2--4~AU).

\begin{figure}[thb]
  \centering
  \resizebox{\hsize}{!}{\includegraphics{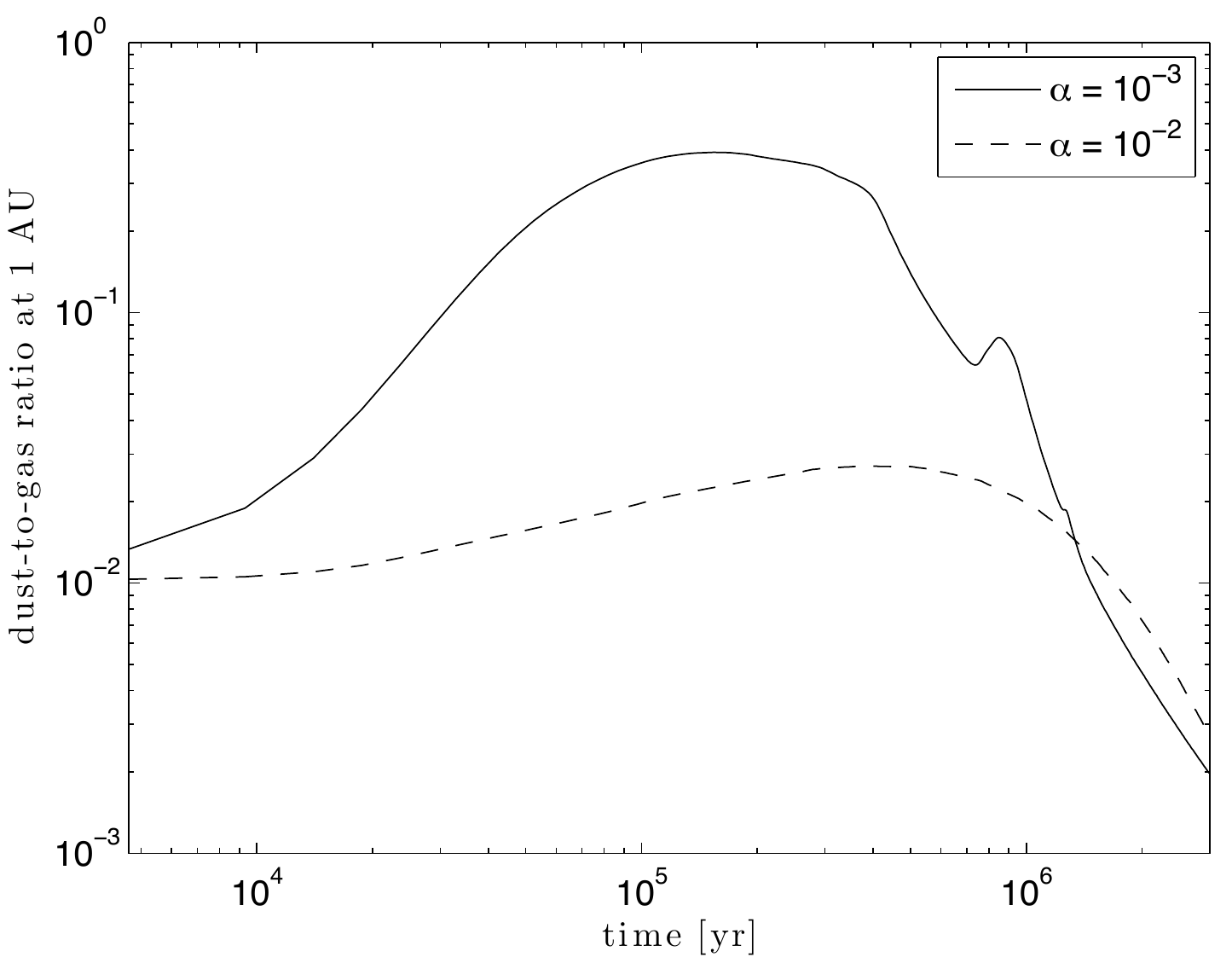}}
  \caption{Dust-to-gas ratio at a distance of 1~AU from the central star as a function of time for the case of low ($\alpha=10^{-3}$, solid line) and high ($\alpha=10^{-2}$, dashed line) turbulence.}
  \label{fig:model:d2g_at_1AU}
\end{figure}

This effect is not as significant in the case of stronger turbulence, where the maximum dust-to-gas ratio is around 0.027. The evolution of the dust-to-gas ratio at a distance of 1~AU from the star is plotted for both cases in Figure~\ref{fig:model:d2g_at_1AU} (the minor bump is an artifact of the initial condition).

The reason for this difference lies in the locations of the drift and fragmentation barriers. The approximate position of the fragmentation barrier is represented by the dot-dashed line in Figure~\ref{fig:model:snapshots_enhancement}. The radial drift barrier cannot be defined as sharply, however radial drift is strongest at a Stokes number of unity, which corresponds to the solid line in Figure~\ref{fig:model:snapshots_enhancement}. An increase of $\alpha$ by one order of magnitude lowers the fragmentation barrier by about one order of magnitude in grain size.

In the lower turbulence case, the fragmentation barrier lies close to $a_\mathrm{St=1}$. Most particles are therefore drifting inwards before they are large enough to experience the fragmentation barrier. Hence, the outer parts of the disk are significantly depleted in small grains.

In contrast to this case, fragmentation is the stronger barrier for grain growth throughout the disk in the high turbulence simulation (right column in Figure~\ref{fig:model:snapshots_enhancement}). It can be seen that particles of smaller sizes are constantly being replenished by fragmentation.

\begin{figure*}
  \centering
  \includegraphics[height=0.7\vsize]{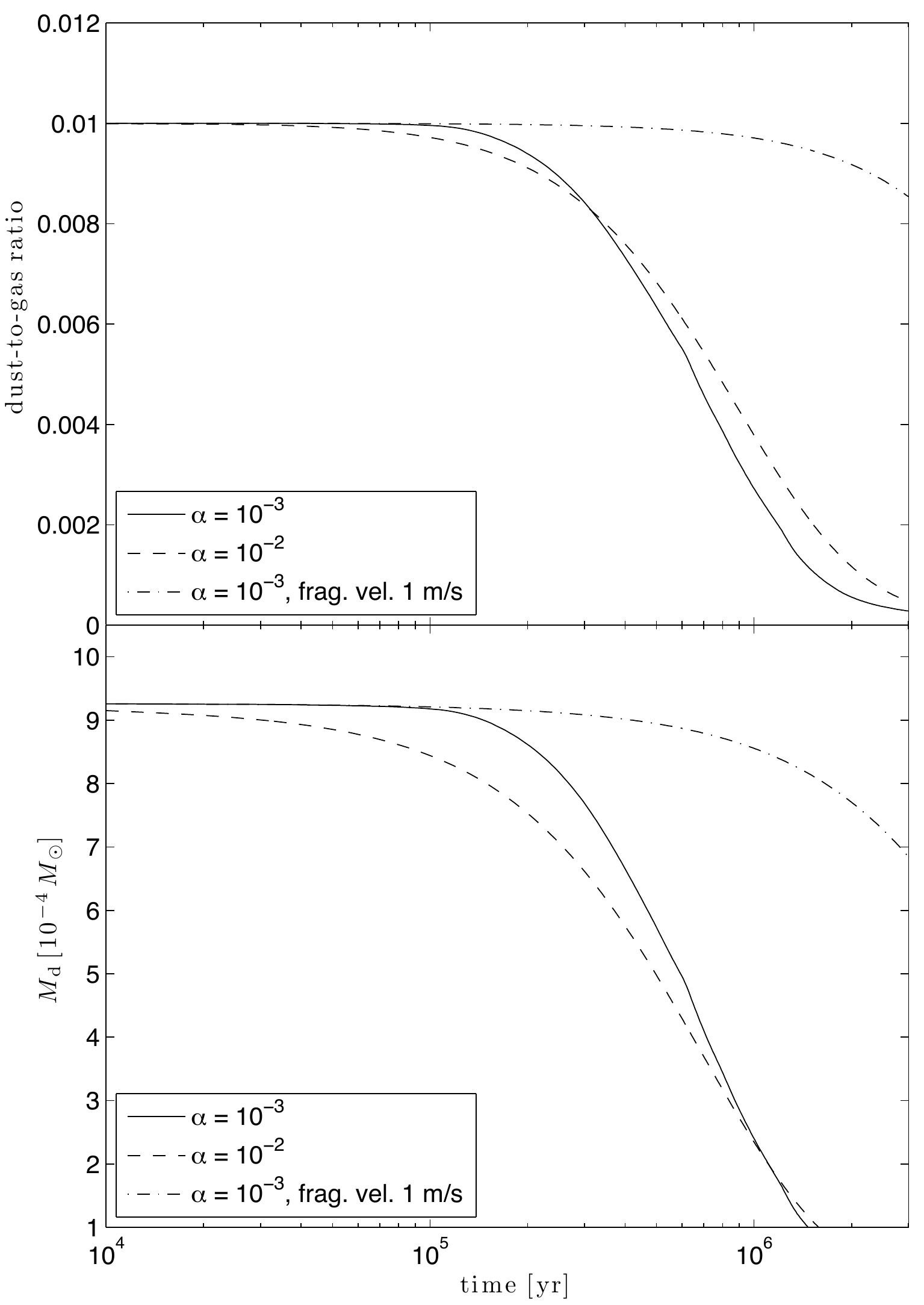}
  \caption{Evolution of the global dust-to-gas ratio (top panel) and the dust disk mass (bottom panel) for the simulations shown in Figure~\ref{fig:model:snapshots_enhancement}. The solid and dashed lines correspond to the low and high turbulence case, respectively. The dash-dotted line shows the evolution of the disk if a low fragmentation velocity is assumed throughout the disk.}
  \label{fig:model:dust_mass_comparison}
\end{figure*}

With these results in mind, the evolution of the disk mass (bottom panel of Figure~\ref{fig:model:dust_mass_comparison}) seems counter-intuitive: the mass of the high turbulence dust disk is decaying faster than in the low turbulence case. This can be understood by looking at the \emph{global} dust-to-gas ratio of the disks (top panel of Figure~\ref{fig:model:dust_mass_comparison}) which does not differ much in both cases. This means that the increased dust mass loss in the high turbulence disk is due to the underlying evolution of the gas disk. Particles in the high turbulence disk may have smaller Stokes numbers (causing drift to be less efficient), however the inward dragging by the accreting gas is stronger in this case.

To show how much the dust evolution depends on the fragmentation velocity, we included the case of a lower fragmentation velocity throughout the disk in Figure~\ref{fig:model:dust_mass_comparison}. It can be seen that the dust mass is retained at its initial value for much longer timescales.

\section{Discussion and conclusions}\label{sec:model:discussion}

We constructed a new model for growth and fragmentation of dust in circumstellar disks.
We combined the (size and radial) evolution of dust of \citetalias{Brauer:2008p215} with a viscous gas evolution code which takes into account the spreading and accretion, irradiation and viscous heating of the gas disk. The dust model includes the growth/fragmentation, radial drift/drag and radial mixing of the dust. We re-implemented and substantially improved the numerical treatment of the Smoluchowski equation of \citetalias{Brauer:2008p215} to solve for the combined size and radial evolution of dust in a fully implicit, un-split scheme (see Chapter~\ref{chapter:algorithm}).
In addition to that, we also included more physics such as relative azimuthal velocities, radial dependence of fragmentation critical velocities and the Stokes drag regime.
The code has been tested extensively and was found to be very accurate and mass-conserving (see Section~\ref{sec:algorithm:tests}).

We compared our results of grain growth in evolving protoplanetary disks to those of steady state disk simulations, similar to \citetalias{Brauer:2008p215}. In spite of many differences in details, we confirm the most general result of \citetalias{Brauer:2008p215}: radial drift and particle fragmentation set strong barriers to particle growth. If fragmentation is included in the calculations, then it poses the strongest obstacle for the formation of planetesimals. Very low turbulence ($\alpha\lesssim 10^{-5}$) and fragmentation velocities of more than a few m~s$^{-1}$ are needed to be able to overcome the fragmentation barrier in the case of turbulent relative velocities.

This model includes also the initial build-up phase of the disk, which is still a very poorly understood phase of disk evolution. We use the Shu-Ulrich infall model which represents a strong simplification. However, the following novel findings of this work do not or only weakly depend on the build-up phase of the disk:
\begin{itemize}

\item Apart from an increased dust-to-gas ratio \citepalias[as in][]{Brauer:2008p215}, other mechanisms such as streaming instabilities or a decreased temperature may be able to weaken this barrier by decreasing the efficiency of radial drift. We found that in simulations without fragmentation the radial drift efficiency needs to be reduced by 80\% to produce particles which crossed the meter-size barrier and are large enough to resist radial drift.

\item If the gas surface density is above a certain threshold (in our simulations about 140~g~cm$^{-2}$ at 1~AU or 330~g~cm$^{-2}$ at 5~AU, see Eq.~\ref{eq:model:sig_threshold}) then the drag force which acts on the dust particles has to be calculated according to the Stokes drag law, instead of the Epstein drag law. The drift barrier in this drag regime is shifting to smaller sizes for smaller radii (independent on the radial profile of the surface density) which means that pure radial drift can already transport dust grains over the drift barrier or at least to larger Stokes numbers even without simultaneous grain growth. In this case, the drift efficiency needs to be reduced only by about 25\% to produce large bodies.

\item If relative azimuthal velocities are included, then grains with $\St>1$ are constantly 'sand-blasted' by small grains (if they are present) which (in our prescription of fragmentation) causes erosion and stops grain-growth even in the case of low turbulence. Only decreasing the radial pressure gradient significantly weakens both relative azimuthal and radial velocities. Low turbulence and a small radial pressure gradient together are needed to allow larger bodies to form. These conditions may be fulfilled at the inner edge of dead zones (\citealp{Brauer:2008p212}; \citealp{Kretke:2007p697}; see also \citealp{Dzyurkevich:2010p11360}). Future work needs to investigate the disk evolution and grain growth of disks with dead zones.
Our prescription of fragmentation and erosion may also need rethinking since \citet{Wurm:2005p1855} and \citet{Teiser:2009p7785} find net-growth by high velocity impacts of small particles onto larger bodies.

\item Higher tensile strengths of particles outside the snow-line allows particles to grow to larger sizes, which are more strongly affected by radial drift. Particles therefore drift from outside the snow-line to smaller radii where they fragment and almost stop drifting (since their radial velocity is decreased by almost two orders of magnitude). This can cause an increase the dust-to-gas ratio inside the snow-line by more than 1.5 orders of magnitude.

\item The critical fragmentation velocity and its radial dependence (and to a lesser extent the level of turbulence) is a very important parameter determining if the dust disk is drift or fragmentation dominated. A drift dominated disk is significantly depleted in small grains and only a fragmentation dominated disk can retain a significant amount of dust for millions of years as is observed in T Tauri disks.
\end{itemize}

The following results depend on the build-up phase of the disk. However unless the collapse of the parent cloud is not inside-out or so fast to cause disk fragmentation, we expect only slight alteration of the results:
\begin{itemize}
\item Disk spreading causes small particles ($\lesssim 10$~$\mu$m) to be transported outward at radii of $\sim$60--190~AU even in 1~Myr old disks.

\item Small particles provided by infalling material are not effectively swept up by large grains if the bulk of the dust mass has already grown to larger sizes. 

\item In an inside-out build-up of circumstellar disks, grains growth is very fast (time scales of some 100~years) because densities are high and orbital time scales are small. Large grains are quickly lost due to drift towards the star if fragmentation is neglected. Fragmentation is firstly needed to keep grains small enough to be able to transport a significant amount of dust to large radii by disk spreading and secondly to retain it in the disk by preventing strong radial inward drift.

\end{itemize}

\ifthenelse{\boolean{chapterbib}}
{
    \clearpage
    \bibliographystyle{aa}
    \bibliography{/Users/til/Documents/Papers/bibliography}
}
{}
\chapter{The Algorithm}\label{chapter:algorithm}
\chapterprecishere{\centering \mbox{Based on \citealp*[][A\&A, 513, 79, Appendix]{Birnstiel:2010p9709}}}
\section{Numerical schemes}\label{sec:num_schemes}
In the next two sections, we will first discuss how the equations of radial evolution of gas (Eq. \ref{eq:model:ssd}) and dust (Eq. \ref{eq:model:dustequation}) as well as the coagulation/fragmentation of dust (Eq. \ref{eq:model:smolu}) are solved separately. In Section \ref{sec:algorithm:alg_both} we will then describe how this model treats the radial and the size evolution of dust in an un-split, fully implicit way.

\subsection{Advection-diffusion Algorithm}\label{sec:algorithm:alg_advdif}
To be able to also model both, the evolution of dust and gas implicitly, we constructed a scheme which solves a general form of an advection-diffusion equation,
\begin{equation}
\ddel{\N}{t} + \ddel{}{x} \left( \N \cdot u\right) -  \ddel{}{x} \left[ h \cdot D_\text{d} \cdot \ddel{}{x} \left(g\cdot \frac{\N}{h}\right) \right] = K + L \cdot \N
\label{eq:algorithm:adv_diff}
\end{equation}
which can be adapted to both, Eq. \ref{eq:model:ssd} and Eq. \ref{eq:model:dustequation} by proper choice of parameters $u$, $D_\text{d}$, $g$, $h$, $K$ and $L$.

We use a flux-conserving donor-cell scheme which is implicit in $\N$.
The time derivative in Eq. \ref{eq:algorithm:adv_diff}, written in a discretized way becomes
\begin{equation}
\ddel{N}{t} \hat{=} \frac{\N^{i+1}_n - \N^{i}_n}{t_{i+1}-t_{i}}
\end{equation}
where $i$ denotes time-dimension and $n$ denotes space-dimension.

The advective part is discretized as
\begin{equation}
\ddel{}{x} \left( \N \cdot u\right) = \frac{F_{n+\half}^{i+1} \cdot S_{n+\half}}{V_n} - \frac{F_{n-\half}^{i+1} \cdot S_{n-\half}}{V_n}
\label{eq:algorithm:advection_num}
\end{equation}
where $F^{i+1}_{n+\half}$ and $S_{n+\half}$ are the future flux and the surface between cell $n$ and cell $n+1$ and $V_n$ is the volume of cell $n$.
The advective interface fluxes can then be written as
\begin{eqnarray}
F_{n+\half}^{i+1} &=& \N_n \cdot \max(0,u_{n+\half}) + \N_{n+1} \cdot \min(0,u_{n+\half}) \\
F_{n-\half}^{i+1} &=& \N_{n-1} \cdot \max(0,u_{n-\half}) + \N_n \cdot \min(0,u_{n-\half}) 
\end{eqnarray}
The diffusive interface flux becomes
\begin{eqnarray}
F_{d,n+\half}^{i+1} &=& D_{\text{d},n+\half} \: h_{n+\half} \: \frac{\frac{g_{n+1}}{h_{n+1}}\N_{n+1} - \frac{g_{n}}{h_{n}}\N_{n}}{x_{n+1}-x_{n}}\\
F_{d,n-\half}^{i+1} &=& D_{\text{d},n-\half} \: h_{n-\half} \: \frac{\frac{g_{n}}{h_{n}}\N_{n} - \frac{g_{n-1}}{h_{n-1}}\N_{n-1}}{x_{n}-x_{n-1}}
\end{eqnarray}
Working out these equations and separating the values of $\N$ leads to
\begin{equation}
\N_n^i = A_n \cdot \N_{n-1}^{i+1} + B_n \cdot \N_{n}^{i+1} + C_n \cdot \N_{n+1}^{i+1} + D_n
\label{eq:algorithm:matrix_r}
\end{equation}
with the coefficients
\begin{equation}
\begin{array}{lll}
A_n &=& -\frac{\Delta t}{V_n} \left( \max(0,u_{n-\half}) \cdot S_{n-\half} + \frac{D_{\text{d},n-\half} \cdot h_{n-\half} \cdot S_{n-\half} \cdot g_{n-1}}{(x_n-x_{n-1}) h_{n-1}} \right)\\
B_n &=& 1- \Delta t  L_n + \frac{\Delta t}{V_n} \Bigg( \max(0,u_{n+\half}) \cdot S_{n+\half} \\
&&- \min(0,u_{n-\half})\cdot S_{n-\half}\\
&&+\frac{D_{\text{d},n+\half}\cdot h_{n+\half} \cdot S_{n+\half} \cdot g_n}{(x_{n+1}-x_n) h_n} + \frac{D_{\text{d},n-\half} \cdot h_{n-\half}\cdot S_{n-\half} \cdot g_n}{(x_n-x_{n-1}) h_n} \Bigg)\\
C_n &=& \frac{\Delta t}{V_n} \left( \min(0,u_{n+\half})\cdot S_{n+\half} - \frac{D_{\text{d},n+\half}\cdot h_{n+\half}\cdot S_{n+\half}\cdot g_{n+1}}{(x_{n+1}-x_n) h_{n+1}} \right)\\
D_n &=& -\Delta t \cdot K_n.
\end{array}
\label{eq:algorithm:coefficients_r}
\end{equation}
Eq. \ref{eq:algorithm:matrix_r} can now be solved by any matrix-solver, but since it is a tri-diagonal matrix, the fastest analytical way to solve it is by \emph{forward elimination/backward substitution}.

It should be noted that Eq. \ref{eq:algorithm:adv_diff} is implicit only in $\N$ which means that the equations we solve are only implicit in the surface density. In the case of the viscous accretion disk, described by Eq. \ref{eq:model:ssd}, we face the problem that also the turbulent gas viscosity $\nug$ depends on the temperature which (in the case of viscous heating) depends on the surface density. This can cause numerical instabilities to develop.

To stabilize the code, we use a scheme which estimates the temperature in several predictor steps. The actual time step is then done with the predicted temperature.

\subsection{Coagulation Algorithm}\label{sec:algorithm:alg_coag}
Discretizing Eq. \ref{eq:model:smolu} on a mass grid $m_i$ gives
\begin{equation}
\ddel{}{t} n_k(r,z) = \sum_{ij} M_{ijk} \,n_i(r,z)\, n_j(r,z),
\end{equation}
where the dust particle number density is
\begin{equation}
n_i(r,z) = \int_{m_{i-1/2}}^{m_{i+1/2}} n(m,r,z)\, \dx m
\end{equation}
If we assume that the coagulation and fragmentation kernels are constant in $z$ and that the vertical distribution of grains is a Gaussian with a scale height according to Eq. \ref{eq:model:h_dust},
\begin{equation}
n_k(r,z) = \frac{N_k(r)}{\sqrt{2\pi}h_k(r)} \cdot \exp\left(-\frac{z^2}{2h_k(r)^2} \right),
\end{equation}
we can vertically integrate the coagulation/fragmentation equation.
\begin{equation}
\begin{split}
\ddel{}{t} N_k(r)
&= \int_{-\infty}^\infty \dot n_k \: \text{d}z\\
&= \sum_{ij} M_{ijk} \: N_i(r) \: N_j(r) \: \frac{1}{2\: \pi \: h_i \: h_j} \times\\
&\qquad\times\int_{-\infty}^\infty \exp{\left(-\frac{h_i^2+h_j^2}{2h_i^2h_j^2} \cdot z^2\right)} \text{d}z\\
&= \sum_{ij}  \widetilde M_{ijk} \: N_i(r) \: N_j(r),
\end{split}
\label{eq:algorithm:smolu_mdiscrete}
\end{equation}
where
\begin{equation}
\widetilde M_{ijk} = \frac{1}{\sqrt{2\:\pi\:\left(h_i^2+h_j^2\right)}}\: M_{ijk}.
\end{equation}
Discretizing Eq. \ref{eq:algorithm:smolu_mdiscrete} also in radial direction and rewriting it in terms of the quantity
\begin{equation}
\N_{kl} = N_k(r_l)\cdot r_l
\end{equation}
yields
\begin{equation}
\ddel{}{t} \N_{kl} = {\displaystyle \sum_{ij}}  \frac{\widetilde M_{ijk}}{r_l} \cdot \N_{il}\cdot \N_{jl} := {S}_{kl}.
\label{eq:algorithm:smolu_in_u}
\end{equation}
where the vector $\mathbf{S}=\{S_{kl}\}_{k=1,m}$ is the source function for each of the $m$ mass bins.

The numerical change of $\N_{kl}$ within a time step $\Delta t = t_i-t_{i-1}$ is $\Delta \N_{kl} = \N_{kl}^{i+1}-\N_{kl}^{i}$. The time-discretized version of Eq. \ref{eq:algorithm:smolu_in_u} then becomes (omitting second order terms)
\begin{equation}
\begin{split}
\frac{\Delta \N_{kl}}{\Delta t} &= \sum_{ij}  \frac{\widetilde M_{ijk}}{r_l} \: \left(\N_{il} + \Delta \N_{il}\right) \: \left(\N_{jl} + \Delta \N_{jl}\right)\\
&=\sum_{ij} \frac{\widetilde M_{ijk}}{r_l} \: \left( \N_{il} \N_{jl} + \Delta \N_{il} \N_{jl} + \N_{il} \Delta\N_{jl} + \cancel{\Delta \N_{il} \Delta \N_{jl}}\right).
\end{split}
\end{equation}
Since the first term on the right hand side is the explicit source function, we can write
\begin{equation}
\begin{split}
\frac{\Delta \N_{kl}}{\Delta t} &= S_{kl} + \sum_{ij} \frac{\widetilde M_{ijk} + \widetilde M_{jik}}{r_l}\cdot \N_{jl} \cdot \Delta \N_{il}.
\end{split}
\end{equation}
Using the vectors $\Delta \mathbf{N}= \{\Delta \N\}_{k=1,n_m}$ and $\mathbf{N}= \{\N\}_{k=1,n_m}$, this can be rewritten in matrix notation,
\begin{equation}
\left( \frac{\mathbbm{1}}{\Delta t} -\mathbf{J} \right) \Delta \mathbf{N} = \mathbf{S}.
\label{eq:algorithm:coag_matrix_equation}
\end{equation}
Were
\begin{equation}
J_{ki} = \sum_{j} \frac{\widetilde{M}_{ijk} + \widetilde{M}_{jik} }{r_l}\:\N_{jl}
\end{equation}
denotes the Jacobian of the source function and $\mathbbm{1}$ the unity matrix.
The solution for the future values can now be derived by inverting the matrix in Eq. \ref{eq:algorithm:coag_matrix_equation},
\begin{equation}
\begin{split}
\mathbf{N}^{i+1} &= \mathbf{N}^{i}+\Delta \mathbf{N} \\
&= \mathbf{N}^{i}+ \left( \frac{\mathbbm{1}}{\Delta t} -\mathbf{J} \right)^{-1} \cdot \mathbf{S}\\
\end{split}
\end{equation}

\subsection{The fragmentation matrix}
The discretization of the coagulation/fragmentation equation contains some pit-falls, which are mostly related to numerical rounding errors: already the growth from 0.1~$\mu$m to just a decimeter sized object spans a mass range of 18 orders of magnitude, which is beyond the numerical precision of a \emph{double precision} floating point number. Problems arise for example if a large body grows by accumulating very small grains. The numerical scheme should thus be written in a way where addition of very different sized numbers or subtraction of almost-equal sized numbers are avoided. In \citet{Brauer:2008p215}, the coagulation matrix was rewritten in order to conserve the total mass, in this subsection, we will show, how the fragmentation matrix needs to be handled.
 
The fragmentation part of the Smoluchowski equation is written in a descretized way as
\begin{equation}
  \dot n_k = \frac{1}{2} \sum_{ij} \frac{m_i+m_j}{m_k} L_{ij} n_i n_j S_{kij} - \sum_i L_{ij} n_k n_i,
  \label{eq:algorithm:fragmentation}
\end{equation}
where $L_{ij}$ is the fragmentation kernel, $S_{kij}$ is the distribution of fragments to bin $k$, which are produced by collisions of $m_i$ and $m_j$. The fraction $(m_i+m_j)/m_k$ converts the source terms from number densities to masses since we want to conserve the total mass.
We again face the problem of numerically subtracting large, almost equal numbers. To be as accurate as possible, it is best to subtract these numbers analytically, if possible, since the computer variables we use have ``only'' 16 decimals of precision.
To be able to do that, we have to rewrite Eq.~\ref{eq:algorithm:fragmentation} keeping in mind that the fragmentation matrix is symmetric ($L_{ij}=L_{ji}$)
\small{
\begin{equation}
\begin{split}
\dot n_k &= \frac{1}{2} \sum_{ij} \frac{m_i+m_j}{m_k} L_{ij} n_i n_j S_{kij} - \sum_i L_{ij} n_k n_i\\
&= \frac{1}{2} \sum_{ij} \frac{m_i+m_j}{m_k} L_{ij} n_i n_j S_{kij} - \frac{1}{2}\left( \sum_i L_{ij} n_k n_i + \sum_j L_{ij} n_k n_j \right)\\
&= \frac{1}{2} \sum_{ij} \frac{m_i+m_j}{m_k} L_{ij} n_i n_j S_{kij} - \frac{1}{2}\left( \sum_{ij} L_{ij} n_k n_i \delta_{kj} + \sum_{ij} L_{ij} n_k n_j \delta_{ki} \right)\\
&= \sum_{ij} L_{ij} n_i n_j \left[ \frac{1}{2} \frac{m_i+m_j}{m_k} \underbrace{\left(S_{kij} -  \frac{m_k}{m_i+m_j} \left( \delta_{ki} + \delta_{kj}\right) \right)}_{\tilde{S}_{kij}}\right]\\
\end{split}
\end{equation}
}

Now in our prescription of cratering, masses $m_\mathrm{big}$ and $m_\mathrm{break}$ collide. $m_\mathrm{break}$ excavates $\chi$ times its own mass from $m_\mathrm{big}$ and fragments itself if it is one order of magnitude smaller in mass. The total mass of fragments is $m_\mathrm{frag} = (1+\chi)\:m_\mathrm{break}$, which is distributed according to the power-law described in Eq.~\ref{eq:model:frag_powerlaw} while the larger mass looses $\chi m_\mathrm{break}$, which leaves
\begin{equation}
  m_\mathrm{rest} = m_\mathrm{big}- \chi m_\mathrm{break}.
\end{equation}
Numerically, $m_\mathrm{rest}$ lies somewhere between $m_\mathrm{big}$ and the next smaller bin, which we will denote as $m_\mathrm{big-1}$. The grid distance is defined as
\begin{equation}
  \Delta m = m_\mathrm{big}- m_\mathrm{big-1}.
\end{equation}
Therefore, $m_\mathrm{rest}$ will be redistributed into the adjacent grid points according to the Podolak prescription. We need to take care since if $\chi$ or the grid resolution is is large, then $m_\mathrm{rest}$ might be smaller than $m_\mathrm{big-1}$. For typical cases of $\chi$ and the grid resolution, this is rarely the case. $(1-\epsilon) \cdot m_\mathrm{rest}$ will be added to $m_\mathrm{big}$ and $\epsilon \cdot m_\mathrm{rest}$ will be added to $m_\mathrm{big-1}$, where $\epsilon$ is defined as
\begin{equation}
  \epsilon = \frac{\chi \; m_\mathrm{break}}{\Delta m}.
\end{equation}
Now in the cases where $m_k = m_\mathrm{big}$ we can write the contributions to $m_\mathrm{big}$, normalized to $m_\mathrm{total} = m_\mathrm{big} + m_\mathrm{break}$ as
\begin{equation}
  \begin{split}
\tilde{S}_{kij} &= (1-\epsilon) \frac{m_\mathrm{rest}}{m_\mathrm{total}} - \frac{m_\mathrm{big}}{m_\mathrm{total}}\\
&= \frac{m_\mathrm{big}-\chi m_\mathrm{break}-\epsilon
m_\mathrm{rest}-m_\mathrm{big}}{m_\mathrm{total}}\\ &= -\chi\frac{m_\mathrm{break}}{m_\mathrm{totla}} - \epsilon \frac{m_\mathrm{rest}}{m_\mathrm{total}}.
\end{split}
\end{equation}
By separating of this special case in the code, the mass is conserved several orders of magnitude better than without rewriting it.

\subsection{Fully implicit scheme for radial motion and coagulation}\label{sec:algorithm:alg_both}

\begin{figure}[t]
  \centering
  \resizebox{0.65\hsize}{!}{\includegraphics{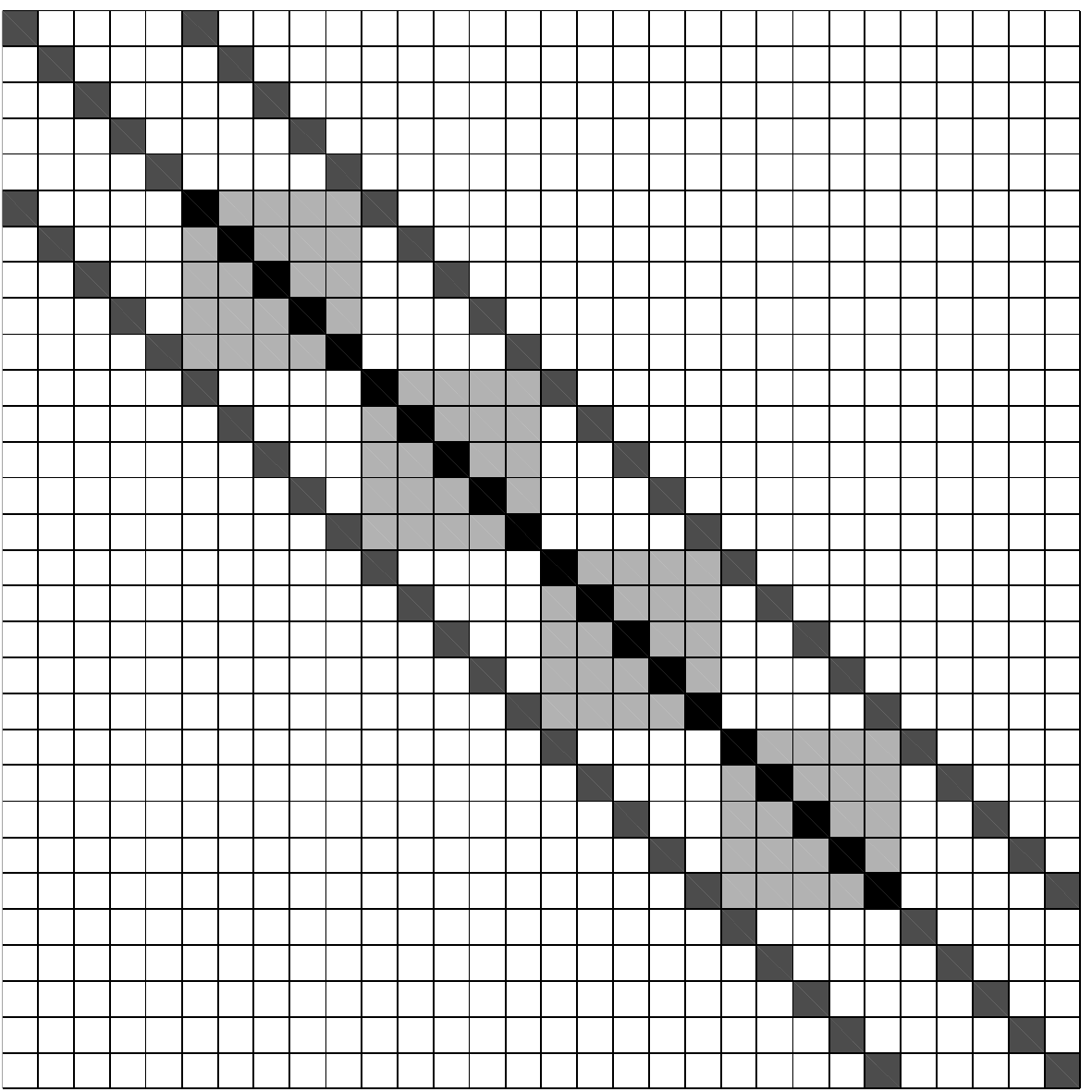}}
  \caption{Pictographic representation of the matrix on the left hand side of Eq.~\ref{eq:algorithm:combined_matrix} with six radial and five mass grid points. White elements are zero, dark grey elements contain contributions from radial transport of dust ($\mathbf{J}$), light gray elements contain contributions from the coagulation/fragmentation ($\mathbf{\hat M}$), and black elements contain both contributions. The upper left and lower 5 rows represent the boundary conditions where coagulation/fragmentation is not taken into account. The matrix in the simulations would typically have a size of 15000$\times$15000.}
  \label{fig:algorithm:matrix}
\end{figure}

To be able to solve the radial motion and the Smoluchowski equation fully implicitly, we rewrite Equation \ref{eq:algorithm:matrix_r} as
\begin{equation}
\mathbf{M} \cdot \mathbf{N}^{i+1}= \mathbf{N}^i- \mathbf{D},
\label{eq:algorithm:radial1}
\end{equation}
where $\mathbf{M}$ is the tri-diagonal matrix with entries $A,B$ and $C$  and source term $\mathbf{D}$ which are given by Eq. \ref{eq:algorithm:coefficients_r}.
$\mathbf{M}$ is now rewritten by separating off the diagonal terms and by dividing by $\Delta t$,
\begin{equation}
\mathbf{M} = \Delta t\cdot \mathbf{\hat M} + \mathbbm{1}
\end{equation}
which brings Eq. \ref{eq:algorithm:radial1} in a form similar to Eq. \ref{eq:algorithm:coag_matrix_equation},
\begin{equation}
\left( \frac{\mathbbm{1}}{\Delta t} + \mathbf{\hat M} \right) \cdot \Delta \mathbf{N} = -\mathbf{\hat M} \cdot \mathbf{N}^i - \mathbf{D}/\Delta t
\end{equation}
The coagulation/fragmentation is now determined by the matrix $\mathbf{J}$ and the source vector $\mathbf{S}$ and similarly, the radial motion is determined by matrix
$\mathbf{\hat M}$ and source vector
\begin{equation}
\mathbf{R} = -\mathbf{\hat M} \cdot \mathbf{N}^i - \mathbf{D}/\Delta t.
\end{equation}
This allows us to combine both operators into one matrix equation,
\begin{equation}
\left( \frac{\mathbbm{1}}{\Delta t} + \mathbf{\hat M} -\mathbf{J} \right) \cdot \Delta \mathbf{N} = \mathbf{R} + \mathbf{S}.
\label{eq:algorithm:combined_matrix}
\end{equation}
In principle, to solve for the vector $\Delta \mathbf{N}$, the matrix on the
left hand side of Eq.~\ref{eq:algorithm:combined_matrix} has to be inverted. This is a
numerically challenging task since the inverse matrix in our simulations would have about 150--500 million elements. The matrix on the left hand side of Eq.~\ref{eq:algorithm:combined_matrix}, however is a sparse matrix (schematically depicted in Figure~\ref{fig:algorithm:matrix}).

We can therefore solve Eq.~\ref{eq:algorithm:combined_matrix} by an iterative incomplete LU decomposition solver for sparse matrices provided by the Sixpack library\footnote{available from \url{www.engineers.auckland.ac.nz/~snor007}} of S. E. Norris.

\begin{figure}[t]
  \centering
  \resizebox{0.7\hsize}{!}{\includegraphics{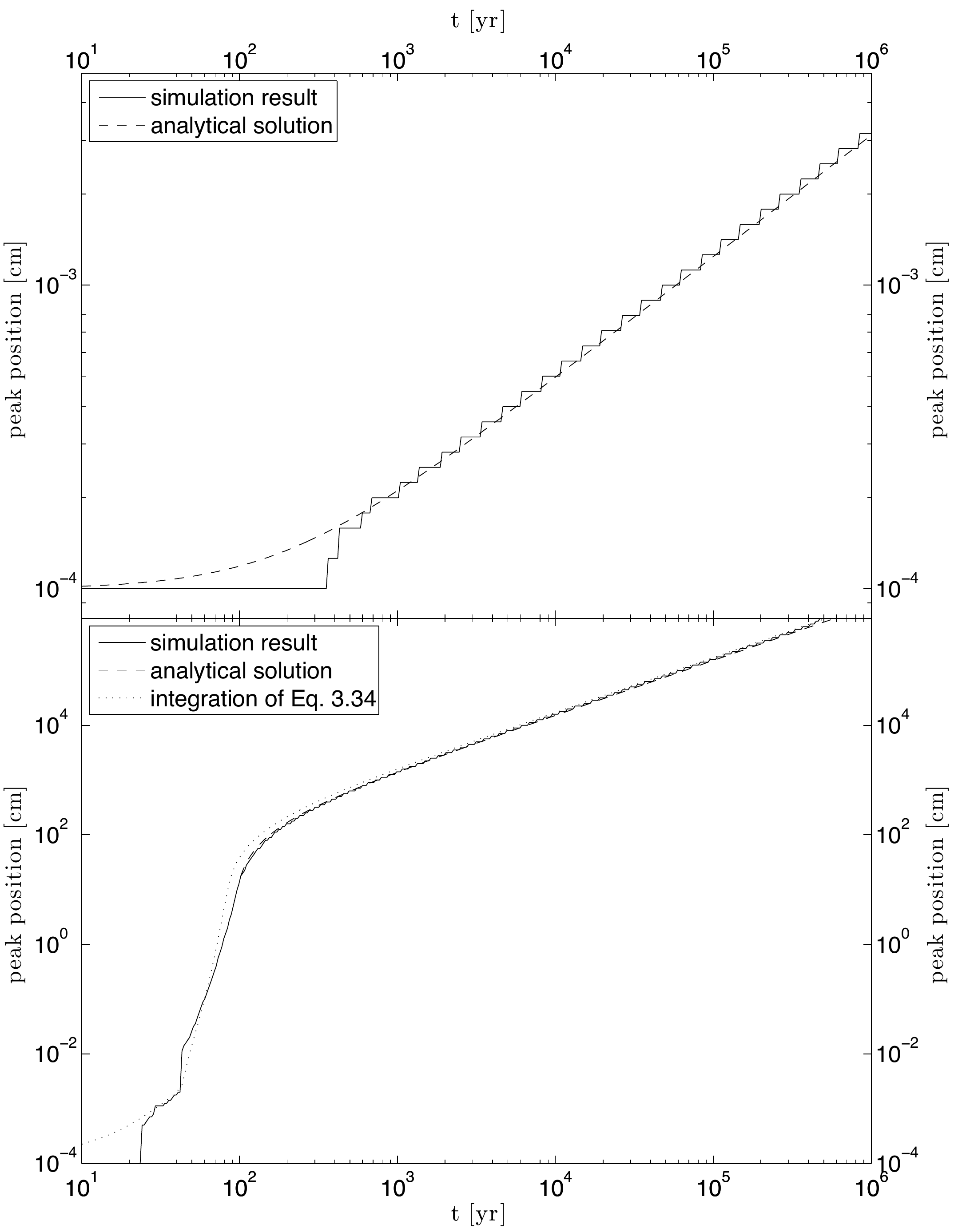}}
  \caption{Comparison between simulation result and analytical solution for growth of equal sized particles. The solid lines denote the position of the peak of the grain size distribution. In the top panel, only Brownian motion is considered as source of relative particle velocities, in the bottom panel, turbulent relative velocities are considered as well. The parameters of these simulation are $T=196$~K, $\rhos$=1.6~$\text{g/cm}^3$, $\Siggas=18$~$\text{g/cm}^2$, $\Sigdust=0.18\text{~g/cm}^2$ and $\alpha=10^{-3}$.}
  \label{fig:algorithm:growthtest}
\end{figure}

\section{Test cases}\label{sec:algorithm:tests}
\subsection{Growth timescales}
To check if the numerical implementation presented above accurately solves Eq. \ref{eq:algorithm:combined_matrix}, we compare results of the simulation to some analytical solutions:
The growth rates of particles can be approximated if we assume the grain size distribution to be a delta function and the sticking probability to be unity. The increase of mass per collision is then given by the mass of the particle, $m$ divided by the time between two collisions, $\tau$, which can be written as

\begin{equation}
\frac{\Del m}{\Del t} = \frac{m}{\tau}.
\end{equation}
Using $\Del m={4\pi} \rhos a^2 \Del a$ and $\tau = m/(\rhodust \;\sigma \; \Delta u)$, we derive
\begin{equation}
\frac{\Del a}{\Del t} = \frac{\rhodust}{\rhos} \; \Delta u,
\label{eq:algorithm:simple_growth}
\end{equation}
where $\Delta u$ is the relative velocity, $\rhodust$ the dust density, $\rhos$ the solid density of the dust particles and $\sigma = 4\pi a^2$ the cross section of the collision.

\begin{figure}[t]
  \resizebox{\hsize}{!}{\includegraphics{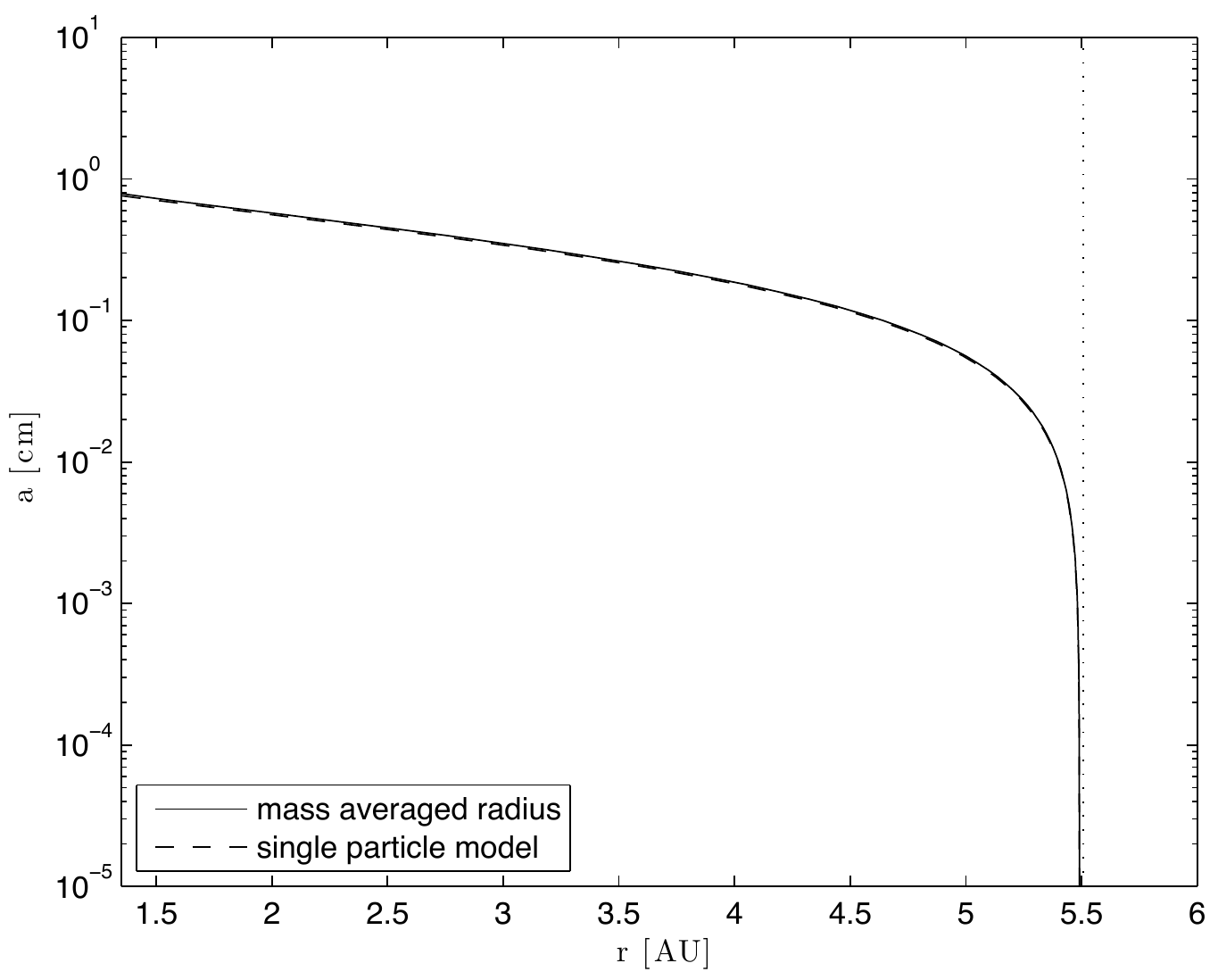}}
  \caption{Test case for only radial drift, without coagulation. The solid line denotes the mass-averaged position of the radial distribution of grains for each grain size after $10^3$~years. The dashed line is the expected solution for a single particle, the dotted line denotes the initial radius of the particles.}
  \label{fig:algorithm:motiontest}
\end{figure}

With this formula, we can estimate the growth time scales if the relative velocities of the dust grains is given. For equal sized particles and Brownian motion, we get
\begin{equation}
\Delta u_\text{BM} = \sqrt{\frac{16 \; \kb T}{\pi \; m}},
\end{equation}
for turbulent motion, the relative velocities are \citep[see][]{Ormel:2007p801}
\begin{equation}
\Delta u_\text{TM} \approx \left\{ \begin{array}{ll}
\csound\sqrt{2\;\alpha \; \St}      &   \text{for }\St \ll 1\\
\csound \sqrt{\frac{2 \alpha}{\St}} &   \text{for }\St \gg 1\\
\end{array}\right..
\end{equation}
Integrating Eq. \ref{eq:algorithm:simple_growth} gives
\begin{equation}
a(t) = \left( \frac{5}{2} \frac{\rho_d}{\rho_s \: \pi} \sqrt{\frac{12 \kb T}{\rhos}} \left( t - t_0\right) + a_0^{\frac{5}{2}} \right)^{\frac{2}{5}}
\label{eq:algorithm:BMgrowth}
\end{equation}
for Brownian motion and
\begin{equation}
a(t) \approx \left\{\begin{array}{ll}
a_0\cdot\exp\left(\frac{\Siggas}{\Sigdust} \frac{\pi}{\sqrt{2}} \cdot(t-t_0)\right) &   \text{for }\St \lesssim 1\\
\frac{\Sigdust \Ok}{2 \rhos \sqrt{\pi}} \cdot (t - t_0) + a_0       &   \text{for }\St > 1\\
\end{array}\right.
\end{equation}
for relative velocities due to turbulent motion.

A comparison between analytical solution and simulation result for Brownian motion growth is shown in the top panel of Figure \ref{fig:algorithm:growthtest}. It can be seen that the position of the peak of the grain size distribution (solid line in Figure \ref{fig:algorithm:growthtest}) follows the analytical result of Eq. \ref{eq:algorithm:BMgrowth}.

A similar comparison for the case of relative velocities due to turbulent motion is not as straight-forward since both the turbulent relative velocities and the dust scale height (Eq. \ref{eq:model:h_dust}) are sub-divided into several regimes. We therefore integrated Eq. \ref{eq:algorithm:simple_growth} numerically for the case of Brownian motion and turbulent relative velocities; the results are shown in the bottom panel of Figure \ref{fig:algorithm:growthtest}. As before, we see that -- after the initial condition is overcome -- the simulation result follows closely the mono-disperse approximation. For grains larger than $\St=1$, the analytical solution is also plotted in the bottom panel of Fig. \ref{fig:algorithm:growthtest}, almost coinciding with the simulation results.

\subsection{Radial motion}
The radial motion of dust particle was tested in a similar fashion: starting from a grain distribution at a radius of 5.5~AU, we let the particles drift (taking the gas drag and the radial drift into account, see Eq. \ref{eq:model:u_r_dust}) without coagulation. We compare the results to results of a numerical integration of the equation of motion for a single particle. The results are shown in Fig. \ref{fig:algorithm:motiontest}.

We find that the size distribution behaves as expected: small particles are well coupled to the gas, they almost retain their initial position since the radial motion due to gas drag is in the order of 0.01~AU. Larger particles, having a larger Stokes number drift towards the star on shorter timescales. Particles of a few centimeters (corresponding to \St=1) are already lost after about 700~years.

The mass in all test cases was found to be conserved on the order of $10^{-11}$\% of the initial value.

\subsection{Temperature structure}
We also compared the results of our temperature equation (cf.~Eq.~\ref{eq:model:t_mid}) to a non-isothermal vertical disk profile, which will be described in the following.
In the non-isothermal case, the source terms are $z$-dependent:
\begin{equation}
\begin{split}
q_\text{+}^\text{visc}(z) &= \frac{9}{4}  \alpha \frac{\kb T(z)}{\mu \mpr} \rho(z) \: \Ok \\
q_\text{+}^\text{irrad}(z) &= \half \rho(z) \: \kappla(z) \: \frac{L_\star}{4 \pi \: R^2}  \cdot \exp\left(-\frac{\tau(z)}{\varphi}\right)
\end{split}
\label{eq:algorithm:qplus}
\end{equation}
Where $\tau$ is the Rosseland optical depth and $\varphi$ the flaring angle, which is usually also dependent on radius. Here we take $\varphi$ to be a constant with the value $0.05$. The factor of $1/2$ in Eq. \ref{eq:algorithm:qplus} comes from the fact that one half of the radiation is radiated away from -- the other half into the disk \citep[see][]{Dullemond:2002p399}.

\begin{figure*}
   \centering
   \includegraphics[width=\columnwidth]{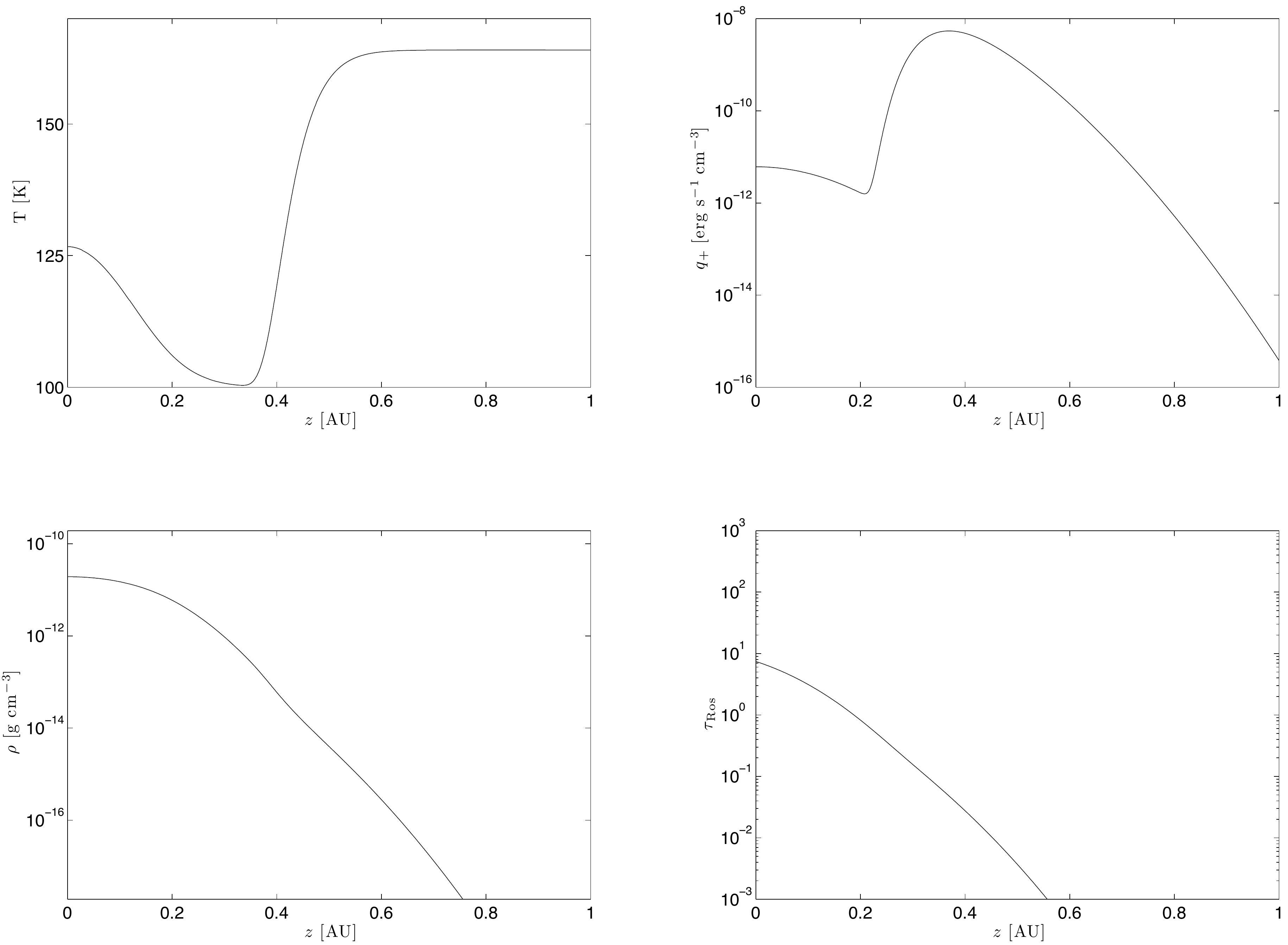}
   \caption{Vertical Structure at 2 AU with a surface density of 94 g cm$^{-2}$. Shown is the temperature profile (top left panel), the heat sources (top right panel), the gas density structure (bottom left) and the Rosseland optical depth (bottom right).}
   \label{fig:algorithm:vertical_struct}
\end{figure*}

Given the source terms, we can solve for the intensity by using the moment equations and the Eddington approximation: if the intensity $I_\nu$ is nearly isotropic,
\begin{equation}
I(\tau,\mu) = a(\tau) + b(\tau) \: \mu
\end{equation}
the moment equations become
\begin{eqnarray}
J &=& \half \int_{-1}^1 I d\mu = a \\
H &=& \half \int_{-1}^1 I \cdot \mu d\mu = \frac{b}{3} \\
K &=& \half \int_{-1}^1 I \cdot \mu^2 d\mu = \frac{a}{3} 
\end{eqnarray}
which leads to the closure relation
\begin{equation}
K = \frac{1}{3} J
\end{equation}
and from the second moment follows
\begin{equation}
\ddel{K}{\tau} = H = \frac{1}{3} \ddel{J}{\tau}.
\label{eq:algorithm:closure}
\end{equation}
Since \citep[see][A.9]{Dullemond:2002p399}
\begin{equation}
\ddel{H}{z} = \frac{q_+(z)}{4 \pi},
\end{equation}
$H$ can be calculated by integrating over the sources starting from the boundary condition $H(0) = 0$. Knowing $H$, we can integrate Eq. \ref{eq:algorithm:closure} to get $J$. The necessary boundary condition at $\tau = 0$ can be determined in the \textit{two-stream-approximation} \citep[see][]{Rybicki:1991p12115}. The temperature follows from \citet{Dullemond:2002p399}, Eq.~A.13, where we use $\kappla ( (J \pi / \sigb )^{1/4} )$ instead of $\kappa_J$. It should also be noted that the last term in A.13 of \citet{Dullemond:2002p399} should be $q/(4\pi \rho)$.

With a given temperature, the vertical hydrostatic equilibrium can be calculated. From $P=\rho \: \csound^2$ follows
\begin{equation}
\frac{\del \left(\rho T\right)}{\del z} = -\rho z \Ok^2 \cdot \frac{\mu \mpr}{\kb},
\end{equation}
which can be done numerically by using $\rho_1 = 1$ and by evaluating the right hand side as the average value between the grid centers. The resulting profile can then be normalized to a given surface density.

\begin{figure}
   \centering
   \includegraphics[width=0.65\columnwidth]{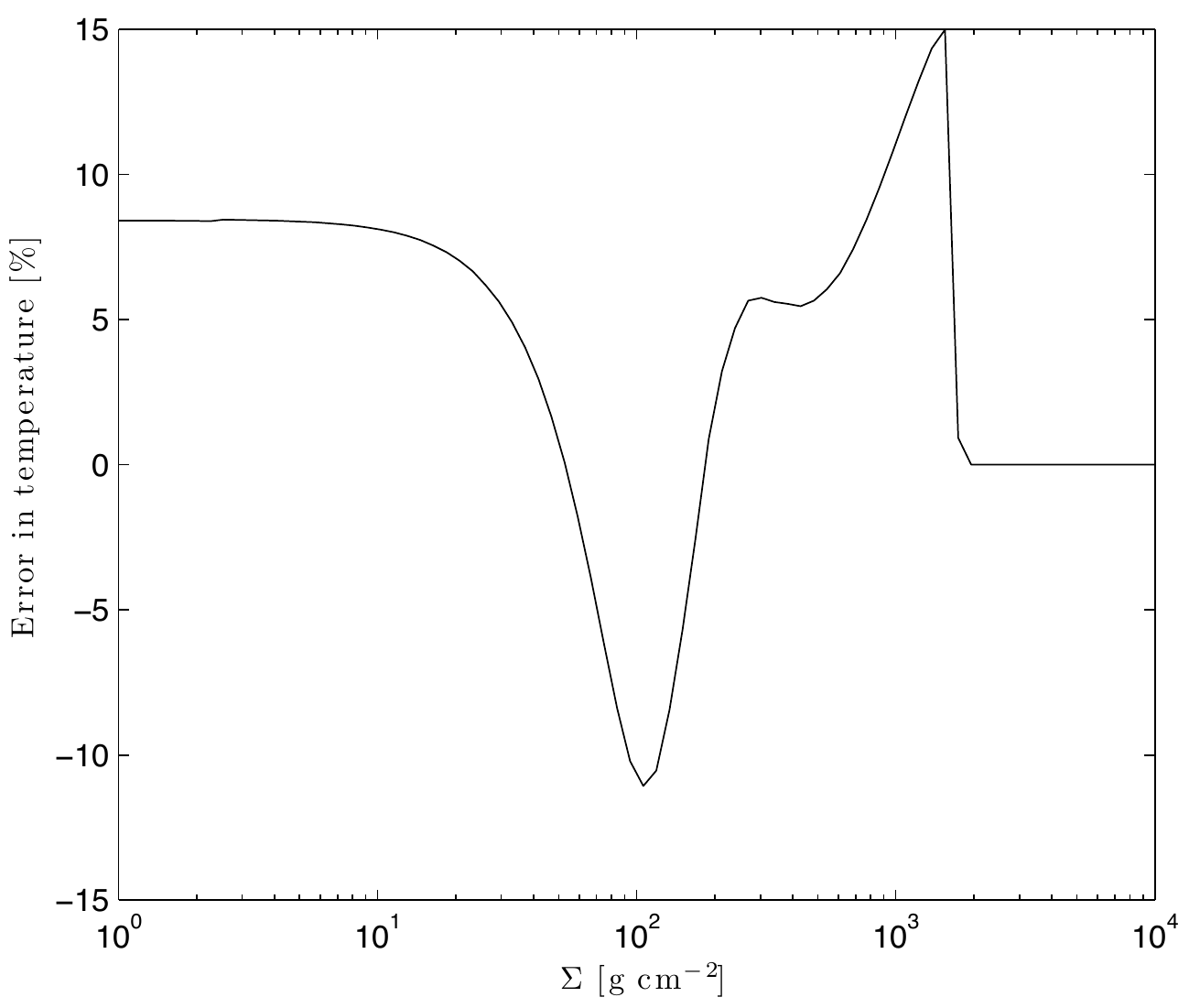}
   \caption{Difference between the approximate mid-plane temperature, calculated from Eq.~\ref{eq:model:t_mid} and the non-isothermal vertical structure, as described in this subsection.}
   \label{fig:algorithm:t_comparison}
\end{figure}

The above procedure has to be done iteratively since the temperature profile depends on the density profile and vice versa. We repeat the $T$ and $\rho$ calculation until the temperature at any point does not change by more than 1\%. An exemplary result for the vertical structure of the disk is shown in Fig.\ref{fig:algorithm:vertical_struct}. An comparison between the results of Eq.~\ref{eq:model:t_mid} and this non-isothermal case is shown in Fig.~\ref{fig:algorithm:t_comparison}.

\ifthenelse{\boolean{chapterbib}}
{
    \clearpage
    \bibliographystyle{aa}
    \bibliography{/Users/til/Documents/Papers/bibliography}
}
{}
\chapter{Dust size distributions in coagulation/fragmentation equilibrium}
\chapterprecishere{\centering From Birnstiel, Ormel, \& Dullemond 2010, submitted to A\&A}
\label{chapter:distri}

\begin{abstract}
Grains in circumstellar disks are believed to grow by mutual collisions and subsequent sticking due to surface forces. Many fields of research involving circumstellar disks, such as radiative transfer calculations, disk chemistry, magneto-hydrodynamic simulations and others largely depend on the resulting grain size distribution.

As detailed calculations of grain growth and fragmentation are both numerically challenging and computationally expensive, we aim to find simple recipes and analytical solutions for the grain size distribution in circumstellar disks for a scenario in which grain growth is limited by fragmentation and radial drift can be neglected.

We generalize previous analytical work on self-similar steady-state grain distributions. Numerical simulations are carried out to identify under which conditions the grain size distributions can be understood in terms of a combination of power-law distributions. A physically motivated fitting formula for grain size distributions is derived using our analytical predictions and numerical simulations. We find good agreement between analytical results and numerical solutions of the Smoluchowski equation for simple shapes of the kernel function. The results for more complicated and realistic cases can be fitted with a physically motivated ``black box'' recipe presented in this chapter.
\end{abstract}

\section{Introduction}\label{sec:distri:introduction}
Dust size distributions are a fundamental ingredient for many astrophysical models in the context of circumstellar disks and planet formation: whenever dust is present, it dominates the opacity of the disk, thereby influencing the temperature and consequently also the vertical structure of the disk \citep[e.g.,][]{Dullemond:2004p390}. Small grains effectively sweep up electrons and therefore strongly affect the chemistry and the ionization fraction (also via grain surface reactions, see Vasyunin et al., in prep.) and thereby also the angular momentum transfer of the disk \citep[e.g.,][]{Wardle:1999p9890,Sano:2000p9889}.

Today, it is well established that the dust distributions in asteroid belts and debris disks are governed by a so-called ``collision cascade'' \citep[see][]{Williams:1994p9637}: larger bodies in a gas free environment exhibit high velocity collisions ($\gtrsim$ km s$^{-1}$), far beyond their critical fragmentation threshold which lead to cratering or even complete shattering of these objects. The resulting fragments in turn suffer the same fate, thus producing ever smaller grains down to sizes below about a few micrometers where Poynting-Robertson drag removes the dust particles \citep[e.g.,][]{Wyatt:1999p10097}. The grain number density distribution in the case of such a fragmentation cascade has been derived by \citet{Dohnanyi:1969p7994} and \citet{Williams:1994p9637} and was found to follow a power-law number density distribution $n(m)\propto m^{-\alpha}$ with index $\alpha=\frac{11}{6}$ (which is equivalent to $n(a)\propto a^{-3.5}$), with very weak dependence on the mechanical parameters of the fragmentation process.
\citet{Tanaka:1996p2320}, \citet{Makino:1998p8778} and \citet{Kobayashi:2010p9774} showed that this result is exactly independent of the adopted model and that the resulting slope $\alpha$ is only determined by the mass-dependence of the collisional cross-section if the model of collisional outcome is self-similar.
The value of $\frac{11}{6}$ agrees well with the size distributions of asteroids \citep[see][]{Dohnanyi:1969p7994} and of grains in the interstellar medium \citep[MRN distribution, see][]{Mathis:1977p789,Pollack:1985p804} and is thus widely applied, even at the gas-rich stage of circumstellar disks.

However, in protoplanetary disks gas drag damps the motions of particles. Very small particles are tied to the gas and, as a result, have a relative velocity low enough to make sticking feasible. The size distribution therefore deviates from the MRN power-law. Theoretical models of grain growth indicate that particles can grow to sizes much larger than a few $\mu$m \citep[see][]{Nakagawa:1981p4533,Weidenschilling:1980p4572,Weidenschilling:1984p4590,Weidenschilling:1997p4593,Dullemond:2005p378,Tanaka:2005p6703,Brauer:2008p215,Birnstiel:2009p7135}. Indeed, the observational evidence suggests that growth up to cm-sizes is possible \citep[e.g.,][]{Testi:2003p3390,Natta:2004p3169,Rodmann:2006p8905,Ricci:2010p9423}.

However, as particles grow, they become more loosely coupled to the gas. This results in an increase in the relative velocity between the particles, a common feature of most sources of particles' relative velocity (i.e., turbulence, radial drift, and settling motions). For this reason, we expect that the perfect sticking assumption will break down at some point and other collisional outcomes, e.g., bouncing, erosion, and catastrophic disruption become possible \citep[see][]{Blum:2008p1920,Guttler:2010p9745}. It is expected, then, that at a certain point growth will cease for the largest particles in the distribution.  Collisions involving these particles result in fragmentation, thus replenishing the small grains. On the other hand mutual collisions among small particles still result in coagulation.  As a result, a steady-state emerges.  This situation is referred to as a fragmentation-coagulation equilibrium.

The situation in protoplanetary disks differs, therefore, from that in debris disks. In the latter only fragmentation operates. The mass distribution still proceeds towards a steady state but, ultimately, mass is removed from the system due to, radiation pressure or Poynting-Robertson drag.

In this chapter, we consider the situation that the total mass budget in the system is conserved. For simplicity, we will ignore motions due to radial drift in this study.  This mechanism effectively removes dust particles from the disk as well as providing particles with a large relative motion.  However, the derived presence of mm-size dust particles in protoplanetary disks is somewhat at odds with the usual (laminar) prescriptions for the radial drift rate \citep{Weidenschilling:1977p865}. Recently, it was shown that mm observations of protoplanetary disks can be explained by steady-state size distributions if radial drift is inefficient \citep{Birnstiel:2010p12008}. On the other hand, if radial drift would operate as effectively as the laminar theory predicts, then the observed populations of mm-sized particles at large radii cannot be sustained \citep{Brauer:2007p232}. Possible solutions to reduce the drift rate include bumps in the radial pressure profile \cite[see][]{Brauer:2008p212,Cossins:2009p3794} or zonal flows \citep[see][]{Johansen:2009p7441}.

In this work we analytically derive steady-state distributions of grains in the presence of both coagulation and fragmentation. The analytical predictions are compared to numerical simulations and applied to grain size distributions in turbulent circumstellar disks. Both theory and simulation results presented in this work are used to derive a fitting formula for steady-state grain size distributions in circumstellar disks.  

This chapter is outlined as follows: in Sect.~\ref{sec:distri:theory}, we briefly summarize and then generalize previous results by \citet{Tanaka:1996p2320} and \citet{Makino:1998p8778}. In Sect.~\ref{sec:distri:simresults}, we test our theoretical predictions and their limitations by a state-of-the-art grain evolution code (see \citealp{Birnstiel:2010p9709} and Chapter~\ref{chapter:algorithm}). Grain size distributions in circumstellar disks are discussed in Sect.~\ref{sec:distri:diskdistris}. In Sect.~\ref{sec:distri:recipe}, we present a fitting recipe for these distributions that can easily be used in models where grain properties are important. Our findings are summarized in Sect.~\ref{sec:distri:conclusions}.

\section{Power-law solutions for a coagulation-fragmentation equilibrium}\label{sec:distri:theory}
In this section, we begin by summarizing some of the previous work on analytical self-similar grain size distributions on which our subsequent analysis is based. We will then extend this to include both coagulation and fragmentation processes in a common framework. Under the assumption that the relevant quantities, i.e., the collisional probability between particles, the distribution of fragments, and the size distribution, behave like power-laws, we will solve for the size distribution in coagulation-fragmentation equilibrium. For simplicity, we consider a single monomer size only of mass $m_0$. This is therefore the smallest mass in the distribution. 

Another key assumption of our analytical model is that we assume the existence of a sharp threshold mass $\mf$, above which collisions always result in fragmentation and below which collision always result in coagulation.  As explained above, the physical motivation for this choice is that relative velocities increase with mass. This also means that in our model we neglect collision outcomes like bouncing or erosion.

\begin{figure}
  \begin{center}
    \includegraphics[width=0.75\hsize]{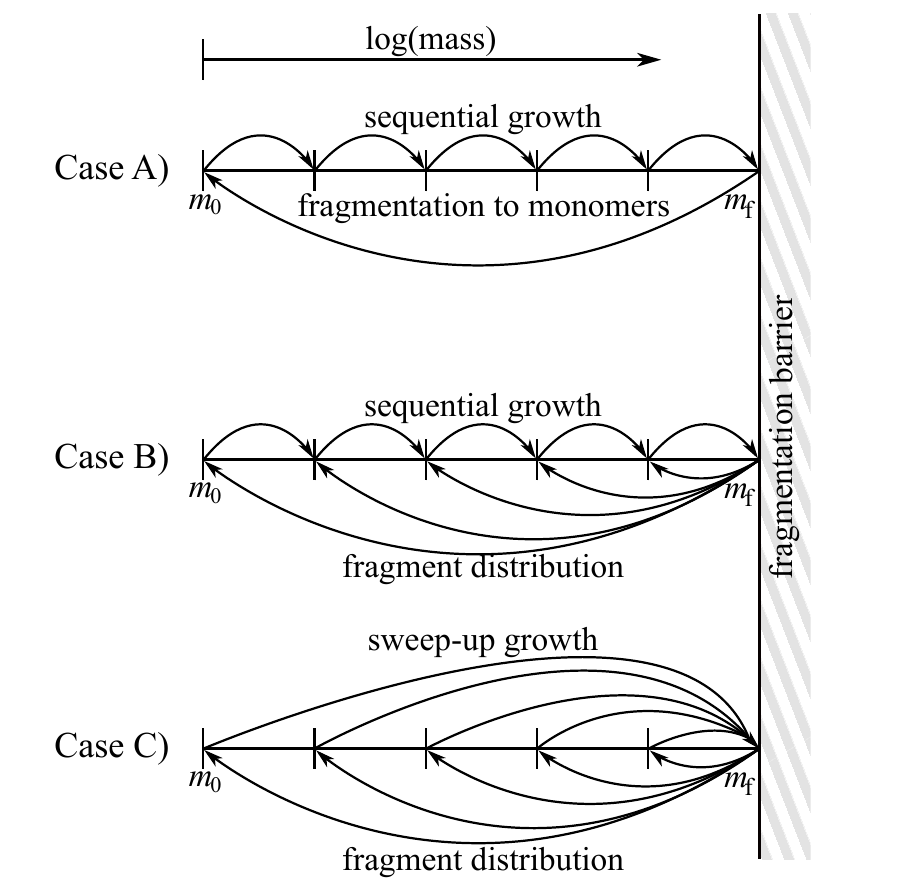}
    \caption{Illustration of the three different regimes for which analytical solutions have been derived. Case A represents the growth cascade discussed in Sect.~\ref{sec:distri:theory_coag}, case B the intermediate regime (see Sect.~\ref{sec:distri:theory_intermediate}) and case C the fragmentation dominated regime which is discussed in Sect.~\ref{sec:distri:theory_fragment}. Particles are shattered once they reach the fragmentation barrier \mf since collision velocities for particles $>\mf$ exceed the fragmentation threshold velocity.}
    \label{fig:distri:sketch}
  \end{center}
\end{figure}

In our analysis, we will identify three different regimes, which are symbolically illustrated in Fig.~\ref{fig:distri:sketch}:
\begin{itemize}
\item 
Regime A represents a case where grains grow sequentially (i.e. hierarchically) by collisions with similar sized grains until they reach an upper size limit and fragment back to the smallest sizes. The emerging power-law slope of the size distribution depends only on the shape of the collisional kernel.
\item 
Regime B is similar to regime A, however in this case the fragmented mass is redistributed over a range of sizes and, thus, influencing the out-coming distribution.
\item 
In Regime C, the upper end of the distribution dominates grain growth at all sizes. Smaller particles are swept up by the upper end of the distribution and are replenished mostly by redistributed fragments of the largest particles. The resulting distribution function depends strongly on how the fragmented mass is distributed after a disruptive collision.
\end{itemize}
For each of these regimes, we derive the parameter ranges in which they apply and also the slopes of the resulting grain size distribution.  

\subsection{The growth cascade}\label{sec:distri:theory_coag}
The fundamental quantity that governs the time-evolution of the dust size distribution is the collision kernel $C_{m_1,m_2}$. It is defined such that
\begin{equation}
C_{m_1,m_2}\cdot n(m_1)\cdot n(m_2)\, \dx{m_1} \dx{m_2}
\label{eq:distri:collisions1}
\end{equation}
gives the number of collisions per unit time per unit volume between particles in mass interval $[m_1, m_1+\dx{m_1}]$ and $[m_2,m_2+\dx{m_2}]$, where $n(m)$ is the number density distribution. Once specified, it determines the collisional evolution of the system. In the case that the number density $n(m)$ does not depend on position, $C_{m_1,m_2}$ is simply the product of the collision cross section and the relative velocity of two particles with the masses $m_1$ and $m_2$. 

We use the same Ansatz as \citet{Tanaka:1996p2320}, assuming that the collision kernel is given in the self-similar form
\begin{equation}
  C_{m_1,m_2} = m_1^\nu \: h\left(\frac{m_2}{m_1}\right),
  \label{eq:distri:kernel_form}
\end{equation}
where $h$ is any function which depends only on the masses through the ratio of $m_2/m_1$ and $\nu$ is called the index of the kernel or the degree of homogeneity.
As we will see in the following, $\nu$ is one of the most important parameters determining the resulting size distribution. Different physical environments are represented by different values of $\nu$. Examples include the constant kernel (i.e., $\nu=0$, mass independent), the geometrical kernel (i.e., $\nu=2/3$, velocity independent) or the linear kernel (i.e., $\nu=1$, as for grains suspended in turbulent gas).
 
It is further assumed that the number density distribution of dust particles follows a power-law,
\begin{equation}
  n(m) = A \cdot  m^{-\alpha}.
  \label{eq:distri:massdistribution}
\end{equation}

The time evolution of the mass distribution obeys the equation of mass conservation,
\begin{equation}
  \frac{\partial m n(m)}{\partial t} + \frac{\partial F(m)}{\partial m} = 0,
  \label{eq:distri:massconservation}
\end{equation}
where the flux $F(m)$ does not represent a flux in a typical continuous way since coagulation is non-local in mass space (each mass can interact with each other mass) but rather an integration of all growth processes which produce a particle with mass greater than $m$ out of a particle that was smaller than $m$ (i.e. collisions of $m_1<m$ with any other mass $m_2$ such that $m_1+m_2>m$). The flux in the case of pure coagulation was derived by \citet{Tanaka:1996p2320}
\begin{equation}
  F(m)   =  \int_{m_0}^m \mathrm{d}m_1 \int_{m-m_1}^{\mf} \mathrm{d}m_2 \, m_1 \, C_{m_1,m_2} \, n(m_1) \, n(m_2),
  \label{eq:distri:tanaka_flux_1}
\end{equation}
where we changed the lower bound of the integration over $m_1$ to start from the mass of monomers $m_0$ (instead of 0) and the upper bound of the integral over $m_2$ to a finite upper end $\mf$ (instead of infinity).

Substituting the definitions of above and using the dimensionless variables $x_1 = m_1/m$, $x_2 = m_2/m$, $x_0 = m_0/m$, and $x_\mathrm{f} = \mf/m$ one obtains
\begin{equation}
  F(m)   =  m^{\nu-2\alpha+3} \underbrace{\int_{x_0}^1 \mathrm{d}x_1\int_{1-x_1}^{x_\mathrm{f}} \mathrm{d}x_2 x_1^{\nu+1-\alpha} x_2^{-\alpha} h\left(\frac{x_2}{x_1}\right)}_\text{:= K},
  \label{eq:distri:tanaka_flux_2}
\end{equation}
where $K$ integrates to a constant.

Postulation of a steady state (i.e. setting the time derivative in Eq.~\ref{eq:distri:massconservation} to zero), leads to the condition
\begin{equation}
  F(m) \propto m^{\nu-2\alpha+3} \stackrel{!}{=} const.,
\end{equation}
from which it follows, that the slope of the distribution is
\begin{equation}
  \alpha = \frac{\nu+3}{2}.
  \label{eq:distri:coag_cascade}
\end{equation}
This result was already derived for the case of fragmentation by \citet{Tanaka:1996p2320} and \citet{Dohnanyi:1969p7994} and for the coagulation by \citet{Klett:1975p3936} and \citet{Camacho:2001p3816}.
The physical interpretation of this is a ``reversed'' fragmentation cascade: instead of a resupply of large particles which produces ever smaller bodies, this represents a constant resupply of monomers which produce ever larger grains (cf. case A in Fig.~\ref{fig:distri:sketch}).

\subsection{Coagulation fragmentation equilibrium}\label{sec:distri:theory_intermediate}
As \citet{Tanaka:1996p2320} and \citet{Makino:1998p8778} pointed out, the previous result is independent of the model of collisional outcomes as long as this model is self-similar (Eq.~\ref{eq:distri:kernel_form}). However this is no longer the case if we consider both coagulation and fragmentation processes happening at the same time.

We will now consider the case with a constant resupply of matter, due to the particles that fragment above \mf.  We assume these fragments obey a power-law mass distribution and are produced at a rate
\begin{equation}
  \dot n_\mathrm{f}(m) = N \cdot m^{-\xi},
  \label{eq:distri:frag_powerlaw}
\end{equation}
where $\xi$ reflects the shape of the fragment distribution.

If there is a constant flux of particles $F(m)$ as defined above, then the flux of fragmenting particles (i.e., the flux produced by particles that are growing over the fragmentation threshold) is given by
\begin{equation}
  F(\mf) =  K\cdot \mf^{\nu-2\alpha+3}
  \label{eq:distri:flux_mf}
\end{equation}
where $K$ is the integral defined in Eq.~\ref{eq:distri:tanaka_flux_2}.

The resulting (downward) flux of fragments $F_\text{f}(m)$ can then be derived by inserting Eq.~\ref{eq:distri:frag_powerlaw} into the equation of mass conservation (Eq.~\ref{eq:distri:massconservation}),
\begin{equation}
  \frac{\partial F_\text{f}(m)}{\partial m} = - m \cdot \dot n_\text{f}.
\end{equation}
Integration from monomer size $m_0$ to $m$ yields
\begin{equation}
  F_\text{f}(m) = - N \cdot \frac{1}{2-\xi} \cdot \left( m^{2-\xi} - m_0^{2-\xi} \right).
  \label{eq:distri:flux_downward}
\end{equation}
The normalization factor $N$ can be determined from the equilibrium condition that the net flux vanishes,
\begin{equation}
  F_\text{f}(\mf) = - F(\mf),
\end{equation}
and was found to be
\begin{equation}
N = \left(2-\xi\right) \, K \, \frac{\mf^{\nu-2\alpha+3}}{\mf^{2-\xi}-m_0^{2-\xi}}.
\label{eq:distri:normalization}
\end{equation}
In Eq.~\ref{eq:distri:flux_downward}, we can distinguish two cases:
\begin{itemize}
  \item If $\xi > 2$, the contribution of $m_0$ dominates the term in brackets. This means that
  most of the fragment mass is redistributed to monomer sizes and the situation is the same
  as in the pure coagulation case (cf. Case A in Fig.~\ref{fig:distri:sketch}). The steady-state condition
  $F(m)+F_\text{f}(m)=0$ (i.e., the net flux is zero) yields that
  \begin{equation}
    K\cdot m^{\nu-2\alpha+3}= - N \cdot \frac{1}{2-\xi} \cdot m_0^{2-\xi},
  \end{equation}
  is constant, which leads to the same condition as Eq.~\ref{eq:distri:coag_cascade}. Intuitively, this is clear since the majority of the redistributed mass ends up at $m \sim m_0$.
  \item If $\xi<2$, the $m$-dependence dominates the term in brackets in Eq.~\ref{eq:distri:flux_downward} and postulation of a steady-state,
  \begin{equation}
    K\cdot m^{\nu-2\alpha+3} \simeq N \cdot \frac{1}{2-\xi} \cdot m^{2-\xi},
  \end{equation}
  leads to an exponent
  \begin{equation}
    \alpha = \frac{\nu+\xi+1}{2},
    \label{eq:distri:intermediate_slope}
  \end{equation}
  less than Eq.~\ref{eq:distri:coag_cascade}.
  In this case, the slope of the fragment distribution matters. This scenario is represented as case B in Fig.~\ref{fig:distri:sketch}. 
\end{itemize}

\subsection{Fragment dominated regime}\label{sec:distri:theory_fragment}
The result obtained in the previous section may seem to be quite general. However, it does not hold for low $\xi$-values as we will show in the following.

In our case, the integrals do not diverge due to the finite integration bounds, however \citet{Makino:1998p8778} used $0$ and $\infty$ as lower and upper bounds for the integration and thus needed to investigate the convergence of the integral. They derived the following conditions:
\begin{align}
  \nu-\gamma-\alpha+1 &< 0 \label{eq:distri:makino_convergence1} \\
  \gamma - \alpha + 2 &> 0 \label{eq:distri:makino_convergence2}. 
\end{align}
where $\gamma$ gives the dependence of $m_2/m_1$ in the $h$-function of Eq.~\ref{eq:distri:gamma_makino} \citep[see][]{Makino:1998p8778}:
\begin{equation}
  h \left( \frac{m_2}{m_1} \right) = h_0 \cdot \left\lbrace
  \begin{array}{ll}
\left( \frac{m_2}{m_1} \right) ^{\gamma}      & \text{for }\frac{m_2}{m_1}\ll 1\\
\\
\left( \frac{m_2}{m_1} \right) ^{\nu-\gamma}  & \text{for }\frac{m_2}{m_1}\gg 1
\end{array}\right.
\label{eq:distri:gamma_makino}
\end{equation}

The first condition (Eq.~\ref{eq:distri:makino_convergence1}) considers the divergence towards the upper masses, whereas the second condition pertains the lower masses.  We assume that Eq.~\ref{eq:distri:makino_convergence2} is satisfied and consider the case of decreasing $\xi$ for $\alpha$ given by Eq.~\ref{eq:distri:intermediate_slope} which results in a steeper size distribution (where the mass is concentrated close to the upper end of the distribution). We see that for $\xi<1+\nu-2\gamma$, Eq.~\ref{eq:distri:makino_convergence1} is no longer fulfilled. The behavior of the flux integral changes qualitatively: growth is no longer hierarchical, but it becomes dominated by contributions by the upper end of the integration bounds.

Physically, this means that the total number of collisions of any grain is determined by the largest grains in the upper end of the distribution. Hence, smaller sized particles are predominantly refilled by fragmentation events of larger bodies instead of coagulation events from smaller bodies and they are  predominantly removed by coagulation events with big particles (near the threshold \mf) instead of similar-sized particles.  This corresponds to Case C in Fig.~\ref{fig:distri:sketch}.

To determine the resulting power-law distribution, we again focus on Eq.~\ref{eq:distri:tanaka_flux_1}. The double integral of the flux $F(m)$ can now be split into three separate integrals,
\begin{align}
F(m) &=& \int_{m_0}^{\frac{m}{2}} \dx{m_1} \int_{m-m_1}^{\mf}  \dx{m_2} \, m_1 \, C_{m_1,m_2} \, n(m_1) \, n(m_2)\nonumber\\
& &+\int_{\frac{m}{2}}^{m}   \dx{m_1} \int_{m-m_1}^{m_1}  \dx{m_2} \, m_1 \, C_{m_1,m_2} \, n(m_1) \, n(m_2)\label{eq:distri:split_integral}\\
& &+\int_{\frac{m}{2}}^{m}   \dx{m_1} \int_{m_1}^{\mf}  \dx{m_2} \, m_1 \, C_{m_1,m_2} \, n(m_1) \, n(m_2)\nonumber
\end{align}
according to whether $m_2$ is larger or smaller than $m_1$.

It can be derived (see Appendix~\ref{sec:distri:appendix}) that if the condition above (Eq.~\ref{eq:distri:makino_convergence1}) is violated and if $\mf\gg m$, then the first and the third integral in Eq.~\ref{eq:distri:split_integral} dominate the flux due to the integration until \mf. In this cases, the flux $F(m)$ is proportional to $m^{2+\gamma-\alpha}$.

A stationary state in the presence of fragmentation (Eq.~\ref{eq:distri:flux_downward}) is reached if the fluxes cancel out, which leads to the condition
\begin{equation}
  \alpha = \xi + \gamma.
  \label{eq:distri:fragment_regime}
\end{equation}
This is the sweep-up regime where small particles are cleaned out by big ones (cf. Case C in Fig.~\ref{fig:distri:sketch}). 
\subsection{Summary of the regimes}\label{sec:distri:theory_summary}

Summarizing these findings, we find that the resulting distribution is described by three scenarios (depicted in Fig.~\ref{fig:distri:sketch}), depending on the slope of the fragment distribution:
\begin{eqnarray}
\begin{array}{lll} 
  \text{\small Case A (growth cascade):}      &\xi>2               & \alpha = \frac{\nu+3}{2}    \\
  \\
  \text{\small Case B (intermediate regime):} &\nu-2\gamma+1<\xi<2 & \alpha = \frac{\nu+\xi+1}{2} \\
  \\
  \text{\small Case C (fragment dominated):}  &\xi < \nu-2\gamma+1 & \alpha = \xi+\gamma
\end{array}
\label{eq:distri:nu_regimes}  
\end{eqnarray} 

\section{Simulation results interpreted}\label{sec:distri:simresults}
In this section, we will test the analytical predictions of the previous section by a coagulation/fragmentation code (\citealp{Birnstiel:2010p9709}, see also \citealp{Brauer:2008p215}). The code solves for the time evolution of the grain size distribution using an implicit integration scheme. This enables us to find the steady-state distribution by using large time steps. In this way, the time evolution is not resolved, but the steady-state distribution is reliably and very quickly derived.

We start out with the simplest case of a constant kernel and then -- step by step -- approach a more realistic  scenario (in the context of a protoplanetary disk). In Sect.~\ref{sec:distri:diskdistris}, we will then consider a kernel taking into account relative velocities of Brownian motion and turbulent velocities and also a fragmentation probability which depends on the masses and the relative velocities of the colliding particles.

\subsection{Constant kernel}\label{sec:distri:simresults_constant}
In the following section, we consider still the case of a constant kernel, but we will include fragmentation above particle sizes of 1~mm because this represents an instructive test case.

\begin{figure}[thb]
  \begin{center}
\includegraphics[width=0.75\columnwidth]{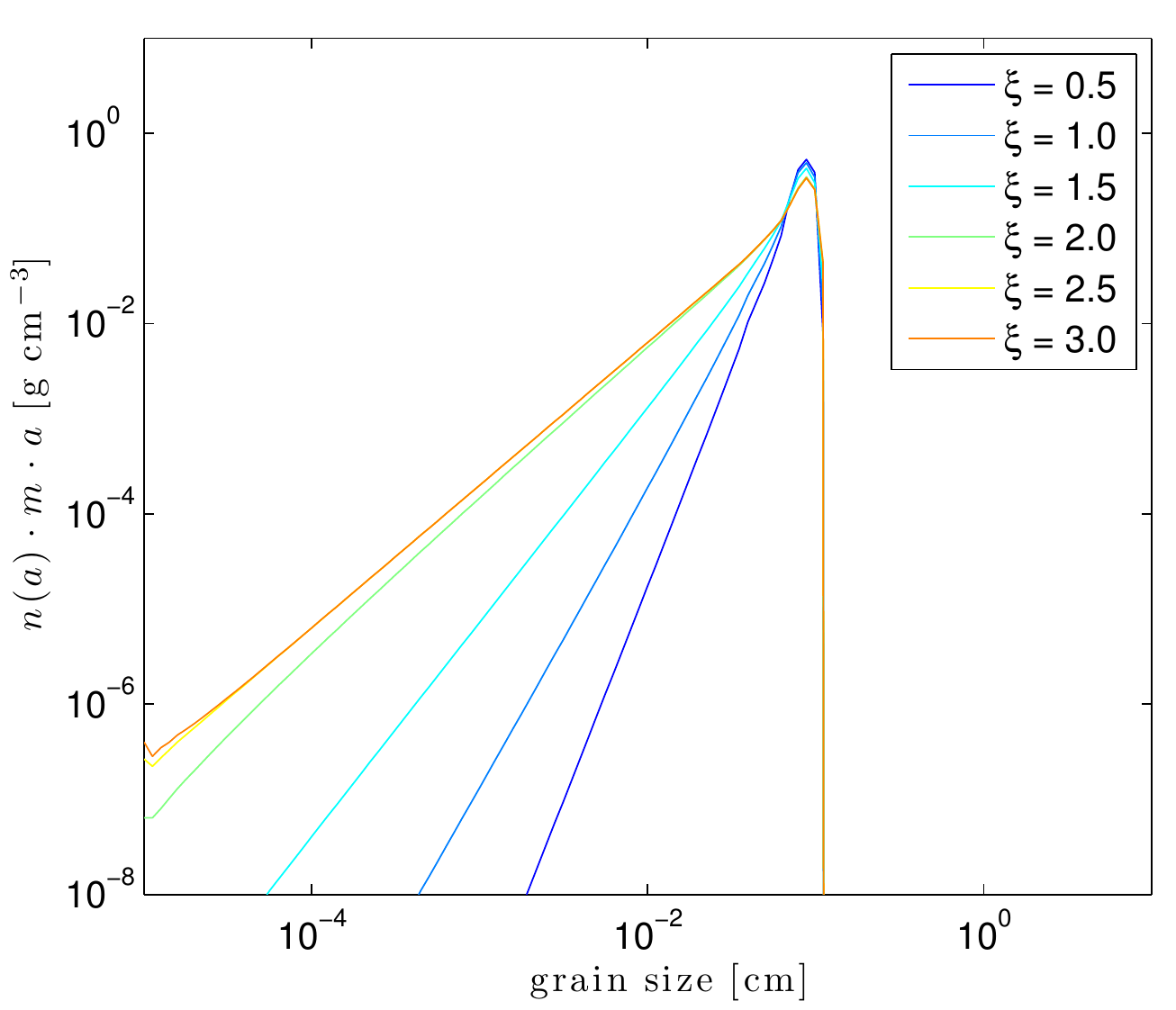}
\caption{Grain size distributions\protect\footnotemark{} for a constant kernel (i.e., $\nu=0$) and different distributions of fragments. The peak towards the upper end of the distribution is due to the fragmentation barrier (explanation in the text). The slope of the mass distribution corresponds to $6-3\alpha$.}
\label{fig:distri:spectra}
\end{center}
\end{figure}
\footnotetext{It should be noted that in this work we will plot the distributions typically in terms of $n(a)\cdot m\cdot a$ which is proportional to the distribution of mass. The advantage of plotting it this way instead of plotting $n(a)$ is the following: when $n(m)$ follows a power-law $m^{-\alpha}$, then the grain size distribution $n(a)$ describes a power-law with exponent $2-3\alpha$ and the mass distribution exponent is $6-3\alpha$. For typical values of $\alpha$, this is less steep and differences between a predicted and a real distribution are more prominent.}

We iteratively solve for the steady-state size distribution between coagulation and fragmentation. The outcome of these simulations for a constant kernel (i.e., $\nu=0$) are power-law distributions where the slope of the distribution depends on the fragmentation law (the slope $\xi$, see Eq.~\ref{eq:distri:frag_powerlaw}). 
Figure~\ref{fig:distri:spectra} shows the corresponding size distributions for some of the different fragment distributions: the steepest distribution corresponds to the case of $\xi=0.5$. For larger $\xi$-values, the slope of the mass distribution flattens. In all cases, a bump develops towards the upper end of the distributions. The reason for this is the following: grains typically grow mostly through collisions with similar-sized or larger particles. Since the distribution is truncated at the upper end (defined as \amax, see also Eq.~\ref{eq:distri:a_max}), particles close to the upper end lack larger collision partners, the growth rate at these sizes would decrease if the distribution keeps its power-law nature. This, in turn means that the flux could not be constant below \amax. To keep a steady state, the number of particles at that point has to increase in order to replace the missing collision partners at larger sizes.   

\begin{figure}[htp]
  \begin{center}
\includegraphics[width=0.75\columnwidth]{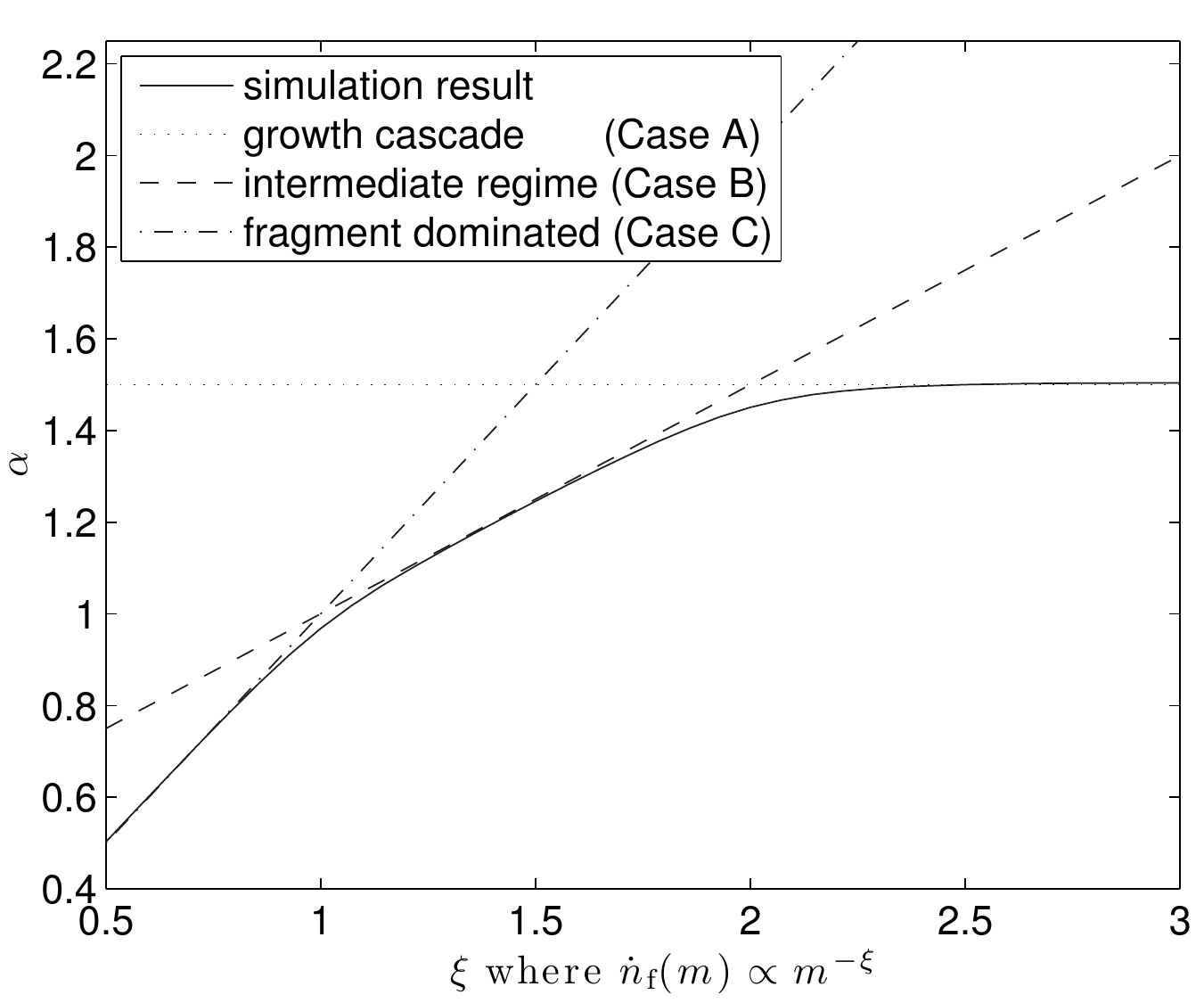}
\caption{Exponent of grain size distributions for a constant kernel (i.e. $\nu=0$) and different distributions of fragments (solid line) and the analytic solution for the growth cascade (Case A, dotted line), the intermediate regime (Case B, dashed line), and the fragment dominated regime (Case C, dash-dotted line).}
\label{fig:distri:exponent}
\end{center}
\end{figure}

Figure~\ref{fig:distri:exponent} shows how the slope of the resulting distribution depends on the fragmentation slope $\xi$, where the three previously discussed regimes can be identified:
\begin{itemize}
  \item
  Case A, growth cascade: as predicted, this scenario holds for values of $\xi \gtrsim 2$ where most of the fragmenting mass is redistributed to fragments. This case corresponds to a ``reversed'' collisional cascade (just the direction in mass space is reversed as collisions mostly lead to growth instead of shattering). 
  
  \item
  Case B, the intermediate regime: when the fragment mass is more or less equally distributed over all sizes, mass gain by redistribution of fragments and the mass loss due to coagulation have to cancel each other.
  
  \item
  Case C: most of the fragmented mass remains at large sizes. Therefore the mass distribution is dominated by the largest particles. In other words, growth is not hierarchical anymore. In this test case $\gamma=0$, and from Eq.~\ref{eq:distri:makino_convergence1} it follows that the transition between the intermediate and fragment-dominated regime lies at $\xi = 1$, which can be seen in Fig.~\ref{fig:distri:exponent}.
\end{itemize}
The measured slopes for $m\ll\mf$ are in excellent agreement with the model outlined in Sect.~\ref{sec:distri:theory}.
\subsection{Including settling effects}\label{sec:distri:simresults_settling}
The simplest addition to this model is settling: as grains become larger, they start to settle towards the mid-plane. However, turbulent mixing counteracts this systematic motion. The vertical distribution of dust in a settling-mixing equilibrium can (close to the mid-plane) be estimated by a Gaussian distribution with a size-dependent dust scale height \Hd.
Smaller particles are well enough coupled to the gas to have the same scale height as the gas, \Hp, while larger particles decouple and their scale height is decreasing with grain size,
\begin{equation}
\frac{\Hd}{\Hp} = \sqrt{\frac{\alphat}{\St}},\qquad\mathrm{for }\qquad \St>\alphat
\label{eq:distri:h_dust}
\end{equation}
\citep[see e.g.,][]{Dubrulle:1995p300,Schrapler:2004p2394,Youdin:2007p2021} where the Stokes number \St is a dimensionless quantity which describes the dynamic properties of a suspended particle. Very small particles have a small Stokes number and are therefore well coupled to the gas. Particles which have different properties (e.g., size or porosity) but the same \St behave aerodynamically the same. In our prescription of turbulence, \St is defined as the product of the particles stopping time $\tau_\mathrm{st}$ and the orbital frequency $\Omega_\mathrm{k}$. We focus on the Epstein regime where the Stokes number can be approximated by
\begin{equation}
  \St = \Ok \cdot \tau_\mathrm{st} \simeq \frac{a \, \rhos}{\Siggas} \frac{\pi}{2}
  \label{eq:distri:st}
\end{equation}
with \Siggas being the gas surface density and \rhos being the solid density of the particles which relates mass and size via $m=4\pi/3\rhos a^3$,

Settling starts to play a role as soon as the Stokes number becomes larger than the turbulence parameter $\alphat$ which can be related to a certain size
\begin{equation}
  \asett = \frac{2\alphat \, \Siggas}{\pi \rhos}.
  \label{eq:distri:asett}
\end{equation}
Eq.~\ref{eq:distri:asett} only holds within about one gas pressure scale height because the dust vertical structure higher up in the disk deviates from the a Gaussian profile \citep[see also][]{Dubrulle:1995p300,Schrapler:2004p2394,Dullemond:2004p390}.

The mass-dependent dust scale height causes the number density distribution $n(m)$ to depend on the vertical height, $z$. In disk-like configurations, it is therefore customary to consider the column density,
\begin{equation}
N(m) = \int_{-\infty}^{\infty} n(m,z) \,\dx{z}.
\end{equation}
Similar to Eq.~\ref{eq:distri:collisions1}, we can write the vertically integrated number of collisions as
\begin{equation}
\tilde{C}_{m_1,m_2}\cdot N(m_1)\cdot N(m_2) \,\dx{m_1}\, \dx{m_2},
\label{eq:distri:collsions2}
\end{equation}
which gives the total number of collisions that take place over the entire column of the disks.

The dependence of the collisional probability on scale height, is now reflected in the modified kernel $\tilde{C}_{m_1,m_2}$ (see Section~\ref{sec:algorithm:alg_coag} for a derivation):
\begin{equation}
  \tilde{C}_{m_1,m_2} = \frac{C_{m_1,m_2}}{\sqrt{2\pi \left(H_1^2+H_2^2\right)}}.
  \label{eq:distri:cmod_1}
\end{equation}

The point to realize here is that, due to the symmetry between Eq.~\ref{eq:distri:collisions1} and Eq.~\ref{eq:distri:collsions2}, the analysis in Sect.~\ref{sec:distri:theory} holds also for disk-like configuration, if the kernel is now replaced by $\tilde{C}_{m_1,m_2}$. The resulting exponent $\alpha$ then concerns the column density.

If we consider the case that $\St>\alphat$ and substitute Eq.~\ref{eq:distri:h_dust} and Eq.~\ref{eq:distri:st} into Eq.~\ref{eq:distri:cmod_1}, we find that 
\begin{equation}
  \begin{split}
\tilde{C}_{m_1,m_2} &= \frac{C_{m_1,m_2}}{\sqrt{2\pi \left(H_1^2+H_2^2\right)}}\\
&= C_{m_1,m_2}\cdot H_1^{-1} \cdot \left(1+\frac{H_2^2}{H_1^2}\right)^{-\frac{1}{2}}\\
&= C_{m_1,m_2}\cdot m_1^{1/6} \cdot h\left(\frac{m_2}{m_1}\right),\\
\end{split}
\end{equation}
has an index $\nu = 1/6$ for grain sizes larger than $a_\mathrm{sett}$ and $\nu=0$ otherwise (it should be noted that $H_1$ and $H_2$ are the dust scale heights whereas $h(m_2/m_1)$ represents the function defined in Eq.~\ref{eq:distri:kernel_form}).

\begin{figure}[htp]
  \centering
  \includegraphics[width=0.75\columnwidth]{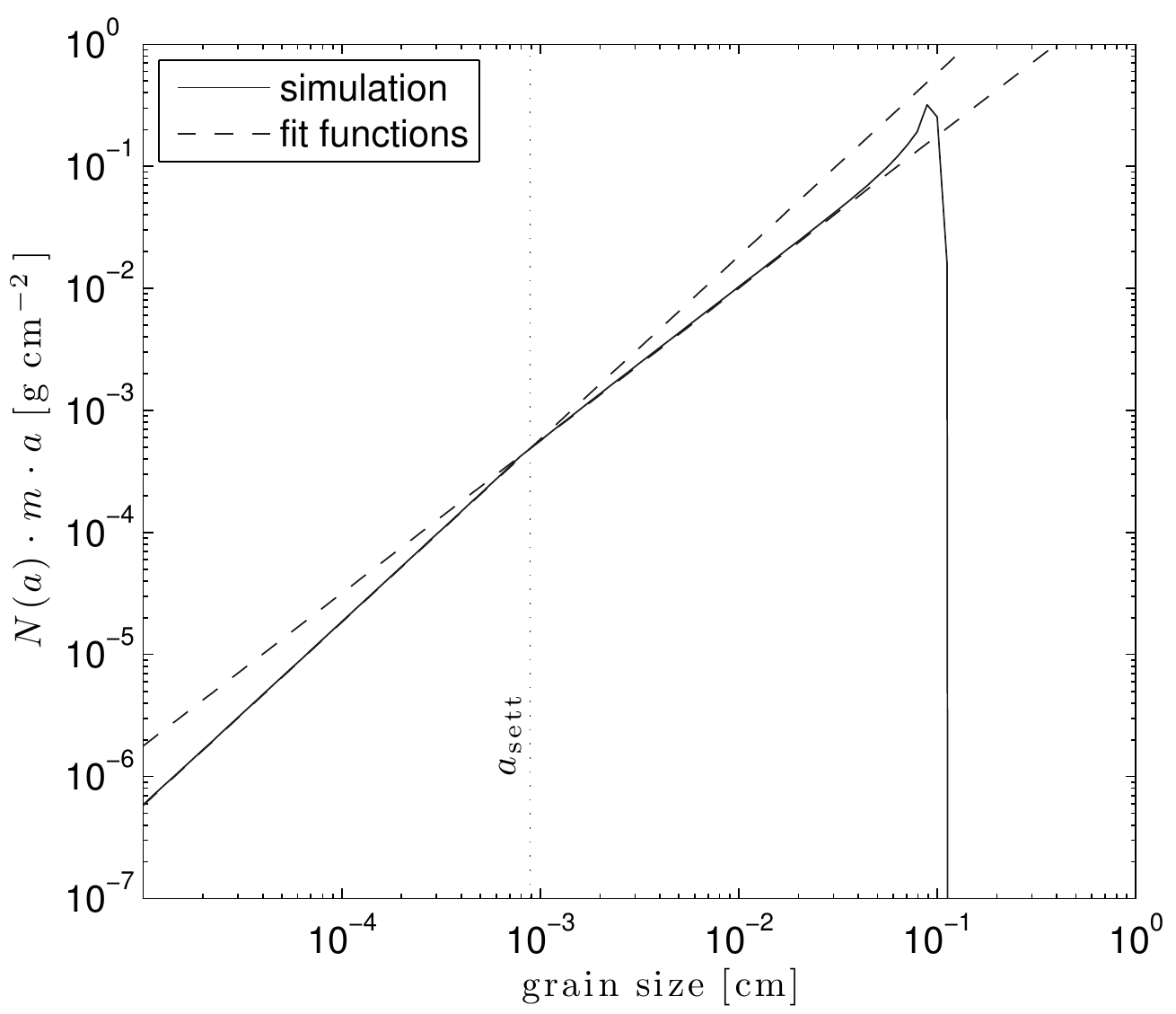}
  \caption{Simulation result (solid line) and a ``two power-law fit'' to the vertically integrated dust distribution for a constant kernel with settling included. The dotted, vertical line denotes the grain size above which grains are affected by settling.}
  \label{fig:distri:settling_fit}
\end{figure}

The theory described in Sect.~\ref{sec:distri:theory} is strictly speaking only valid for a constant $\nu$-index, but if this index is constant over a significant range of masses, then the local slope of the distribution will still adapt to this index. In the case of settling, we can therefore describe the distribution with two power-laws as can be seen in Fig.~\ref{fig:distri:settling_fit}.

The fact that the distribution will locally follow a power-law is an important requirement for being able to construct fitting formulas which reproduce the simulated grain size distributions. It allows us \emph{in some cases} to explain the simulation outcomes with the local kernel index (although coagulation and fragmentation are non-local processes in mass space, since each mass may interact with each other mass). A physically motivated recipe to fit the numerically derived distribution functions for the special case of $\xi=11/6$ is presented in Sect.~\ref{sec:distri:recipe}.

\subsection{Non-constant kernels}\label{sec:distri:simresults_nonconstant}
\begin{figure*}
  \begin{center}
\includegraphics[width=0.46\columnwidth]{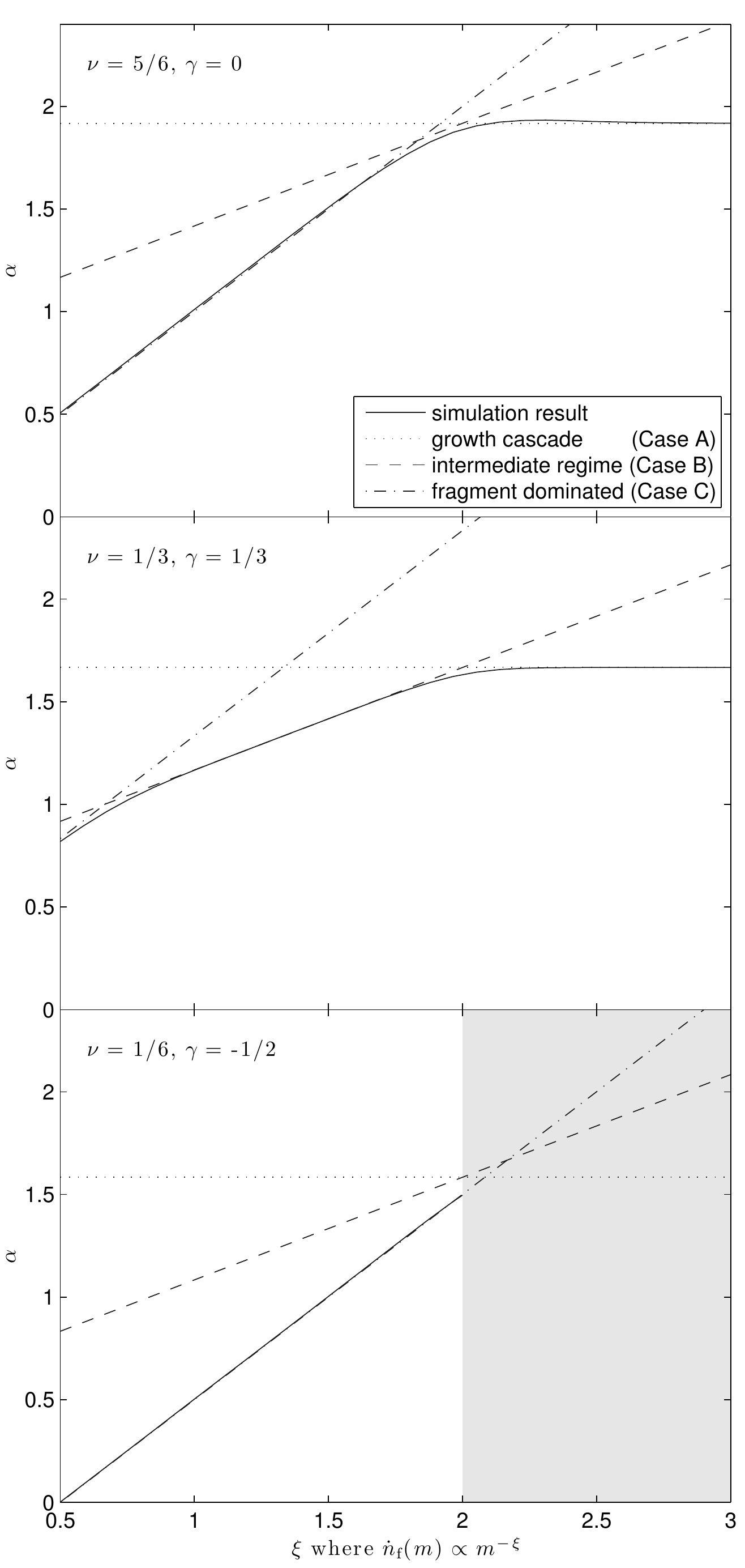}
\caption{Exponents of the grain size distributions for three different kernels as function of the fragment distribution index. The plot shows the simulation results (solid lines), the analytic solution for the growth cascade (Case A, dotted lines), the intermediate regime (Case B, dashed lines), and the fragment dominated regime (Case C, dash-dotted lines). The upper panel was calculated with a  $(\nu = 5/6,\gamma=0)$-kernel (i.e. turbulent relative velocities, see Sect.~\ref{sec:distri:diskdistris_vrel}), the middle panel with a $(\nu = 1/3, \gamma=1/3)$-kernel and the lower panel with a $(\nu = 1/6, \gamma=-1/2)$-kernel (i.e. Brownian motion relative velocities). In the grey shaded area, Eq.~\ref{eq:distri:makino_convergence2} is not fulfilled, and the simulation result is not a power-law like distribution.}
\label{fig:distri:exponents_both}
\end{center}
\end{figure*}

We performed the same tests as in Sect.~\ref{sec:distri:simresults_constant} also for non-constant kernels, i.e. kernels with $\nu>0$. The measured slopes for the cases of $(\nu = 5/6, \gamma = 0)$ (corresponding to the second turbulent regime, see Sect.~\ref{sec:distri:diskdistris_vrel}), $(\nu = 1/3, \gamma = 1/3)$, and $(\nu = 1/6, \gamma = -1/2)$ (i.e., Brownian motion, see Sect.~\ref{sec:distri:diskdistris_vrel}) are shown in Fig.~\ref{fig:distri:exponents_both}. Similar to Fig.~\ref{fig:distri:exponent}, the distribution always follows the minimum index of the three different regimes (Cases A, B, and C). It should be noted that for all indices $\xi$ of the fragmentation law, all processes -- coagulation, fragmentation, and re-distribution of fragments -- take place, but the relative importance of them is what determines the resulting slope.

In Fig.~\ref{fig:distri:exponents_both}, it can be seen that Case B (defined in Sect.~\ref{sec:distri:theory_intermediate}) vanishes for the kernel in the upper panel, while it is present for a large range of $\xi$ for the kernel in the middle panel. This can be explained by the definitions of the three regimes which were summarized in Eq.~\ref{eq:distri:nu_regimes}: with $(\nu = 5/6, \gamma=0)$ (cf. upper panel in Fig.~\ref{fig:distri:exponents_both}), Case B is confined between $\xi=11/6$ and 2. The grain size distribution, therefore, switches almost immediately from being growth dominated (Case A) to fragmentation dominated (Case C). In the case of a kernel with $\nu=1/3$ and $\gamma=1/3$, this range spans from $2/3$ to 2, as can be seen in the central panel of Fig.~\ref{fig:distri:exponents_both}.

The lower panel shows the distribution for a Brownian motion kernel (i.e., $\nu = 1/6$ and $\gamma=-1/2$). The grey shaded area highlights the range of $\xi$ values where our predictions do not apply, that is, Eq.~\ref{eq:distri:makino_convergence2} is not fulfilled any more. The resulting steady-state distributions are not power-law distributions anymore.

\section{Grain size distributions in circumstellar disks}\label{sec:distri:diskdistris}
In this section, we will leave the previous ``clean'' kernels and focus on the grain size distribution in circumstellar disks including relative velocities due to Brownian and turbulent motion of the particles, settling effects, and a fragmentation probability as function of particle mass and impact velocity.

Combining all these effects makes it impossible to find simple analytical solutions in the case of a coagulation-fragmentation equilibrium. We will therefore use a coagulation fragmentation code to find the steady-state solutions and to show how the steady-state distributions in circumstellar disks depend on our input parameters.

We will first discuss the model ``ingredients'', i.e. the relative velocities and the prescription for fragmentation/sticking. Afterwards, we will define characteristic sizes at which the shape of the size distribution changes due to the underlying physics. In the last subsection, we will then show the simulation results and discuss the influence of different parameters. 

\subsection{Relative velocities}\label{sec:distri:diskdistris_vrel}
We will now include the effects of relative velocities due to Brownian motion, and due to turbulent mixing.
The Brownian motion relative velocities are given by
\begin{equation}
  \Delta u_\mathrm{BM}=\sqrt{\frac{8 \kb T \;(m_1+m_2)}{\pi \, m_1\, m_2}}
  \label{eq:distri:dv_BM}
\end{equation}
where $\kb$ is the Boltzmann constant and $T$ the mid-plane temperature of the disk.
\citet{Ormel:2007p801} have derived closed form expressions for particle collision velocities induced by turbulence. They also provided easy-to-use approximations for the different particle size regimes which we will use in the following. Small particles (i.e. stopping time of particle $\ll$ eddy crossing time) belong to the first regime of \citet{Ormel:2007p801} with velocities proportional to
\begin{equation}
  \Delta u_\mathrm{I} \propto \left|\St_1-\St_2\right|.
  \label{eq:distri:dv_TM1}
\end{equation}
The relative velocities of particles in the second turbulent regime of \citet{Ormel:2007p801} are given by
\begin{equation}
  \Delta u_\mathrm{II} \propto \sqrt{\St_\mathrm{max}},
  \label{eq:distri:dv_TM2}
\end{equation}
where $\St_\mathrm{max}$ is the larger of the particles Stokes numbers. Velocities in this regime show also a weak dependence on the ratio of the Stokes numbers which we will neglect in the following discussion.

Together with the geometrical cross section $\sigma_\mathrm{geo} = \pi (a_1+a_2)^2$, it is straight-forward to estimate the indices of the kernel, $\nu$ and $\gamma$, as defined in Eq.~\ref{eq:distri:kernel_form} and \ref{eq:distri:gamma_makino} for all these regimes, without settling (assuming that only Brownian motion or turbulent motion dominates the relative velocities). The indices for these three sources of relative particle motion are summarized in Table~\ref{tab:distri:nu_and_gamma}.
\begin{table}
  \caption{Kernel indices for the different regimes without settling.}
  \centering
  \begin{tabular}{l|ccl}
Regime&                  $\nu$&         $\gamma$&           upper end\\
\hline
Brownian motion regime&  $\frac{1}{6}$& $-\frac{1}{2}$&     \aBT\\[0.1cm]
Turbulent regime I&      1&             0&                  \aonetwo\\[0.1cm]
Turbulent regime II&     $\frac{5}{6}$& 0&                  \amax\\[0.1cm]
\end{tabular}
\label{tab:distri:nu_and_gamma}
\end{table}
If settling is to be included, then the $\nu$ index for particle sizes above $a_\mathrm{sett}$ has to be increased by $\frac{1}{6}$ (see Sect.~\ref{sec:distri:simresults_settling}) while $\gamma$ remains the same. The $\nu$ and $\gamma$ indices for all three regimes can be found in Table~\ref{tab:distri:nu_and_gamma}.

\begin{figure*}
  \begin{center}
\includegraphics[width=\hsize]{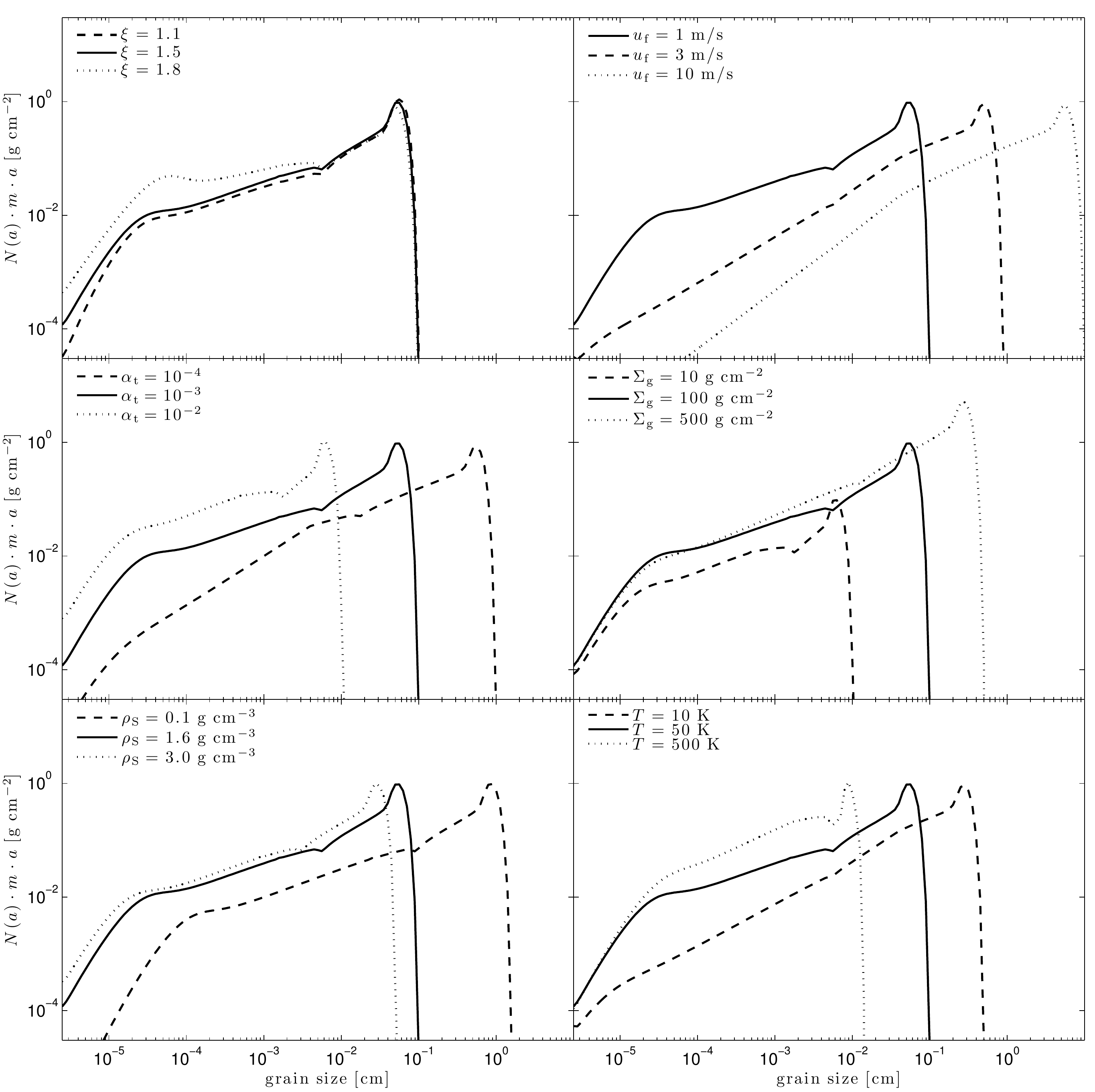}
\caption{Fiducial model and variations of the most important parameters: fragmentation power-law index $\xi$, critical fragmentation velocity \uf, turbulence parameters \alphat, surface density \Siggas, particle solid density \rhos, and mid-plane temperature $T$. The shape of the vertically integrated size distributions does not depend on the stellar mass or the distance to the star (only via the radial dependence of the parameters above).}
\label{fig:distri:variations}
\end{center}
\end{figure*}

\subsection{Fragmentation and cratering}\label{sec:distri:diskdistris_frag_crater}
We introduce fragmentation and cratering according to the recipe of \citet{Birnstiel:2010p9709}: whether a collision leads to sticking or to fragmentation/cratering is determined by the fragmentation probability
\begin{equation}
p_\text{f} = \left\{
\begin{array}{ll}
0&                              \text{if } \Delta u < \uf - \delta u\\
\\
1&                              \text{if } \Delta u > \uf\\
\\
1-\frac{\uf-\Delta u}{\delta u}&    \text{else}
\end{array}
\right.
\end{equation}
which means that all impacts with velocities above the critical break-up threshold \uf lead to fragmentation or cratering while impacts with velocities below $\uf-\delta u$ lead to sticking. The width of the linear transition region $\delta u$ is chosen to be $0.2 \, \uf$.

The mass ratio of the two particles determines whether the impact completely fragments the larger body (masses within one order of magnitude) or if the smaller particle excavates mass from the larger body (masses differ by more than one order of magnitude).

In the case of fragmentation, the whole mass of both collision partners is redistributed to all masses smaller than the larger body according to Eq.~\ref{eq:distri:frag_powerlaw}.

In the case of cratering, it is assumed that the smaller particle (with mass $m_\mathrm{imp}$) excavates its own mass from the larger body. The mass of the impactor as well as the excavated mass is then redistributed to masses smaller than the impactor mass according to Eq.~\ref{eq:distri:frag_powerlaw}. Thus, the total redistributed mass equals
\begin{equation}
\int_{m_0}^{\mf} n(m) \cdot m \, \dx{m} = 2\, m_\mathrm{imp}
\end{equation}
and the mass of the larger body is reduced by $2\, m_\mathrm{imp}$.

Most parameters such as the fragmentation velocity or the amount of excavated material during cratering are not yet well enough constrained (for the most recent experimental work, see \citealp{Blum:2008p1920}, \citealp{Guttler:2010p9745}, and references therein). Experiments suggest fragmentation velocities of a few m s$^{-1}$ and fragment distributions with $\xi$ between 1 and 2 (\citealp{Guttler:2010p9745} find $\xi$ values between 1.07 and 1.37 for SiO$_2$ grains). Simulations of silicate grain growth around 1~AU show that also bouncing (i.e. collisions without sticking or fragmentation) can play an important role \citep[see][]{Weidling:2009p9833,Zsom:2010p9746}. However, changes in material composition such as organic or ice mantles or the monomer size are expected to change this picture. As there is still a large parameter space to be explored, we continue with the rather simple recipe of sticking, fragmentation and cratering outlined above.

\subsection{Regime boundaries}\label{sec:distri:diskdistris_boundaries}
From Eq.~\ref{eq:distri:nu_regimes}, we can calculate the slope of the distribution in the different regimes if we assume that the slope of the distribution at a given grain size always follows from the kernel index $\nu$. To construct a whole distribution consisting of several power-laws for each regime, we need to know where each of the different relative velocity regime applies.

It is important to note that relative velocities due to Brownian motion decrease with particle size whereas the relative velocities induced by turbulent motion increase with particle size (up to $\St=1$). Therefore, Brownian motion dominates the relative velocities for small particles, while for larger particles, turbulence dominates. From numerical simulations, we found that at those sizes where the highest turbulent relative velocities (i.e. collisions with the smallest grains) start to exceed the smallest Brownian motion relative velocities (i.e. collisions with similar sized grains), the slope of the distribution starts to be determined by the turbulent kernel slope. By equating the approximate relative velocity of \citet{Ormel:2007p801} and Eq.~\ref{eq:distri:dv_BM}, the according grain size can be estimated to be
\begin{equation}
a_\mathrm{BT} \approx \left[\frac{8 \Siggas}{\pi \rhos} \cdot \mathrm{Re}^{-\frac{1}{4}} \cdot \sqrt{\frac{\mu \, \mpr}{3\pi \, \alphat}} \cdot \left(\frac{4\pi}{3}\rhos\right)^{-\frac{1}{2}}  \right]^{\frac{2}{5}},
\label{eq:distri:a_01}
\end{equation}
where we approximate the Reynolds number (i.e., the ratio of turbulent viscosity $\nu_\mathrm{t} = \alpha \csound \Hp$ over molecular viscosity) near the disk mid-plane by
\begin{equation}
\Rey \approx \frac{\alphat \, \Siggas \, \sighyd}{2 \, \mu \, \mpr}.
\label{eq:distri:reynolds}
\end{equation}
Here, \sighyd is the cross section of molecular hydrogen (taken to be $2\times~10^{-15}$~cm$^2$) and $\mu=2.3$ is the mean molecular weight in proton masses $\mpr$.

Turbulent relative velocities strongly increase for grains with a stopping time that is larger or about the turn-over time of the smallest eddies. More specifically, the Stokes number of the particles at this change in the relative velocity is
\begin{equation}
\St_\mathrm{12} = \frac{1}{y_a} \, \frac{t_\eta}{t_\mathrm{L}} = \frac{1}{y_a} \, \Rey^{-\frac{1}{2}},
\end{equation}
where $t_\mathrm{L}=1/\Ok$ and $t_\eta = t_\mathrm{L} \cdot \Rey^{-\half}$ are the turn-over times of the largest and the smallest eddies, respectively, and $\Ok$ is the Kepler frequency. $y_a$ is a factor which was approximated by \citet{Ormel:2007p801} to be about 1.6. The corresponding grain size in the Epstein regime is therefore given by
\begin{equation}
\aonetwo = \frac{1}{y_a} \, \frac{2 \Siggas}{\pi \, \rhos} \cdot \Rey^{-\frac{1}{2}},
\label{eq:distri:a12}
\end{equation}
As mentioned above, the Brownian motion relative velocities of small grains decrease with their size. For larger sizes, the relative velocities due to turbulent motion are gaining importance, which are increasing with size until a Stokes number of unity. For typical values of the sound speed
\begin{equation}
\csound=\sqrt{\frac{\kb\, T}{\mu \mpr}}
\label{eq:distri:csound}
\end{equation}
and the turbulence parameter $\alphat$, the largest turbulent relative velocity $\Delta u_\mathrm{max} \approx  \sqrt{\alphat}\,\csound$ exceed the critical collision velocity of the grains (which is in the order of a few m s$^{-1}$) and therefore leads to fragmentation of the dust particles. In the case of very quiescent environments and/or larger critical collision velocities, particles do not experience this fragmentation barrier and can continue to grow. Hence, a steady state is never reached.
The work presented here focuses on the former case where
\begin{equation}
\Delta u_\mathrm{max} > \uf,
\label{eq:distri:frag_condition}
\end{equation}
and grain growth is always limited by fragmentation.

As relative turbulent velocities are (in our case) increasing with grain size, we can relate the maximum turbulent relative velocity and the critical collision velocity to derive the approximate maximum grain size which particles can reach \cite[see][]{Birnstiel:2009p7135}
\begin{equation}
a_\mathrm{max} \simeq \frac{2\Siggas }{\pi \alphat \rho_\mathrm{s}} \cdot \frac{\uf^2}{\csound^2}.
\label{eq:distri:a_max}
\end{equation}

\subsection{Resulting steady-state distributions}\label{sec:distri:diskdistris_results}
The  parameter space is too large to even nearly discuss all possible outcomes of steady state grain size distributions. We will therefore focus on a few examples and rather explain the basic features and the most general results only. For this purpose, we will adopt a fiducial model and consider the influences of several parameters on the resulting grain size distribution: $\xi$, \uf, \alphat, \Siggas, \rhos, and $T$ (see Fig.~\ref{fig:distri:variations}).

The fiducial model (see the solid black line in Fig.~\ref{fig:distri:variations}) shows the following features:
steep increase from the smaller sizes until a few tenth of a micrometer. This relates to the regime dominated by Brownian motion relative velocities. The upper end of this regime can be approximated by Eq.~\ref{eq:distri:a_01}. The flatter part of the distribution is caused by a different kernel index $\nu$ in the parts of the distribution which are dominated by turbulent relative velocities. The dip at about 60~$\mu$m (cf. Eq.~\ref{eq:distri:a12}) is due to the jump in relative velocities as the stopping time of particles above this size exceeds the shortest eddy turn-over time \citep[see][]{Ormel:2007p801}.

The upper end of the distribution is approximately at \amax. The increased slope of the distribution and the bump close to the upper end are caused by two processes. Firstly, a boundary effect: grains mostly grow by collisions with similar or larger sized particles. Grains near the upper end of the distribution lack larger sized collision partners and therefore the number density needs to increase in order to keep the flux constant with mass (i.e. in order to keep a steady-state).
Secondly, the bump is caused by cratering: impacts of small grains onto the largest grains do not cause growth or complete destruction of the larger bodies, instead they erode them. Growth of these larger bodies is therefore slowed down and, similar to the former case, the mass distribution needs to increase in order to fulfill the steady-state criterion (``pile-up effect'').

The upper left panel in Fig.~\ref{fig:distri:variations} shows the influence of the distribution of fragments after a collision event: larger values of $\xi$ mean that more of the fragmented mass is redistributed to smaller sizes. Consequently, the mass distribution at smaller sizes increases relative to the values of smaller $\xi$ values.

The strong influence of the fragmentation threshold velocity \uf can be seen in the upper right panel in Fig.~\ref{fig:distri:variations}: according to Eq.~\ref{eq:distri:a_max}, an order of magnitude higher \uf leads to a 100 times larger maximum grain size.

The grain size distributions for different levels of turbulence are shown in the middle left panel of Fig.~\ref{fig:distri:variations}. The effects are two-fold: firstly, an increased \alphat leads to increased turbulent relative velocities, thus, moving the fragmentation barrier \amax to smaller sizes (cf. Eq.~\ref{eq:distri:a_max}).
Secondly, a larger \alphat shifts the transition within the turbulent regime, \aonetwo, to smaller sizes. Consequently, the second turbulent regime gains importance as \alphat is increased since its upper end lower boundary extend ever further.

The middle right panel in Fig.~\ref{fig:distri:variations} displays the influence of an decreased gas surface density \Siggas (assuming a fixed dust-to-gas ratio). It can be seen that not only the total mass is decreased due to the fixed dust-to-gas ratio but also the upper size of the distribution \amax decreases. This is due to the coupling of the dust to the gas: with larger gas surface density, the dust is better coupled to the gas. This is described by a decreased Stokes number (see Eq.~\ref{eq:distri:st}) which, in turn, leads to smaller relative velocities and hence a larger \amax.
Interestingly, the shape of the grain size distribution does not depend on the total dust mass, but on the total gas mass, as long as gas is dynamically dominating (i.e., $\Siggas \gg \Sigdust$). If more dust were to be present, grains would collide more often, thus, a steady-state would be reached faster and the new size distribution would be a scaled-up version of the former one. However the velocities at which grains collide are determined by the properties of the underlying gas disk. In this way, the dust grain size distribution is not only a measure of the dust properties, but also a measure of the gas disk physics like the gas density and the amount of turbulence.

The shape of the size distribution for different grain volume densities \rhos does not change significantly. However most regime boundaries (\asett, \aonetwo, and \amax) are inversely proportional to \rhos (because of the coupling to the gas, described by the Stokes number). A decrease (increase) in \rhos therefore shifts the whole distribution to larger (smaller) sizes as can be seen in the lower left panel in Fig.~\ref{fig:distri:variations}.

The upper end of the distribution, \amax, is inversely proportional to the mid-plane temperature $T$ (as in the case of the turbulence parameter) whereas the transition between the different turbulent regimes \aonetwo does not. Therefore, increasing the temperature decreases \amax in the same way as decreasing \Siggas does. However \aonetwo is not influenced by temperature changes, therefore the shape of the size distribution changes in a different way than in the case of changing \Siggas as can be seen by comparing the middle right and the  lower right panels of Fig.~\ref{fig:distri:variations}. 
\section{Fitting formula for steady-state distributions}\label{sec:distri:recipe}
In this section, we will describe a simple recipe which allows us to construct vertically integrated grain size distributions which fit reasonably well to the simulation results presented in the previous section.


The recipe does not directly depend on the radial distance to the star or on the stellar mass. A radial dependence only enters via radial changes of the input parameters listed in Table~\ref{tab:distri:glossary}. This recipe has been tested for a large grid of parameter values\footnote{See 
\href{http://www.mpia.de/distribution-fits}{www.mpia.de/distribution-fits}},
however there are some restrictions.

\begin{figure}[tbh]
  \begin{center}
\includegraphics[width=0.75\hsize]{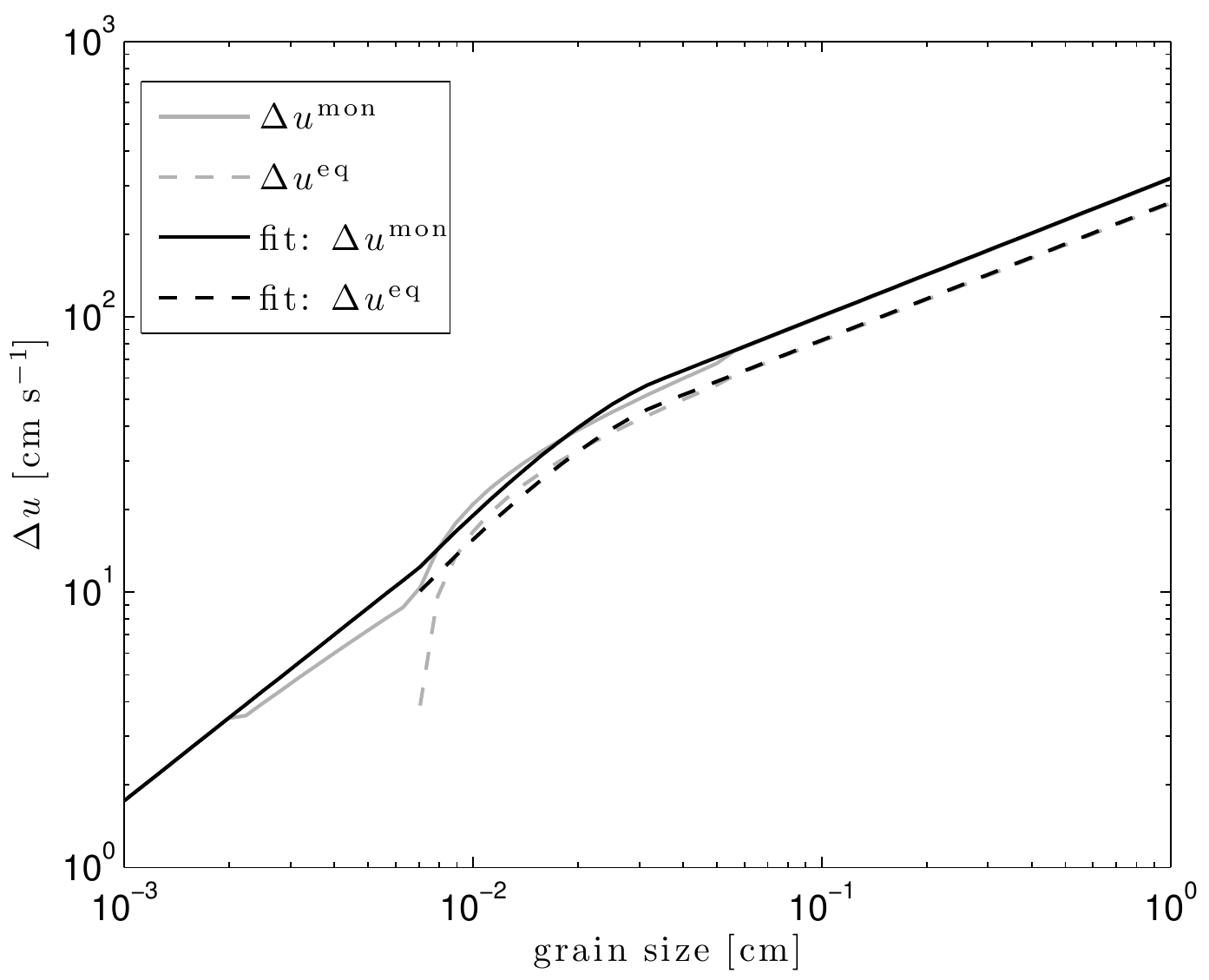}
\caption{Comparison of the turbulent relative velocities of \citet{Ormel:2007p801} to the fitting formula used in the recipe (see Eqs.~\ref{eq:distri:du_eq} and \ref{eq:distri:du_mon}) for grain size distributions. The largest error in the resulting upper grain size $a_\mathrm{P}$ derived from the fitting formula is about 25\%.}
\label{fig:distri:dv_fit}
\end{center}
\end{figure}

\begin{figure*}
  \begin{center}
\includegraphics[width=\hsize]{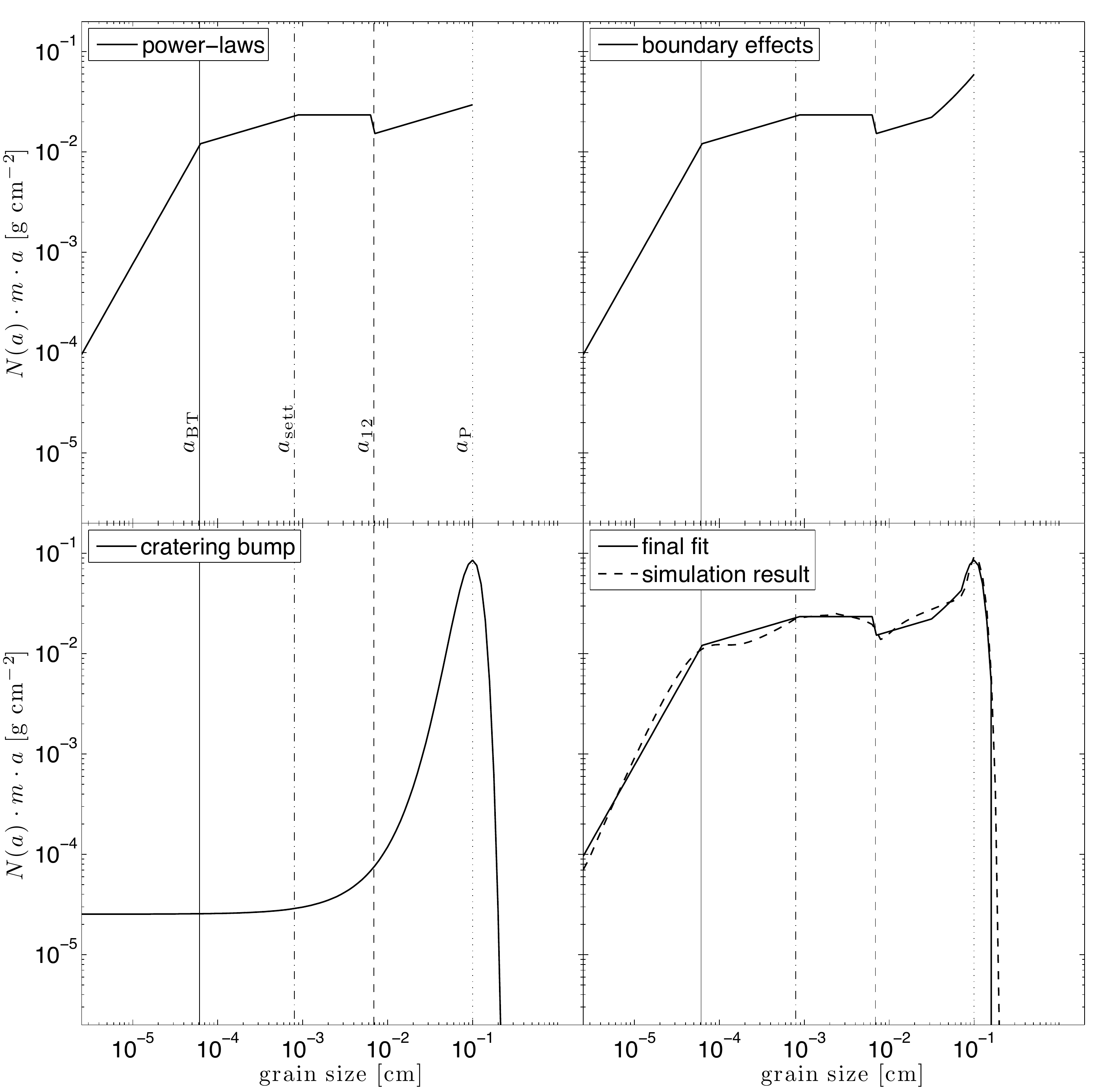}
\caption{Step-by-step construction of the fit distribution for the following parameters: $\Siggas=20$ g cm$^{-2}$, $\Sigdust=0.2$ g cm$^{-2}$, $\alphat=1\e{-4}$, $\uf=1$ m s$^{-1}$, $\xi=1.833$, $T=50$ K, $\rhos = 1.6$ g cm$^{-3}$. \textit{The upper left panel} shows the distribution after step 5: each interval obeys a different power-law, the distribution is continuous apart from a jump at \aonetwo. \textit{The upper right panel} displays the fit including and increase at the upper end, according to step 6. The bump caused by cratering (cf. Eq.~\ref{eq:distri:bump}) is shown in \textit{the bottom left panel} while \textit{the bottom right panel} compares the final fit distribution (solid curve) to the simulation result (dashed curve). The vertical lines correspond to the regime boundaries: \aBT (solid), \asett (dash-dotted), \aonetwo (dashed) and $a_\mathrm{P}$ (dotted).}
\label{fig:distri:app_recipe}
\end{center}
\end{figure*}

\subsection{Limitations}
These fits strictly apply only for the case of $\xi=11/6$. In this case, the slopes of the distribution agree well with the predictions of the intermediate regime (Case B, defined in Sec.~\ref{sec:distri:theory_intermediate}). For smaller values of $\xi$, the slopes do not strictly follow the analytical predictions. This is due to the fact that we include cratering, which is not covered by our theory. Erosion is therefore an important mode of fragmentation: it dominates over complete disruption through the high number of small particles \citep[see also][]{Kobayashi:2010p9774} and it is able to redistribute significant amounts of mass to the smallest particle sizes.

One important restriction for this recipe is the upper size of the particles \amax: it needs to obey the condition of Eq.~\ref{eq:distri:frag_condition}, since otherwise, particles will not experience the fragmenting high-velocity impacts and a steady state will never be reached since particles can grow unhindered over the meter-size barrier.

There are also restrictions to a very small \amax: if \amax is close to or even smaller than \aonetwo, then the fit will not represent the true simulation outcome very well. In this case, the upper end of the distribution, and in more extreme cases the whole distribution will look much different. Thus, the sizes should obey the condition
\begin{equation}
5\,\mu\text{m} < \aonetwo < \amax,
\end{equation}
where each inequality should be within a factor of a few.

\subsection{Recipe}
The following recipe calculates the vertically integrated mass distribution of dust grains in a turbulent circumstellar disk within the the above mentioned limitations. The recipe should be applied on a logarithmic grain size grid $a_i$ with a lower size limit of 0.025 $\mu$m and a fine enough size resolution ($a_{i+1}/a_i\lesssim 1.12$). 
For convenience, all variables are summarized in Table~\ref{tab:distri:glossary}.
The steps to be performed are as follows:
\begin{enumerate}
  \item Calculate the grain sizes which represent the regime boundaries \aBT, \aonetwo, \asett which are given by Eqs.~\ref{eq:distri:a_01}, \ref{eq:distri:a12}, and \ref{eq:distri:asett}.
  \item Calculate the turbulent relative velocities for each grain size. For this, we approximate the equations which are given by \citet{Ormel:2007p801}. Collision velocities with monomers are given by
  \begin{equation}
  \Delta u_i^\mathrm{mon} = \ugas\cdot \, \left\{
  \begin{array}{lll}
                    \Rey^\frac{1}{4}\cdot(\St_i-\St_0)       &\mathrm{for}& a_i<\aonetwo\\ \\
    (1-\epsilon)  \cdot\Rey^\frac{1}{4}\cdot(\St_i-\St_0)  &\mathrm{for}& \aonetwo<a_i<5\,\aonetwo\\
    \phantom{(1}+ \epsilon\phantom{)} \cdot\sqrt{3\cdot\St_i}\\ \\
    \sqrt{3\cdot\St_i}                                      &\mathrm{for}& a_i>5\,\aonetwo\\ \\
  \end{array}\right.
  \label{eq:distri:du_mon}
  \end{equation}
  where \Rey is the Reynolds number (see Eq.~\ref{eq:distri:reynolds}), and $\St_i$ and $\St_0$ are the Stokes numbers of $a_i$ and monomers, respectively (cf. Eq.~\ref{eq:distri:st}). \ugas is given by
  \begin{equation}
    \ugas = \csound \, \sqrt{\frac{3}{2}\alphat},
  \end{equation}
  and the interpolation parameter $\epsilon$ is defined as
  \begin{equation}
    \epsilon = \frac{a_i-\aonetwo}{4\,\aonetwo}.
  \end{equation}
  and collisions with similar sized bodies are approximated as
  \begin{equation}
  \Delta u_i^\mathrm{eq} = \left\{
  \begin{array}{lll}
    0                        &\mathrm{for}& a_i<\aonetwo\\ \\
    \sqrt{\frac{2}{3}}\cdot\Delta u_i^\mathrm{mon}  &\mathrm{for}& a_i>\aonetwo\\
  \end{array}\right.
  \label{eq:distri:du_eq}
  \end{equation}
  A comparison between these approximations and the formulas of \citet{Ormel:2007p801} is shown in Fig.~\ref{fig:distri:dv_fit}.
  \item Using Eqs.~\ref{eq:distri:du_mon} and \ref{eq:distri:du_eq} for the relative velocities, and the transition width $\delta u=0.2\uf$, find the grain sizes which correspond to the following conditions:
  \begin{itemize}
    \item  $a_\mathrm{L}$: particles above this size experience impacts with monomers with velocities of $\Delta u^\mathrm{mon}_i\geq\uf-\delta u$ (i.e., cratering starts to become important).
    \item $a_\mathrm{P}$: particles above this size experience impacts with equal sized grains with velocities of $\Delta u^\mathrm{eq}_i\geq\uf-\delta u$ (i.e., fragmentation becomes important).
    \item $a_\mathrm{R}$: particles above this size experience impacts with similar sized grains with velocities of $\Delta u^\mathrm{eq}_i\geq\uf$ (i.e. every impact causes fragmentation/cratering).
  \end{itemize}
  \item Calculate the factor $J$ according to the recipe
  \begin{align}
    V &= \csound \, \left(\frac{8 \, \mu \mpr \, \Siggas}{\alphat\,\sighyd}\right)^{\frac{1}{4}}
    \sqrt{\frac{3}{4}\, \frac{\alphat}{\Siggas \, \, y_a}}\\
    J &= \left(2.5^{-9}+\left(1.1^9+\left(1+2\,\sqrt{3}\, \frac{V}{\uf}\right)^9\right)^{-1}\right)^{-\frac{1}{9}},
  \end{align}
  where $\mu=2.3$, $\sighyd=2\e{-15}$~cm$^2$ and $y_a= 1.6$.
  \item The power-law indices of the mass distribution $\delta_i$ for each interval between the regime boundaries (\aBT, \aonetwo, \asett, $a_\mathrm{P}$) are calculated according to the intermediate regime (cf.~Eq.~\ref{eq:distri:intermediate_slope}).
  The slopes have to be chosen according to the kernel regime (Brownian motion, turbulent regime 1 or 2) and according to whether the regime is influenced by settling or not (i.e., if $a$ is larger or smaller than \asett). The resulting slopes of the mass distribution are given in Table~\ref{tab:distri:slopes}. The first version of the fit $f(a_i)$ (where $a_i$ denotes the numerical grid point of the particles size array) is now constructed by using power-laws ($\propto a_i^{\delta_i}$) in between each of the regimes, up to $a_\mathrm{P}$. The fit should be continuous at all regime boundaries except a drop of 1/J at the transition at \aonetwo. An example of this first version is shown in the top left panel of Fig.~\ref{fig:distri:app_recipe}.
  \item Mimic the cut-off effects which cause an increase in the distribution function close to the upper end: linearly increase the distribution function for all sizes $a_\mathrm{inc} < a < a_\mathrm{P}$: 
  \begin{equation}
    f(a_\mathrm{i}) \rightarrow f(a_\mathrm{i})\cdot \left(2-\frac{a_\mathrm{i}-a_\mathrm{P}}{a_\mathrm{inc}-a_\mathrm{P}}\right),
  \end{equation}
  with
  \begin{equation}
    a_\mathrm{inc} = 0.3\,a_\mathrm{P}.
  \label{eq:distri:a_inc}
  \end{equation}
  The resulting fit after this step is shown in the upper right panel of Fig.~\ref{fig:distri:app_recipe}. 
  \item The bump due to cratering is mimicked by a Gaussian,
  \begin{equation}
    b(a_i) = 2\cdot f(a_\mathrm{L})\cdot \exp\left(-\frac{\left(a_i-a_\mathrm{P}\right)^2}{\sigma^2}\right),
    \label{eq:distri:bump}
  \end{equation}
  where $\sigma$ is defined as
  \begin{equation}
    \sigma = \frac{\min\left(\left| a_\mathrm{R}-a_\mathrm{P}\right|,\left|a_\mathrm{L}-a_\mathrm{P}\right| \right)}{\sqrt{\log(2)}},
  \end{equation}
  but should be limited to be at least
  \begin{equation}
    \sigma > 0.1\,a_\mathrm{P}.
  \end{equation} 
  \item The fit $\mathcal{F}(a_i)$ is now constructed by using the maximum of $f(a_i)$ and $b(a_i)$ in the following way:
  \begin{equation}
    \mathcal{F}(a_i) = \left\{\begin{array}{lll}
    f(a_i)&                         \mathrm{if }&   a_i \leq a_\mathrm{L}\\
    \\
    \max\left(f(a_i),b(a_i)\right)& \mathrm{if }&   a_\mathrm{L} \leq a_i \leq a_\mathrm{P}\\
    \\
    b(a_i)&                         \mathrm{if }&   a_\mathrm{P} \leq a_i \leq a_\mathrm{R}\\
    \\
    0&                              \mathrm{else}
    \end{array}
    \right.
  \end{equation}
  \item Finally, the fit needs to be normalized to the dust surface density at the given radius. $\mathcal{F}$ is a (yet un-normalized) vertically integrated mass distribution (shown in the bottom right panel of Fig.~\ref{fig:distri:app_recipe}). To translate this mass distribution to a vertically integrated number density distribution $N(a)$, we need to normalize it as 
  \begin{equation}
    N(a) = \frac{\Sigdust}{\int_{a_0}^{\infty}\mathcal{F}(a)\,\dx{\text{ln}a}} \cdot \frac{\mathcal{F}(a)}{m\cdot a}.
\end{equation}
\end{enumerate}

\begin{table}
  \caption{Power-law exponents of the distribution $n(m)\cdot m\cdot a$. The slopes were calculated using the formula for a coagulation/fragmentation equilibrium (Eq.~\ref{eq:distri:intermediate_slope}). Within each regime of relative velocities, it has to be differentiated whether grains are influenced by settling or not.}
  \begin{center}
\begin{tabular}{p{5cm}|c|c}
\hline
\hline
\multirow{2}{*}{Regime} & \multicolumn{2}{c}{$\delta_i$}\\
\cline{2-3}
& $a_i<\asett$&  $a_i>\asett$\\
\hline
&&\\[-0.3cm]
Brownian motion regime&  $\frac{3}{2}$& $\frac{5}{4}$\\[0.1cm]
Turbulent regime I&      $\frac{1}{4}$& $0$\\[0.1cm]
Turbulent regime II&     $\frac{1}{2}$& $\frac{1}{4}$\\[0.1cm]
\hline
\end{tabular}
\label{tab:distri:slopes}
\end{center}
\end{table}

\begin{table*}
  \caption{Definition of the variables used in this chapter. The variables grouped as ``input variables'' are the parameters of the fitting recipe in Sect.~\ref{sec:distri:recipe}. ``Other variables'' summarizes the definitions of all other variables used in this work.}
  \begin{center}
  \begin{scriptsize}
  \setlength{\extrarowheight}{2pt}
\begin{tabularx}{\linewidth}{l|l>{\raggedright\arraybackslash}Xl}
\hline
\hline
\multirow{10}{0.5cm}{
    \rotatebox{90}{
    Input variables
    }}
    & \multirow{2}{*}{variable} & \multirow{2}{*}{definition} & \multirow{2}{*}{unit}\\
    \\
\cline{2-4}
    & \alphat &          Turbulence strength parameter &                              -\\
    & \uf &              Fragmentation threshold velocity &                           cm s$^{-1}$\\
    & \Siggas &          Gas surface density &                                        g cm$^{-2}$\\
    & \Sigdust &         Dust surface density &                                       g cm$^{-2}$\\
    & $\xi$ &            Power-law index of the mass 
                         distribution of fragments, see
                         Eq.~\ref{eq:distri:frag_powerlaw} &                          -\\
    & $T$ &              Mid-plane temperature &                                      K\\
    & \rhos &            Volume density of a dust particle&                           g cm$^{-3}$\\
\hline
    \multirow{30}{0cm}{\rotatebox{90}{ \centering Other variables}}
    & $\alpha$ &                  Slope of the number density
                                  distribution $n(m)\propto m^{-\alpha}$, 
                                  see Eq.~\ref{eq:distri:massdistribution} &          -\\
    & \aBT &                      Eq.~\ref{eq:distri:a_01} &                          cm\\
    & \aonetwo &                  Eq.~\ref{eq:distri:a12} &                           cm\\
    & \amax &                     Eq.~\ref{eq:distri:a_max} &                         cm\\
    & $a_\mathrm{L}$ &            Left boundary of the bump function
                                  $b(a)$, see Sect.~\ref{sec:distri:recipe},
                                  paragraph~3 &                                       cm\\
    & $a_\mathrm{P}$ &            Peak size of the bump function $b(a)$,
                                  see Sect.~\ref{sec:distri:recipe}, paragraph~3 &    cm\\
    & $a_\mathrm{R}$ &            Right boundary of the bump function
                                  $b(a)$, see Sect.~\ref{sec:distri:recipe},
                                  paragraph~3 &                                       cm\\
    & \asett &                    Eq.~\ref{eq:distri:asett} &                         cm\\
    & $a_\mathrm{inc}$ &          Eq.~\ref{eq:distri:a_inc} &                         cm\\
    & $b(a)$ &                    Bump function, see Eq.~\ref{eq:distri:bump} &       -\\
    & $C_{m_1,m_2}$ &             Collision kernel, see
                                  Eq.~\ref{eq:distri:kernel_form} &                   cm$^{3}$ s$^{-1}$\\
    & $\tilde{C}_{m_1,m_2}$ &     Collision kernel including settling effects,
                                  see Eq.~\ref{eq:distri:kernel_form} &               cm$^{2}$ s$^{-1}$\\
    & \csound &                   sound speed, see Eq.~\ref{eq:distri:csound} &       cm s$^{-1}$\\
    & $\delta_i$ &                Slopes of the fit-function,
                                  see Table~\ref{tab:distri:slopes} &                 -\\
    & $\delta u$ &                Width of the transition between sticking
                                  and fragmentation, taken to be 0.2 \uf &            cm s$^{-1}$\\
    & $\gamma$ &                  Power-law index of the function
                                  $h(m_2/m_1)$ for large ratios of $m_2/m_1$,
                                  as defined in Eq.~\ref{eq:distri:gamma_makino} &    -\\
    & $K$ &                       Integral defined in
                                  Eq.~\ref{eq:distri:tanaka_flux_2} &                 -\\
    & \kb &                       Boltzmann constant &                                erg K$^{-1}$\\
    & $m_0$ &                     Monomer mass &                                      g\\ 
    & $m_1$ &                     Mass above which particles fragment &               g\\ 
    & $\mu$ &                     Mean molecular weight in
                                  proton masses, taken to be 2.3 &                    -\\
    & $\mpr$ &                    Proton mass &                                       g\\
    & $n(m)$ &                    Number density distribution &                       g cm$^{-3}$\\
    & $N(m)$ &                    Vertically integrated
                                  number density distribution &                       g cm$^{-2}$\\
    & $\nu$ &                     Degree of homogeneity of the kernel as
                                  defined in Eq.~\ref{eq:distri:kernel_form} &        -\\
    & \Rey &                      Reynolds number, see Eq.~\ref{eq:distri:reynolds} & -\\
    & \St &                       Mid-plane Stokes number in the
                                  Epstein regime, see Eq.~\ref{eq:distri:st} &        -\\
    & \sighyd&                    Cross-section of molecular hydrogen &               cm$^2$\\
    & $\Delta u_i^\mathrm{mon}$ & Relative velocities between monomers 
                                  and grains of size $a_i$,
                                  see Eq.~\ref{eq:distri:du_mon} &                    cm s$^{-1}$\\
    & $\Delta u_i^\mathrm{eq}$ &  Relative velocities between grains
                                  of size $a_i$, see Eq.~\ref{eq:distri:du_eq} &      cm s$^{-1}$\\
\hline
\hline
\end{tabularx}
\label{tab:distri:glossary}
\end{scriptsize}
\end{center}
\end{table*}

\section{Conclusions}\label{sec:distri:conclusions}
In this work, we generalize the analytical findings of previous works to the case of grain size distributions in a coagulation/fragmentation equilibrium. Under the assumption that all grains above a certain size, \amax, fragment into a power-law distribution $n_\text{f}(m)\propto m^{-\xi}$, we derived analytical steady-state solutions for self-similar kernels and determined three different cases (see \ref{sec:distri:theory_summary}).

Results show that dust size distributions in circumstellar disks do not necessarily follow the often adopted MRN power-law distribution of $n(a)\propto a^{-3.5}$ \citep[see][]{Mathis:1977p789,Dohnanyi:1969p7994,Tanaka:1996p2320,Makino:1998p8778,Garaud:2007p405} when both coagulation and fragmentation events operate. We performed detailed simulations of grain growth and fragmentation to test the analytical predictions and found very good agreement between the theory and the simulation results.

We applied the theory to the gaseous environments of circumstellar disks. Unlike the models of \citet{Garaud:2007p405}, the upper end of the size distribution is not limited by the growth timescale but by fragmentation as relative velocities increase with grain size and reach values large enough to fragment grains. The shape of the dust distribution is determined by the gaseous environment (e.g., gas surface density, level of turbulence, temperature and others) since the gas is dynamically dominating as long as the gas surface density significantly exceeds the dust surface density. The total dust mass merely provides the normalization of the distribution and the timescale in which an equilibrium is reached. The results presented in this work show that the physics of growth and fragmentation directly link the upper and the lower end of the dust distribution in circumstellar disks.

A ready-to-use recipe for deriving vertically integrated dust size distributions in circumstellar disks for a fixed value of $\xi=11/6$ is presented in Sect.~\ref{sec:distri:recipe}.  Although the collision kernel in circumstellar disks is complicated, we found good agreement with our fitting recipe for a fragment distributions with $\xi=11/6$. The recipe can readily be used for further modeling such as disk chemistry or radiative transfer calculations.

\ifthenelse{\boolean{chapterbib}}
{
    \clearpage
    \bibliographystyle{aa}
    \bibliography{/Users/til/Documents/Papers/bibliography}
}
{}
\begin{subappendices}
\section{Derivation of the fragment dominated size distribution}\label{sec:distri:appendix}
In this section, we will derive the slope of the size distribution which is dominated by the largest particles. Since in our scenario, the integrals are confined between the monomer mass $m_0$ and the largest particles at the fragmentation barrier \mf, the integrals do not diverge as in the scenario of \citet{Tanaka:1996p2320} and \citet{Makino:1998p8778}, who consider integration bounds of zero and infinity. However, if the first condition of \cite{Makino:1998p8778} is not fulfilled, then the mass flux (cf. Eq.~\ref{eq:distri:tanaka_flux_1}) is dominated by the upper bound of the integral. This is the case which we will consider in the following.

As noted in Sect.~\ref{sec:distri:theory_fragment}, the flux integral (Eq.~\ref{eq:distri:tanaka_flux_1}) can be split into three separate integrals, in order to distinguish cases of $m_2 > m_1$ or $m_2 < m_1$,
\begin{equation}
F(m) \equiv I_1 + I_2 + I_3,
\end{equation}
where $I_1$, $I_2$, and $I_3$ correspond (from left to right) to the three integrals defined in  Eq.~\ref{eq:distri:split_integral}.
We will now evaluate these integrals in the limits of $m_0 \ll m \ll \mf$, using the limiting behavior of $h(m_2/m_1)$ as given in Eq.~\ref{eq:distri:gamma_makino}.

\subsection{First integral}
Carrying out the integration over $m_2$ in $I_1$ leads to
\begin{equation}
\begin{aligned}
I_1 = & \frac{A^2 \cdot h_0}{\nu-\gamma-\alpha+1} \cdot \int_{m_0}^{m/2} \dx{m_1} \, m_1^{\gamma - \alpha +1} \cdot \\
      & \quad   \left[ \mf^{\nu-\gamma-\alpha+1} - \left(m-m_1\right)^{\nu-\gamma-\alpha+1} \right], 
\end{aligned} 
\end{equation}
from which we can derive Eq.~\ref{eq:distri:makino_convergence1}, the first convergence criterion of \citet{Makino:1998p8778}.

We consider the case where this condition does not hold, i.e.,
\begin{equation}
\nu-\gamma-\alpha+1>0.
\end{equation}
Then, the \mf term in brackets dominates over the other term. Thus, the term in brackets is constant and carrying out the integration yields
\begin{equation}
I_1 = \frac{A^2\cdot h_0 \cdot \mf^{\nu-\gamma-\alpha+1}}{(\nu-\gamma-\alpha+1)\cdot (\gamma-\alpha+2)} \cdot \left[\left(\frac{m}{2}\right)^{\gamma-\alpha+2}  - m_0^{\gamma-\alpha+2}\right].
\end{equation}
Now, if the second condition, Eq.~\ref{eq:distri:makino_convergence2} holds, using $m_0 \ll m$, we derive
\begin{equation}
I_1 = \frac{A^2\cdot h_0 \cdot \mf^{\nu-\gamma-\alpha+1}}{(\nu-\gamma-\alpha+1)\cdot (\gamma-\alpha+2)} \cdot \left(\frac{m}{2}\right)^{\gamma-\alpha+2}.
\label{eq:distri:I1}
\end{equation}

\subsection{Second integral}
\begin{figure}[t!bh]
  \begin{center}
\includegraphics[width=0.75\hsize]{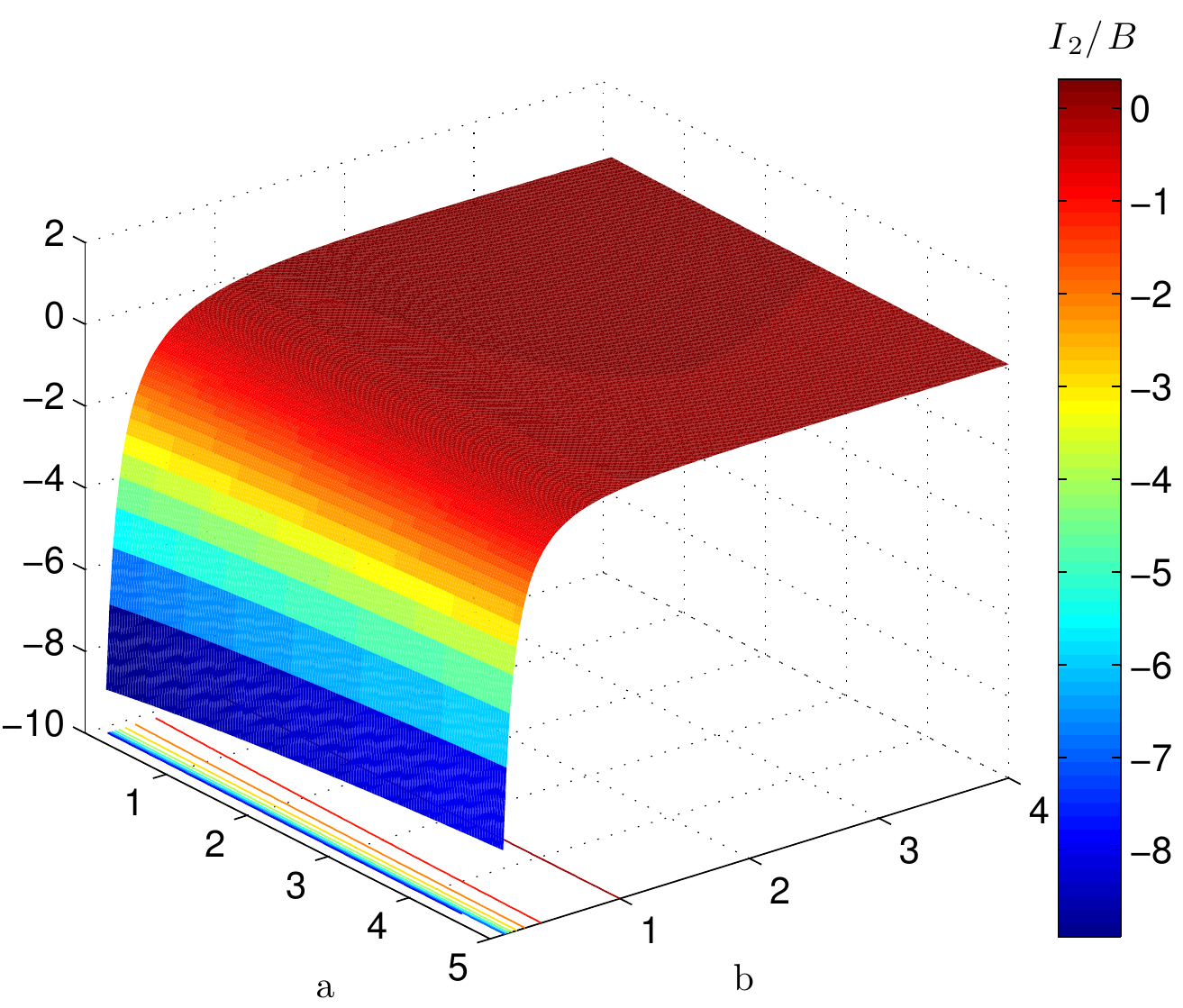}
\caption{Integral $I_2/D$ as function of $a$ and $b$.}
\label{fig:distri:I2}
\end{center}
\end{figure}

We rewrite $I_2$ using the dimensionless variables $x_1 = m_1/m$ and $x_2 = m_2/m$ which yields
\begin{equation}
I_2 = A^2 \cdot h_0 \cdot m^{3+\nu-2\alpha} \int_\frac{1}{2}^1 \dx{x_1} \, \int_{1-x_1}^{x_1} \dx{x_2} \, x_1^{1+\nu-\gamma-\alpha} \cdot x_2^{\gamma-\alpha}.
\end{equation}
By integrating over $x_2$, we derive
\begin{equation}
I_2=\underbrace{\frac{A^2 \cdot h_0\cdot m^{3+\nu-2\alpha}}{\gamma-\alpha+1}}_{:=D} \cdot \int_\frac{1}{2}^{1}\dx{x_1} x_1^{1+\nu-\gamma-\alpha} \cdot \left[x_1^{\gamma-\alpha+1}-\left(1-x_1\right)^{\gamma-\alpha+1} \right].
\end{equation}
The term in square brackets can be split into a sum of integrals, the first of which is straight-forward to evaluate as
\begin{equation}
\frac{I_2}{D} = \frac{1-\left(\frac{1}{2}\right)^{3+\nu-2\alpha}}{\nu-2\alpha+3} -  \int_\frac{1}{2}^{1}\dx{x_1} \, x_1^{\overbrace{1+\nu-\gamma-\alpha}^{a-1}} \cdot \left(1-x_1\right)^{\overbrace{\gamma-\alpha+1}^{b-1}},
\end{equation}
while the second can be identified as a sum of a Beta function $B(a,b)$ and an incomplete Beta function $B_\frac{1}{2}(a,b)$,
\begin{equation}
\frac{I_2}{D} = \frac{1-\left(\frac{1}{2}\right)^{a+b-1}}{a+b-1} -  \left( B(a,b) - B_\frac{1}{2}(a,b) \right).
\end{equation}
The conditions from above,
\begin{alignat}{2}
a&=2+\nu-\gamma-\alpha\quad  &>0 \label{eq:distri:new_condition1} \\
b&=2+\gamma-\alpha    \quad  &>0 \label{eq:distri:new_condition2}
\end{alignat}
assure that the Beta functions, and thus $I_2$, are real. The numerical value of $I_2/D$ for a range of values of $a$ and $b$ are shown in Fig.~\ref{fig:distri:I2}.

\subsection{Third integral}
Similarly, we can rewrite $I_3$ as
\begin{align}
I_3 &= A^2 \cdot h_0 \cdot \int_\frac{1}{2}^m \dx{m_1} \int_{m_1}^{\mf}\dx{m_2}\, m_1^{1+\gamma-\alpha} \cdot m_2^{\nu-\gamma-\alpha}\\
    &= \frac{A^2 \cdot h_0}{\nu-\gamma-\alpha+1} \cdot \int_\frac{m}{2}^m \dx{m_1} \, m_1^{1+\gamma-\alpha} \cdot \left( \mf^{\nu-\gamma-\alpha+1} - \left(\frac{m}{2}\right)^{\nu-\gamma-\alpha+1} \right). \nonumber
\end{align}
If Eq.~\ref{eq:distri:new_condition1} holds, then from $\mf \gg m$ follows
\begin{equation}
I_3 = \frac{A^2 \cdot h_0\cdot \mf^{\nu-\gamma-\alpha+1}}{(\nu-\gamma-\alpha+1)\cdot(\gamma-\alpha+2)} \cdot \left( 1 - \left(\frac{1}{2}\right)^{\gamma-\alpha+2} \right) \cdot m^{\gamma-\alpha+2}.
\end{equation}

\subsection{Deriving the steady-state distribution}
The first and the third integrand $I_1$ and $I_3$ show the same mass dependence and can be summed up to
\begin{equation}
\frac{I_1+I_3}{A^2\cdot h_0} \propto
  \mf^{\nu-\gamma-\alpha+1}\cdot m^{\gamma-\alpha+2}
\label{eq:distri:I13}
\end{equation}
while $I_2$ is proportional to
\begin{equation}
  \begin{aligned}
  \frac{I_2}{A^2 \cdot h_0} &\propto m^{3+\nu-2\alpha} \\
                            &\propto \left(\frac{m}{\mf}\right)^{\nu-\alpha-\gamma+1} \, \mf^{\nu-\gamma-\alpha+1} \,m^{\gamma-\alpha+2} ,
  \end{aligned}
\end{equation}
and therefore
\begin{equation}
\frac{I_2}{I_1+I_3} \propto \left(\frac{m}{\mf}\right)^{\nu-\alpha-\gamma+1}.
\end{equation}
Since the constants of proportionality are factors of order unity and $m<\mf$, the integrals $I_1+I_3$ are much larger than $I_2$, therefore, the flux $F(m)$ is proportional to $m^{\gamma-\alpha+2}$.

In the case of a steady state, this flux and the downward flux of fragments (which is proportional to $m^{2-\xi}$) have to cancel each other. Therefore, the exponents of the mass dependence need to cancel out, i.e.
\begin{equation}
\alpha = \gamma + \xi,
\end{equation}
which is the slope of the steady-state condition in the fragmentation dominated regime (Case C in Fig.~\ref{fig:distri:sketch}).
\end{subappendices}
\chapter{Dust retention in protoplanetary disks}\label{chapter:dustretention}
\chapterprecishere{\centering Published as \citealp*[A\&A, 503, L5]{Birnstiel:2009p7135}}

\begin{abstract}
  {Protoplanetary disks are observed to remain dust-rich for
  up to several million years. Theoretical modeling, on the other hand,
  raises several questions. Firstly, dust coagulation occurs so rapidly, that if
  the small dust grains are not replenished by collisional fragmentation of
  dust aggregates, most disks should be observed to be dust poor, which is
  not the case. Secondly, if dust aggregates grow to sizes of the order of
  centimeters to meters, they drift so fast inwards, that they
  are quickly lost.}
  {We attempt to verify if collisional fragmentation
  of dust aggregates is effective enough to keep disks `dusty' by replenishing
  the population of small grains and by preventing excessive radial drift.}
  {With a new and sophisticated implicitly integrated
  coagulation and fragmentation modeling code, we solve the combined problem
  of coagulation, fragmentation, turbulent mixing and radial drift
  and at the same time solve for the 1-D viscous gas disk
  evolution.}
  {We find that for a critical collision velocity of 1~m/s, as
  suggested by laboratory experiments, the fragmentation is so
  effective, that at all times the dust is in the form of relatively small
  particles. This means that radial drift is small and that large amounts
  of small dust particles remain present for a few million years, as
  observed. For a critical velocity of 10~m/s, we find that particles
  grow about two orders of magnitude larger, which leads again to significant
  dust loss since larger particles are more strongly affected by radial drift.}
\end{abstract}
\clearpage
\raggedbottom
\section{Introduction}\label{sec:dustretention:introduction}
The quest for a comprehensive understanding of the formation of planets in
the dusty circumstellar disks around many young stars has gained significant
impetus. This is partly because of developments in the numerical
modeling of these processes, but also because of the high-quality infrared and
\mbox{(sub-)}millimeter data that are now available of hundreds of T Tauri and
Herbig Ae star disks from observatories such as the Spitzer Space Telescope
\citep[e.g.][]{Furlan:2006p4444,KesslerSilacci:2006p4450}, the Very
Large Telescope \citep[e.g.][]{vanBoekel:2004p4708}, the Submillimeter Array
\citep[e.g.][]{Andrews:2007p3380} and the Plateau de Bure Interferometer
\citep[e.g.][]{Pietu:2007p4501}. The challenge for theoretical astrophysicists is now not only to come up with a plausible model of how to overcome the various problems in our current understanding of how planets form, but also to ensure that new models are consistent with the observational data obtained from these planetary birthplaces.

One of the challenges of theories of planet formation is what is often called the `meter-size barrier'.
It was shown by \citet{Weidenschilling:1977p865} that as dust particles grow by coagulation, they acquire increasingly large relative velocities with respect to other particles in their vicinity, as well as a systematic inward drift velocity as originally described by \citet{Whipple:1972p4621}. The high relative velocities lead to destructive collisions \citep{Blum:2008p1920}, and thereby limit the growth of the aggregates to some maximum size. The radial drift also leads to a loss of solids toward the star well before large bodies can even form \citep[][hereafter \citetalias{Brauer:2008p215}]{Brauer:2008p215}.

\citet{Dullemond:2005p378} showed that if, hypothetically, coagulation can proceed without limit, i.e.,\ if no fragmentation occurs and no radial drift is present, the growth is so efficient that within $10^5$ years very few small ($a\lesssim 1$mm) opacity bearing grains remain in the disk. The disk becomes optically thin and the mid- to far-infrared flux of the disk drops to levels well below that observed in the majority of gas-rich circumstellar disks around pre-main sequence stars. By excluding other mechanisms, \citet{Dominik:2008p4626} argue that fragmentation is the most promising process for maintaining a high abundance of fine-grained dust in evolved disks.

However, it has never been demonstrated explicitly that fragmentation is efficient enough to retain the population of small grains at the required levels. Moreover, if we include the effect of radial drift, it remains to be seen whether this radial drift is suppressed sufficiently by keeping the grains small through fragmentation. It is the goal of this chapter to investigate this issue and to determine how efficient fragmentation should be to agree with observations.

\section{Model}\label{sec:dustretention:model}
\flushbottom
The model presented here is a combination of a 1D viscous gas disk evolution code and a dust evolution code treating the turbulent mixing, radial drift, coagulation and fragmentation of the dust. A detailed description of the model can be found in Chapters~\ref{chapter:model} and \ref{chapter:algorithm}. Although the evolution of dust can influence the gas disk evolution, we neglect this effect in our model.

The gas disk model is similar to that described in \citet{Hueso:2005p685}. The mid-plane temperature of the disk is approximated by following \citet{Nakamoto:1994p798}, taking irradiation by the central star and viscous heating into account. The viscous evolution of the disk surface density $\Sigma_\mathrm{g}$ is described by \citep[see][]{LyndenBell:1974p1945},
\begin{equation}
\frac{\del \Sigma_\mathrm{g} }{\del t} - \frac{1}{r}\frac{\del}{\del r}\left( \Sigma_\mathrm{g} \, r \, u_\mathrm{g}\right) = 0,
\end{equation}
where the gas radial velocity $u_\mathrm{g}$ is given by
\begin{equation}
u_\mathrm{g} = - \frac{3}{\Sigma_\mathrm{g}\sqrt{r}} \frac{\del}{\del r} \left( \Sigma_\mathrm{g} \nu_\mathrm{g} \sqrt{r} \right).
\end{equation}
$\nu_\mathrm{g}$ is the gas turbulent viscosity,
\begin{equation}
\nu_\mathrm{g} = \alpha \, \csound \, \Hp,
\end{equation}
\csound denotes the sound speed, \Hp the pressure scale height, and $\alpha$ is the turbulence parameter \citep{Shakura:1973p4854}.

Grains are affected by a systematic radial inward drift related to headwind caused by the pressure-supported gas \citep{Weidenschilling:1977p865},
\begin{equation}
u_\text{drift} = \frac{1}{\St^{-1}+\St}\cdot\frac{ \del_{r}{P_\text{g}}}{\rho_
\text{g} \: \Omega_\text{k}},
\label{eq:dustretention:v_drift}
\end{equation}
where \St is the Stokes number of the particle (a dimensionless representation of the particle size), $P_\text{g}$ is the gas pressure, $\rho_\text{g}$ is the gas volume density, and $\Omega_\text{k}$ is the Kepler frequency.
In the Epstein regime, \St is given by
\begin{equation}
\St = \frac{a \rho_\mathrm{s}}{\Sigma_\mathrm{g}} \frac{\pi}{2},
\label{eq:dustretention:stokesnumber}
\end{equation}
where $\rho_\mathrm{s}$ is the solid density of the dust grains.

The second contribution to radial dust velocities is the gas drag due to the radial motion of the gas,
\begin{equation}
u_\text{drag} = \frac{u_\text{g}}{\Sc},
\label{eq:dustretention:v_drag}
\end{equation}
where $\Sc=1 + \St^2$ is the Schmidt number, following \citet{Youdin:2007p2021}.

The coupling to the gas turbulence leads to turbulent mixing of each dust species with a diffusion constant that is taken to be
\begin{equation}
D_\text{d} = \frac{\nu_\mathrm{g}}{\Sc}.
\end{equation}

The dust grain number density $n(m,r,z)$ is a function of mass $m$, radius $r$, height above the mid-plane $z$, and time. If we define
\begin{equation}
\sigma(m,r) \equiv \int_{-\infty}^{\infty} n(m,r,z) \, m^2 \dx z,
\end{equation}
the vertically integrated time-evolution of this distribution can now be described by a general two-body process and an advection-diffusion equation,
\begin{equation}
\begin{array}{lll}
\frac{\partial \sigma(m,r)}{\partial t} 
&=&\iint_{0}^{\infty} K(m,m',m'') \, \sigma(m',r) \, \sigma(m'',r)\, \dx m'\, \dx m''\\
\\
&&-\frac{1}{r}\frac{\partial}{\partial r}
\left[ r \cdot \sigma(m,r) \cdot (u_\mathrm{drag}+u_\mathrm{drift}) \right]\\
\\
&&+\frac{1}{r}\frac{\partial}{\partial r} \left[ r \cdot D_\mathrm{d} \cdot \frac{\partial}{\partial r} \left( \frac{\sigma(m,r)}{\Sigma_\mathrm{g}}\right) \cdot \Sigma_\mathrm{g} \right].
\end{array}
\label{eq:dustretention:smolu}
\end{equation}
Here, the right-hand side terms (from top to bottom) correspond to co\-agu\-lation/frag\-mentation, advection, and turbulent mixing.

Since the detailed version of the coagulation/\-fragmentation equation is lengthy
(\citealp{Weidenschilling:1980p4572}; \citealp{Nakagawa:1981p4533}; \citealp{Dullemond:2005p378}; \citetalias{Brauer:2008p215}), we include here only the general integral of a two-body process (first term on the right hand side). For a detailed description of the physics of coagulation/fragmentation, we refer to \citetalias{Brauer:2008p215}; the numerical implementation is discussed in Chapter~\ref{chapter:algorithm}.

The physical effects that produce the relative particle velocities considered here are Brownian motion, relative radial motion, vertical settling \citepalias[see][]{Brauer:2008p215}, and turbulent motion \citep[see][]{Ormel:2007p801}.

If particles collide with a relative velocity higher than the critical velocity \uf, they either fragment into a power-law size distribution of fragments \citepalias[i.e., $n(m)\propto m^{-1.83}$, see][]{Brauer:2008p215} or the smaller body excavates mass from the larger one (cratering). We assume that the amount of excavated mass equals the mass of the smaller body. The fragmentation velocity \uf is a free parameter in our model, and unless otherwise noted is taken to be 1~m/s, as suggested by experiments \citep[e.g.,][]{Blum:1993p4324} and by theoretical modeling \citep[e.g.,][]{Stewart:2009p5281}.

We note that the velocities and therefore also the relative velocities are dependent on the particle size. If we assume that relative particle velocities are dominated by turbulent motion,
\begin{equation}
\Delta u_\mathrm{turb} \propto \sqrt{\alpha\:\St} \: \csound,
\label{eq:dustretention:v_turb}
\end{equation}
then we can estimate the Stokes number of the largest particles by equating the relative velocity and fragmentation velocity,
\begin{equation}
\St_\mathrm{max} \simeq \frac{\uf^2}{\alpha\:\csound^2}.
\label{eq:dustretention:st_max}
\end{equation}
With Eq. \ref{eq:dustretention:stokesnumber}, we can relate this maximum `dimensionless size' $\St_\mathrm{max}$ to a particle size,
\begin{equation}
a_\mathrm{max} \simeq \frac{2\Sigma_\mathrm{g} }{\pi \alpha \rho_\mathrm{s}} \cdot \frac{\uf^2}{\csound^2}.
\label{eq:dustretention:a_max}
\end{equation}
It should be noted that Eq. \ref{eq:dustretention:v_turb} is an approximation of the equations derived in \citet{Ormel:2007p801}. The detailed relative velocities of particles depend on the Stokes numbers of both particles. Hence, Eq. \ref{eq:dustretention:a_max} is an order of magnitude estimate.

The combined equations for coagulation, fragmentation, and radial motion due to drift, drag, and mixing are solved using the technique of implicit integration \citepalias[similar to][]{Brauer:2008p215}. We use the SixPACK F-90 linear algebra package\footnote{available from \url{www.engineers.auckland.ac.nz/~snor007}} to solve the sparse matrix equation.

\begin{figure*}
  \centering
   \includegraphics[width=\hsize]{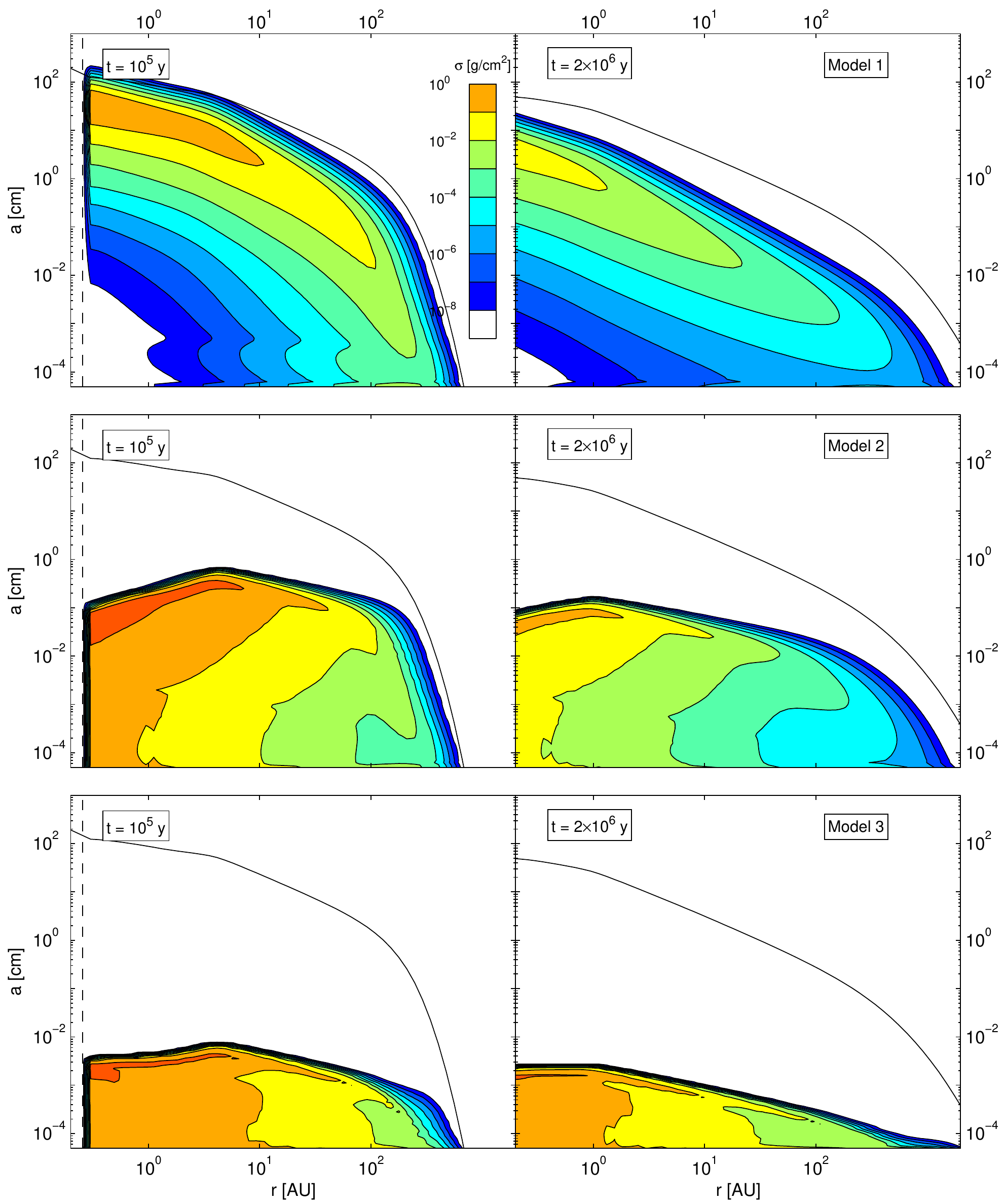}
   \caption{Snapshots of the vertically integrated dust density distribution of model~1 (only coagulation), model~2 (fragmentation at 10~m/s) and model~3 (fragmentation at 1~m/s). The grain size is given by $a=(3m/4\pi\rho_\mathrm{s})^{1/3}$. The dashed line shows the evaporation radius within which no coagulation is calculated. The solid line shows the particle size corresponding to a Stokes number of unity.}
   \label{fig:dustretention:snapshots}
\end{figure*}

\section{Results}\label{sec:dustretention:results}
The simulation results presented here begin with a 0.07~$M_\odot$ disk of power-law surface density $\Sigma(r)\propto r^{-1}$ (from 0.05 to 150 AU) around a 0.5~$M_\odot$ star. Unless otherwise noted, we used a turbulence parameter of $10^{-2}$. The three models differ in the prescription of fragmentation. In model~1, we accounted only for growth without fragmentation. In model~2, particles fragment at $\uf=10$ m/s, whereas in model~3, the fragmentation speed is taken to be 1~m/s. Selected  results of all three models are depicted in Figure \ref{fig:dustretention:snapshots}.

The top row shows the evolution in the surface density distribution of dust because of coagulation, radial mixing, and radial drift in an evolving protoplanetary disk, corresponding to model~1.
The dashed line denotes the grain evaporation radius, which moves inwards as the temperature falls (lower surface density caused by accretion produces less viscous heating).
The evaporation radius in our model is defined as the radius where the temperature rises above 1500~K assuming that all particles evaporate at this temperature. The solid line denotes the dust grain size, which corresponds to a Stokes number of unity.  It can be seen that particles grow to decimeter size and then quickly drift inside the evaporation radius.

The results of simulations that also include grain fragmentation are plotted in the second and third row of Figure~\ref{fig:dustretention:snapshots}.
In the innermost regions of model 2, relative particle velocities are high enough (as a result of shorter dynamical times) to fragment particles already at relatively small Stokes number. However, the maximum size of the grain distribution is still strongly affected by radial drift, especially at larger radii. Larger particles are subject to radial drift and therefore drift to smaller radii, where they are pulverized.

In contrast, fragmentation becomes the main limiting factor for particle growth throughout the disk if the critical fragmentation velocity is 1~m/s (model~3). The result is that particles fragment at a small Stokes number. Particles with a Stokes number less than about $10^{-3}$ are still strongly coupled to the gas and not subject to strong radial drift \citep[see][]{Brauer:2007p232}. Therefore, significant amounts of dust can be retained for several million years if fragmentation is included in the calculations.

\begin{figure}[htp]
  \centering
   \includegraphics[width=0.9\textwidth]{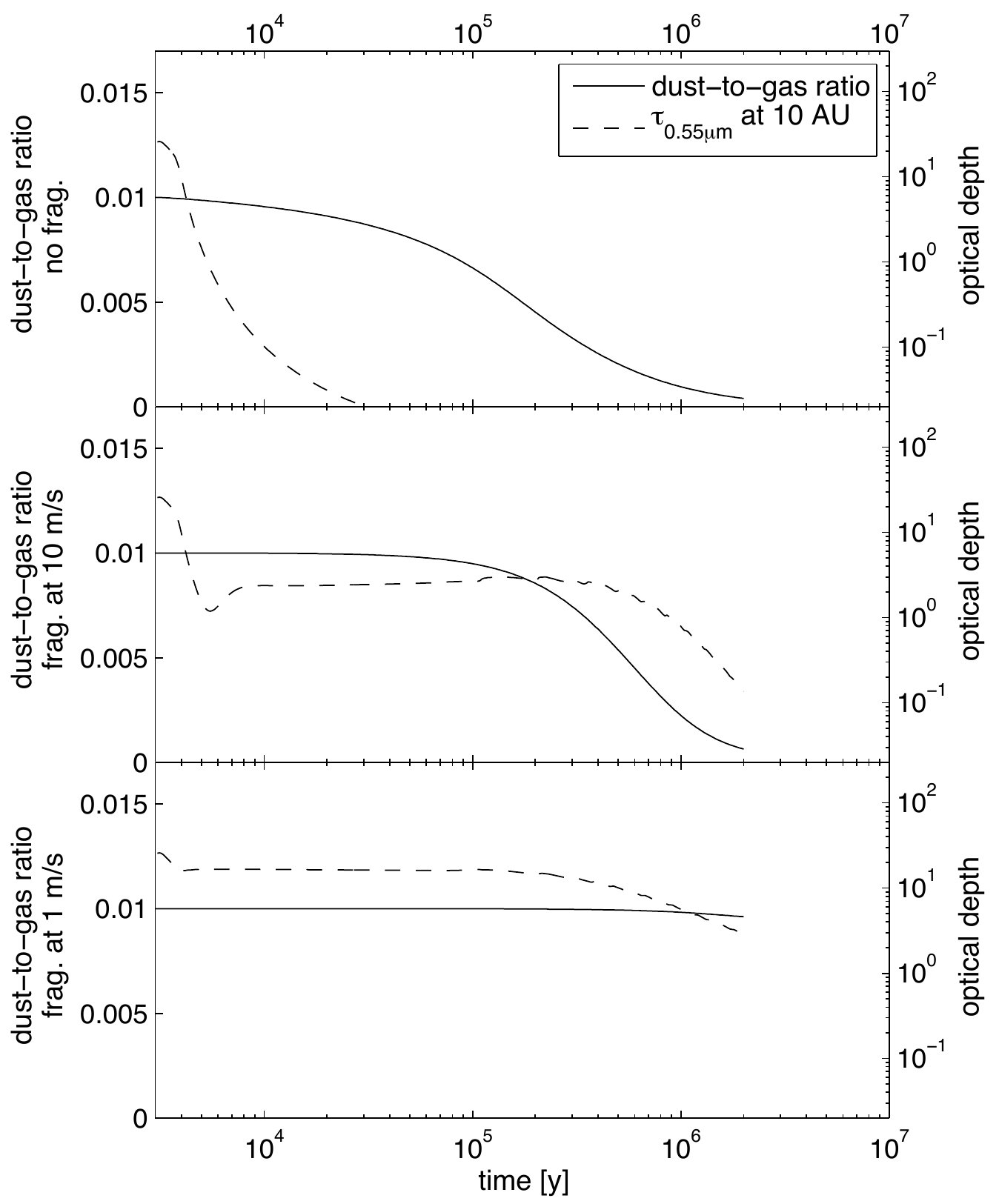}
   \caption{Dust-to-gas ratio (solid line, linear scale) and 0.55 $\mu$m optical depth at 10 AU (dashed line, logarithmic scale) as function of time for model~1 (top), model~2 (center) and model~3 (bottom).}
   \label{fig:dustretention:d2g_1}
\end{figure}

Figure~\ref{fig:dustretention:d2g_1} compares the evolution in the dust-to-gas ratio (solid line) and the 0.55 $\mu$m optical depth at 10 AU (dashed line) of all three simulations.
The optical depth is defined as
\begin{equation}
\tau_\nu = \int_{0}^\infty \sigma(m) \, \kappa_\nu(m) \dx m.
\end{equation}
It can be seen that the dust-to-gas ratio declines dramatically in the first two cases. The optical depth in model~1 drops on even shorter timescales than the dust-to-gas ratio because not only the dust loss but also the dust growth causes the optical depth to decrease.

Model~2 can retain the optical depth longer because small particles remain (due to fragmentation), providing enough surface to keep the disk optically thick for about 0.9 Myr. The initial dip in the optical depth is caused by the initial condition: since initially, all mass is in small grains, the optical depth decreases as particles grow to larger sizes until fragmentation sets in, which increases the optical depth until a semi-equilibrium of growth and fragmentation is reached.

If particles are already fragmented at 1 m/s (model~3), the dust-to-gas ratio declines only slightly, after 2 My, only about 4\% of the dust being lost (compared to 96\% and 94\% in models~1 and 2). The optical depth decreases on longer timescales than in model~2 since the dust mass is lost only on the viscous timescale.

This mechanism of grain retention relies on the particles remaining small due to fragmentation. From Eq. \ref{eq:dustretention:st_max} and \ref{eq:dustretention:a_max}, it follows that the maximum size or Stokes number depends strongly on the critical velocity \uf and less strongly on $\alpha$. If \uf is 10 m/s instead of 1 m/s, the corresponding particle size is 100 times larger meaning that the largest particles are now also experiencing the radial drift barrier and therefore drifting quickly towards the central star.

\section{Discussion and conclusions}\label{sec:dustretention:discussion}
We have shown that grain fragmentation plays an important role, not only in keeping the disk optically thick (by replenishing small grains) but also in maintaining a relatively high abundance of solids of all sizes.
If particles fragment at relatively small sizes, they can stay below the size-regime where they are affected by radial drift.

The simulation results presented here suggest that fragmentation of relatively low critical velocity (a few m/s) is needed to retain the dust required for planet formation, although the formation of planetesimals probably requires periods (and/or regions) of quiescence such as dead zones or pressure bumps \citep[see][]{Kretke:2007p697,Brauer:2008p212}. These locally confined quiescent environments could decrease both the relative and radial velocities of particles, thus allowing particles possibly to either grow to larger sizes or to accumulate until gravitational instabilities become important \citep[see][]{Johansen:2007p4788,Lyra:2009p4812}. These effects are not yet included in this model and will be the subject of future research.

Assuming that the proposed mechanism of dust retention is effective in protoplanetary disks, we can constrain the critical fragmentation velocity to be smaller than 10 m/s since velocities $\gtrsim$ 10 m/s cannot explain the observed lifetimes of disks even in the case of small $\alpha$.

\ifthenelse{\boolean{chapterbib}}
{
    \clearpage
    \bibliographystyle{aa}
    \bibliography{/Users/til/Documents/Papers/bibliography}
}
{}
\chapter
[Testing the theory of grain growth and fragmentation by millimeter observations of protoplanetary disks]
[Testing the theory of grain growth and fragmentation]
{Testing the theory of grain growth and fragmentation by millimeter observations of protoplanetary disks}
\label{chapter:mmobs}
\chapterprecishere{\centering Published as \citealp*[A\&A, 503, L5]{Birnstiel:2010p12008}}

\begin{abstract}
{Observations at sub-millimeter and mm wavelengths will in the near future be able to resolve the radial dependence of the mm spectral slope in circumstellar disks with a resolution of around a few~AU at the distance of the closest star-forming regions.}
{We aim to constrain physical models of grain growth and fragmentation by a large sample of (sub-)mm observations of disks around pre-main sequence stars in the Taurus-Auriga and Ophiuchus star-forming regions.}
{State-of-the-art coagulation/fragmentation and disk-structure codes are coupled to produce steady-state grain size distributions and to predict the spectral slopes at (sub-)mm wavelengths.}
{This work presents the first calculations predicting the  mm spectral slope based on a physical model of grain growth. Our models can quite naturally reproduce the observed mm-slopes, but a simultaneous match to the observed range of flux levels can only be reached by a reduction of the dust mass by a factor of a few up to about 30 while keeping the gas mass of the disk the same. This dust reduction can either be caused by radial drift at a reduced rate or during an earlier evolutionary time (otherwise the predicted fluxes would become too low) or due to efficient conversion of dust into larger, unseen bodies.}
\end{abstract}

\section{Introduction}                  \label{sec:mmobs:intro}
Circumstellar disks play a fundamental role in the formation of stars as most of the stellar material is believed to be transported through the disk before being accreted onto the star \citep{LyndenBell:1974p1945}. At the same time circumstellar disks are thought to be the birth places of planets. Understanding the physics of circumstellar disks is therefore the key to some of the most active fields of astrophysical research today.

However, observing these disks in order to learn about the physical processes taking place in their interior is a challenging task.
\citet{Strom:1989p9475} and \citet{Beckwith:1990p3768} were the first to use observations at mm-wavelengths to confirm that many of the observed pre-main sequence (PMS) stars showed excess radiation above the spectrum of a T Tauri star.
While these single dish observations provided valuable insight into dust masses (because mm observations probe not only the thin surface layers, but the bulk of the dust mass in the mid-plane), recent developments in the field of \mbox{(sub-)mm} interferometry allow one to constrain models of disk structure and evolution of protoplanetary disks by fitting parametric models to the observed radial profiles \citep[e.g.,][]{Andrews:2009p7729,Isella:2009p7470}. Spatially resolving the disks is important because it ensures that low millimeter spectral slopes are not just an artifact of high optical depth.

Today, mm spectral slopes are known for quite a number of disks, and spatially resolved observations indicate that the low values measured in these samples are related to grain growth \citep[e.g.,][]{Testi:2003p3390,Natta:2004p3169,Rodmann:2006p8905}. Grains are believed to collide and stick together by van der Waals forces, thus forming larger and larger aggregates \citep{Dominik:1997p9440,Poppe:2000p9447,Blum:2008p1920}. Due to this loose binding, collisions with velocities in excess of a few m~s$^{-1}$ may lead to fragmentation of the aggregates.

Larger samples of radially resolved mm spectral slopes are expected in the near future, but so far no study interpreted mm observations using simulated grain size distributions. Instead simple parametric power-law distributions were used with an upper size cut-off. In this work, we use a state of the art dust grain evolution code \citep[similar to][]{Brauer:2008p215,Birnstiel:2010p9709} to derive steady-state grain distributions where grain growth and fragmentation effects balance each other. We self-consistently solve for the grain size distributions and the disk structure to predict fluxes at mm wavelengths and the radial dependence of the mm spectral index. Comparing these results with observed values in the Taurus and Ophiuchus star-forming regions allows us to test the predictions of the theory of grain growth/fragmentation and to infer constraints on grain properties such as the critical collision velocity and the distribution of fragments produced in collision events.

Grains orbiting at the Keplerian velocity in a laminar gas disk feel a constant head wind (caused by the gas rotating slightly sub-keplerian), which forces them to spiral inwards \citep{Weidenschilling:1977p865}. If this drag is as efficient as the laminar theory predicts, all dust particles that are necessary to explain the observed spectral indices would quickly be removed \citep[see][]{Brauer:2007p232}. We therefore assume that the radial drift is halted by an unknown mechanism. Under this assumption, we find that the low values of the mm spectral index can be explained by the theory. We show that in order to explain the observed flux levels, the amount of observable\footnote{by ``visible'' or ``observable'' dust we mean the dust particles which are responsible for most of the thermal continuum emission at \mbox{(sub-)mm} wavelengths, which are typically smaller than a few centimeter in radius.} dust needs to be reduced by either reducing the dust-to-gas ratio (perhaps by radial drift at a intermediate efficiency or during an earlier evolutionary epoch) or by dust particle growth beyond centimeter sizes. 

\section{Model description}             \label{sec:mmobs:model}
\subsection{Disk model}                 \label{sec:mmobs:model_disk}
We consider disks around a PMS star with a mass of 0.5~$M_{\odot}$, bolometric luminosity of 0.9~$L_{\odot}$ and effective temperature of 4000~K, at a distance of 140~pc, which are typical values for the sample of low-mass PMS stars studied in \citet[][hereafter \citetalias{Ricci:2010p9423}]{Ricci:2010p9423}. To derive the disk structure we adopted a modified version of the two-layer models of passively irradiated flared disks developed by \citet{Dullemond:2001p9307} (following the schematization by \citealp{Chiang:1997p1986}), in which we have relaxed the common assumption that dust grain properties are constant throughout the disk. For the disk surface density we adopted the self-similar solution for a viscous disk \citep[see][]{LyndenBell:1974p1945} with parameters lying in the ranges observationally constrained by \citet{Andrews:2009p7729}. The surface density gradient $\gamma$ and the characteristic radius $R_\mathrm{c}$ \citep[for the definitions, see][]{Hartmann:1998p664} are assumed to be $\gamma=1$ and $R_\mathrm{c}=60$~AU, respectively. Throughout this work we assume a constant dust-to-gas mass ratio of 1\%.

\begin{table}
\caption{Parameters of the model grid: $M_\text{disk}$ is the total disk mass, $\alphat$ is the turbulence parameter, \uf is the critical collision velocity, $f_\text{vac}$ is the grain volume fraction of vacuum, and $\xi$ is the index of the distribution of fragments (see Eq.~\ref{eq:mmobs:n_frag}). The parameters of the fiducial model are highlighted in bold face.}
\label{tab:mmobs:model_grid}
\centering
\begin{tabular}{lr|llll}
\hline\hline
parameter & \multicolumn{4}{c}{values}\\
\hline\\[-0.25cm]
    $M_\mathrm{disk}$ &$[M_\odot]$ & 5\e{-3}                 & $\mathbf{1\e{-2}}$          & 5\e{-2} & 1\e{-1}\\
    $\alphat$         &            & $\mathbf{5\e{-4}}$      & 1\e{-3}                     & 5\e{-3}                    & -      \\
    \uf               &[m/s]       & 1                       & \textbf{3}                  & 10                         & -      \\
    $f_\text{vac}$    &[\% vol.]   & \textbf{10}             & 30                          & 50                         & -      \\
    $\xi$             &            & 1.0                     & \textbf{1.5}                & 1.8                        & -      \\
\hline
\end{tabular}
\end{table}

\subsection{Dust model}                 \label{sec:mmobs:model_dust}
We use a coagulation/fragmentation code as described in \citet{Brauer:2008p215} and \citet{Birnstiel:2010p9709} to simulate the growth of dust particles. Particles grow through mutual collisions (induced by Brownian motion and by turbulence, see \citealp{Ormel:2007p801}) and subsequent sticking by van der Waals forces. We assume the dust particles to be spheres of internal density \rhos and vary \rhos to account for porosity effects. However, we do not employ a dynamic porosity model \citep[see][]{Ormel:2007p7127,Zsom:2008p7126}.

With increasing collision velocity $\Delta u$, the probability of sticking decreases and fragmentation events start to become important. Here, we use the fragmentation probability  
\begin{equation}
p_\text{f} = \left\{
\begin{array}{ll}
0&                              \text{if } \Delta u < \uf - \delta u\\
1&                              \text{if } \Delta u > \uf\\
1-\frac{\uf-\Delta u}{\delta u}&    \text{else}
\end{array}
\right.,
\end{equation}
where \uf is the collision velocity above which particles are assumed to fragment, and $\delta u$ is the transition width between coagulation and fragmentation (taken to be 0.2 \uf). Recent studies of collision experiments \citep{Guttler:2010p9745} and numerical simulations \citep{Zsom:2010p9746} indicate that there is also a regime in which particles may bounce. However, because this topic is still not well understood, we will omit these effects here. 

Radial drift is an as yet unsolved problem \citep{Brauer:2007p232,Birnstiel:2009p7135}. Still there are several effects such as spiral wave structure \citep[e.g.,][]{Cossins:2009p3794}, density sinks \citep[e.g.,][]{Brauer:2008p212}, or zonal flows \citep[e.g.,][]{Johansen:2009p7441} which may reduce the effectiveness of radial drift. We assume that radial drift is ineffective because we focus on the question whether observations can be explained through the physics of grain growth and fragmentation. The question to answer in this case is not how to retain particles at these radii, but rather how to create them there in the first place.

To investigate this problem, we simulate the physics of particle growth and fragmentation until a steady state between both processes develops. Because the relative velocities for particles typically increase with grain radius, we can relate the fragmentation velocity to a certain grain size \citep[which defines the ``fragmentation barrier'', see][]{Birnstiel:2009p7135}
\begin{equation}
a_\mathrm{max} \simeq \frac{2\Siggas }{\pi \alphat \rho_\mathrm{s}} \cdot \frac{\uf^2}{\csound^2},
\label{eq:mmobs:a_max}
\end{equation}
above which particles fragment (with \Siggas, $\alphat$ and \csound being the gas surface density, the turbulence parameter, and the sound speed, respectively).
\uf and \alphat are assumed to be radially constant with \alphat values within a range expected from theoretical \citep[see][]{Johansen:2005p8425,Dzyurkevich:2010p11360} and observational works \citep[see][]{Andrews:2009p7729}.
Grains which reach \amax will experience high velocity collisions, causing them to be eroded or even completely fragmented. The resulting fragments can again contribute to growth processes at smaller sizes, and the grain size distribution will at some point reach a steady state where gain and loss terms caused by coagulation and by fragmentation cancel out at all sizes.

Particles will need a certain time to grow to the fragmentation barrier. The time to reach the steady state will therefore be several of these growth time scales. Depending on the distance to the central star, the steady state is typically reached after a few thousand years at 1~AU up to about 1~Myr at 100~AU. The mean ages of the sources in our sample are $\approx 2$~Myr and $\approx 0.5-1$~Myr for the Taurus and Ophiuchus PMS stars, respectively. Since radii around $40-80$~AU dominate the observed emission at \mbox{(sub-)mm} wavelengths, we expect most of the samples to be in or at least close to a steady state. 

If the highest collision velocity that turbulent motion induces (depending on $\alphat$ and \csound) is lower than the critical collision velocity \uf, then (at least some) particles do not fragment (i.e. the break through the fragmentation barrier) and a steady state is never reached. Owing to this scenario, some of the possible combinations of the parameter values (see Table~\ref{tab:mmobs:model_grid}) do not reach a steady state and are therefore not included in the results. 

The shape of the steady-state grain size distributions is influenced mainly by five parameters: the previously mentioned $\alphat$, \uf, \Siggas, the temperature $T$ (through the sound speed \csound), and by the prescription of fragmentation. In our models, we assume the distribution of fragments to follow a power-law number density distribution,
\begin{equation}
n(m) \propto m^{-\xi},
\label{eq:mmobs:n_frag}
\end{equation} 
with an upper end at $m_\mathrm{f}$. We consider fragmentation and cratering, as described in \citet{Birnstiel:2010p9709}.
Recent experiments suggest $\xi$ values between 1.07 and 1.37 \citep[see][]{Guttler:2010p9745}. In this work, we consider $\xi$ values between $1.0$ and $1.8$.

To calculate the dust opacity of a given grain size distribution we adopted the same dust grain model as in \citetalias{Ricci:2010p9423}, i.e. porous composite spherical grains made of astronomical silicates, carbonaceous materials and water ices (see \citetalias{Ricci:2010p9423} for the references to the optical constants). The ratio between the fractional abundance of each species comes from \citealp{Semenov:2003p9622}, and models with three different porosities have been considered in this chapter (see Table~\ref{tab:mmobs:model_grid}).
We used the Bruggeman mixing theory to combine the refractive indices of the different materials and to calculate the dust opacity of the composite grains.
The opacity induces probably the largest uncertainties in our calculations because grain composition, grain structure and temperature effects may lead to largely different opacities \citep[see, for example][]{Henning:1996p4181}.

\subsection{Comparison to observations} \label{sec:mmobs:model_comparison}
We compare the \mbox{(sub-)mm} SED generated by our models with observational data of \citetalias{Ricci:2010p9423} and Ricci et al. (in prep), more specifically the flux at 1~mm (\Fmm) and the spectral index between 1 and 3~mm ($F(\lambda) \propto \lambda^{-\alphamm}$). The samples considered include all class II disks in the Taurus-Auriga and $\rho$-Oph star-forming regions for which both the central PMS star and the disk are observationally well characterized through optical-NIR spectroscopy/photometry and \mbox{(sub-)mm} photometry/interferometry. To calculate the dust opacity as a function of wavelengths and radius, and the temperature in the disk mid-plane, we iterated the two-layer disk model (keeping the profile of $\Siggas$ constant in time) with the dust model described above until convergence is reached. Once the physical structure of the disk is determined, the two-layer disk models return the disk SED, which can be compared with the observations.

The influence of the different parameters on the calculated \alphamm values can mostly be understood by a simple model for a dust distribution, as used in \citetalias{Ricci:2010p9423} (cf. Fig.~3 in \citetalias{Ricci:2010p9423}):
for maximum grain sizes much smaller than the observed wavelengths, the spectral index of the dust opacity \betamm ($\kappa(\lambda) \propto \lambda^{-\betamm}$) is constant, while it decreases for maximum particle sizes larger than sub-mm. In between (at a few tenth of a mm), there is a peak which is caused by an increased opacity of grains with sizes similar to the observed wavelength. The relation between $\alphamm$ and $\betamm$ depends on the emitting spectrum and the optical depth. For a completely optically thin disk in the Rayleigh-Jeans regime $\betamm=\alphamm-2$. However, if the emitted spectrum deviates from the Rayleigh-Jeans limit, then $\betamm \gtrsim \alphamm-2$. In our models, $\alphamm-\betamm$ turns out to be typically between $1.4$ and $1.7$ if \amax is outside of the peak of opacity.

\section{Results}                       \label{sec:mmobs:results}
\subsection{Sub-mm fluxes and spectral indices}                 \label{sec:mmobs:results_fluxes}
For all possible combinations of the parameters shown in Table~\ref{tab:mmobs:model_grid}, we solved for the steady-state grain size distributions and derived the \alphamm and \Fmm values. As noted before, some of the models do not result in a steady state and are therefore not shown here.

\begin{figure}[t]
  \centering
  \resizebox{0.9\hsize}{!}{\includegraphics{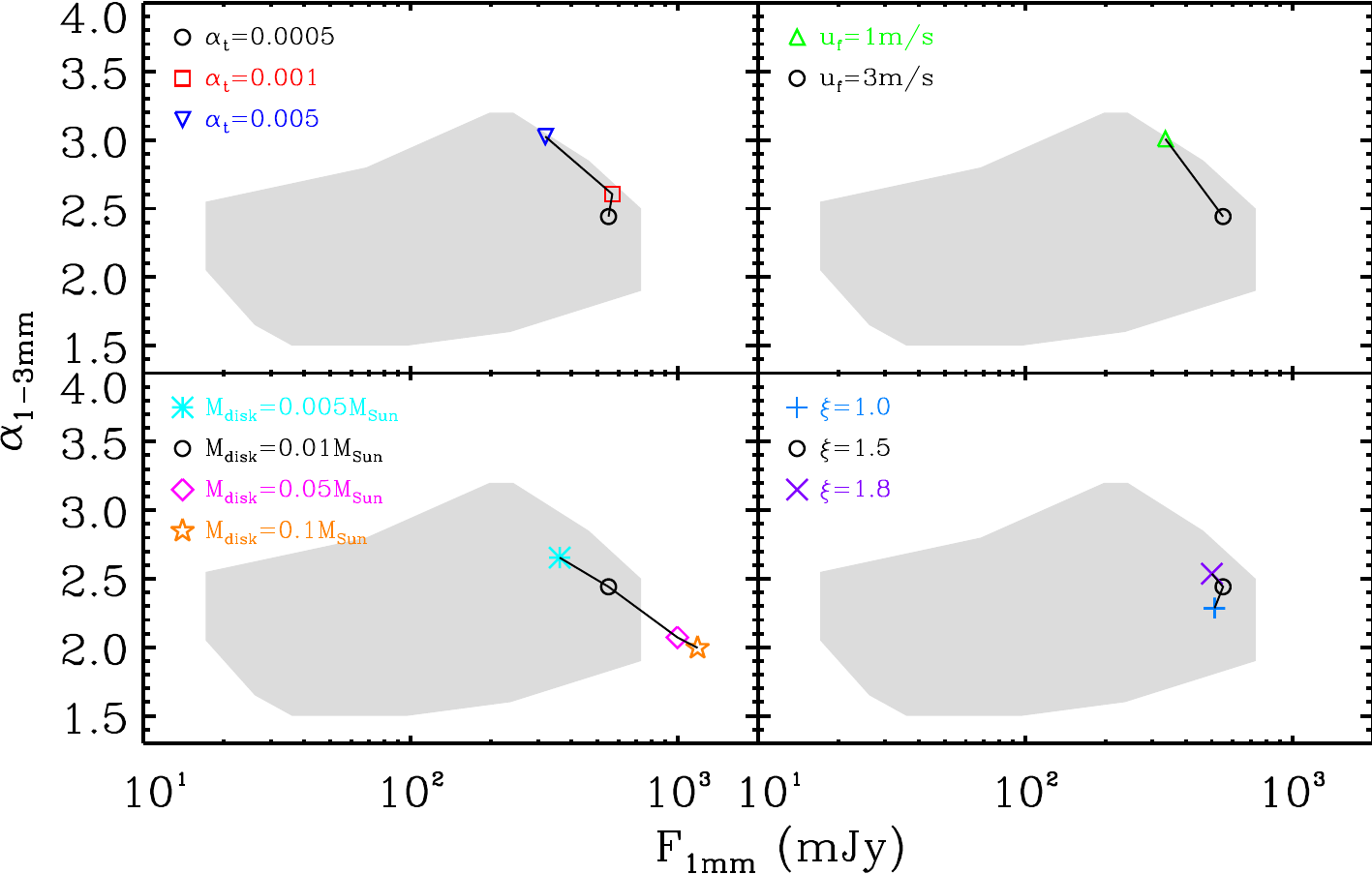}}
  \caption{Influence of the parameters \alphat (top left), fragmentation velocity (top right), disk mass (bottom left) and grain porosity (bottom right) on the observed fluxes and spectral indices. The black circle denotes the fiducial model whose parameters are given in Table~\ref{tab:mmobs:model_grid}. The grey area represents the region in which the observed sources lie (see Fig.~\ref{fig:mmobs:clouds}).}
  \label{fig:mmobs:multiplot}
\end{figure}

The top left panel of Fig.~\ref{fig:mmobs:multiplot} shows the influence of the turbulence parameter \alphat. According to Eq.~\ref{eq:mmobs:a_max}, the maximum grain size increases if \alphat decreases. Depending on where \amax lies with respect to the opacity peak (see \citetalias{Ricci:2010p9423}, Fig.~3), \alphamm can increase or decrease with increasing \alphat. In the simulations presented here, \amax is typically so large that increasing \alphat predicts steeper spectral slopes.

\amax is more sensitive to \uf (cf. top right panel in Fig.~\ref{fig:mmobs:multiplot}): the maximum grain size $a_\text{max}$ is proportional to $\uf^2$, therefore a change of \uf by a factor of about 3 significantly changes \alphamm by increasing the grain size by about one order of magnitude.
However many models with a fragmentation velocity of 10~m/s never reach a steady state. It is therefore not possible to explain lower \alphamm values by a further increase of \uf alone.

The influence of $M_\text{disk}$ on \Fmm and \alphamm is twofold. Firstly, a decrease in $M_\text{disk}$ (assuming a constant dust-to-gas ratio and a fixed shape of the disk surface density, i.e. not varying $R_\mathrm{c}$ and $\gamma$, see Section~\ref{sec:mmobs:model_disk}) reduces the amount of emitting dust and thus \Fmm. Secondly, such a reduction in gas mass also reduces \amax (Eq.~\ref{eq:mmobs:a_max}), which tends to increase \alphamm. This combined trend is seen in Fig.~\ref{fig:mmobs:multiplot}. Hence, in order to explain faint sources with low \alphamm, the amount of emitting dust has to be reduced while the disk gas mass stays large. This effect could be achieved in two ways: the amount of dust could be reduced by radial drift at a reduced rate (full radial drift would quickly remove all mm-sized grains, see \citealp{Brauer:2007p232}) or only the ``visible'' amount of dust is reduced if some of the dust is already contained in larger bodies. This latter case is predicted
by our non-steady-state distribution models and will be discussed in more detail in a forthcoming
paper.

\begin{figure}[b!t]
  \centering
  \resizebox{0.9\hsize}{!}{\includegraphics{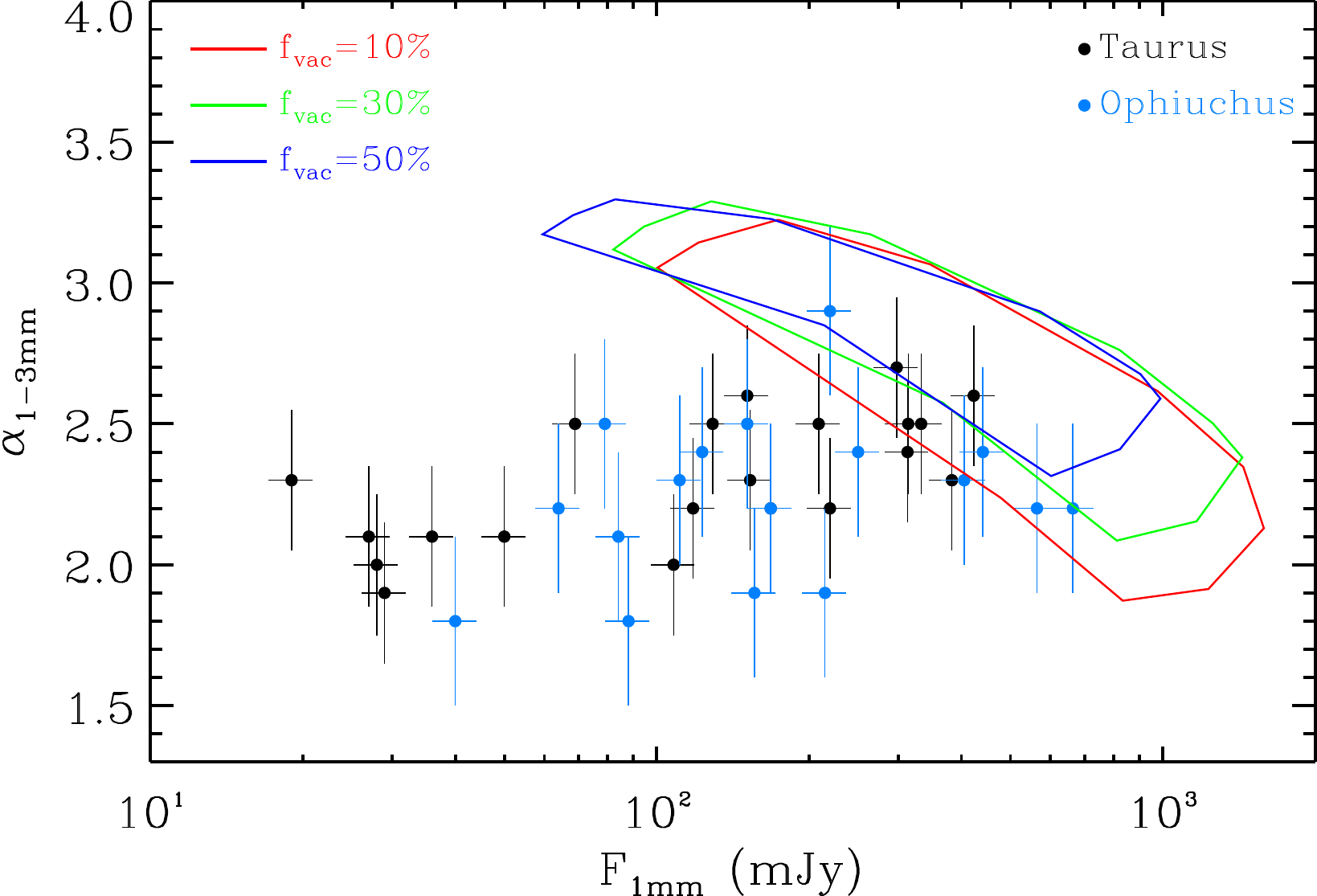}}
  \caption{Observed fluxes at mm-wavelengths of the Taurus (black dots) and the Ophiuchus (blue dots) star-forming regions (see \citetalias{Ricci:2010p9423} and Ricci et al. 2010, in prep) and the areas covered by the simulation results for different vacuum fractions of the grains (varying all other parameters according to Table~\ref{tab:mmobs:model_grid}).}
  \label{fig:mmobs:clouds}
\end{figure}

In general, lower values of $\xi$ translate to shallower grain-size distribution, which results in lower values of \betamm \citep[see][]{Draine:2006p9433}.
The lower right panel in Fig.~\ref{fig:mmobs:multiplot} does not seem to indicate a strong dependence on $\xi$, but lower values of $\xi$ (around ~ 1) seem to be closer to the observations especially at high fluxes.

Figure~\ref{fig:mmobs:clouds} shows the areas which are covered by our sets of simulations for different porosities in comparison to the observational samples. It can be seen that only the brightest sources are covered by the simulations. The trend of larger \alphamm for a larger vacuum fraction seems to be in contradiction with Eq.~\ref{eq:mmobs:a_max}, because smaller grain volume density leads to larger \amax. However in this case, the opacity is much more affected by changing the grain structure: reducing the grains' vacuum fraction increases the spectral index at sub-mm, while it is reduced for longer wavelengths. Therefore opacity effects outweigh the smaller changes in \amax.
A more thorough analysis of opacity effects is beyond the scope of this chapter, but it seems implausible that the large spread in the observations could be explained by different kinds of grains alone \citep[see][]{Draine:2006p9433}.

\subsection{Radial profiles of the dust opacity index}                 \label{sec:mmobs:results_beta}
\begin{figure}[bt]
  \centering
\resizebox{0.9\hsize}{!}{\includegraphics{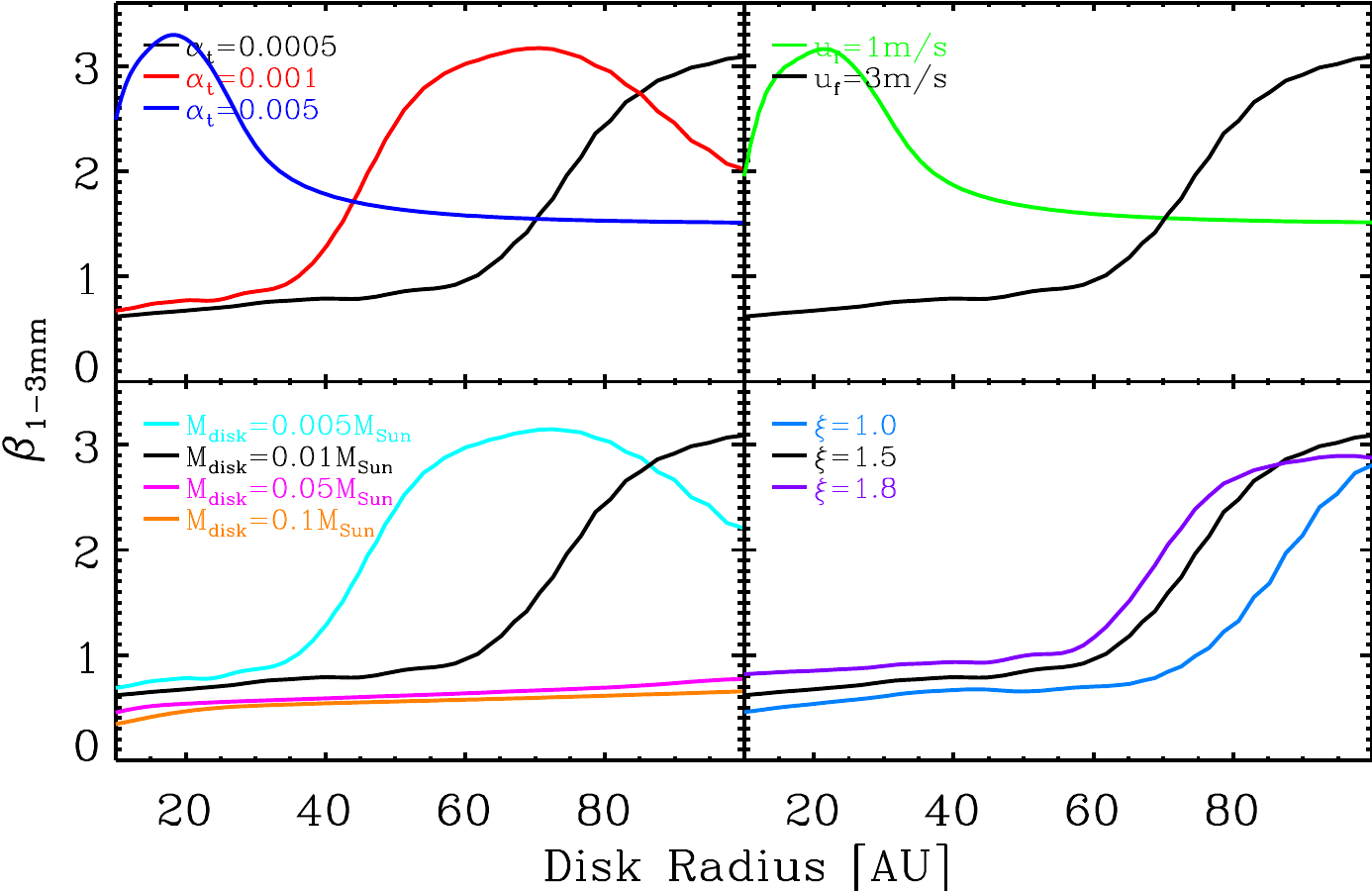}}
  \caption{Predicted profiles of the dust opacity index at mm-wavelengths for different variations of the fiducial model. The colors correspond to the parameters shown in Fig.~\ref{fig:mmobs:multiplot}.}
  \label{fig:mmobs:betamultiplot}
\end{figure}

The presented models also compute \alphamm as function of radius. From the point of view of a comparison with the observations, this is somewhat premature because observational methods are not yet able to provide reliable radial profiles of \alphamm (e.g., \citealp{Isella:2010p9438}, Banzatti et al., in prep.). 
Yet, the predicted radial dependence of \betamm (shown in Fig.~\ref{fig:mmobs:betamultiplot}) agrees with the observations so far. It can be seen that the shape of most models looks similar, slightly increasing from \betamm-values around 0.5 at 10~AU up to around 1.5 at 100~AU. The reason for this is that \amax depends on the ratio of surface density over temperature. Under typical assumptions, \amax will decrease with radius. An upper grain size, which is decreasing with radius and stays outside the peak in the opacity, results in \betamm increasing with radius (cf. Fig.~3 in \citetalias{Ricci:2010p9423}). If the radially decreasing upper grain size \amax reaches sizes just below mm, then the peak in opacity will produce also a peak in the radial profile of \betamm (the size of which depends much on the assumed opacity), which can be seen in Fig.~\ref{fig:mmobs:betamultiplot}. Thus, even though \amax is monotone in radius, \betamm does not need to be monotone. 

\section{Discussion and conclusions}                   \label{sec:mmobs:conclusions}
We present the first in-depth comparison of simulated grain size distributions and observed mm spectral indices of YSOs in the Taurus and the Ophiuchus star-forming regions. Additionally we present the first predictions of the radial profile of the dust opacity index at mm wavelength that are consistent with the limits set by \citet{Isella:2010p9438}.

Low values of the observed mm-slopes are quite naturally reproduced by our models, favoring low values of $\xi$ and \alphat as well as fragmentation threshold velocities above 1~m~s$^{-1}$. However, a simultaneous match to the observed range of flux levels requires a reduction of the dust mass by a factor of a few up to about 30. This over-prediction of fluxes cannot be fixed by simply reducing the disk mass because the predicted \alphamm would be too large for smaller disk masses.
Opacities induce a large uncertainty in the flux levels. However, considering the results of \citet{Draine:2006p9433}, it seems implausible that the large spread in observed fluxes for different disks with similar \alphamm (which is probably even larger as very faint disks are not contained in the sample) can be explained by different grain mineralogy alone.

The aforementioned reduction of observable dust could be due to radial drift at a reduced rate or during an earlier epoch (drift has been artificially suppressed in this work in order to explain the low values of \alphamm by $\gtrsim 1$~mm sized grains). Another possible explanation is grain growth to even larger sizes, as these bodies have a small opacity coefficient per unit mass.

Finally, a different dependence between \alphamm and the observed flux \Fmm might also originate from disk surface densities profiles that differ from what we have assumed in this work. This possibility, as well as the effect of a different dust composition, will be considered in a future work.
\ifthenelse{\boolean{chapterbib}}
{
    \clearpage
    \bibliographystyle{aa}
    \bibliography{/Users/til/Documents/Papers/bibliography}
}
{}
\chapter{Summary and outlook}\label{chapter:outlook}
In this thesis, we investigated how dust in circumstellar disks evolves due to radial transport within the gaseous disk and due to grain growth and fragmentation. The physics of the underlying gas disk is modeled as well using a viscous accretion disk model including irradiation by the central star and viscous heating.

The first two chapters of this thesis focused on the various physical aspects of the evolution of dust and the gas in these disks using a newly developed code which can integrate the gas disk evolution and at the same time the coupled equations of advection, diffusion and the Smoluchowski equation including fragmentation of grains. Additionally, the solver of the Smoluchowski equation can in the future be extended to also solve for the mean porosity of each grain size, as derived and demonstrated by \citet{Okuzumi:2009p9772}.

Our results show that the radial inward drift and the fact that larger grains are less effectively mixed and carried outwards causes a strong radial dependence of the dust-to-gas ratio. Only if grain fragmentation effectively grinds down larger grains, the dust can be relatively equally be distributed with the gas. Observations in the near future (e.g., ALMA) should be sensitive enough to detect the different sizes of the gas and the dust disk (Pani\'c \& Birnstiel, in prep.). 

We also investigated the two main barriers preventing particles to accumulate to planetesimals: radial drift and fragmentation. Our results show that there are several mechanisms which are able slow down radial drift to an extend where pure sticking would allow particles to overcome the radial drift barrier. However, fragmentation seems to be a much stronger obstacle, especially in the inner hot parts of the disk. A very low degree of turbulence is needed to keep the relative velocities of the particles low enough to avoid shattering. Even in this case, radial and azimuthal relative velocities are typically still large enough to fragment the aggregates. Therefore, we propose that both a low degree of turbulence and a low radial pressure gradient are necessary to allow further growth by sticking collisions.

\begin{figure}[htb]
  \centering
  \resizebox{0.65\hsize}{!}{\includegraphics{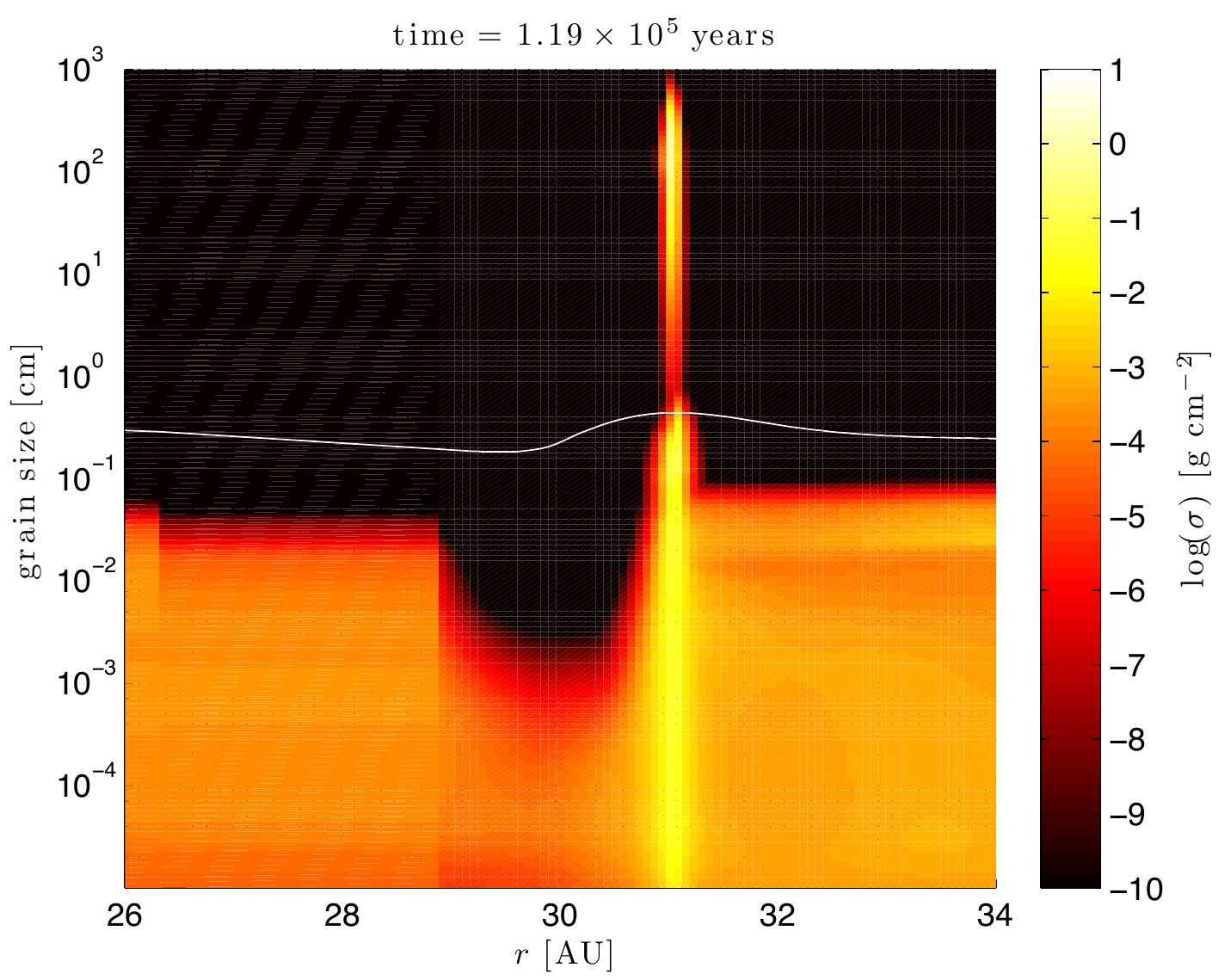}}
  \caption{An underlying change in the turbulence parameter \alphat (from $10^{-4}$ inside 30~AU to $10^{-5}$) induces a surface density and thus pressure maximum. Dust accumulates in this pressure bump and -- because radial and azimuthal velocities vanish there -- dust can continue to grow to larger sizes. The white line denotes the grain size which corresponds to a Stokes number of unity and is proportional to the gas surface density.}
  \label{fig:outlook:bump}
\end{figure}

Dead zones seem to be such an environment. \citet{Dzyurkevich:2010p11360} found that indeed a pressure bump can exist at the inner boundary of a dead zone. \citet{Brauer:2008p212} showed that particles can break through the growth barrier at a pressure bump. Preliminary simulations of such a pressure bump, caused by a radial variation in the amount of turbulence in an evolving gas disk, show that radially drifting grains accumulate in the center of the pressure peak where the physical conditions (low turbulence, vanishing radial and azimuthal relative velocities) are such that larger grains grow fast enough to overcome the growth barrier. A snapshot of such a simulation is shown in Fig.~\ref{fig:outlook:bump}. Future work in this direction has to take into account more detailed profiles of the relative velocities and of the gas surface density, such as produced by MHD simulations \citet{Dzyurkevich:2010p11360}. The code presented in this thesis can possibly be extended to a layered disk scenario. These simulations could give some first results about the enrichment in dust and the dust size evolution within a dead zone. 

A second step in this direction would be to use the code to solve not the radial but instead the vertical transport of dust caused by settling and turbulent mixing. This could be coupled to the already available vertical temperature and density profile of the disk to investigate the observable dust distribution in the upper layers of disks. Together with a dust-dependent recipe for the MRI activity (available from Neal Turner), this would allow us to study the interplay between MRI activity (which in turn depends on the amount of small dust) and the dust evolution (which depends on the MRI activity as source of turbulent velocities).

Another result of this thesis was that if radial drift is ineffective or if particles stay small enough (due to the fragmentation), then grain growth and fragmentation balance each other. This balance determines the size distribution of dust in circumstellar disks. Therefore, in Chapter~\ref{chapter:distri}, we tried to analytically understand the dust distribution in such a case. For this purpose, we generalized previous analytical results for collisional cascades to include both processes, coagulation and fragmentation at the same time. We could derive power-law distributions for three different regimes, depending on which of these effects dominates the shape of the distribution.

For simple collision kernels, we found very good agreement between the analytical results and the numerically simulated steady-state distributions. For the more complicated collision kernel which represents the condition in a turbulent circumstellar disk, our analytical results helped us to understand the general shape of the resulting distributions. We were able to find a recipe which can reproduce the distributions we derived numerically. This recipe can readily be used for further modeling, such as grain surface chemistry (Vasyunin et al., in prep.), disk heating-cooling and FUV photoevaporation, radiative transfer modeling of disks (Mulders \& Birnstiel, in prep.; Sauter \& Birnstiel, in prep.) and many more. 

As mentioned before, the dust distribution is an important ingredient to many models of circumstellar disks and is thus a key to understanding the astrophysics of circumstellar disk and of planet formation. Chapters~\ref{chapter:dustretention} and \ref{chapter:mmobs} are an example of this: in Chapter~\ref{chapter:dustretention}, we showed how grain fragmentation and cratering could solve the riddle why disks are observed to be dusty for several million years, while theory predicts the dust to grow and to disappear quickly due to radial drift: if fragmentation is happening at an collision velocity of around 1~m~s$^{-1}$, as suggested by experiments \citep{Blum:2008p1920}, then the maximum grain size which can exist is small enough to be unaffected by radial drift. Particles larger than this size experience larger impact velocities and are quickly ground to smaller sizes.

\begin{figure}[htb]
  \centering
  \resizebox{0.75\hsize}{!}{\includegraphics{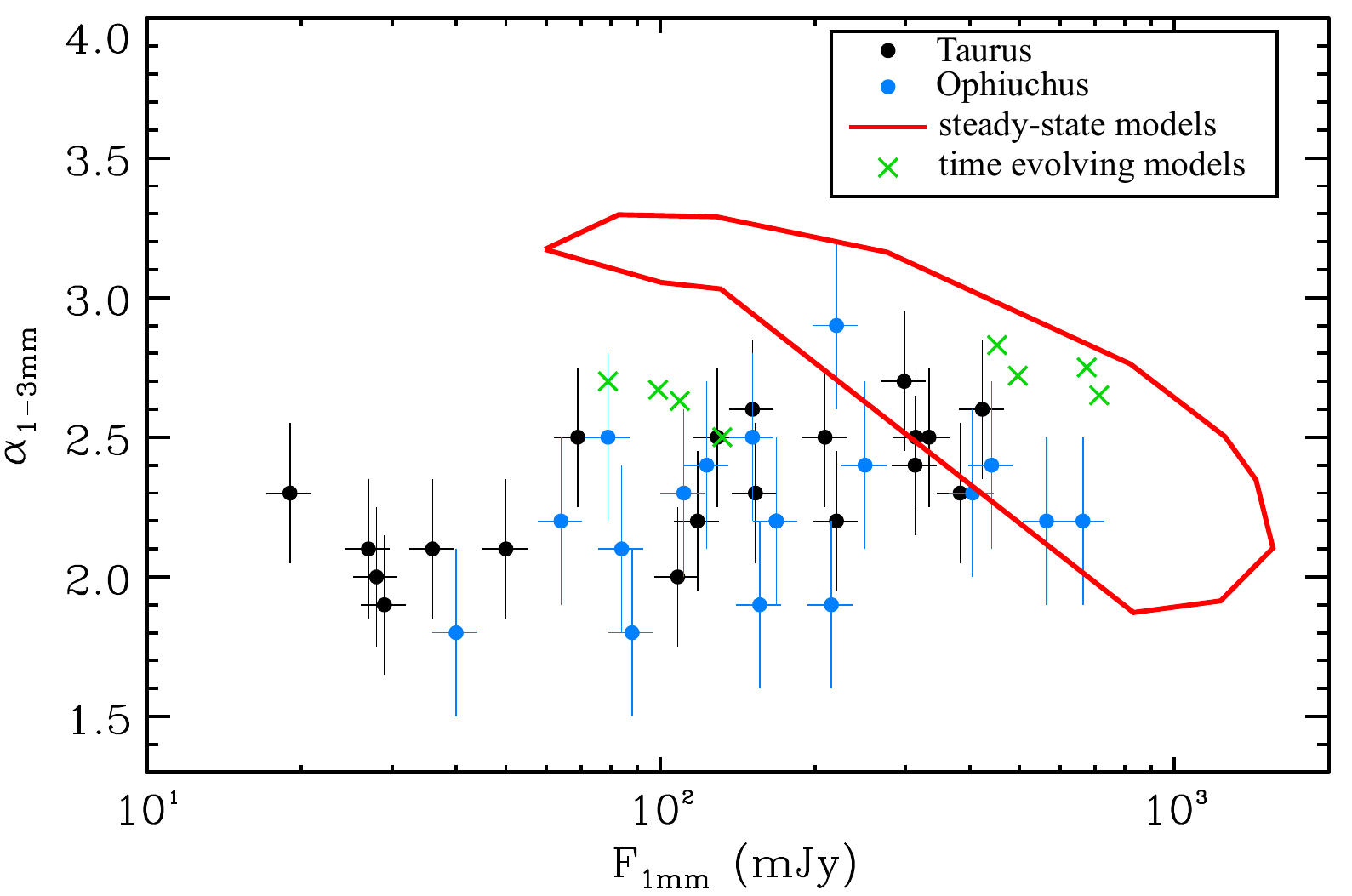}}
  \caption{Preliminary results: comparison of fluxes and spectral indices of observational data (black and blue dots for Taurus and Ophiuchus, respectively), steady-state solutions from Chapter~\ref{chapter:mmobs} (the area inside the red polygon) and time evolving simulations (green crosses). The snapshots are taken at $10^{5}$ years of evolution. Higher-mass disks are already inside of the steady-state area, while lower mass disks reproduce some of the lower-flux values.}
  \label{fig:outlook:luca}
\end{figure}

Another link to observations was established in Chapter~\ref{chapter:mmobs}: we argued that some mechanism must exist to stop radial drift. Otherwise, all larger grains in the outer disk would disappear quickly, as shown by \citet{Brauer:2007p232}. This assumption allows us to investigate the question whether grains can grow at these radii to the observed sizes on a reasonable timescale. We found that the grain size distribution reaches a steady-state after about a million years, even at 100~AU. The largest grain size is therefore not limited by the growth time scale, but by fragmentation. We calculated disk models for a large set of models to determine the flux and the spectral index predicted by the theory and compared it to a sample of observations by \citet{Ricci:2010p9423}. We found that the models indeed reproduce the low spectral indices, however the fluxes are only reproduced for the brightest objects. This indicates a reduced dust-to-gas ratio (or decreased dust opacity) since increasing both the dust \emph{and} gas mass would increase the spectral index. Future work in this direction will focus on the impact of using other opacity models and on time evolving simulations of grain growth and fragmentation. Preliminary results in this direction show that disk at earlier evolution times produce lower flux values, as shown in Fig.~\ref{fig:outlook:luca}.

It has now been mentioned several times, that planet(-esimal) formation heavily depends on this initial phase of grain growth. We will therefore use the code presented in this thesis also to study the inward mass flux of dust and the maximum size of the drifting particles (Birnstiel \& Klahr, in prep.). The reasons for this are the following: if planetesimals are to be formed by gravoturbulent mechanisms \citep[see][]{Johansen:2007p4788}, then the initial conditions for these scenarios need to be met. The conditions are firstly that grains need to have the right sizes \citep[for example about decimetre sized particles for the scenario of][]{Johansen:2007p4788} and secondly that there needs to be enough of these particles (i.e., a strongly enhanced dust-to-gas ratio). Our model can calculate self-consistently, which particles are produced under which condition and at the same time the radial drift and possible accumulation of the dust. This drift rate and the particle sizes are also important for other alternative planet formation scenarios, such as dust capturing in vortices \citep{Barge:1995p11993,Klahr:2006p7719}.    
    
\ifthenelse{\boolean{chapterbib}}
{
    \clearpage
    \bibliographystyle{aa}
    \bibliography{/Users/til/Documents/Papers/bibliography}
}
{}
\backmatter
\ifthenelse{\boolean{chapterbib}}
{}
{
    \bibliographystyle{aa}
    \bibliography{bibliography}

\begin{thebibliography}{184}
\expandafter\ifx\csname natexlab\endcsname\relax\def\natexlab#1{#1}\fi

\bibitem[{Alexander {et~al.}(2006)Alexander, Clarke, \&
  Pringle}]{Alexander:2006p136}
Alexander, R.~D., Clarke, C., \& Pringle, J.~E. 2006, MNRAS, 369, 216

\bibitem[{Andrews \& Williams(2007)}]{Andrews:2007p3380}
Andrews, M.~R. \& Williams, J.~P. 2007, ApJ, 671, 1800

\bibitem[{Andrews \& Williams(2005)}]{Andrews:2005p3779}
Andrews, S.~M. \& Williams, J.~P. 2005, ApJ, 631, 1134

\bibitem[{Andrews {et~al.}(2009)Andrews, Wilner, Hughes, Qi, \&
  Dullemond}]{Andrews:2009p7729}
Andrews, S.~M., Wilner, D.~J., Hughes, A.~M., Qi, C., \& Dullemond, C.~P. 2009,
  ApJ, 700, 1502

\bibitem[{Armitage(2009)}]{Armitage:2009p10482}
Armitage, P. 2009, Astrophysics of Planet Formation‎

\bibitem[{Armitage {et~al.}(2001)Armitage, Livio, \&
  Pringle}]{Armitage:2001p993}
Armitage, P.~J., Livio, M., \& Pringle, J.~E. 2001, MNRAS, 324, 705

\bibitem[{Balbus \& Hawley(1991)}]{Balbus:1991p4932}
Balbus, S.~A. \& Hawley, J.~F. 1991, ApJ, 376, 214

\bibitem[{Balbus \& Hawley(1998)}]{Balbus:1998p7450}
Balbus, S.~A. \& Hawley, J.~F. 1998, Rev. Mod. Phys., 70, 1

\bibitem[{Banerjee \& Pudritz(2006)}]{Banerjee:2006p8491}
Banerjee, R. \& Pudritz, R.~E. 2006, ApJ, 641, 949

\bibitem[{Barge \& Sommeria(1995)}]{Barge:1995p11993}
Barge, P. \& Sommeria, J. 1995, A{\&}A, 295, L1

\bibitem[{Beckwith {et~al.}(1990)Beckwith, Sargent, Chini, \&
  Guesten}]{Beckwith:1990p3768}
Beckwith, S. V.~W., Sargent, A.~I., Chini, R.~S., \& Guesten, R. 1990, AJ, 99,
  924

\bibitem[{Birnstiel {et~al.}(2009)Birnstiel, Dullemond, \&
  Brauer}]{Birnstiel:2009p7135}
Birnstiel, T., Dullemond, C.~P., \& Brauer, F. 2009, A{\&}A, 503, L5

\bibitem[{Birnstiel {et~al.}(2010{\natexlab{a}})Birnstiel, Dullemond, \&
  Brauer}]{Birnstiel:2010p9709}
Birnstiel, T., Dullemond, C.~P., \& Brauer, F. 2010{\natexlab{a}}, A{\&}A, 513,
  79

\bibitem[{Birnstiel {et~al.}(2010{\natexlab{b}})Birnstiel, Ricci, Trotta,
  Dullemond, Natta, Testi, Dominik, Henning, Ormel, \&
  Zsom}]{Birnstiel:2010p12008}
Birnstiel, T., Ricci, L., Trotta, F., {et~al.} 2010{\natexlab{b}}, A{\&}A, 516,
  L14

\bibitem[{Blum \& Muench(1993)}]{Blum:1993p4324}
Blum, J. \& Muench, M. 1993, Icarus, 106, 151

\bibitem[{Blum \& Wurm(2000)}]{Blum:2000p8099}
Blum, J. \& Wurm, G. 2000, Icarus, 143, 138

\bibitem[{Blum \& Wurm(2008)}]{Blum:2008p1920}
Blum, J. \& Wurm, G. 2008, ARA{\&}A, 46, 21

\bibitem[{Blum {et~al.}(2000)Blum, Wurm, Kempf, Poppe, Klahr, Kozasa, Rott,
  Henning, Dorschner, Schr{\"a}pler, Keller, Markiewicz, Mann, Gustafson,
  Giovane, Neuhaus, Fechtig, Gr{\"u}n, Feuerbacher, Kochan, Ratke, Goresy,
  Morfill, Weidenschilling, Schwehm, Metzler, \& Ip}]{Blum:2000p8110}
Blum, J., Wurm, G., Kempf, S., {et~al.} 2000, Phys. Rev. Lett., 85, 2426

\bibitem[{Bodenheimer(2006)}]{Bodenheimer:2006p11151}
Bodenheimer, P. 2006, Planet Formation

\bibitem[{Boss(1997)}]{Boss:1997p11650}
Boss, A.~P. 1997, Science, 276, 1836

\bibitem[{Boss(2000)}]{Boss:2000p11660}
Boss, A.~P. 2000, ApJ, 536, L101

\bibitem[{Boss(2007)}]{Boss:2007p11741}
Boss, A.~P. 2007, ApJ, 661, L73

\bibitem[{Bouvier {et~al.}(2007)Bouvier, Alencar, Harries, Johns-Krull, \&
  Romanova}]{Bouvier:2007p10356}
Bouvier, J., Alencar, S. H.~P., Harries, T.~J., Johns-Krull, C.~M., \&
  Romanova, M.~M. 2007, PP V, 479

\bibitem[{Bouwman {et~al.}(2001)Bouwman, Meeus, de~Koter, Hony, Dominik, \&
  Waters}]{Bouwman:2001p8118}
Bouwman, J., Meeus, G., de~Koter, A., {et~al.} 2001, A{\&}A, 375, 950

\bibitem[{Brauer {et~al.}(2008{\natexlab{a}})Brauer, Dullemond, \&
  Henning}]{Brauer:2008p215}
Brauer, F., Dullemond, C.~P., \& Henning, T. 2008{\natexlab{a}}, A{\&}A, 480,
  859

\bibitem[{Brauer {et~al.}(2007)Brauer, Dullemond, Johansen, Henning, Klahr, \&
  Natta}]{Brauer:2007p232}
Brauer, F., Dullemond, C.~P., Johansen, A., {et~al.} 2007, A{\&}A, 469, 1169

\bibitem[{Brauer {et~al.}(2008{\natexlab{b}})Brauer, Henning, \&
  Dullemond}]{Brauer:2008p212}
Brauer, F., Henning, T., \& Dullemond, C.~P. 2008{\natexlab{b}}, A{\&}A, 487,
  L1

\bibitem[{Brush(1996)}]{Brush:1996p10354}
Brush, S.~G. 1996, Fruitful encounters: the origin of the solar system and of
  the moon from Chamberlin to Apollo

\bibitem[{Burkhardt {et~al.}(2008)Burkhardt, Kleine, Bourdon, Palme, Zipfel,
  Friedrich, \& Ebel}]{Burkhardt:2008p10908}
Burkhardt, C., Kleine, T., Bourdon, B., {et~al.} 2008, GCA, 72, 6177

\bibitem[{Calvet {et~al.}(2002)Calvet, D'Alessio, Hartmann, Wilner, Walsh, \&
  Sitko}]{Calvet:2002p10424}
Calvet, N., D'Alessio, P., Hartmann, L., {et~al.} 2002, ApJ, 568, 1008

\bibitem[{Calvet {et~al.}(2000)Calvet, Hartmann, \& Strom}]{Calvet:2000p1083}
Calvet, N., Hartmann, L., \& Strom, S.~E. 2000, PP IV, 377

\bibitem[{Camacho(2001)}]{Camacho:2001p3816}
Camacho, J. 2001, Phys. Rev. E, 63, 46112

\bibitem[{Chiang \& Goldreich(1997)}]{Chiang:1997p1986}
Chiang, E. \& Goldreich, P. 1997, ApJ, 490, 368

\bibitem[{Chiang \& Youdin(2010)}]{Chiang:2010p11570}
Chiang, E. \& Youdin, A.~N. 2010, Ann. Rev. Earth Planet. Sci., 38, 493

\bibitem[{Clarke {et~al.}(2001)Clarke, Gendrin, \& Sotomayor}]{Clarke:2001p969}
Clarke, C., Gendrin, A., \& Sotomayor, M. 2001, MNRAS, 328, 485

\bibitem[{Cossins {et~al.}(2009)Cossins, Lodato, \& Clarke}]{Cossins:2009p3794}
Cossins, P., Lodato, G., \& Clarke, C.~J. 2009, MNRAS, 393, 1157

\bibitem[{Cumming {et~al.}(2008)Cumming, Butler, Marcy, Vogt, Wright, \&
  Fischer}]{Cumming:2008p11954}
Cumming, A., Butler, R.~P., Marcy, G.~W., {et~al.} 2008, PASP, 120, 531

\bibitem[{Cuzzi {et~al.}(2001)Cuzzi, Hogan, Paque, \&
  Dobrovolskis}]{Cuzzi:2001p2167}
Cuzzi, J.~N., Hogan, R.~C., Paque, J.~M., \& Dobrovolskis, A.~R. 2001, ApJ,
  546, 496

\bibitem[{Davis \& Ryan(1990)}]{Davis:1990p7995}
Davis, D.~R. \& Ryan, E.~V. 1990, Icarus, 83, 156

\bibitem[{Dohnanyi(1969)}]{Dohnanyi:1969p7994}
Dohnanyi, J.~S. 1969, J. Geophys. Res., 74, 2531

\bibitem[{Dominik \& Dullemond(2008)}]{Dominik:2008p4626}
Dominik, C. \& Dullemond, C.~P. 2008, A{\&}A, 491, 663

\bibitem[{Dominik \& Tielens(1997)}]{Dominik:1997p9440}
Dominik, C. \& Tielens, A. G. G.~M. 1997, ApJ, 480, 647

\bibitem[{Draine(2006)}]{Draine:2006p9433}
Draine, B.~T. 2006, ApJ, 636, 1114

\bibitem[{Drake(2003)}]{Drake:2003p10346}
Drake, S. 2003, Galileo at Work: His Scientific Biography

\bibitem[{Dubrulle {et~al.}(1995)Dubrulle, Morfill, \&
  Sterzik}]{Dubrulle:1995p300}
Dubrulle, B., Morfill, G.~E., \& Sterzik, M.~F. 1995, Icarus, 114, 237

\bibitem[{Dullemond \& Dominik(2004)}]{Dullemond:2004p390}
Dullemond, C.~P. \& Dominik, C. 2004, A{\&}A, 421, 1075

\bibitem[{Dullemond \& Dominik(2005)}]{Dullemond:2005p378}
Dullemond, C.~P. \& Dominik, C. 2005, A{\&}A, 434, 971

\bibitem[{Dullemond {et~al.}(2001)Dullemond, Dominik, \&
  Natta}]{Dullemond:2001p9307}
Dullemond, C.~P., Dominik, C., \& Natta, A. 2001, ApJ, 560, 957

\bibitem[{Dullemond {et~al.}(2007)Dullemond, Hollenbach, Kamp, \&
  D'Alessio}]{Dullemond:2007p336}
Dullemond, C.~P., Hollenbach, D., Kamp, I., \& D'Alessio, P. 2007, PP V, 555

\bibitem[{Dullemond {et~al.}(2002)Dullemond, van Zadelhoff, \&
  Natta}]{Dullemond:2002p399}
Dullemond, C.~P., van Zadelhoff, G.~J., \& Natta, A. 2002, A{\&}A, 389, 464

\bibitem[{Dzyurkevich {et~al.}(2010)Dzyurkevich, Flock, Turner, Klahr, \&
  Henning}]{Dzyurkevich:2010p11360}
Dzyurkevich, N., Flock, M., Turner, N.~J., Klahr, H., \& Henning, T. 2010,
  A{\&}A, 515, 70

\bibitem[{Fischer \& Valenti(2005)}]{Fischer:2005p11968}
Fischer, D.~A. \& Valenti, J. 2005, ApJ, 622, 1102

\bibitem[{Furlan {et~al.}(2006)Furlan, Hartmann, Calvet, D'Alessio,
  Franco-Hern{\'a}ndez, Forrest, Watson, Uchida, Sargent, Green, Keller, \&
  Herter}]{Furlan:2006p4444}
Furlan, E., Hartmann, L., Calvet, N., {et~al.} 2006, ApJS, 165, 568

\bibitem[{Gammie(1996)}]{Gammie:1996p1515}
Gammie, C.~F. 1996, ApJ, 457, 355

\bibitem[{Gammie(2001)}]{Gammie:2001p6435}
Gammie, C.~F. 2001, ApJ, 553, 174

\bibitem[{Garaud(2007)}]{Garaud:2007p405}
Garaud, P. 2007, ApJ, 671, 2091

\bibitem[{Gaukroger(2006)}]{Gaukroger:2006p10348}
Gaukroger, S. 2006, The emergence of a scientific culture: science and the
  shaping of modernity

\bibitem[{Goldreich \& Ward(1973)}]{Goldreich:1973p11184}
Goldreich, P. \& Ward, W.~R. 1973, ApJ, 183, 1051

\bibitem[{Goodman {et~al.}(1993)Goodman, Benson, Fuller, \&
  Myers}]{Goodman:1993p8362}
Goodman, A.~A., Benson, P.~J., Fuller, G.~A., \& Myers, P.~C. 1993, ApJ, 406,
  528

\bibitem[{Gorti {et~al.}(2009)Gorti, Dullemond, \&
  Hollenbach}]{Gorti:2009p8414}
Gorti, U., Dullemond, C.~P., \& Hollenbach, D. 2009, ApJ, 705, 1237

\bibitem[{G{\"u}ttler {et~al.}(2010)G{\"u}ttler, Blum, Zsom, Ormel, \&
  Dullemond}]{Guttler:2010p9745}
G{\"u}ttler, C., Blum, J., Zsom, A., Ormel, C.~W., \& Dullemond, C.~P. 2010,
  A{\&}A, 513, 56

\bibitem[{Haisch {et~al.}(2001)Haisch, Lada, \& Lada}]{Haisch:2001p10426}
Haisch, K.~E., Lada, E.~A., \& Lada, C.~J. 2001, ApJ, 553, L153

\bibitem[{Hartmann {et~al.}(1998)Hartmann, Calvet, Gullbring, \&
  D'Alessio}]{Hartmann:1998p664}
Hartmann, L., Calvet, N., Gullbring, E., \& D'Alessio, P. 1998, ApJ, 495, 385

\bibitem[{Heim {et~al.}(1999)Heim, Blum, Preuss, \& Butt}]{Heim:1999p11580}
Heim, L.-O., Blum, J., Preuss, M., \& Butt, H.-J. 1999, Phys. Rev. Lett., 83,
  3328

\bibitem[{Henning \& Stognienko(1996)}]{Henning:1996p4181}
Henning, T. \& Stognienko, R. 1996, A{\&}A, 311, 291

\bibitem[{Hollenbach {et~al.}(1994)Hollenbach, Johnstone, Lizano, \&
  Shu}]{Hollenbach:1994p1186}
Hollenbach, D., Johnstone, D., Lizano, S., \& Shu, F.~H. 1994, ApJ, 428, 654

\bibitem[{Hueso \& Guillot(2005)}]{Hueso:2005p685}
Hueso, R. \& Guillot, T. 2005, A{\&}A, 442, 703

\bibitem[{Ida \& Makino(1993)}]{Ida:1993p11613}
Ida, S. \& Makino, J. 1993, Icarus, 106, 210

\bibitem[{Inutsuka {et~al.}(2010)Inutsuka, Machida, \&
  Matsumoto}]{Inutsuka:2010p11359}
Inutsuka, S., Machida, M.~N., \& Matsumoto, T. 2010, ApJL, 718, L58

\bibitem[{Isella {et~al.}(2009)Isella, Carpenter, \&
  Sargent}]{Isella:2009p7470}
Isella, A., Carpenter, J.~M., \& Sargent, A.~I. 2009, ApJ, 701, 260

\bibitem[{Isella {et~al.}(2010)Isella, Carpenter, \&
  Sargent}]{Isella:2010p9438}
Isella, A., Carpenter, J.~M., \& Sargent, A.~I. 2010, ApJ, 714, 1746

\bibitem[{Johansen \& Klahr(2005)}]{Johansen:2005p8425}
Johansen, A. \& Klahr, H. 2005, ApJ, 634, 1353

\bibitem[{Johansen {et~al.}(2006)Johansen, Klahr, \&
  Henning}]{Johansen:2006p7466}
Johansen, A., Klahr, H., \& Henning, T. 2006, ApJ, 636, 1121

\bibitem[{Johansen {et~al.}(2007)Johansen, Oishi, Low, Klahr, Henning, \&
  Youdin}]{Johansen:2007p4788}
Johansen, A., Oishi, J.~S., Low, M.-M.~M., {et~al.} 2007, Nature, 448, 1022

\bibitem[{Johansen {et~al.}(2009)Johansen, Youdin, \&
  Klahr}]{Johansen:2009p7441}
Johansen, A., Youdin, A., \& Klahr, H. 2009, ApJ, 697, 1269

\bibitem[{Kalas {et~al.}(2008)Kalas, Graham, Chiang, Fitzgerald, Clampin, Kite,
  Stapelfeldt, Marois, \& Krist}]{Kalas:2008p11975}
Kalas, P., Graham, J.~R., Chiang, E., {et~al.} 2008, Science, 322, 1345

\bibitem[{Kessler-Silacci {et~al.}(2006)Kessler-Silacci, Augereau, Dullemond,
  Geers, Lahuis, Evans, van Dishoeck, Blake, Boogert, Brown, J{\o}rgensen,
  Knez, \& Pontoppidan}]{KesslerSilacci:2006p4450}
Kessler-Silacci, J., Augereau, J.-C., Dullemond, C.~P., {et~al.} 2006, ApJ,
  639, 275

\bibitem[{Klahr \& Bodenheimer(2006)}]{Klahr:2006p7719}
Klahr, H. \& Bodenheimer, P. 2006, ApJ, 639, 432

\bibitem[{Klahr \& Bodenheimer(2003)}]{Klahr:2003p10848}
Klahr, H.~H. \& Bodenheimer, P. 2003, ApJ, 582, 869

\bibitem[{Kleine {et~al.}(2009)Kleine, Touboul, Bourdon, Nimmo, Mezger, Palme,
  Jacobsen, Yin, \& Halliday}]{Kleine:2009p10853}
Kleine, T., Touboul, M., Bourdon, B., {et~al.} 2009, GCA, 73, 5150

\bibitem[{Klett(1975)}]{Klett:1975p3936}
Klett, J.~D. 1975, Journal of Atmospheric Sciences, 32, 380

\bibitem[{Kley \& Lin(1992)}]{Kley:1992p7134}
Kley, W. \& Lin, D. N.~C. 1992, ApJ, 397, 600

\bibitem[{Kobayashi \& Tanaka(2010)}]{Kobayashi:2010p9774}
Kobayashi, H. \& Tanaka, H. 2010, Icarus, 206, 735

\bibitem[{Kokubo \& Ida(2000)}]{Kokubo:2000p11616}
Kokubo, E. \& Ida, S. 2000, Icarus, 143, 15

\bibitem[{Kretke \& Lin(2007)}]{Kretke:2007p697}
Kretke, K.~A. \& Lin, D. N.~C. 2007, ApJ, 664, L55

\bibitem[{Larson(1969)}]{Larson:1969p2574}
Larson, R.~B. 1969, MNRAS, 145, 271

\bibitem[{Larson(1985)}]{Larson:1985p10355}
Larson, R.~B. 1985, Royal Astronomical Society, 214, 379

\bibitem[{Larson(2003)}]{Larson:2003p3025}
Larson, R.~B. 2003, Rep. Prog. Phys, 66, 1651

\bibitem[{Leinhardt \& Stewart(2009)}]{Leinhardt:2009p5282}
Leinhardt, Z.~M. \& Stewart, S.~T. 2009, Icarus, 199, 542

\bibitem[{Lesur \& Papaloizou(2010)}]{Lesur:2010p10824}
Lesur, G. \& Papaloizou, J. C.~B. 2010, A{\&}A, 513, 60

\bibitem[{Lissauer(1993)}]{Lissauer:1993p11600}
Lissauer, J.~J. 1993, ARA{\&}A, 31, 129

\bibitem[{Lynden-Bell \& Pringle(1974)}]{LyndenBell:1974p1945}
Lynden-Bell, D. \& Pringle, J.~E. 1974, MNRAS, 168, 603

\bibitem[{Lyra {et~al.}(2009)Lyra, Johansen, Klahr, \&
  Piskunov}]{Lyra:2009p4812}
Lyra, W., Johansen, A., Klahr, H., \& Piskunov, N. 2009, A{\&}A, 493, 1125

\bibitem[{Makino {et~al.}(1998)Makino, Fukushige, Funato, \&
  Kokubo}]{Makino:1998p8778}
Makino, J., Fukushige, T., Funato, Y., \& Kokubo, E. 1998, New Astronomy, 3,
  411

\bibitem[{Marois {et~al.}(2008)Marois, Macintosh, Barman, Zuckerman, Song,
  Patience, Lafreni{\`e}re, \& Doyon}]{Marois:2008p11978}
Marois, C., Macintosh, B., Barman, T., {et~al.} 2008, Science, 322, 1348

\bibitem[{Mathis {et~al.}(1977)Mathis, Rumpl, \& Nordsieck}]{Mathis:1977p789}
Mathis, J.~S., Rumpl, W., \& Nordsieck, K.~H. 1977, ApJ, 217, 425

\bibitem[{Mayor \& Queloz(1995)}]{Mayor:1995p10615}
Mayor, M. \& Queloz, D. 1995, Nature, 378, 355

\bibitem[{McKee \& Ostriker(2007)}]{McKee:2007p2813}
McKee, C.~F. \& Ostriker, E.~C. 2007, ARA{\&}A, 45, 565

\bibitem[{Meru \& Bate(2010)}]{Meru:2010p11987}
Meru, F. \& Bate, M.~R. 2010, MNRAS, 858

\bibitem[{Mordasini {et~al.}(2009)Mordasini, Alibert, Benz, \&
  Naef}]{Mordasini:2009p11966}
Mordasini, C., Alibert, Y., Benz, W., \& Naef, D. 2009, A{\&}A, 501, 1161

\bibitem[{Murdin(2001)}]{Murdin:2001p10349}
Murdin, P. 2001, Encyclopedia of astronomy and astrophysics‎

\bibitem[{Myers(2005)}]{Myers:2005p4950}
Myers, P.~C. 2005, ApJ, 623, 280

\bibitem[{Nakagawa {et~al.}(1981)Nakagawa, Nakazawa, \&
  Hayashi}]{Nakagawa:1981p4533}
Nakagawa, Y., Nakazawa, K., \& Hayashi, C. 1981, Icarus, 45, 517

\bibitem[{Nakagawa {et~al.}(1986)Nakagawa, Sekiya, \&
  Hayashi}]{Nakagawa:1986p2048}
Nakagawa, Y., Sekiya, M., \& Hayashi, C. 1986, Icarus, 67, 375

\bibitem[{Nakamoto \& Nakagawa(1994)}]{Nakamoto:1994p798}
Nakamoto, T. \& Nakagawa, Y. 1994, ApJ, 421, 640

\bibitem[{Natta {et~al.}(2004)Natta, Testi, Neri, Shepherd, \&
  Wilner}]{Natta:2004p3169}
Natta, A., Testi, L., Neri, R., Shepherd, D.~S., \& Wilner, D.~J. 2004, A{\&}A,
  416, 179

\bibitem[{Okuzumi(2009)}]{Okuzumi:2009p7473}
Okuzumi, S. 2009, ApJ, 698, 1122

\bibitem[{Okuzumi {et~al.}(2009)Okuzumi, Tanaka, \& aki
  Sakagami}]{Okuzumi:2009p9772}
Okuzumi, S., Tanaka, H., \& aki Sakagami, M. 2009, ApJ, 707, 1247

\bibitem[{O'neil {et~al.}(2008)O'neil, Carlson, Francis, \&
  Stevenson}]{Oneil:2008p10905}
O'neil, J., Carlson, R., Francis, D., \& Stevenson, R. 2008, Science, 321, 1828

\bibitem[{Ormel \& Cuzzi(2007)}]{Ormel:2007p801}
Ormel, C.~W. \& Cuzzi, J.~N. 2007, A{\&}A, 466, 413

\bibitem[{Ormel {et~al.}(2010{\natexlab{a}})Ormel, Dullemond, \&
  Spaans}]{Ormel:2010p11595}
Ormel, C.~W., Dullemond, C.~P., \& Spaans, M. 2010{\natexlab{a}}, arXiv,
  astro-ph.EP

\bibitem[{Ormel {et~al.}(2010{\natexlab{b}})Ormel, Dullemond, \&
  Spaans}]{Ormel:2010p11601}
Ormel, C.~W., Dullemond, C.~P., \& Spaans, M. 2010{\natexlab{b}}, ApJL, 714,
  L103

\bibitem[{Ormel \& Klahr(2010)}]{Ormel:2010p11591}
Ormel, C.~W. \& Klahr, H.~H. 2010, arXiv, astro-ph.EP

\bibitem[{Ormel {et~al.}(2009)Ormel, Paszun, Dominik, \&
  Tielens}]{Ormel:2009p8002}
Ormel, C.~W., Paszun, D., Dominik, C., \& Tielens, A. G. G.~M. 2009, A{\&}A,
  502, 845

\bibitem[{Ormel {et~al.}(2007)Ormel, Spaans, \& Tielens}]{Ormel:2007p7127}
Ormel, C.~W., Spaans, M., \& Tielens, A. G. G.~M. 2007, A{\&}A, 461, 215

\bibitem[{Paszun \& Dominik(2009)}]{Paszun:2009p8871}
Paszun, D. \& Dominik, C. 2009, A{\&}A, 507, 1023

\bibitem[{Penston(1969)}]{Penston:1969p2601}
Penston, M.~V. 1969, MNRAS, 145, 457

\bibitem[{Pi{\'e}tu {et~al.}(2007)Pi{\'e}tu, Dutrey, \&
  Guilloteau}]{Pietu:2007p4501}
Pi{\'e}tu, V., Dutrey, A., \& Guilloteau, S. 2007, A{\&}A, 467, 163

\bibitem[{Pollack {et~al.}(1996)Pollack, Hubickyj, Bodenheimer, Lissauer,
  Podolak, \& Greenzweig}]{Pollack:1996p11635}
Pollack, J.~B., Hubickyj, O., Bodenheimer, P., {et~al.} 1996, Icarus, 124, 62

\bibitem[{Pollack {et~al.}(1985)Pollack, McKay, \&
  Christofferson}]{Pollack:1985p804}
Pollack, J.~B., McKay, C.~P., \& Christofferson, B.~M. 1985, Icarus, 64, 471

\bibitem[{Poppe {et~al.}(2000)Poppe, Blum, \& Henning}]{Poppe:2000p9447}
Poppe, T., Blum, J., \& Henning, T. 2000, ApJ, 533, 454

\bibitem[{Pringle(1981)}]{Pringle:1981p805}
Pringle, J.~E. 1981, ARA{\&}A, 19, 137

\bibitem[{Rafikov(2005)}]{Rafikov:2005p11787}
Rafikov, R.~R. 2005, ApJ, 621, L69

\bibitem[{Ricci {et~al.}(2010)Ricci, Testi, Natta, Neri, Cabrit, \&
  Herczeg}]{Ricci:2010p9423}
Ricci, L., Testi, L., Natta, A., {et~al.} 2010, A{\&}A, 512, 15

\bibitem[{Rice {et~al.}(2005)Rice, Lodato, \& Armitage}]{Rice:2005p11667}
Rice, W. K.~M., Lodato, G., \& Armitage, P.~J. 2005, MNRAS, 364, L56

\bibitem[{Rodmann {et~al.}(2006)Rodmann, Henning, Chandler, Mundy, \&
  Wilner}]{Rodmann:2006p8905}
Rodmann, J., Henning, T., Chandler, C.~J., Mundy, L.~G., \& Wilner, D.~J. 2006,
  A{\&}A, 446, 211

\bibitem[{Ruden \& Pollack(1991)}]{Ruden:1991p1806}
Ruden, S.~P. \& Pollack, J.~B. 1991, ApJ, 375, 740

\bibitem[{Russell {et~al.}(2006)Russell, Hartmann, Cuzzi, Krot, Gounelle, \&
  Weidenschilling}]{Russell:2006p11856}
Russell, S.~S., Hartmann, L., Cuzzi, J., {et~al.} 2006, Meteorites and the
  Early Solar System II, 233

\bibitem[{Rybicki \& Lightman(1991)}]{Rybicki:1991p12115}
Rybicki, G.~B. \& Lightman, A.~P. 1991, Radiative processes in astrophysics‎

\bibitem[{Safronov(1969)}]{Safronov:1969p11177}
Safronov, V.~S. 1969, Evolution of the protoplanetary cloud and formation of
  the earth and planets. English translation (1972)

\bibitem[{Sano {et~al.}(2000)Sano, Miyama, Umebayashi, \&
  Nakano}]{Sano:2000p9889}
Sano, T., Miyama, S.~M., Umebayashi, T., \& Nakano, T. 2000, ApJ, 543, 486

\bibitem[{Sch{\"a}fer {et~al.}(2007)Sch{\"a}fer, Speith, \&
  Kley}]{Schafer:2007p7468}
Sch{\"a}fer, C., Speith, R., \& Kley, W. 2007, A{\&}A, 470, 733

\bibitem[{Schneider(2008)}]{Schneider:2008p10345}
Schneider, P. 2008, Einf{\"u}hrung in die extragalaktische Astronomie und
  Kosmologie‎

\bibitem[{Schr{\"a}pler \& Henning(2004)}]{Schrapler:2004p2394}
Schr{\"a}pler, R. \& Henning, T. 2004, ApJ, 614, 960

\bibitem[{Scott(2007)}]{Scott:2007p10936}
Scott, E. R.~D. 2007, Ann. Rev. Earth Planet. Sci., 35, 577

\bibitem[{Semenov {et~al.}(2003)Semenov, Henning, Helling, Ilgner, \&
  Sedlmayr}]{Semenov:2003p9622}
Semenov, D., Henning, T., Helling, C., Ilgner, M., \& Sedlmayr, E. 2003,
  A{\&}A, 410, 611

\bibitem[{Shakura \& Sunyaev(1973)}]{Shakura:1973p4854}
Shakura, N.~I. \& Sunyaev, R.~A. 1973, A{\&}A, 24, 337

\bibitem[{Shu(1977)}]{Shu:1977p843}
Shu, F.~H. 1977, ApJ, 214, 488

\bibitem[{Sicilia-Aguilar {et~al.}(2006{\natexlab{a}})Sicilia-Aguilar,
  Hartmann, Calvet, Megeath, Muzerolle, Allen, D'Alessio, Mer{\'\i}n, Stauffer,
  Young, \& Lada}]{SiciliaAguilar:2006p10359}
Sicilia-Aguilar, A., Hartmann, L., Calvet, N., {et~al.} 2006{\natexlab{a}},
  ApJ, 638, 897

\bibitem[{Sicilia-Aguilar {et~al.}(2006{\natexlab{b}})Sicilia-Aguilar,
  Hartmann, F{\"u}r{\'e}sz, Henning, Dullemond, \&
  Brandner}]{SiciliaAguilar:2006p10361}
Sicilia-Aguilar, A., Hartmann, L.~W., F{\"u}r{\'e}sz, G., {et~al.}
  2006{\natexlab{b}}, AJ, 132, 2135

\bibitem[{Simon \& Prato(1995)}]{Simon:1995p11800}
Simon, M. \& Prato, L. 1995, ApJ, 450, 824

\bibitem[{Sirono(2004)}]{Sirono:2004p8225}
Sirono, S.-I. 2004, Icarus, 167, 431

\bibitem[{Stewart \& Leinhardt(2009)}]{Stewart:2009p5281}
Stewart, S.~T. \& Leinhardt, Z.~M. 2009, ApJ, 691, L133

\bibitem[{Stone \& Balbus(1996)}]{Stone:1996p10488}
Stone, J.~M. \& Balbus, S.~A. 1996, ApJ, 464, 364

\bibitem[{Strom {et~al.}(1989)Strom, Strom, Edwards, Cabrit, \&
  Skrutskie}]{Strom:1989p9475}
Strom, K.~M., Strom, S.~E., Edwards, S., Cabrit, S., \& Skrutskie, M.~F. 1989,
  AJ, 97, 1451

\bibitem[{Struve(1952)}]{Struve:1952p10743}
Struve, O. 1952, The Observatory, 72, 199

\bibitem[{Tanaka {et~al.}(2005)Tanaka, Himeno, \& Ida}]{Tanaka:2005p6703}
Tanaka, H., Himeno, Y., \& Ida, S. 2005, ApJ, 625, 414

\bibitem[{Tanaka {et~al.}(1996)Tanaka, Inaba, \& Nakazawa}]{Tanaka:1996p2320}
Tanaka, H., Inaba, S., \& Nakazawa, K. 1996, Icarus, 123, 450

\bibitem[{Teiser \& Wurm(2009)}]{Teiser:2009p7785}
Teiser, J. \& Wurm, G. 2009, MNRAS, 393, 1584

\bibitem[{Terquem(2008)}]{Terquem:2008p7731}
Terquem, C. E. J. M. L.~J. 2008, ApJ, 689, 532

\bibitem[{Testi {et~al.}(2003)Testi, Natta, Shepherd, \&
  Wilner}]{Testi:2003p3390}
Testi, L., Natta, A., Shepherd, D.~S., \& Wilner, D.~J. 2003, A{\&}A, 403, 323

\bibitem[{Thommes \& Duncan(2006)}]{Thommes:2006p11628}
Thommes, E.~W. \& Duncan, M.~J. 2006, Planet Formation

\bibitem[{Thommes {et~al.}(2003)Thommes, Duncan, \&
  Levison}]{Thommes:2003p11622}
Thommes, E.~W., Duncan, M.~J., \& Levison, H.~F. 2003, Icarus, 161, 431

\bibitem[{Toomre(1964)}]{Toomre:1964p1002}
Toomre, A. 1964, ApJ, 139, 1217

\bibitem[{Turner \& Drake(2009)}]{Turner:2009p11341}
Turner, N.~J. \& Drake, J.~F. 2009, ApJ, 703, 2152

\bibitem[{Ulrich(1976)}]{Ulrich:1976p856}
Ulrich, R.~K. 1976, ApJ, 210, 377

\bibitem[{Urpin(1984)}]{Urpin:1984p1473}
Urpin, V.~A. 1984, Soviet Astron., 28, 50

\bibitem[{van Boekel {et~al.}(2004)van Boekel, Min, Leinert, Waters, Richichi,
  Chesneau, Dominik, Jaffe, Dutrey, Graser, Henning, de~Jong, K{\"o}hler,
  de~Koter, Lopez, Malbet, Morel, Paresce, Perrin, Preibisch, Przygodda,
  Sch{\"o}ller, \& Wittkowski}]{vanBoekel:2004p4708}
van Boekel, R., Min, M., Leinert, C., {et~al.} 2004, Nature, 432, 479

\bibitem[{van Boekel {et~al.}(2003)van Boekel, Waters, Dominik, Bouwman,
  de~Koter, Dullemond, \& Paresce}]{vanBoekel:2003p8117}
van Boekel, R., Waters, L. B. F.~M., Dominik, C., {et~al.} 2003, A{\&}A, 400,
  L21

\bibitem[{Visser {et~al.}(2009)Visser, Dishoeck, Doty, \&
  Dullemond}]{Visser:2009p9087}
Visser, R., Dishoeck, E. F.~V., Doty, S.~D., \& Dullemond, C.~P. 2009, A{\&}A,
  495, 881

\bibitem[{Wada {et~al.}(2008)Wada, Tanaka, Suyama, Kimura, \&
  Yamamoto}]{Wada:2008p4903}
Wada, K., Tanaka, H., Suyama, T., Kimura, H., \& Yamamoto, T. 2008, ApJ, 677,
  1296

\bibitem[{Wada {et~al.}(2009)Wada, Tanaka, Suyama, Kimura, \&
  Yamamoto}]{Wada:2009p8776}
Wada, K., Tanaka, H., Suyama, T., Kimura, H., \& Yamamoto, T. 2009, ApJ, 702,
  1490

\bibitem[{Wardle \& Ng(1999)}]{Wardle:1999p9890}
Wardle, M. \& Ng, C. 1999, MNRAS, 303, 239

\bibitem[{Weidenschilling(1977)}]{Weidenschilling:1977p865}
Weidenschilling, S.~J. 1977, MNRAS, 180, 57

\bibitem[{Weidenschilling(1980)}]{Weidenschilling:1980p4572}
Weidenschilling, S.~J. 1980, Icarus, 44, 172

\bibitem[{Weidenschilling(1984)}]{Weidenschilling:1984p4590}
Weidenschilling, S.~J. 1984, Icarus, 60, 553

\bibitem[{Weidenschilling(1995)}]{Weidenschilling:1995p11583}
Weidenschilling, S.~J. 1995, Icarus, 116, 433

\bibitem[{Weidenschilling(1997)}]{Weidenschilling:1997p4593}
Weidenschilling, S.~J. 1997, Icarus, 127, 290

\bibitem[{Weidling {et~al.}(2009)Weidling, G{\"u}ttler, Blum, \&
  Brauer}]{Weidling:2009p9833}
Weidling, R., G{\"u}ttler, C., Blum, J., \& Brauer, F. 2009, ApJ, 696, 2036

\bibitem[{Wetherill \& Stewart(1989)}]{Wetherill:1989p11603}
Wetherill, G.~W. \& Stewart, G.~R. 1989, Icarus, 77, 330

\bibitem[{Whipple(1972)}]{Whipple:1972p4621}
Whipple, F.~L. 1972, From Plasma to Planet, 211

\bibitem[{Whitehouse \& Bate(2006)}]{Whitehouse:2006p8532}
Whitehouse, S.~C. \& Bate, M.~R. 2006, MNRAS, 367, 32

\bibitem[{Wilde {et~al.}(2001)Wilde, Valley, Peck, \&
  Graham}]{Wilde:2001p10906}
Wilde, S.~A., Valley, J.~W., Peck, W.~H., \& Graham, C.~M. 2001, Nature, 409,
  175

\bibitem[{Williams \& Wetherill(1994)}]{Williams:1994p9637}
Williams, D.~R. \& Wetherill, G.~W. 1994, Icarus, 107, 117

\bibitem[{Wilner {et~al.}(2005)Wilner, D'Alessio, Calvet, Claussen, \&
  Hartmann}]{Wilner:2005p3383}
Wilner, D.~J., D'Alessio, P., Calvet, N., Claussen, M.~J., \& Hartmann, L.
  2005, ApJ, 626, L109

\bibitem[{Wolk \& Walter(1996)}]{Wolk:1996p11850}
Wolk, S.~J. \& Walter, F.~M. 1996, AJ, 111, 2066

\bibitem[{Wolszczan \& Frail(1992)}]{Wolszczan:1992p10543}
Wolszczan, A. \& Frail, D.~A. 1992, Nature, 355, 145

\bibitem[{Wurm {et~al.}(2005)Wurm, Paraskov, \& Krauss}]{Wurm:2005p1855}
Wurm, G., Paraskov, G., \& Krauss, O. 2005, Icarus, 178, 253

\bibitem[{Wyatt {et~al.}(1999)Wyatt, Dermott, Telesco, Fisher, Grogan, Holmes,
  \& Pi{\~n}a}]{Wyatt:1999p10097}
Wyatt, M.~C., Dermott, S.~F., Telesco, C.~M., {et~al.} 1999, ApJ, 527, 918

\bibitem[{Youdin(2010)}]{Youdin:2010p10939}
Youdin, A.~N. 2010, EAS Publications Series, 41, 187

\bibitem[{Youdin \& Goodman(2005)}]{Youdin:2005p11585}
Youdin, A.~N. \& Goodman, J. 2005, ApJ, 620, 459

\bibitem[{Youdin \& Lithwick(2007)}]{Youdin:2007p2021}
Youdin, A.~N. \& Lithwick, Y. 2007, Icarus, 192, 588

\bibitem[{Zsom \& Dullemond(2008)}]{Zsom:2008p7126}
Zsom, A. \& Dullemond, C.~P. 2008, A{\&}A, 489, 931

\bibitem[{Zsom {et~al.}(2010)Zsom, Ormel, G{\"u}ttler, Blum, \&
  Dullemond}]{Zsom:2010p9746}
Zsom, A., Ormel, C.~W., G{\"u}ttler, C., Blum, J., \& Dullemond, C.~P. 2010,
  A{\&}A, 513, 57

\end{thebibliography}
}
\chapter{Acknowledgements}
\setlength{\parindent}{0pt}
\setlength{\parskip}{0.24cm}
Firstly, I would like to thank my advisor, Kees Dullemond. Not only did he put enough trust in me and gave me the opportunity to do my PhD at the MPIA, but also did he spend a lot of time patiently explaining things to me, discussing ideas with me, opening up new opportunities for me. Additionally, he had plenty of useful advices about how to structure and present talks, about good paper writing, or about how to get organized, and much more.

Of similar importance to this work was a colleague and friend, Frithjof Brauer, who died on September 19th, 2009. He always found the time to explain the physical and numerical concepts of coagulation and fragmentation to me and also cheered me up during several frustrating periods of my work. This work had not been possible without his contribution. Thank you Frithjof!

I am indebted to Gerhard Hoffman who made my code run so fast on different machines.

Many thanks go to all my colleagues for contributing to my work and/or for fruitful discussions:
Carsten Dominik,
Natalia Dzyurkevich,
Mario Flock,
Thomas Henning,
Hubert Klahr,
Christoph Mordasini,
Gijs Mulders,
Antonella Natta,
Olja Pani\'c,
Luca Ricci,
J\"urgen Sauter,
Leonardo Testi,
Francesco Trotta,
Anton Vasyunin,
Svitlana Zhukovska,
Andras Zsom,
and probably many more. I especially like to thank Chris Ormel who always pushed me to ge a bit further, which significantly improved my work.

I thank Uma Gorti and David Hollenbach for hosting me at SETI Institute. 

I am also thankful to my committee members (Kees Dullemond, Ralf Klessen, Thomas Henning and Mario Trieloff) for finding the time for my exam and for reading my thesis.

There are many more other people and institutions at the MPIA than I can acknowledge here. To mention just a few: the theory lunch group for lots of interesting scientific and non-scientific discussions, the student coffee round, the Ernst-Patzer Foundation, and the PhD Advisory Committee. I also enjoyed the PSF and student retreats and like to thank all people who contributed.

Most importantly, I would like to thank my parents for their endless support, for believing in me and for making me who I am.    

Finally, I thank Anna for her strong support, her cheerful company and for everything else during the last three years and hope for many more years to come.
\end{document}